\documentclass[11pt,aps,amsfonts,showpacs,nofootinbib,longbibliography,superscriptaddress,onecolumn,tightenlines]{revtex4-1}
\usepackage{graphicx} 
\usepackage{amsmath}
\usepackage{amssymb}
\usepackage{amsthm}
\usepackage{mathrsfs}
\usepackage{mathtools}
\usepackage{extarrows} 
\usepackage{xcolor}
\usepackage[margin=0.8 in]{geometry}
\usepackage{tikz-cd}
\usetikzlibrary{decorations.pathmorphing}
\usepackage[normalem]{ulem}
\usepackage{scalerel}
\usepackage{bbm}
\usepackage[hypertexnames=false]{hyperref}
\usepackage{framed}
\usepackage{cleveref}
\usepackage{enumerate}
\usepackage[all]{xy}
\usepackage{todonotes}

\usetikzlibrary{patterns}

\definecolor{detector}{RGB}{20,116,132}
\definecolor{excitation}{RGB}{183,65,71}
\definecolor{annotation}{RGB}{84,88,92}

\usepackage{array}   
\newcolumntype{L}{>{$}r<{$}} 

\newcolumntype{C}{>{$}c<{$}} 

\newcommand{\Z}{\mathbb Z}
\newcommand{\C}{\mathbb C}
\newcommand{\F}{\mathbb F}

\newcommand{\T}{\mathcal T}

\newcommand{\R}{\mathbb R}
\newcommand{\Q}{\mathbb Q}
\newcommand{\N}{\mathbb N}

\newcommand{\br}{\mathbf r}

\renewcommand{\ol}{\overline}

\newcommand{\im}{\operatorname{im}}
\newcommand{\Hom}{\operatorname{Hom}}

\newcommand{\Ext}{\operatorname{Ext}}

\newcommand{\ExtRn}{\operatorname{Ext}_{R_n}}
\newcommand{\Tor}{\operatorname{Tor}}

\newcommand{\cR}{\mathcal{R}}

\newcommand{\cC}{\mathcal{C}}

\newcommand{\can}{\operatorname{can}}
\newcommand{\scan}{\operatorname{scan}}
\newcommand{\coker}{\operatorname{coker}}
\newcommand{\RQZ}{R_{\mathbb{Q}/\mathbb{Z}}}
\newcommand{\RRZ}{R_{\mathbb{R}/\mathbb{Z}}}
\newcommand{\Cone}{\operatorname{Cone}}
\newcommand{\inv}{\operatorname{inv}}
\newcommand{\Mod}{\operatorname{Mod}}

\newcommand{\FC}{\mathcal{F}}

\newtheorem{theorem}{Theorem}[section]

\newtheorem{proposition}[theorem]{Proposition}
\newtheorem{prop}[theorem]{Proposition}
\newtheorem{lemma}[theorem]{Lemma}

\newtheorem{corollary}[theorem]{Corollary}

\newtheorem{conjecture}[theorem]{Conjecture}
\newtheorem{question}[theorem]{Question}

\theoremstyle{definition}
\newtheorem{defn}[theorem]{Definition}
\newtheorem{remark}[theorem]{Remark}
\newtheorem{example}[theorem]{Example}
\newtheorem{property}[theorem]{Property}
\newtheorem{warning}[theorem]{Warning}

\newtheorem{convention}[theorem]{Convention}

\newcommand{\id}{\mathrm{id}}

	\makeatletter
\newsavebox{\@brx}
\newcommand{\llangle}[1][]{\savebox{\@brx}{\(\m@th{#1\langle}\)}%
  \mathopen{\copy\@brx\kern-0.5\wd\@brx\usebox{\@brx}}}
\newcommand{\rrangle}[1][]{\savebox{\@brx}{\(\m@th{#1\rangle}\)}%
  \mathclose{\copy\@brx\kern-0.5\wd\@brx\usebox{\@brx}}}

\makeatother

\newcommand{\EW}[1]{\textcolor{blue}{#1}}

\newcommand{\Supp}{\operatorname{Supp}}
\newcommand{\supp}{\operatorname{supp}}
\newcommand{\Ass}{\operatorname{Ass}}
\newcommand{\Spec}{\operatorname{Spec}}

\newcommand{\Ann}{\operatorname{Ann}}
\newcommand{\RHom}{\operatorname{RHom}}
\newcommand{\trunc}{\operatorname{trunc}}

\newcommand{\depth}{\operatorname{depth}}
\newcommand{\height}{\operatorname{ht}}

\newcommand{\op}{\mathrm{op}}
\newcommand{\Ch}{\mathrm{Ch}}
\newcommand{\cD}{\mathcal{D}}
\newcommand{\fD}{\mathfrak{D}}
\newcommand{\hR}{\widehat{R}}
\newcommand{\cS}{\mathcal{S}}

\DeclareFontFamily{T1}{cbgreek}{}
\DeclareFontShape{T1}{cbgreek}{m}{n}{<-6>  grmn0500 <6-7> grmn0600 <7-8> grmn0700 <8-9> grmn0800 <9-10> grmn0900 <10-12> grmn1000 <12-17> grmn1200 <17-> grmn1728}{}
\DeclareSymbolFont{quadratics}{T1}{cbgreek}{m}{n}
\DeclareMathSymbol{\qoppa}{\mathord}{quadratics}{19}
\DeclareMathSymbol{\Qoppa}{\mathord}{quadratics}{21}

\newcommand{\K}{\mathrm{K}}
\newcommand{\D}{\mathrm{D}}
\newcommand{\ev}{\operatorname{ev}}

\makeatletter
\def\@bibstyle{apsrev4-1-fs}
\makeatother

\begin{document}
\linespread{1.08}\selectfont

\title{Algebraic theory of abelian fractons and beyond}

\author{Wilbur Shirley}
\thanks{These authors contributed equally to this work.}
\affiliation{Leinweber Institute for Theoretical Physics, University of Chicago, Chicago, IL 60637, USA}
\affiliation{School of Natural Sciences, Institute for Advanced Study, Princeton, NJ 08540, USA}
\author{Evan Wickenden}
\thanks{These authors contributed equally to this work.}
\affiliation{Department of Physics, University of Colorado, Boulder, CO 80309, USA}
\affiliation{Center for Theory of Quantum Matter, University of Colorado, Boulder, CO 80309, USA}
\author{Bowen Yang}
\affiliation{Center of Mathematical Sciences and Applications, Harvard University, Cambridge, MA 02138, USA}
\author{Agn\`es Beaudry}
\affiliation{Department of Mathematics, University of Colorado, Boulder, CO 80309, USA}
\author{Michael Hermele}
\affiliation{Department of Physics, University of Colorado, Boulder, CO 80309, USA}
\affiliation{Center for Theory of Quantum Matter, University of Colorado, Boulder, CO 80309, USA}
\date{\today}

\begin{abstract}
We develop an algebraic theory of fusion, mobility, and statistics in abelian gapped fracton orders and topological phases. The framework generalizes the algebraic description of abelian anyons to systems with restricted-mobility and spatially extended excitations. We consider translation-invariant bosonic systems whose pointlike and extended excitations have finite fusion order. The basic objects are \emph{braided fusion complexes} (BFCs): chain complexes encoding fusion and mobility, equipped with a braiding chain map encoding statistics. Their equivalence classes define \emph{braided fusion theories} (BFTs). Our framework encompasses arbitrary translation-invariant Pauli codes, as well as all abelian anyon theories in two spatial dimensions, including chiral theories. We give evidence for the conjecture that every nondegenerate BFT can be realized as the excitation theory of a gapped phase and that the BFT describing such a phase is a phase invariant. We derive constraints on pointlike fusion and mobility, showing, for example, that neither the Pauli $X$ nor $Z$ sectors of the X-cube and anisotropic models can occur alone in a nondegenerate BFT. For nondegenerate theories with only pointlike excitations, we extract a simpler \emph{theory of mutual statistics}, consisting of a fusion module and a map to its \emph{statistical dual}. This theory of mutual statistics is an invariant of the BFT and, at odd fusion order, gives an equivalent description of the full braided fusion theory. We interpret remote detection as the generalization of mutual statistics to fracton orders. In the BFC setting, we give a precise formulation of the excitation-detector principle: every nontrivial excitation can be detected by an operator supported at spatial infinity, and every nontrivial detector detects some excitation. Assuming a unique infinite-space ground state, we prove that this principle is equivalent to nondegeneracy.
\end{abstract}

\maketitle

\tableofcontents

\newpage

\section{Introduction}
\label{sec:introduction}

\subsection{Background and motivation}
\label{subsec:background-motivation}

The defining characteristic of fracton quantum matter is the presence of restricted-mobility excitations.  Early interest in these fascinating states centered on their slow relaxation to equilibrium \cite{Chamon_2005} and their resulting properties as candidate self-correcting quantum memories \cite{Haah_2011, Bravyi_2011, Bravyi_2013}. The topological quantum order and restricted-mobility excitations of Chamon’s model \cite{Chamon_2005} were subsequently analyzed \cite{Bravyi_2011_top_order}.  Later on, following the identification of simpler fracton models \cite{Vijay_2015, Vijay_2016, Pretko_2017}, broader interest grew in fracton systems as a class of quantum phases lying beyond theoretical frameworks available at the time \cite{Nandkishore_2019, Pretko_2020_review, Gromov_2024, You_2025}.  More than 20 years after the initial discovery of fractons, despite the development of significant new ideas, construction of many example models, and progress on classification in particular settings \cite{Yoshida_2013, Haah_2013, Haah_2014, Williamson_2016, Prem_2017, Ma_2017, Vijay_2017, Halasz_2017, Slagle_2017, Shirley_2018, Pretko_2018_elasticity, You_2018, Devakul_2018_diagnostics, Slagle_2019, Pai_2019,Shirley_2019,Song_2019,Bulmash_2019_braiding,Prem_2019, Bulmash_2019, Gromov_2019, You_2020, You_2020_CS,Dua_2020, Aasen_2020, Devakul_2020, Wen_2020, Sala_2020, Khemani_2020, Gromov_2020_hydrodynamics, Seiberg_2020, Seiberg_2021,Gorantla_2021,Tantivasadakarn_2021, Hsin_2021, Wang_2023,Song_2024,Gorantla_2025,Gorantla_2026, Wickenden_2026}, the general classification of fracton phases remains poorly understood.  Moreover, characterization in terms of universal properties that are readily compared between models remains incomplete, although there has been progress in this direction \cite{Shirley_2019, Wickenden_2025}.   Our understanding of both classification and characterization remains
limited even in perhaps the simplest setting: gapped fracton phases in
zero-temperature equilibrium, with no microscopic symmetries imposed
except perhaps lattice translation.

It has been a longstanding goal to develop an algebraic theory of excitations for gapped fracton phases. This should encode fusion and statistics of the restricted-mobility excitations, in the spirit of the algebraic theory of anyons in 2D \cite{Moore_1989, Kitaev_2006}, while also capturing excitation mobility.  In this paper, we propose such an algebraic framework as a general theory of excitations in abelian fracton orders and abelian topological phases.
For systems with only pointlike excitations, we use \emph{abelian}
to mean that the type of a composite excitation is uniquely determined
by the types of its constituents. Defining abelianity in the presence
of extended excitations is more subtle; we return to this issue in
Section~\ref{sec:discussion}.
 We focus throughout on bosonic systems with all excitations of finite fusion order, \emph{i.e.} for every excitation there is a positive integer $n$ such that fusing $n$ copies of the excitation is trivial.

The algebraic theory of anyons greatly simplifies in the abelian case  \cite{Belov_2005, Stirling_2008, Kapustin_2011, Galindo_2016, Lee_2018}.  The data of the theory is a pair $({\cal A}, \vartheta)$, where ${\cal A}$ is a finite abelian group whose elements are anyon types, and $\vartheta : {\cal A} \to \Q / \Z$ is a nondegenerate quadratic form encoding topological spins as $e^{2\pi i \vartheta (x)}$ for $x \in {\cal A}$.  The mutual statistics between $x$ and $y$ anyons is given by $e^{2\pi i b_a(x,y) }$, where $b_a(x,y) = \vartheta(x+y) - \vartheta(x) - \vartheta(y)$. While the theory we develop is both richer and more complex, it should be viewed as a generalization of this very simple algebraic theory.

Our theory is built on encoding the fusion and mobility of excitations in certain chain complexes, called \emph{fusion complexes}.  The terms of a fusion complex have direct physical interpretations, and elements of their negative-degree homology modules -- called fusion modules -- correspond to distinct types of pointlike and extended excitations.  Universal fusion and mobility information is encoded in the quasi-isomorphism type of a fusion complex; this is an equivalence class of fusion complexes referred to as a fusion theory.  (A quasi-isomorphism is a chain map that induces isomorphisms on homology.)

Excitation mobility enters our theory through lattice translation symmetry: the modules above are modules over a Laurent polynomial ring encoding translations \cite{Haah_2013}. Yet we believe fracton phases should not be viewed as translation-symmetry-enriched phases. We resolve this tension in two steps. First, we study translation-invariant phases, imposing translation as an ordinary symmetry. We then allow finite enlargements of the crystalline unit cell (“coarsening”) \cite{Haah_2013} and certain deformations of the underlying lattice \cite{Hermele_UQM_talk, Hermele_UQM_AM_talk}. We refer to the operations thus generated as lattice restructuring, and the resulting equivalence classes of translation-invariant phases are called \emph{lattice-restructured phases}. In this approach, translation supplies a spatial homogeneity condition rather than a conventional symmetry: no fixed translation operator is required to commute with the Hamiltonian as the lattice is restructured. For most purposes of this paper, only the first step is required, with coarsening entering only occasionally.  Full lattice restructuring plays no role in this paper, but 
lattice-restructured phases will be necessary in future applications of our framework to classification and characterization; we defer a full treatment to those works \cite{Wickenden_invariants_inprep, Hermele_restructured_inprep}.

Our description of fusion incorporates and extends key earlier results.  For gapped abelian phases with only pointlike excitations, there is a single fusion module $S$, also often denoted as $Q_0$ \cite{Haah_2013, Pai_2019}.  Elements of $S$ are superselection sectors of pointlike excitations; these are equivalence classes of excitations, where two excitations are equivalent if one can be transformed into the other by acting with a local operator.  Within the setting of Pauli codes, further progress was made by recognizing that there are multiple fusion modules $Q_i$, with $i = 0, \dots, d-1$ \cite{Ruba_2024}.  For certain codes called fully mobile models, elements of $Q_i$ were shown to be types of $i$-dimensional extended excitations, \emph{e.g.} the $i = 1$ flux lines of the 3D toric code.  In our framework, the fusion modules $Q_i$ arise as the homology modules of the fusion complex in negative degree.  When only pointlike excitations are present, $S = Q_0$ completely determines the fusion theory.  More generally, while the fusion modules $Q_i$ are important fusion invariants, they do not fully determine the fusion theory (Example~\ref{ex:hybrid-Xcube}), and thus our description of fusion goes beyond the earlier results.

This distinction between a fusion theory and its homology modules is naturally understood in the \emph{derived category}, where quasi-isomorphisms become invertible maps. The derived category plays an important role in our theory. Because the surrounding constellation of mathematical ideas is not familiar within condensed matter theory, we make use of these ideas but do not embrace a fully derived perspective in the main text. In the appendices, we provide some background on the derived category and make free use of its tools to prove supporting technical results.

Compared to fusion and mobility, the general theory of statistics of restricted-mobility excitations has been less developed in prior work.  Planons (mobile within 2D planes), lineons (mobile along lines), and fractons (immobile when isolated from other excitations) can participate in statistical processes that generalize more familiar braiding and exchange \cite{Pai_2019, Shirley_2019,Song_2019,Bulmash_2019_braiding,You_2020, You_2020_CS, Hsin_2021, Song_2024}.  A general framework to describe such processes was presented in \cite{Pai_2019}, but that approach does not provide an algebraic theory that, for instance, is capable of encoding all consistent statistical data for a given fusion theory.  More recently, such an algebraic theory was developed for planon-only fracton orders \cite{Wickenden_2026}, \emph{i.e.} those with only planon excitations, and a partial theory of statistics was introduced for p-modular fracton orders \cite{Wickenden_2025}, where every excitation can be remotely detected by braiding with a planon.  The statistics in these theories involves braiding and exchange of planons, and is thus based directly  on the familiar algebraic theory of 2D abelian anyons.  Within the setting of Pauli codes with fully mobile excitations, descriptions of mutual and self statistics have also been developed using homological methods \cite{Ruba_2026}.

In the approach of this paper, statistics is described by a chain map from a fusion complex to a dual complex, referred to as the \emph{braiding}.  The resulting structure is called a \emph{braided fusion complex} (BFC), and equivalence classes are \emph{braided fusion theories} (BFTs).  The terminology reflects the analogous role played by braided fusion categories in describing anyon fusion and statistics.  The braiding is equivalent to a kind of antisymmetric form defined on the fusion complex, which, in part, can be understood as encoding the commutators between operators that move excitations around in space.  

A BFC is called \emph{nondegenerate} when its braiding chain map is a quasi-isomorphism and when the infinite-space ground state is unique.\footnote{Here and later, by infinite space we mean infinite Euclidean space, which is not the same as taking the large-system limit of a finite periodic lattice.  Infinite-space ground state degeneracy in systems without microscopic symmetries is synonymous with fine-tuning.  Robust ground state degeneracy in finite periodic systems is a distinct phenomenon.}  Nondegenerate BFCs encompass Lagrangian Pauli codes (those with a unique infinite-space ground state), as well as all theories of abelian anyons in 2D.  Moreover, we prove that odd-fusion-order nondegenerate BFTs in $d=2$, after suitable coarsening of translation symmetry, are in bijective correspondence with abelian anyon theories of odd fusion order, and we believe this is likely to hold irrespective of the fusion order.  These and other results lead us to the following:
\begin{conjecture}[Physical]  \label{conj:nondeg-BFC}
A  BFT in spatial dimension $d$ describes the excitations of some gapped phase of an essentially $d$-dimensional bosonic system if and only if the BFT is nondegenerate.  Whenever such a description exists, the BFT is uniquely determined by the phase.
\end{conjecture}
\noindent  Here, an {\bf essentially} $d${\bf -dimensional} system is either a $d$-dimensional system with a tensor product Hilbert space, or a gapped $d$-dimensional boundary of a $(d+1)$-dimensional gapped bulk with a tensor product Hilbert space and such that the bulk does not support fractional excitations.  The BFT does not in general completely determine the phase:  it describes excitation data, and in two dimensions does not retain the chiral central charge beyond its value modulo eight.

The proposed criterion of nondegeneracy is supported by Theorem~\ref{thm:bfc-excitation-detector}, which shows that a BFC with a unique infinite-space ground state is nondegenerate if and only if it satisfies the \emph{excitation-detector principle}.  This principle, introduced in \cite{Wickenden_2026} as an extension of the principle of remote detectability \cite{Kitaev_2006, Lan_2018}, is the statement that every nontrivial excitation is detectable remotely, and every nontrivial detector detects something.  It was argued to be a necessary condition for physical realizability of a gapped state, and this was proved in the context of planon-only fracton orders \cite{Wickenden_2026}.  See Remark~\ref{rmk:conj-overview} for a more complete overview of evidence supporting the conjecture, with references to later sections.

Not every gapped phase can be described by a braided fusion theory.  For instance, non-abelian theories of anyons in 2D certainly do not admit such a description.  We thus ask:
\begin{question}[Physical]  \label{Q:BFC-generality}
Which gapped phases of bosonic essentially $d$-dimensional systems admit a nondegenerate BFT description of their excitations?
\end{question}
\noindent  We expect the answer to this question is ``gapped abelian phases,'' with a suitable definition of abelianity when extended excitations are present. We discuss this question further in Section~\ref{sec:discussion}.

Beyond developing the general theory of fusion complexes and BFCs, we obtain a number of specific consequences of the theory.  Many of these are summarized at the beginning of Section~\ref{subsec:outline}; we highlight one particular such result here.  The presentation of a BFC requires a considerable amount of data, especially when compared to a theory of abelian anyons $({\cal A}, \vartheta)$.  However, much of the data is nonuniversal, and in specific cases the braided fusion theory equivalence classes can be described in more concrete and simpler terms.  

We carry out such a simplification, which captures some but in general not all of the statistical data, in systems with only pointlike excitations. Among abelian fracton orders in $d=3$ spatial dimensions, this includes Chamon's model, Haah's cubic code, the X-cube model, and many other examples, excluding hybrid fracton orders \cite{Tantivasadakarn_2021}.  This results in an algebraic theory of mutual statistics for such systems that is analogous to the pair $({\cal A}, b_a)$ in abelian anyons.  

Given a nondegenerate BFC describing such a pointlike system, we extract a pair $(S,b)$, where $S$ is the fusion module and $b : S \to S^\vee$, called the \emph{homological braiding}, is an isomorphism to the \emph{statistical dual} $S^\vee$, a module that depends only on $S$.  The pair $(S,b)$ is called a pointlike theory of mutual statistics,
and under a suitable equivalence it is an invariant of the BFT and, assuming Conjecture~\ref{conj:nondeg-BFC}, a universal property of the phase.  Moreover, when $S$ is of odd fusion order, we prove that $(S,b)$ is a complete BFC invariant, meaning that it encodes all the universal fusion, mobility and statistics information.  At even fusion order, additional data is needed to describe self statistics, and incorporating this into the simplified description provided by $(S,b)$ remains open.

\subsection{Key results and outline}
\label{subsec:outline}

Given the considerable length of this paper, before giving an outline, we first highlight some of the key results obtained from our general framework, beyond those already described above.

\begin{itemize}
\item We construct a nondegenerate pointlike BFC realizing every theory
of abelian anyons in two spatial dimensions,
including chiral theories (Section~\ref{subsec:anyon-realization}).
The construction explicitly recovers both the mutual statistics and
the topological spins of the input theory.

\item We characterize the fusion modules $S$ of pointlike nondegenerate  BFCs, \emph{i.e.} those with only pointlike excitations.  In spatial dimension $d$, $S$ is pure $(d-2)$-dimensional Cohen-Macaulay (Corollary~\ref{cor:pfm-characterization}).  In addition, the fusion module admits an antisymmetric isomorphism $S \cong S^\vee$ to its statistical dual (Corollary~\ref{cor:full-fm-characterization}).  At odd fusion order, these properties provide a complete characterization:  any fusion module satisfying them is the fusion module of some pointlike nondegenerate BFC.

\item Our characterization places strong constraints on fusion modules of pointlike fracton orders. 
We use the preceding results to show that the fracton sector of the $\Z_n$ X-cube model, under arbitrary coarsening of translation symmetry, cannot arise on its own as the fusion module of any pointlike nondegenerate BFC (Section \ref{subsec:statistical-duality-examples}).  We thus expect there is no pointlike abelian fracton order whose fusion module is that of the X-cube model fracton sector. The same result holds for the $\Z_n$ X-cube model lineon sector (Section \ref{subsec:statistical-duality-examples}), as well as the $X$ and $Z$ sectors of the $\Z_n$ anisotropic model (Section \ref{subsec:statistical-duality-examples}).

\item We give a physical interpretation of the higher fusion modules $Q_i$ ($i > 0$) in general BFCs, including Pauli codes, that is not limited to fully mobile models (Section \ref{subsec:hfm-interpretation}).  In a certain sense, each $Q_i$ continues to describe $i$-dimensional extended excitations, but the interpretation is more subtle than examining the support of excitations in space.

\item We define proper and complete fusion complexes (Definition \ref{defn:propercomplete}), and prove that these are characterized by bounds on the Krull dimensions of the fusion modules that agree with those obtained in \cite[Proposition~25]{Ruba_2024} for general and Lagrangian Pauli codes, respectively (Theorem \ref{theorem:dimension-bounds}).  The fusion complex of a nondegenerate BFC is always complete.  These results imply that nondegenerate BFCs in $d=2$ have only fully mobile pointlike excitations (Corollary~\ref{cor:2d-mobile-pointlike}).

\item  We derive constraints relating pointlike and extended excitations in nondegenerate BFCs in three and four spatial dimensions (Theorem~\ref{example:proper-d3} and Theorem~\ref{example:proper-d4}). These constrain which excitation types can occur together and how they can detect one another. In $d=3$, a nontrivial line excitation requires the presence of a nontrivial point excitation. This applies, in particular, to Lagrangian Pauli codes (Corollary~\ref{cor:3d-no-loop-only}).

\item We present examples of \emph{degenerate} BFCs in $d=1$ that describe gapped boundaries of 2D anyon systems (Section \ref{subsec:example1D}).  The fusion complexes of these examples are proper but not complete.

\item  We state the excitation-detector principle for BFCs and Pauli codes (Section \ref{sec:rdms}).  There are detection modules $D_i$ with $i \geq 0$, and bilinear pairings $\theta_i : D_i \times Q_i  \to \Q/\Z$.  Elements of $D_0$ can be understood as locally canceling infinite products of stabilizers, with $\theta_0$ encoding the action of these extended operators on pointlike excitations.  We also provide a physical interpretation of the higher detection modules and pairings (Remark~\ref{rmk:physical-higher-ed-pairings}).  The excitation-detector principle is the statement that each $\theta_i$ is two-sided nondegenerate, and as noted above we prove this is equivalent to nondegeneracy of the underlying BFC under the assumption of a unique infinite-space ground state (Theorem \ref{thm:bfc-excitation-detector}).

\item  We explain why remote detection and mutual statistics should be viewed as synonymous, and use the excitation-detector pairing $\theta_0$ to explain how the theory of mutual statistics $(S,b)$ actually encodes mutual statistics (Section \ref{sec:rdms}).  For anyons in 2D, from $\theta_0$ we extract the usual symmetric bilinear form encoding mutual statistics (Section \ref{subsec:2d-detector-particles}).

\end{itemize}

We now give a more detailed but brief outline of the remainder of the paper.  Section~\ref{subsec:fc1} gives physical motivation for the definitions of fusion complex and fusion theory.  In Section~\ref{subsec:bfc1}, beginning with physical motivation, we define the dual $D(F)$ of a fusion complex $F$, and introduce the braiding as an antisymmetric chain map $\lambda : F \to D(F)$ that is also constant-alternating in degree zero.  A BFC is defined as a pair $(F, \lambda)$.  We discuss when two BFCs should be viewed as equivalent and define braided fusion theories as the equivalence classes.

We construct BFCs from Pauli codes in Section~\ref{sec:Pauli}. We first review the algebraic description of translation-invariant Pauli codes in Section~\ref{subsec:Pauli-review}. In Section~\ref{subsec:Pauli-BFC}, we construct the associated fusion complex and braiding, and show that Lagrangian Pauli codes give nondegenerate BFCs. Section~\ref{subsec:hfm-interpretation} gives a physical interpretation of the higher fusion modules in terms of extended excitations, using selection criteria that specify when an excitation type can be assigned to a state.

Examples of BFCs that cannot be realized by a Pauli code in the same spatial dimension are presented in Section~\ref{sec:examples}. In Section~\ref{subsec:example1D}, we construct one-dimensional BFCs describing the gapped boundaries of the toric code and double semion models. Section~\ref{subsec:example2D} gives two-dimensional examples describing a $\Z_2$ boson, fermion, and chiral semion, illustrating how different statistics can arise for the same fusion theory.  Section~\ref{subsec:anyon-realization} then gives a general construction
of a nondegenerate pointlike BFC realizing any prescribed theory of
abelian anyons in two dimensions.

Section~\ref{sec:dual_complex} introduces proper and complete fusion complexes and discusses their physical significance, including the interpretation of homology of the dual complex in terms of objects that can be remotely detected. We characterize properness and completeness by bounds on the Krull dimensions of the fusion modules (Theorem~\ref{theorem:dimension-bounds}). We then examine the relations between fusion modules and dual homology in two, three, and four spatial dimensions, obtaining constraints on the excitation content of nondegenerate BFCs.

Turning to pointlike excitations, in Section~\ref{sec:pfm-char} we study fusion theories for which both the fusion complex and its dual have homology only in degree $-1$. We characterize their fusion modules in terms of purity and the Cohen--Macaulay property, obtaining necessary conditions on fusion modules of pointlike nondegenerate BFCs. We also establish the equivalence between zero Krull dimension and full mobility, and conclude that complete fusion theories in two dimensions have only fully mobile pointlike excitations.

The simplified algebraic description of pointlike excitations in terms of theories of mutual statistics $(S,b)$ is developed in Section~\ref{sec:homological-braiding}. Section~\ref{subsec:derived-duality} introduces derived duality and canonical evaluation, providing the tools needed to define statistical duality. In Section~\ref{subsec:tpe}, we define the statistical dual $S^\vee$ and extract the homological braiding $b : S \to S^\vee$ from a pointlike BFC. We prove that $(S,b)$ is an invariant of a nondegenerate pointlike BFT and, at odd fusion order, a complete invariant that can always be realized by a BFC (Theorem~\ref{thm:Sb-invariant}). Section~\ref{subsec:statistical-duality-examples} applies statistical duality to the X-cube and anisotropic models, showing that their separate excitation sectors cannot occur on their own as fusion modules of pointlike nondegenerate BFCs, even after coarsening.

In Section~\ref{sec:rdms}, we develop the relation between remote detection and mutual statistics. Section~\ref{subsec:ed-phys} gives a physical discussion of the excitation-detector principle in Pauli codes, illustrated by the two-dimensional toric code. In Section~\ref{subsec:ed-bfc}, we define detection modules and excitation-detector pairings for general BFCs, explain their physical interpretation, and relate the pointlike pairing to the homological braiding. We prove that the excitation-detector principle is equivalent to nondegeneracy when the infinite-space ground state is unique (Theorem~\ref{thm:bfc-excitation-detector}). Section~\ref{subsec:2d-detector-particles} specializes to two dimensions, identifying detectors with particle types and recovering the usual symmetric bilinear form describing anyon mutual statistics.

We conclude with a brief summary of our results in Section~\ref{sec:discussion}.

Appendix~\ref{subsec:notation} collects notation, conventions, and definitions used throughout the paper. 
Appendices~\ref{app:derived_background}--\ref{app:extended-evaluation-proof}
provide background on the derived category, commutative algebra, and duality,
and develop a derived formulation of fusion complexes and braidings.
They establish the relation between this formulation and the tame complexes
used in the main text, and supply proofs of supporting technical results.


\section{Fusion complexes and braided fusion complexes}
\label{sec:first}

In this section, we give the definitions of fusion complex and braided fusion complex (BFC).  We also define equivalence of fusion complexes and BFCs, referring to the equivalence classes as fusion theories and braided fusion theories (BFTs), respectively.  Section~\ref{subsec:fc1} begins with physical motivation before presenting the formal definition of fusion complex, and Section~\ref{subsec:bfc1} follows the same pattern for BFCs.

Our mathematical notation, conventions and basic definitions are collected
in Appendix~\ref{subsec:notation}. Readers are encouraged to consult this appendix as
needed. In particular, it specifies our conventions for dual modules and
sesquilinear pairings; chain-complex conventions are given in Appendix~\ref{dra:new-conventions}.

\subsection{Fusion complexes}
\label{subsec:fc1}

Throughout this paper, we focus on zero-temperature bosonic gapped systems in $d$ spatial dimensions, with all excitations of finite fusion order.  Moreover, we are interested in abelian systems.  As discussed in Sections~\ref{subsec:background-motivation} and \ref{sec:discussion}, it is subtle to give a physical definition of abelianity when extended excitations are present.  We will avoid these subtleties in providing physical motivation for the definitions of fusion complex and BFC by focusing primarily on systems with only pointlike excitations.

We are primarily but not exclusively interested in essentially $d$-dimensional systems (as defined in Section~\ref{subsec:background-motivation}).  We will also consider gapped $d$-dimensional boundaries of essentially $(d+1)$-dimensional bulk systems; an example is a gapped boundary of the 2d toric code.  In such cases, the bulk can support fractional excitations, but we will build a theory that only describes excitations at the boundary.  For now, we do not make any assumptions related to the tensor product structure of Hilbert space.  In principle, the theory of fusion complexes (FCs) and braided fusion complexes (BFCs) should apply to $d$-dimensional defects of arbitrary codimension, but such a general setting is largely beyond the scope of this paper.

We always work in infinite $d$-dimensional Euclidean space.  We assume unbroken lattice translation symmetry, and do not impose any other microscopic symmetries.  We assume a $d$-dimensional hypercubic lattice with translation symmetry group ${\cal T} = \Z^d$; because we do not impose any other spatial symmetries, there is no loss of generality in this choice of lattice geometry.  As noted in Section~\ref{subsec:background-motivation}, we are not interested in translation-symmetry-protected phenomena \emph{per se}, but instead we view lattice translations as a tool ensuring spatial homogeneity and enabling an account of excitation mobility.  However, for the purposes of this paper, it is largely sufficient to focus on translation-invariant phases.

To motivate the definition of fusion complex, it is helpful for the moment to make some simplifying assumptions.  First, we assume that our system has a unique ground state in infinite space.  It is important to emphasize that this is not in conflict with topological order, in the sense of a robust ground state degeneracy in the thermodynamic limit of finite periodic systems.  For instance, the 2d toric code has a fourfold ground state degeneracy on the 2-torus, but a unique ground state on the infinite Euclidean plane.  In infinite space, ground state degeneracy is not associated with topological order, but instead arises from spontaneous symmetry breaking or from fine-tuning (\emph{e.g.} tuning to a first-order phase transition).  We are thus primarily interested in systems with a unique ground state, but fusion complexes do allow for ground state degeneracy.
Second, we assume that only pointlike excitations are present.  That is, for instance, there are no extended line excitations as in the 3d toric code.  Fusion complexes do also encompass systems with extended excitations, as discussed below and in more depth in Section~\ref{sec:Pauli}.

Now we turn to the key object of our discussion, namely the set $S$ of superselection sectors of pointlike excitations.  Elements of $S$ are equivalence classes of excitations of bounded support; that is, we consider states that are locally identical to the ground state as we approach spatial infinity, but may differ from the ground state within a ball of some radius.  Two such excitations are considered equivalent if they are related by the action of some local operator, \emph{i.e.} any operator of bounded spatial support.  We use the term particle type interchangeably with superselection sector; the term topological charge is also sometimes used in the literature.

Given two particle types and spatially separated representative excitations, we can form a composite state that contains both excitations and ask about its type.  This process is called fusion.  We recall that we restricted our attention from the outset to abelian systems; this means, by definition, that fusion gives a well-defined binary operation $S \times S \to S$ that makes $S$ into an abelian group.  In particular, the type of a composite excitation only depends on the type of its constituents; this is not the case in non-abelian systems, including 2d systems of non-abelian anyons.  We use additive notation for the fusion operation on $S$.

While for abelian anyons $S$ is a finite group, in fracton systems $S$ is infinite, and usually it is not even finitely generated.  Therefore the abelian group structure of $S$ alone does not provide much insight into the properties of excitations, and it gives us no insight into their mobility or lack thereof.  To proceed we take advantage of the translation symmetry ${\cal T}$, which acts on superselection sectors, thus making $S$ into a module over the Laurent polynomial ring $R = \mathbb{Z}[{\cal T}] \cong \mathbb{Z}[t_1^{\pm}, \dots, t_d^{\pm}]$ (see Appendix~\ref{subsec:notation}) \cite{Haah_2013, Pai_2019}.
We thus refer to $S$ as the {\bf module of superselection sectors} or the {\bf fusion module}.  One reason $S$ is a nice object to work with is that its isomorphism class is a phase invariant for translation-invariant phases \cite{Haah_2013, Pai_2019}.

We assume that $S$ is finitely generated as an $R$-module.  This amounts to the statement that there are only a finite number of excitation types per unit volume, which must hold in systems with a finite-dimensional local Hilbert space (\emph{i.e.} in spin models), where only a finite number of different excitations can fit into a bounded spatial volume.   We also assume that every element of $S$ is of finite order under fusion, \emph{i.e.} $S$ is $\Z$-torsion.  Because $S$ is finitely generated, this implies there exists a smallest  $n_S \in \N$ such that $n_S x = 0$ for all $x \in S$.  We call $n_S$ the {\bf fusion order} of $S$.    Choosing any $n \in \N$ divisible by $n_S$, we can just as well view $S$ as a module over $R_n =  \Z_n [{\cal T}] \cong \Z_n [t_1^{\pm}, \dots, t_d^{\pm}]$, the ring of Laurent polynomials with $\Z_n$ coefficients.  This holds because $R_n \cong R / (n)$, where $(n) \subset R$ is the ideal generated by $n$, and because $(n)$ annihilates $S$.  Sometimes it will be more convenient to work over $R$, and sometimes more convenient to work over $R_n$; the relationship between these choices is discussed below.  We note that there are interesting fracton orders where $S$ has infinite-order elements \cite{MacDonald_1989,MacDonald_1990, Sondhi_2000,Sondhi_2001,Ma_2022,Chen_2023}; extending the framework developed here to encompass such examples is left for future work.

The rings $R$ and $R_n$ come with an involution, denoted by $r \mapsto \bar{r}$, that reverses translations ($\bar{t} = t^{-1}$ for $t \in \T$) and leaves the $\Z$ or $\Z_n$ coefficients invariant.  For a module $M$, we can always form its conjugate module $\overline M$, which has the same underlying abelian group as $M$, but the ring action is twisted by the involution.  We write $\overline{m} \in \overline{M}$ to denote the element corresponding to $m \in M$, and the ring action is given by $r\overline{m}=\overline{\bar r m}$. An antilinear map $f:M\to N$ satisfies $f(rm)=\bar r f(m)$, or equivalently defines a linear map $\overline M\to N$.  A distinct notion is a module with involution; this is a module $M$ equipped with an antilinear map $M \to M$ whose square is the identity.  The involution on $M$ is also written using a bar, \emph{i.e.} $m \mapsto \overline{m}$.  The notation $\overline{m}$ is thus used in two different ways:  to indicate elements of the conjugate module $\overline{m} \in \overline{M}$, or to indicate the value of the involution $m \mapsto \overline{m}$ when $M$ is a module with involution.  When the meaning of a bar is not clear from context, we clarify it explicitly.  See Appendix~\ref{subsec:notation} for further details.

We begin by working over $R$, and we choose a free resolution of $S$ given by
\begin{equation}
\begin{tikzcd}
\cdots \arrow[r] & F_2 \arrow[r, "d_2"] & F_1 \arrow[r, "d_1"]  &  F_0 \arrow[r, "d_0"]  & F_{-1} \arrow[r, "\pi"]  & S \arrow[r] & 0 
\end{tikzcd} \text{,}  \label{eqn:first-FC}
\end{equation}
where each $F_i$ is a finitely generated free $R$-module.  The nonstandard numbering convention anticipates the definition of BFC, where $F_0$ will play a special role.  We denote by $F$ the bounded-below complex obtained by deleting $S$ and replacing it with an infinite sequence of zeros.  That is, for $i \geq -1$ the modules $F_i$ are as given above, and $F_i = 0$ for $i < -1$.  The complex $F$ is exact except in degree $-1$, where $H_{-1}(F) \cong S$.  In fact, by Lemma~\ref{lemma:bounded-replacement}(c) and the freeness of finitely generated projective $R$-modules used in Lemma~\ref{v2:lemma:tame-in-FC}(c), it is always possible to find a finite free resolution for a finitely generated $R$-module $S$, so we can choose $F$ to be a bounded complex, though we do not necessarily make this restriction.  Looking ahead to Pauli codes, it will be convenient to represent $S$ by a complex $F$ that is not necessarily bounded below.  That is, $F$ may be any complex of finitely generated free $R$-modules with homology concentrated in degree $-1$, and with $H_{-1}(F) \cong S$.

The above construction of a complex $F$ from $S$ can be repeated working over $R_n$, producing a chain complex where each $F_i$ is a free $R_n$-module.  The only difference is that it is no longer always possible to find a bounded $F$.  For instance, if $S \cong \Z_2$, there is no finite free resolution of $S$ by $R_4$-modules.  When working over $R_n$, we also allow for complexes that are not necessarily bounded below.  

At this point it is helpful to give a physical interpretation of the modules $F_i$ and boundary maps $d_i$ in the complex $F$, focusing on the case where $F$ is a free resolution of $S$.  To arrive at a simple interpretation, we will freely make certain reasonable physical assumptions.  These assumptions will be used to motivate the definitions, especially the definition of BFC in Sec.~\ref{subsec:bfc1}, and will provide some intuition for working with BFCs.  Once the definitions are in hand, one can put the motivating physical assumptions aside and ask about the generality of the BFC framework, as stated in Question~\ref{Q:BFC-generality}.

Choosing a basis $\{ \hat{e}_1, \dots, \hat{e}_k \}$ for $F_{-1}$, a general element $e \in F_{-1}$ is of the form $e = \sum_i r_i \hat{e}_i$, where either $r_i \in R$ or $r_i \in R_n$.  Therefore $e$ can be viewed as a configuration on the spatial lattice where we place $k$ different $\Z$-valued (or $\Z_n$-valued) degrees of freedom at each site, and only finitely many of these single-site degrees of freedom are nonzero.  Under the map $\pi$ of Equation~\ref{eqn:first-FC}, elements of $S$ can be viewed as equivalence classes of these configurations.  Therefore it is natural to view elements of $F_{-1}$ as states of the system, and the zero element of $F_{-1}$ represents the ground state.  We refer to elements of $F_{-1}$ as {\bf bounded excitations}; they are spatially bounded due to the finiteness condition on the single-site degrees of freedom.  We often drop the ``bounded'' qualifier when this is clear from context.  The addition operation on $F_{-1}$ represents fusion of excitations, so we are assuming it is possible to lift the module structure on $S$ that encodes fusion and mobility to a set of states that moreover forms a free module.

It is tempting to say that $F_{-1}$ gives a basis for the Hilbert space of the system, up to a choice of ${\rm U}(1)$ phase for each element, but this is not quite correct, in part due to the subtleties of working in infinite space.  First of all, there may be states that cannot be obtained as linear combinations of the states in $F_{-1}$ at all; however, we are not interested in these states, because by construction we do not need them to reach all the superselection sectors.  So we might like to say that $F_{-1}$ forms a basis for a certain interesting subspace of Hilbert space.  However, this still is not correct, because states in different superselection sectors do not live in the same Hilbert space; it is meaningless to construct a linear superposition of such states.  A better statement is that if we take the elements of $F_{-1}$ within any superselection sector, these form a basis for a subspace of ``interesting states'' within that sector.  

Now given $A \in F_0$, we see that $d_0 (A) \in F_{-1}$ is an excitation of trivial particle type.  That is, $d_0(A)$ is a locally creatable excitation.  Exactness at $F_{-1}$ in Equation~\ref{eqn:first-FC} means that every locally creatable excitation can be obtained in this way.  Therefore we view elements  $A \in F_0$ as projective local operators, where acting with $A$ adds the locally creatable excitation $d_0(A)$ to the system.  By projective, we mean that each element of $F_0$ corresponds to an operator only up to an overall ${\rm U}(1)$ phase.  This is consistent with the interpretation of elements of $F_{-1}$ as states, \emph{i.e.} as rays rather than vectors in Hilbert space.  We further observe that $F_0$ is countable, but any set of operators including the overall ${\rm U}(1)$ phase would be uncountable.
More concretely, when we construct fusion complexes and BFCs from Pauli codes in Section~\ref{sec:Pauli}, we see that $F_0$ appears as the projective Pauli group, which is precisely the group of products of Pauli operators modulo ${\rm U}(1)$ phase factors.  Addition in $F_0$ corresponds to (projective) operator multiplication, and $0 \in F_0$ represents the identity operator.  This immediately implies that $A \in F_0$ corresponds to an invertible operator, because $-A \in F_0$ and $A + (-A) = 0$.  For simplicity, we further assume that elements $A \in F_0$ correspond to unitary operators.

We refer to elements of $F_0$ as {\bf movers}, because the way to move an excitation in space is by acting with local operators.  For instance, it is familiar that a fully mobile excitation, such as an anyon in $d=2$, can be moved by acting with a finite string operator.  In fracton systems, where excitations have restricted mobility, more complicated cooperative motions are possible, where an excitation splits into multiple excitations at different spatial positions, where a group of excitations moves together, and so on.  Of course, it can be the case that some elements of $F_0$ do not literally have the effect of moving some excitation.  Introducing the term ``mover'' allows us to use ``operator'' in the standard fashion, \emph{i.e.} to refer to elements of the operator algebra of the underlying physical system we are describing.

Next, for any $f_1 \in F_1$, $d_1(f_1)$ is a mover that does not create any excitations, and all such movers lie in the image of $d_1$ by exactness at $F_0$.  With a nod to Pauli codes, elements of $F_1$ are thus called {\bf stabilizers}, because they represent operators that leave a system's state unchanged.  At this point we have run out of physics-inspired names for the terms in $F$; elements of $F_2$ are local relations among stabilizers, $F_3$ consists of relations among relations, and so on.

If $F$ is not bounded below, the above discussion is almost unchanged.  The one modification is that only the submodule $\operatorname{ker} d_{-1} \subset F_{-1}$ should correspond to a set of states, because this is the module that maps surjectively to $S$ via the projection $\operatorname{ker} d_{-1} \to \ker d_{-1} / \im d_0 \cong S$.  Therefore more generally we refer to elements of $F_{-1}$ as {\bf configurations}, and elements of the submodule $\ker d_{-1}$ as excitations.  Looking ahead to the discussion of Pauli codes, the interpretation of $F_1$ as the module of stabilizers, $F_0$ as movers, and $\ker d_{-1} \subset F_{-1}$ as excitations holds in general beyond the pointlike case.  Elements of $F_{-2}$ will be called {\bf violations}, while elements of larger negative degree are {\bf higher violations}.  This terminology is better justified once we have a braiding, which for instance provides a map $F_{-2} \to D_{\RRZ}(F_2)$, with the dual defined below Equation~\ref{eqn:omega-Omega}.  An element of $F_{-2}$ thus provides a functional on relations.  If we consider $f_{-2} = d_{-1} f_{-1}$ for some $f_{-1} \in F_{-1}$, then this functional tells us about which relations are ``violated'' by the configuration $f_{-1}$.

\begin{defn} \label{def:FC}
A {\bf fusion complex} $F$ is a complex over $R$ with homology concentrated in degrees $i = 0, -1, -2, \dots, -d$.  Moreover, for all $i \in \Z$, $H_i(F)$ is a finitely generated $R$-module and is $\Z$-torsion.  The {\bf fusion order} of a fusion complex is the smallest natural number $n$ that annihilates all of the $H_i(F)$.  We say a fusion complex is trivial if it is acyclic; that is, if $H_i(F) = 0$ for all $i \in \Z$.  

The {\bf fusion modules} of a fusion complex are its negative-degree homology modules and are denoted $Q_i = H_{-(i+1)}(F)$ for $0 \leq i \leq d-1$.  We also write $S = S(F) = Q_0$.  A fusion complex is {\bf pointlike} if its homology is concentrated in degree $-1$, \emph{i.e.} if $H_0(F) = 0$ and $Q_i = 0$ for all $i > 0$.  In the pointlike case we refer to $S$ as \emph{the} fusion module of $F$, and the fusion order of $S$  is the same as that of $F$.  The fusion modules $Q_i$ with $i > 0$ are referred to as the {\bf higher fusion modules}.
\end{defn}

\begin{remark}
Fusion complexes are precisely perfect complexes over $R$ with the additional requirement of $\Z$-torsion homology concentrated in degrees $i=0,-1,\dots,-d$.  (See Definition~\ref{defn:perfect} and Remark~\ref{rmk:bh-perfect} below, and Lemma~\ref{lemma:bounded-replacement}(c).)  We note this fact because the full subcategory of perfect complexes $\Ch^p(R) \subset \Ch(R)$ has many nice properties.
\end{remark}

We work over $R$ in Definition~\ref{def:FC} because this ring always encodes the action of translation, independent of the properties of a specific module $S$; in particular, the definition does not involve an arbitrary \emph{a priori} choice of fusion order.  While we do not assume that $F$ consists of free $R$-modules, any free resolution of $S$ over $R$ or over $R_n$ is a fusion complex, so the definition encompasses the examples discussed above.

As is clear from the terminology, the definition goes beyond the case of a system with only pointlike excitations; the fusion modules $Q_i$ describe  excitations of dimension $i$ (in a sense to be explained in Sec.~\ref{subsec:hfm-interpretation}), so for $i > 0$ these are extended excitations.  For instance, $Q_1$ describes line-like excitations such as the flux lines of the 3d toric code.  This interpretation will be justified in Sec.~\ref{subsec:hfm-interpretation}. For purposes of physical understanding, it will not cause the reader any issues to focus on pointlike fusion complexes until reading that section.  Indeed, many of the main results of this paper are specific to the pointlike case.

The condition that the homology $H_i(F)$ is $\Z$-torsion is a generalization of the same restriction on $S$ for pointlike systems.  There may be interesting fracton orders, likely in spatial dimension $d > 3$, where the higher fusion modules have infinite-order elements.  Once again, extending our framework to encompass such examples, should they exist, is left to future work.

The homology $H_0(F)$ measures ground state degeneracy in infinite space. This can be motivated from the physical discussion above, because nonzero elements of $H_0$ correspond to movers that do not create any excitations, but are not themselves stabilizers.  The ground state eigenvalues of stabilizers are fixed by the system's Hamiltonian, while nonzero elements of $H_0 \neq 0$ correspond to classes of operators that can be simultaneously diagonalized together with all the stabilizers, but whose eigenvalues are not fixed.

Definition~\ref{def:FC} is intentionally very flexible, and in particular there are no conditions placed on the homologies $H_{i}(F)$ except that they are finitely generated and have elements of finite order.  At least for the essentially $d$-dimensional systems of primary interest, this is certainly too general.  For instance, in an essentially $d$-dimensional system, we do not expect it is possible for $S$ to be a finitely generated free $R_n$-module.  This would describe completely immobile particles, which do not have even the cooperative mobility characteristic of fractons.  That is, a particle at a given lattice position cannot be moved to any collection of particles elsewhere in space by any local operator.  On the other hand, such behavior can naturally occur on defects of codimension two.  For instance, we can consider a trivial one-dimensional defect line within a three-dimensional stack of decoupled 2d layers, with the defect line perpendicular to the layers.  If each layer has fusion module $S_{{\rm layer}} \cong \Z_n$, then the fusion module of the defect line is $S \cong \Z_n[t^\pm]$, where the polynomial indeterminate $t$ represents translation normal to the layers.

The remarks above illustrate that different physical settings impose different constraints on the homology modules of a fusion complex.  Indeed, informed by the definition of a braided fusion complex, in Sec.~\ref{sec:dual_complex} we introduce two subclasses of fusion complexes, called \emph{proper} and \emph{complete} (see Definition~\ref{defn:propercomplete}).  Complete fusion complexes are intended to model gapped phases of essentially $d$-dimensional systems, while proper fusion complexes form a broader class that also encompasses gapped boundaries.

Now we discuss the notion of equivalence of fusion complexes.  Clearly there are many different possible choices of the complex $F$ for a given physical system.  The physical interpretation makes it clear that much of the data is non-universal, in the sense that it is not a universal property of a translation-invariant phase.  In the pointlike case, the isomorphism class of $S$ is the only universal information contained in $F$, because we started from $S$ to construct $F$.  This should be reflected in a notion of equivalence that relates different choices of $F$.  In particular, the choice of coefficients should not matter; for instance, in the pointlike case with $S$ of fusion order $n$, it should not matter whether we work with a free resolution of $S$ over $R$ or over $R_n$.

Mathematically, fusion complexes should be the objects of some category, in which equivalence should be isomorphism.  Broadly speaking, there are three options.  First, we can work in the category of chain complexes $\Ch(R)$, where the morphisms are chain maps $f : F \to G$, and a chain isomorphism is a chain map where each  component $f_i : F_i \to G_i$ is a module isomorphism.  This is clearly not the choice we want, because it preserves \emph{all} the non-universal data in $F$, up to a relabeling of elements within the modules $F_i$.

The next option is the homotopy category $\K(R)$, where the morphisms are (chain) homotopy classes of chain maps.  Isomorphisms are homotopy equivalences, which induce isomorphisms on homology.  This is a better choice; for instance, recalling the interpretation of elements of $F_0$ as operators, homotopy equivalences allow us to add and remove ancilla degrees of freedom.  That is, we recover the desired equivalence between a physical system $A$ and the same system $A$ stacked with a trivial system $T$ of ancillas, \emph{i.e.} of decoupled single-site degrees of freedom.  For $\Z_n$ degrees of freedom, one such trivial system can be modeled by the trivial fusion complex $T$:
\begin{equation}
\begin{tikzcd}
0 \arrow[r] & R_n \arrow[r, "\left(\begin{smallmatrix} 1 \\ 0 \end{smallmatrix}\right)"] &
R^2_n \arrow[r, "\left(0 \,\, 1\right)"]  & R_n \arrow[r] & 0 
\end{tikzcd} \text{,}
\end{equation}
where $T_0 = R^2_n$ and the matrices represent $R_n$-linear maps between the free $R_n$-modules.  This fusion complex can be obtained from a trivial $\Z_n$ Pauli code with Hamiltonian $H = - \sum_i( X^{\vphantom\dagger}_i + X^\dagger_i )$, where $X_i$ is the Pauli $X$ operator on lattice site $i$ (see Sec.~\ref{sec:Pauli}).  The complex $T$ is contractible, which means it is homotopy equivalent to the zero complex.  Therefore, $F$ is homotopy equivalent to $F \oplus T$ for any fusion complex $F$.

However, working in the homotopy category still does not give the desired equivalence of fusion complexes.  In particular, there are some systems of ancillas that are not trivial in the homotopy category.  For instance, we consider the infinite fusion complex
\begin{equation}
\begin{tikzcd}
\cdots R^2_4  \arrow[r, "\left(\begin{smallmatrix} 2 \,\, 0 \\ 0 \,\, 2 \end{smallmatrix}\right)"] & R^2_4 \arrow[r, "\left(\begin{smallmatrix} 2 \,\, 0 \\ 0 \,\, 2 \end{smallmatrix}\right)"] &
R^2_4  \arrow[r, "\left(\begin{smallmatrix} 2 \,\, 0 \\ 0 \,\, 2 \end{smallmatrix}\right)"]  & R^2_4  \arrow[r, "\left(\begin{smallmatrix} 2 \,\, 0 \\ 0 \,\, 2 \end{smallmatrix}\right)"] & R^2_4 \cdots
\end{tikzcd} \text{.} \nonumber
\end{equation}
This fusion complex is acyclic, but is not contractible; it is straightforward to see that the identity map on the complex is not null-homotopic.  This complex models a $\Z_4$ Pauli code of ancillas with Hamiltonian $H = - \sum_i( X^2_i + Z^2_i)$.

Another reason the homotopy category is not satisfactory is that it does not allow for change of coefficients.  Suppose that $F$ is a free resolution of $S$ by $R$-modules, and $G$ is a free resolution by $R_n$-modules.  We can view any $R_n$-module as an $R$-module, so if we work over $R$ we can try to relate $F$ and $G$. By the comparison theorem of homological algebra (Theorem 2.2.6 of \cite{Weibel_1994}), there is a chain map $f : F \to G$ that induces an isomorphism on homology.  If $S \neq 0$, $f$ cannot be a homotopy equivalence; that requires a homotopy inverse to exist, but the only chain map $G \to F$ is the zero map.  

 This leads us to the derived category $\D(R)$, which is the third option.  Recall that a quasi-isomorphism between any two chain complexes is a chain map that induces isomorphisms on homology.  For instance, the map $f : F \to G$ discussed immediately above is a quasi-isomorphism.  As that example illustrates, given a quasi-isomorphism, there need not be a chain map in the reverse direction that is also a quasi-isomorphism.  The essential idea of the derived category is to work in a setting where quasi-isomorphisms become isomorphisms.  This is similar in spirit to the idea of localization in commutative algebra, where we select certain elements of a ring, and then construct a new ring (the localization) where our selected elements become units, \emph{i.e.} they become invertible.  Unlike localization, the construction of the derived category is rather complicated, but fortunately one does not need to know the details of the construction to work with the derived category.  For now, a few basic ideas will suffice (see Appendix~\ref{dra:new-derived-category} for further background):  The objects of the derived category are still chain complexes, and every chain map represents a morphism in the derived category.  Moreover, homotopic chain maps represent the same derived morphism; in other words, the derived category is built on the homotopy category.  However, there are derived morphisms that are not represented by any chain map; that is, unlike the homotopy category, morphisms in the derived category are not equivalence classes of chain maps.  Instead we have the following:
\begin{property}[Corollary~\ref{lemma:iso-zigzag}]\label{property}
Two chain complexes $F$ and $G$ are isomorphic in the derived category $\D(R)$, and thus said to be {\bf quasi-isomorphic}, if and only if they are connected by a finite alternating sequence of quasi-isomorphisms, \emph{i.e.} a diagram of the form
\begin{equation}
\begin{tikzcd}
F \arrow[r, "f_1"] & C^1 & C^2 \arrow[l, "s_1"] \arrow[r, "f_2"] & \cdots & \arrow[l, "s_k"] G 
\end{tikzcd} \text{,} \label{eqn:zigzag}
\end{equation}
where the $C^i$ are complexes, and each $f_i$ and $s_i$ is a quasi-isomorphism.  Such a sequence is called a {\bf zigzag} of quasi-isomorphisms.   In particular, if there is a quasi-isomorphism $F \to G$, then $F$ and $G$ are quasi-isomorphic.  So every quasi-isomorphism represents an isomorphism in the derived category.  If $F$ and $G$ are quasi-isomorphic, we write $F \simeq G$.
\end{property}

We note that any of the maps in Equation~\ref{eqn:zigzag} can be the identity, so there is nothing special about the particular alternating pattern chosen for the arrow directions.  For instance, $F \to G$, $F \leftarrow C^1 \leftarrow G$ and $F \leftarrow C^1 \rightarrow G$ are all legitimate zigzags, and can be written in the form of Equation~\ref{eqn:zigzag} by introducing explicit arrows for identity maps.  It is important that both directions of arrows are allowed.

\begin{defn} \label{defn:perfect}
A {\bf perfect} chain complex over a commutative ring ${\mathcal R}$ is one that is quasi-isomorphic to a bounded complex of finitely generated projective ${\mathcal R}$-modules.
\end{defn}
\noindent  The reader should not be disturbed by the sudden appearance of projective modules.  Over the ring $R$, finitely generated projective modules and finitely generated free modules are the same, by the Quillen-Suslin-Swan theorem.

\begin{remark}  \label{rmk:bh-perfect}
Over the ring $R$, any complex with finitely generated homology concentrated in finitely many degrees is perfect.  See Lemma~\ref{lemma:bounded-replacement}(c).
\end{remark}

\begin{defn}[Equivalence of fusion complexes]  \label{def:fc-equiv}
We work in the category $\Ch^p(R)$ of perfect chain complexes over $R$. 
Any two fusion complexes are considered equivalent if they are quasi-isomorphic; in other words, equivalent fusion complexes are connected by a zigzag of quasi-isomorphisms; the intermediate complexes $C^i$ are automatically also fusion complexes.  The equivalence classes are referred to as {\bf fusion theories}.
\end{defn}

\begin{remark}
Equivalently, two fusion complexes are considered equivalent if they are isomorphic as objects of the derived category $\D^p(R)$, the full subcategory of $\D(R)$ whose objects are perfect complexes.  This is how Definition~\ref{def:fc-equiv} would be stated if this paper were written primarily for a mathematics audience; in that case, we would work in the derived category from the beginning.  However, for greater concreteness and accessibility to a physics audience, in the main text we often prefer to work at the chain complex level, but using ideas from the derived category; Definition~\ref{def:fc-equiv} is stated in terms of $\Ch^p(R)$ to reflect this approach. 
In Appendices~\ref{app:derived_background} and~\ref{app:derived-fusion}--\ref{v2:app:tame-BFC} we more fully adopt a derived perspective.
\end{remark}

\begin{example} \label{ex:hybrid-Xcube}
The homology of a fusion complex does not in general determine the fusion theory. To see this, we compare the fractonic hybrid X-cube model~\cite{Tantivasadakarn_2021} with a stack of the X-cube model and the 3D toric code. Let $F_{\mathrm{hyb}}$ be the fusion complex of the lineon and flux-loop sector of the hybrid model, and let $F_{\mathrm{stack}} = F_{\mathrm{lineon}} \oplus F_{\mathrm{loops}}$ be the corresponding fusion complex for the stacked model. Both complexes have $H_{-1} \cong S_{\mathrm{lineon}}$, the fusion module of the X-cube lineon sector, and $H_{-2} \cong \Z_2$ as their only nonzero homology modules. Nevertheless, $F_{\mathrm{hyb}} \not\simeq F_{\mathrm{stack}}$.

Physically, the homology modules of $F_{\mathrm{hyb}}$ describe loop and pointlike excitations separately, and do not capture the fusion of two identical hybrid flux loops into a configuration of lineons (see \cite{Tantivasadakarn_2021} for a discussion of this phenomenon). However, the quasi-isomorphism type of $F_{\mathrm{hyb}}$ retains this information. This inequivalence can also be established by comparing the dual complexes, defined in Section~\ref{subsec:bfc1}, whose homology modules are not isomorphic.  Physically, this difference reflects the fact that the fracton excitations of the hybrid model have order four under fusion, while those of the stacked model have order two.
\end{example}

Given a fusion theory, the choice of $\Z$ or $\Z_n$ coefficients amounts to picking a certain representative fusion complex.  For a pointlike fusion theory of fusion order $n_S$, we can always choose a representative where $F$ consists of free $R_n$-modules provided $n_S$ divides $n$.  The following terminology will be useful:
\begin{defn}
Let $A$ be a chain complex of $R$-modules.  For a fixed positive integer $n$, we say $A$ is {\bf $\Z_n$-tame} if each $A_i$ is a finitely generated free $R_n$-module; no boundedness condition is imposed.  We say $A$ is {\bf $\Z$-tame} if it is bounded and each $A_i$ is a finitely generated free $R$-module.  A complex is {\bf tame} if it is $\Z$-tame or $\Z_n$-tame for some $n$.

A fusion complex is {\bf tame}, {\bf $\Z$-tame}, or {\bf $\Z_n$-tame} when its underlying chain complex has the corresponding property.  For a $\Z_n$-tame fusion complex, we refer to $n$ as the {\bf coefficient order}.
\end{defn}

\begin{remark}
The coefficient order of a $\Z_n$-tame fusion complex is divisible by the fusion order, but the coefficient and fusion orders are not equal in general.  Moreover, by the definition of perfect complex, every fusion complex is equivalent to a $\Z$-tame fusion complex.  Therefore, every fusion theory has a tame representative (Lemma~\ref{v2:lemma:tame-in-FC}(c)).  Tameness depends on the chosen representative and need not be preserved by quasi-isomorphisms.
\end{remark}

The following result brings us back to where we started in the pointlike case.  There, our input to construct a fusion complex was the module $S$, and we should not be able to get out more than what we put in.  More precisely, we have:

\begin{proposition}  \label{prop:S-complete-invt}
The isomorphism class of $S = H_{-1}(F)$ is a complete invariant for pointlike fusion complexes.  That is, two pointlike fusion complexes are equivalent if and only if they have isomorphic fusion modules.  Moreover, given two pointlike fusion complexes $F$ and $G$, and an isomorphism $\alpha : H_{-1}(F) \to H_{-1}(G)$, there exists a unique derived isomorphism $F \to G$ that induces $\alpha$ on homology.  
\end{proposition}

\begin{remark} \label{rmk:S-complete-invt-Zn}
If $F$ and $G$ are $\Z_n$-tame pointlike fusion complexes with an isomorphism $\alpha : H_{-1}(F) \to H_{-1}(G)$, then the corresponding derived isomorphism $F \to G$ can be represented as a zigzag of quasi-isomorphisms between $\Z_n$-tame fusion complexes.
\end{remark}

Proposition~\ref{prop:S-complete-invt} follows from Corollary~\ref{v2:cor:S-complete-invt}, and Remark~\ref{rmk:S-complete-invt-Zn} follows from Lemma~\ref{v2:lem:truncated-gamma}(c).

\subsection{Braided fusion complexes}
\label{subsec:bfc1}

A fusion complex is only a starting point for an algebraic theory of excitations, because it includes no information about statistics.  The next step is thus to understand what additional data we can add to encode statistics; this will lead us to the definition of braided fusion complex.  As in Section~\ref{subsec:fc1}, we will sometimes appeal to the pointlike case for physical motivation, but will give definitions that also apply to systems with extended excitations, with further physical motivation to follow in Section~\ref{sec:Pauli}.  This will lead to a definition of BFC that makes sense for tame fusion complexes, and a notion of equivalence between such tame BFCs.  While this provides a physically motivated working definition of BFC and BFC equivalence, it is not entirely satisfactory,  because the definitions only make sense for tame representatives of a fusion theory.  At odd fusion order, Theorem~\ref{v2:thm:odd-tame-derived-BFC} relates this description to a derived definition that does not suffer from this issue.

Following Ref.~\cite{Pai_2019} (see also Ref.~\cite{Levin_2003}), a statistical process can be understood as a sequence of $k$ steps where excitations are moved over large spatial distances during each step.  While Ref.~\cite{Pai_2019} focused on the pointlike case, the discussion of this paragraph also applies to extended excitations. The system starts in an initial state $| \psi \rangle$, and returns to the same state at the end of the process, up to a ${\rm U}(1)$ phase factor.  As appropriate for restricted-mobility excitations, during each step excitations can be created and destroyed.  Representing the $i$th step by a unitary operator $\hat{A}_i$, the final state is $| \psi' \rangle = \hat{A}_k \cdots \hat{A}_1 | \psi \rangle$, and we have assumed that $| \psi' \rangle = e^{i \theta'} | \psi \rangle$.  The phase $e^{i \theta'}$  is not of interest, however, because it depends on an arbitrary overall phase choice of each operator $\hat{A}_i$.  In order to obtain a well-defined ${\rm U}(1)$ phase, we need to compare to another process represented by $| \psi'' \rangle =  \hat{A}_{\sigma(k)} \cdots \hat{A}_{\sigma(1)} | \psi \rangle$, where $\sigma : \{1, \dots, k\} \to \{1, \dots, k\}$ is a permutation, such that again $| \psi'' \rangle = e^{i \theta''} | \psi \rangle$ for some other phase factor $e^{i \theta''}$.  The relative ${\rm U}(1)$ phase $e^{i (\theta' - \theta'')}$ between the two final states is well-defined, and if  certain other conditions are satisfied it is a statistical phase factor \cite{Pai_2019}.  The intuition is that we want to compare two processes, one where a nontrivial generalized braiding or exchange of excitations occurs, and another where no generalized braiding or exchange occurs, such that the two processes are made out of the same steps but carried out in a different order.  If the pairwise commutator of the operators $\hat{A}_i$ is a ${\rm U}(1)$ phase, \emph{i.e.} if $[\hat{A}_i, \hat{A}_j ] = \hat{A}^{\vphantom 1}_i \hat{A}^{\vphantom 1}_j \hat{A}^{-1}_i \hat{A}^{-1}_j  = e^{2\pi i \phi_{i j}} \mathbbm{1}$, then the statistical phase is determined by the commutator.

Now we consider a tame fusion complex, which need not be pointlike. We assume that the movers (elements of $F_0$) can be viewed as projective unitary operators, with addition in $F_0$ corresponding to operator multiplication modulo ${\rm U}(1)$ phases, as discussed for the pointlike case in Sec.~\ref{subsec:fc1}. This is one of the physical assumptions used to motivate our definitions. Based on this assumption, there is a natural commutator on $F_0$.  Because movers are precisely the local operators that can be used to move excitations around, they are the building blocks for the steps of a statistical process; that is, each $\hat{A}_i$ above should be a product of movers.   Therefore, under these assumptions, any possible statistical phase factor is determined by the commutator on movers.

Let $A,B \in F_0$, and choose operator representatives $\hat{A}$ and $\hat{B}$.  The commutator is defined to be $[A,B] = \hat{A} \hat{B} \hat{A}^{-1} \hat{B}^{-1}$, which is well-defined independent of the choice of representatives.  We note that we always use the notation $[A,B]$ to represent the group-theoretic commutator as defined above; the commutator of operators $\hat{A} \hat{B} - \hat{B} \hat{A}$ is not used in this paper.  We define a function $\omega : F_0 \times F_0 \to \R/\Z$ by
\begin{equation}
\left[A,B\right] = e^{2\pi i \omega(A,B)} \mathbbm{1} \text{.}
\end{equation}
This form holds because the product $\hat{A} \hat{B} \hat{A}^{-1} \hat{B}^{-1}$ corresponds to $A + B - A - B = 0$ in $F_0$.  We also refer to $\omega$ as the commutator.

The commutator satisfies the following properties for all $A,B,C \in F_0$:
\begin{itemize}
\item $\omega(A,A) = 0$.
\item $\omega(A,B) = - \omega(B,A)$.
\item $\omega(A+B,C) = \omega(A,C) + \omega(B,C)$.
\end{itemize} 
These properties imply that $\omega$ is an alternating (and thus also antisymmetric) bilinear form.  Moreover, the commutator is translation invariant, \emph{i.e.} $\omega(A,B) = \omega(t A, t B)$ for all $A,B \in F_0$ and all translations $t \in {\cal T}$.  Finally, because movers are local operators, we should assume that $\omega$ is spatially local, in the sense that movers commute at sufficiently large spatial separation.  Formally we require that for any $A,B \in F_0$, $\omega(A, t B)$ is nonzero for only finitely many $t \in {\cal T}$.  More generally, it could be the case that $\omega$ decays to zero at large spatial separation as a function of $t$, in contrast with the strict spatial locality that we assume.  However, for systems with $\Z$-torsion fusion modules $Q_i$, we expect that a looser notion of spatial locality for the commutator will not lead to new gapped phases.

Following Ref.~\cite{Ruba_2024}, we repackage $\omega$ into an $\RRZ$-valued sesquilinear form $\Omega : F_0 \times F_0 \to \RRZ$, also referred to as the commutator, which is defined as the unique sesquilinear form whose constant component is $\omega$.  Here the involution sends each translation $t$ to $t^{-1}$ and
fixes coefficients, and our sesquilinearity convention is
$\Omega(rA,sB)=\bar r s\,\Omega(A,B)$. The two forms are related via the explicit formula
\begin{equation}
\Omega(A, B) = \sum_{t \in {\cal T}} \omega(t A, B) t \text{,} \label{eqn:omega-Omega}
\end{equation}
which is well-defined by spatial locality of $\omega$.  The value $\Omega(A,B)$ is a polynomial that encodes the commutator of $B$ with all the translates of $A$, and vice versa (by translation-invariance of $\omega$).  It follows from Equation~\ref{eqn:omega-Omega} that $\Omega$ is antisymmetric, \emph{i.e.} $\Omega(B,A) = - \overline{\Omega(A,B)}$.  We also write $\Omega$ for its associated $R$-linear map
$\Omega:F_0\to D_{\RRZ}(F_0)$, defined by $\Omega(B)(\overline{A})=\Omega(A,B)$.
Here we introduce the dual $D_{\RRZ}(M)=\Hom_R(\overline M,\RRZ)$, with scalar action $(rf)(\overline{m})=r f(\overline{m})$; its elements are linear on $\overline M$, or equivalently antilinear on $M$. If $f : M \to N$ is an $R$-linear map, $D_{\RRZ}(f) : D_{\RRZ}(N) \to D_{\RRZ}(M)$ is precomposition by $f$. When no confusion can arise as to the target of the duality (which is $\RRZ$ in most of this section), we sometimes write $M^* =  D_{\RRZ}(M)$ and $f^* = D_{\RRZ}(f)$, or we drop the coefficient module subscript and write $D(M) = D_{\RRZ}(M)$ and $D(f) = D_{\RRZ}(f)$. More generally, a sesquilinear pairing $\eta:M\times N\to\RRZ$ has linear left and right associated maps
\[
\eta_L(m)(\overline{n})=\overline{\eta(m,n)},\qquad
\eta_R(n)(\overline{m})=\eta(m,n).
\]
For a symmetric or antisymmetric form, $\eta_R=\eta_L$ or $\eta_R=-\eta_L$, respectively. We use the right map as the default associated map. See Appendix~\ref{subsec:notation} for our conventions.

\begin{defn}\label{def:constant-alternating}
An $\RRZ$-valued or $\RQZ$-valued sesquilinear form $\eta$ on an $R$-module $M$ is {\bf constant-alternating} if its constant component is alternating, \emph{i.e.} if the constant coefficient of $\eta(a,a)$ vanishes for every $a\in M$.  We use the same terminology for the associated map of the form.
\end{defn}

An important property of the commutator is that stabilizers should commute.  That is, we require that $\Omega(d_1(A), d_1(B)) = 0$ for all $A,B \in F_1$.  In terms of the associated map, this is the same as the condition $d^*_1 \Omega d_1 = 0$.  The reason for this condition is that stabilizers map under $d_1$ to movers that do not create any excitations; thus, acting with an operator representing $d_1(A)$ leaves every state in $F_{-1}$ (\emph{i.e.} every element of $\ker d_{-1}$) invariant.  Within each superselection sector, the stabilizers are thus simultaneously diagonal in the basis given by elements of $\ker d_{-1}$ within that sector.  This is only possible if any pair of stabilizers commutes.

We summarize the properties of the commutator in the following definition:
\begin{defn}
A {\bf commutator} on a fusion complex $F$ is an antisymmetric, constant-alternating $R$-sesquilinear form $\Omega : F_0 \times F_0 \to \RRZ$ satisfying $\Omega(d_1(A), d_1(B)) = 0$ for all $A, B \in F_1$.  
\end{defn}

For a fusion complex of $R_n$-modules we have $n A = 0$ for any $A \in F_0$, which implies $\omega(A,B) \in \Z_n \subset \Q / \Z \subset \R / \Z$.  We can thus view $\Omega$ as an $R_n$-sesquilinear form valued in $R_n$, using the natural inclusions $\Z_n \hookrightarrow \Q / \Z \hookrightarrow \R/\Z$ and $R_n \hookrightarrow \RQZ \hookrightarrow \RRZ$.  That is, in this case we can write $\omega : F_0 \times F_0 \to \Z_n$ and $\Omega : F_0 \times F_0 \to R_n$, with associated map $\Omega : F_0 \to \Hom_{R_n}(\overline{F_0}, R_n)$.  We use the same constant-alternating terminology for $R_n$-valued forms, via the inclusion $R_n\hookrightarrow\RQZ$.

\begin{remark}  \label{rmk:alternating-superfluous}
For a commutator on a fusion complex of $R_n$-modules with $n$ odd, the requirement that $\Omega$ be constant-alternating is superfluous.  In that case, $2$ is invertible in $R_n$, and for any antisymmetric $R_n$-sesquilinear form we have
\begin{equation}
\Omega(A,B) = 2^{-1} \Big[  \Omega(A,B) - \overline{\Omega(B,A)} \Big] \text{.} \nonumber
\end{equation}
It follows immediately that $\Omega$ is constant-alternating.
\end{remark}

While a fusion complex with a commutator gives us all the data we need to describe statistics, we have not quite arrived at the definition of a BFC.  Not only can we put the data into a nicer mathematical package, but not every commutator is guaranteed to be physically sensible.  Some further physical discussion is helpful to motivate the definition.  Now we specialize to a pointlike tame fusion complex, and assume for the moment that $F$ is a free resolution for $S$; this is a simplification because all elements of $F_{-1}$ can be viewed as states.  Given a stabilizer $A \in F_1$ and a representative operator, every state $e \in F_{-1}$ is an eigenstate of the stabilizer.  While \emph{a priori} the eigenvalue is only defined up to a ${\rm U}(1)$ phase, this can be fixed by requiring that the eigenvalue is unity in the ground state.  We thus have a bilinear pairing $\xi : F_1 \times F_{-1} \to \R/\Z$, where $e^{2\pi i \xi(A, e)}$ is the eigenvalue of the stabilizer $A$ acting on the state $e$.  The pairing $\xi$ is clearly translation-invariant and spatially local in the same sense as $\omega$.  From $\xi$ we can thus construct $\Xi : F_1 \times F_{-1} \to \RRZ$, which is the unique $R$-sesquilinear pairing whose constant component is $\xi$, explicitly $\Xi(B,e)=\sum_{t\in\T}\xi(tB,e)t$.

We will see that $\Omega$ and $\Xi$ can naturally be viewed as parts of the same object.  In order to do that, first we will need to define dual chain complexes and establish some basic results on duality.  

\begin{defn} \label{def:dual-complex}
Duality is the functor $D_{\RRZ}=\Hom_R(\overline{(-)},\RRZ):\Ch(R)^\op\to\Ch(R)$.  That is, given a chain complex $F$ with boundary maps $d_i:F_i\to F_{i-1}$, its {\bf dual complex} is $D_{\RRZ}(F)=\Hom_R(\overline F,\RRZ)$ (see Appendix~\ref{dra:new-conventions}), where on the right-hand side $\RRZ$ denotes the chain complex with $\RRZ$ in degree zero and all other terms zero.  In more detail, 
$D_{\RRZ}(F)_i=D_{\RRZ}(F_{-i})$, with differential denoted $d^{D(F)}_i=(-1)^{i-1}d^*_{-(i-1)}$, where we write $D(F)$ in place of $D_{\RRZ}(F)$ in the superscript to avoid notational clutter. The dual complex is thus the following diagram:
\begin{equation}
\begin{tikzcd}[column sep=small]
\cdots \arrow[r] & D_{\RRZ}(F_{-2}) \arrow[r,"-d^*_{-1}"] & D_{\RRZ}(F_{-1}) \arrow[r,"d^*_{0}"] & D_{\RRZ}(F_0) \arrow[r,"-d^*_{1}"] & D_{\RRZ}(F_1) \arrow[r,"d^*_{2}"] & D_{\RRZ}(F_2) \arrow[r,"-d^*_{3}"] & \cdots
\end{tikzcd} \text{.} \nonumber
\end{equation}
Moreover, given a chain map $f:F\to G$, the {\bf dual chain map} $D_{\RRZ}(f):D_{\RRZ}(G)\to D_{\RRZ}(F)$ has components $D_{\RRZ}(f)_i=f^*_{-i}$.  When the target of the duality is clear from context, we often write $D(F) = D_{\RRZ}(F)$ and $D(f) = D_{\RRZ}(f)$.
\end{defn}

The minus signs in the definition of $d^{D(F)}_i$ arise from a standard convention for the boundary maps of the Hom-complex.  While the signs can be unpleasant, and other conventions would allow a definition of duality without minus signs, we are not aware of any convention that allows one to completely avoid  signs in all the constructions we employ in this paper.  Therefore we feel it is safest to follow an established convention in the mathematical literature.  No signs are included in the definition of dual chain map; this ensures the dual of the identity chain map is the identity.

\begin{warning}  \label{warning:dual-complex}
The duality of Definition~\ref{def:dual-complex} is not derived; here we briefly explain this statement.  Very generally, when working in any category, if we make a construction that takes some objects as input and outputs another object, the isomorphism type of the output should only depend on the isomorphism types of the inputs.  Constructions that do not have this property are not natural, and need to be used with caution.  In the derived category, this principle becomes the statement that the quasi-isomorphism type of the output should only depend on the quasi-isomorphism types of the inputs; constructions that have this property are said to be derived.  For the duality as defined, if $F \simeq G$, it may be the case that $D(F) \not\simeq D(G)$.  This issue will be important later on in our treatment, but it does not play a role in this section.

We emphasize that the duality as defined is still useful because it allows us to express statistical information in a physically transparent manner, and thus to introduce the key ideas without too much abstraction.  Moreover, we will define a derived duality (see Appendix~\ref{app:derived_background}) that agrees with Definition~\ref{def:dual-complex} for tame fusion complexes (Theorem~\ref{v2:thm:tame-dual} and Lemma~\ref{v2:lemma:real-coefficients}(b)).  This justifies the use of the dual complex as defined, as long as we start with a tame fusion complex.
\end{warning}

If a chain map $\alpha : F \to G$ is a chain isomorphism, then so is its dual $D(\alpha)$.   This holds because  duality is a functor on the category of chain complexes.  Moreover, if $\alpha$ is a homotopy equivalence, then so is $D(\alpha)$.  To see this, let $\beta : G \to F$ be a homotopy inverse. Then applying duality to the homotopies $\alpha \beta \simeq \operatorname{id}_G$ and $\beta \alpha \simeq \operatorname{id}_F$ gives the desired result.  Finally, if $F$ and $G$ are tame fusion complexes and $\alpha$ is a quasi-isomorphism, then $D(\alpha)$ is also a quasi-isomorphism (Lemma~\ref{v2:lemma:real-coefficients}(c)).   Caution is required, because the same result does not hold if $F$ and $G$ are general complexes of $R$-modules.

Returning to our pointlike fusion complex  $F$, we assemble the map $-\Omega : F_0 \to F^*_0$ and the associated maps $\Xi_L : F_1 \to F^*_{-1}$ and $\Xi_R : F_{-1} \to F^*_1$ into a chain map from $F$ to the dual complex.  That is, the desired chain map is $\lambda : F \to D(F)$ and is represented by the diagram
\begin{equation}
\begin{tikzcd}
\cdots \arrow[r] & F_2 \arrow[r, "d_2"] \arrow[d] & F_1 \arrow[r, "d_1"] \arrow[d, "\Xi_L"] & F_0 \arrow[r, "d_0"] \arrow[d, "-\Omega"] & F_{-1} \arrow[r]  \arrow[d, "\Xi_R"] & 0 \arrow[r] \arrow[d] & \cdots \\
\cdots \arrow[r] & 0 \arrow[r] & F^*_{-1} \arrow[r,  "d^*_{0}"] & F^*_0 \arrow[r, "-d^*_{1}"] & F^*_{1} \arrow[r, "d^*_{2}"] & F^*_2 \arrow[r] & \cdots 
\end{tikzcd} \text{.}  \label{eqn:first-example-lambda}
\end{equation}
Thus we have $\lambda_0 = -\Omega_R=-\Omega$, $\lambda_{1} = \Xi_L$ and $\lambda_{-1} = \Xi_R$, with $\lambda_i = 0$ for other values of $i$. Explicitly,
\[
\lambda_0(A)(\bar A')=-\Omega(A',A),\quad
\lambda_1(B)(\bar e)=\overline{\Xi(B,e)},\quad
\lambda_{-1}(e)(\bar B)=\Xi(B,e).
\]  In order for this to be a chain map, the diagram must commute, and we now give a physical explanation why this holds.  First we recall that $F_2$ consists of relations among stabilizers, so elements of $\im d_2$ correspond to products of stabilizers that are equal to the identity operator.  Such a product thus gives unity acting on any state, and commutativity of both the leftmost and rightmost squares expresses this fact.  Commutativity of the square involving $\Xi_L$ and $\Omega$ is the statement $\Xi(B, d_0(A)) = \Omega(d_1(B), A)$ for any $A \in F_0$ and $B \in F_1$, or equivalently, taking the constant component, $\xi(B, d_0(A)) = \omega(d_1(B), A)$.  Letting $\hat{B}$ and $\hat{A}$ be the operators representing $d_1(B)$ and $A$, respectively, this is simply the statement that the eigenvalue of $\hat{B}$ acting on the state $\hat{A} | \psi_0 \rangle$, where $|\psi_0\rangle$ is the ground state, is the commutator $[\hat{B}, \hat{A}]$.  Commutativity of the square involving $\Xi_R$ and $\Omega$ is the equivalent statement $\overline{ \Xi(B, d_0(A) ) } = - \Omega(A, d_1(B))$.  

Our discussion so far suggests that, in the case of a physical system described by a general fusion complex, statistics can be described by a chain map to the dual complex.  We observe that the chain map we constructed satisfies the following antisymmetry property:
\begin{equation}
\lambda_i(f_i)(\overline{f_{-i}}) = (-1)^{i+1} \overline{ \lambda_{-i}(f_{-i})(\overline{f_i})} \text{,} \label{eqn:lambda-asym}
\end{equation}
for all $i \in \Z$ and any $f_i \in F_i$ and $f_{-i} \in F_{-i}$.  For $i = 0$, this is nothing but antisymmetry of $\Omega$, and for $i = 1$ it expresses the fact that $\Xi_L$ and $\Xi_R$ are the two associated maps of a single sesquilinear pairing.  More generally, any chain map satisfying Equation~\ref{eqn:lambda-asym} encodes a family of sesquilinear pairings $F_i \times F_{-i} \to \RRZ$ for $i = 0,1,\dots$.  While Equation~\ref{eqn:lambda-asym} appears somewhat messy, it can be expressed more elegantly as a natural property of the chain map $\lambda$; in order to do this, a brief mathematical digression is needed.

The first step is to understand how symmetry or antisymmetry of a sesquilinear form is encoded as a property of its associated map.  This is done using the canonical map from an $R$-module to its double dual:
\begin{defn}  \label{def:can-module}
Let $M$ be an $R$-module. Then the canonical map $\can_M : M \to D^2(M) := D( D(M))$ is the twisted evaluation map.  That is, $\can_M(x)(\overline{\varphi}) = \overline{\varphi(\overline{x})}$ for all $x \in M$ and any $\varphi \in D(M)$.  Here, the outer bar on the right-hand side denotes the involution on $\RRZ$, while the others indicate elements of conjugate modules.  When it is clear from the context, we often omit the ``$M$'' subscript and simply write $\can$.
\end{defn}

The following standard result is stated without proof (see Lemma~\ref{v2:lemma:ev-can-formulas}, Corollary~\ref{v2:cor:can-identities}(b), and the degreewise finite-free calculation in Lemma~\ref{v2:lemma:Zn-can}(b), with Lemma~\ref{v2:lemma:real-coefficients}(b)):
\begin{proposition}  
The canonical map $\can_M$ is an $R$-module homomorphism.  Moreover, it is a natural transformation between the identity and double dual functors on the category of $R$-modules.  Finally, if $F$ is a finitely generated free $R_n$-module, then $\can_F$ is an isomorphism, and $\can_F$ is thus a natural isomorphism between the identity and double dual functors on the full subcategory of finitely generated free $R_n$-modules.
\end{proposition}

Now let $\mu : M \times M \to \RRZ$ be a sesquilinear form on an $R$-module $M$.  We assume that $\mu$ is symmetric or antisymmetric, which means $\mu(x, x') = \pm \overline{\mu(x', x)}$ for all $x, x' \in M$.  Of course, the $\pm$ sign is independent of $x$ and $x'$.  In terms of the associated map $\mu : M \to M^*$ we have $\mu(x)(\overline{x'}) = \mu(x', x)$.  Now we observe that using the canonical map, we can construct another map from $M$ to $M^*$, namely $\mu^* \circ \can : M \to M^*$, where $\mu^* : M^{**} \to M^*$ is the dual map.  Evaluating this we have
\begin{equation}
(\mu^* \circ \can)(x)(\overline{x'}) = \can(x)(\overline{\mu(x')}) = \overline{\mu(x')(\overline{x})} = \overline{\mu(x,x')}  \text{.} \nonumber
\end{equation}
Therefore the expression of symmetry/antisymmetry in terms of the associated map is
\begin{equation} \label{eqn:form-sym}
\mu = \pm \mu^* \circ \can  \text{.} 
\end{equation}

This discussion has a parallel for chain maps $\lambda : F \to D(F)$.  First we observe that the double dual complex $D^2(F) := D(D(F))$ is given by
\begin{equation}
\begin{tikzcd}
\cdots \arrow[r] & F^{**}_{2} \arrow[r,"-d^{**}_{2}"] & F^{**}_{1} \arrow[r,"-d^{**}_1"] &  F^{**}_0 \arrow[r,"-d^{**}_0"] &  F^{**}_{-1} \arrow[r,"-d^{**}_{-1}"] &  F^{**}_{-2} \arrow[r,"-d^{**}_{-2}"] & \cdots
\end{tikzcd} \text{.} \nonumber
\end{equation}
We can then define a canonical chain map:
\begin{defn} \label{def:can-RRZ}
Let $F$ be a chain complex of  $R$-modules. Then the canonical chain map $\can_F  : F \to D^2(F)$ is given in terms of its components $\can_i : F_i \to F^{**}_i$ by $\can_i = (-1)^i \can_{F_i}$.
\end{defn}
\noindent The alternating minus signs are required to make the squares of the chain map commute, given the overall minus sign in the boundary maps of $D^2(F)$.  Commutativity of the squares then follows from the fact that $\can_{M}$ is a natural transformation.  If $F$ is a $\Z_n$-tame complex, then $\can_F : F \to D^2(F)$ is a chain isomorphism by the degreewise finite-free calculation in Lemma~\ref{v2:lemma:Zn-can}(b) and Lemma~\ref{v2:lemma:real-coefficients}(b).  On the other hand, if $F$ is a nonzero complex of free $R$-modules, then $\can_F$ is not a chain isomorphism due to size issues, and for a general such complex (e.g.\ $F=R[0]$) it need not even be a quasi-isomorphism. For $\Z$-tame fusion complexes, $\can_F$ is a quasi-isomorphism (Lemma~\ref{v2:lemma:real-coefficients}(d)).
This is a manifestation of the same issue raised in Warning~\ref{warning:dual-complex}, but it will not play a role until later sections (see Sec.~\ref{sec:homological-braiding} and Appendix~\ref{app:derived-fusion}) where it is addressed by working with the derived duality.  A different way of addressing the issue, which we do not employ in this paper but has the virtue of being physically intuitive, is briefly discussed below in Remark~\ref{rmk:tdd}.

\begin{defn}  \label{def:chain-map-sym}
Let $F$ be a chain complex of  $R$-modules, and $\lambda : F \to D(F)$ a chain map.  We say that $\lambda$ is symmetric (resp. antisymmetric) if $\lambda = D(\lambda) \circ \can_F$ (resp. $\lambda = - D(\lambda) \circ \can_F$).  We say that $\lambda$ is {\bf constant-alternating in degree zero} if its degree-zero component $\lambda_0$ is constant-alternating.
\end{defn}
\noindent  The chain map  $\lambda$ for our pointlike fusion complex is thus antisymmetric. Here it is important to note that, while the alternating minus signs are required for $\can_F$ to be a chain map, we do have freedom to choose an overall minus sign.  Changing this overall sign would flip the meaning of symmetric vs. antisymmetric chain maps.  We choose our sign convention by requiring $\can_0 = \can_{F_0}$, which ensures that symmetry of forms agrees with symmetry of chain maps in the following sense:  suppose we have a  module $M$ and a symmetric or antisymmetric sesquilinear form $\mu : M \to M^*$.  We view $M$ as a complex with $F_0 = M$ and $F_i = 0$ for $i \neq 0$.  Then the chain map $\lambda : F \to D(F)$ with $\lambda_0 = \mu$ is merely a repackaging of the form $\mu$.  With our convention, $\mu$ is symmetric (resp. antisymmetric) if and only if $\lambda$ is symmetric (resp. antisymmetric).  

\begin{remark}[Topological double dual]  \label{rmk:tdd}
Here we briefly sketch an alternative construction of a double dual, which is better-behaved on fusion complexes of finitely generated free $R$-modules in the sense that a complex is naturally quasi-isomorphic (in fact, chain isomorphic) to its double dual. We give $\R/\Z$ its standard topology, and give $\RRZ$ a topology by viewing it as a subspace of the product space $\prod_{t \in \T} \R / \Z$.  Giving the terms $F_i$ of a tame fusion complex $F$ the discrete topology, we construct the dual $\Hom^c_R (\overline F, \RRZ)$, where ``$c$'' indicates that we take continuous maps.  Since every map from a discrete space is continuous, this dual agrees with that already defined, but now we give the terms of $\Hom^c_R (\overline F, \RRZ)$ the compact-open topology, as appropriate for a function space.  Dualizing again using $\Hom^c_R(\overline{(-)}, \RRZ)$ results in the topological double dual, which is chain-isomorphic to $F$; the key fact needed to show this is that $\Hom^c_R(\overline{\RRZ}, \RRZ) \cong R$.  We note that replacing the double dual with the topological double dual has no effect on any of the definitions in this section, and the topological double dual will not be used in this paper.  
\end{remark}

We can now state the key definition, focusing on tame fusion complexes; this will be our working definition of BFC for the majority of the paper.  For odd fusion order, this is equivalent to the derived description by Theorem~\ref{v2:thm:odd-tame-derived-BFC}.  When the fusion order is even we do not yet have a satisfactory abstract definition. See Remark~\ref{rem:2rem}.
\begin{defn}\label{defn:BFC}
A tame {\bf braided fusion complex (BFC)} over $F$  is a pair $(F, \lambda)$, where $F$ is a  tame fusion complex and $\lambda : F \to D(F)$ is an antisymmetric chain map that is constant-alternating in degree zero.  The chain map $\lambda$ is referred to as the {\bf braiding}.  A $\Z_n$-tame BFC is one where $F$ is a $\Z_n$-tame fusion complex, and a $\Z$-tame BFC is one where $F$ is a $\Z$-tame fusion complex.  The coefficient order of a $\Z_n$-tame BFC is that of its underlying fusion complex.  If $H_0(F) = 0$ and $\lambda$ is a quasi-isomorphism, we say the BFC is {\bf nondegenerate}; otherwise, it is degenerate.  A pointlike BFC is one where $F$ is a pointlike fusion complex, and a trivial BFC is one where $F$ is a trivial fusion complex.  The fusion order of a BFC is the fusion order of the underlying fusion complex.
\end{defn}

In the main text, all BFCs are understood to be tame, and we often omit this qualifier.  By Remark~\ref{rmk:alternating-superfluous}, for a $\Z_n$-tame BFC with $n$ odd, the condition that $\lambda$ is constant-alternating in degree zero is superfluous.

\begin{remark} \label{rmk:bfc-forms}
In the motivating discussion above, the braiding of a BFC was constructed from the two sesquilinear forms $\Omega$ and $\Xi$.  This holds more generally, in the sense that the braiding $\lambda$ gives a collection of sesquilinear forms $F_i \times F_{-i} \to \RRZ$ for $i \geq 0$.  Remark~\ref{v2:rmk:BFC-as-forms} describes how the entire braiding $\lambda$ can in fact be understood as a graded antisymmetric form.
\end{remark}

\begin{remark}  \label{rmk:conj-overview}
Recalling Conjecture~\ref{conj:nondeg-BFC}, we give a brief overview of supporting evidence presented in later sections.  We first motivate the claim that nondegeneracy is necessary for a BFT to describe the excitations of a gapped essentially $d$-dimensional bosonic system.  First of all, the condition $H_0(F) = 0$ simply says that the infinite-space ground state is unique. A degenerate ground state would mean the system has been tuned to a first-order phase transition or that translation symmetry is spontaneously broken. To motivate the necessity of the condition that $\lambda$ is a quasi-isomorphism, we prove in Section~\ref{sec:rdms} that any BFC with $H_0(F) = 0$ is nondegenerate if and only if it satisfies the excitation-detector principle. The excitation-detector principle was introduced in \cite{Wickenden_2026} and conjectured to be a necessary condition for the physical realizability of a gapped phase; this conjecture was established in the context of planon-only fracton orders.

Our examples are also consistent with the conjectured equivalence between nondegeneracy and physical realizability in gapped essentially $d$-dimensional systems. Every Lagrangian Pauli code (\emph{i.e.} those Pauli codes with a unique infinite-space ground state) gives rise to a nondegenerate BFC (see Section~\ref{subsec:Pauli-BFC}). 
Moreover, Section~\ref{subsec:anyon-realization} constructs a
nondegenerate pointlike BFC for every theory of abelian anyons in two
dimensions, including the chiral semion example of
Section~\ref{subsec:example2D}.
 The assertion that the BFT is uniquely determined by the phase is a separate part of Conjecture~\ref{conj:nondeg-BFC}, motivated by the physical interpretation of BFC equivalence discussed below.
 \end{remark}

While our main interest will be in nondegenerate BFCs, the degenerate case is also physically interesting.  In Section~\ref{subsec:example1D} we show that the BFC describing a gapped boundary of the 2D toric code is degenerate.  Then in Section~\ref{sec:dual_complex}, we define a class of possibly degenerate BFCs that we conjecture can occur at $d$-dimensional boundaries of essentially $(d+1)$-dimensional systems. 

Based on the physical discussion of this subsection, the most important data of a BFC should be its fusion complex $F$ and the commutator $-\lambda_0 : F_0 \to F_0^*$.  This expectation is substantiated by Corollary~\ref{v2:cor:complete-commutator}, from which it follows that if two nondegenerate tame BFCs $\lambda, \lambda' : F \to D(F)$ have the same commutator (\emph{i.e.} $\lambda_0 = \lambda'_0$), then the BFCs are equivalent in the sense of Definition~\ref{def:tame-bfc-equivalence}.  We can also ask whether every commutator $\Omega : F_0 \to F^*_0$ can be realized by a BFC.  In the case that $F$ is a complete tame fusion complex (see Definition~\ref{defn:propercomplete}), the same corollary shows that for any commutator $\Omega : F_0 \to F^*_0$, there exists a braiding $\lambda : F \to D(F)$ with $\lambda_0 = -\Omega$.  We note that the fusion complex of a nondegenerate BFC is always complete.

\begin{remark}[Matrix BFCs]  \label{rmk:matrix-bfcs}
A tame BFC can be completely specified as a collection of matrices, and this is often the best way to present examples.  Since each $F_i$ is either a free $R$-module or a free $R_n$-module, we begin by choosing a basis for each $F_i$.  This allows us to identify either $F_i = R^{k_i}$ or $F_i = R^{k_i}_n$.  The boundary map $d_i : F_i \to F_{i-1}$ is thus a $k_{i-1} \times k_i$ matrix with entries in $R$ or $R_n$.

In the $\Z_n$-tame case, each term $D(F)_i = F^*_{-i}$ is again a free $R_n$-module.  The choice of basis on $F_{i}$ also provides us with a corresponding dual basis on $F^*_{i}$, defined by $e^a(\overline{e_b})=\delta_{ab}$, after identifying the $n$-torsion submodule of $\RRZ$ with $R_n$, so we can identify $F^*_{i} = R^{k_{i}}_n$.  Therefore each map $d^{D(F)}_i$ and $\lambda_i$ is also a matrix with entries in $R_n$.  Moreover, as a matrix $d^*_i = d^\dagger_i$, where ``$\dagger$'' denotes transpose followed by the involution applied to each entry.  Therefore $d^{D(F)}_i = (-1)^{i-1} d^\dagger_{-(i-1)}$.  Antisymmetry of the braiding is equivalent to the family of matrix equations $\lambda_i = (-1)^{i+1} \lambda_{-i}^\dagger$.  In particular, we have $\lambda_0 = - \lambda_0^\dagger$, \emph{i.e.} $\lambda_0$ is an anti-Hermitian matrix.

In the $\Z$-tame case, using the choice of basis on each $F_i$, we identify $F^*_{i} = (\RRZ)^{k_{i}}$.  The map $d^*_i : F^*_{i-1} \to F^*_i$ can be viewed as a  $k_i \times k_{i-1}$ matrix with entries in $R$.  A matrix entry equal to $r \in R$ represents multiplication by $r$; that is, it represents the $R$-module map  $\mu_r : \RRZ \to \RRZ$  that sends $x \mapsto r x$.  With this interpretation, the matrix equation $d^*_i = d^\dagger_i$ again holds.  Similarly, $\lambda_i : F_i \to F^*_{-i}$ is a $k_{-i} \times k_i$ matrix with entries in $\RRZ$.  Each entry represents an $R$-module map $R \to \RRZ$, which is the same as an element of $\RRZ$.  We again have $\lambda_i = (-1)^{i+1} \lambda_{-i}^\dagger$.

\end{remark}

\begin{example}[3D toric code]
	A BFC for the  $\Z_p$ toric code in three spatial dimensions is given by 
	\[
		\begin{tikzcd}[column sep = 3em, row sep = 2.5em, ampersand replacement = \&]
			0 \ar r \&
			R_p \oplus 0 \ar r {\varphi \oplus 0} 
			\ar d {-1}
			\&
			R^3_p\oplus R_p \ar r {\chi \oplus \psi^\dagger} 
			\ar  d {\left[\begin{smallmatrix} 0 & 1 \\ 1 & 0 \end{smallmatrix}\right]}
			\&
			R^3_p \oplus R_p^3 \ar r {\psi \oplus - \chi^\dagger} \ar d 
			{\left[\begin{smallmatrix} 0 & 1 \\ -1 & 0 \end{smallmatrix}\right]}
				\&
			R_p \oplus R_p^3 \ar r {0 \oplus \varphi^\dagger} 
			\ar d {\left[\begin{smallmatrix} 0 & 1 \\ 1 & 0 \end{smallmatrix}\right]}
			\&
			0 \oplus R_p \ar r \ar d {1} \&
			0
			\\
			0 \ar r \&
			0 \oplus R_p
			\ar r {-(0 \oplus \varphi)} \&
			R_p \oplus R_p^3 \ar r {\psi^\dagger \oplus -\chi} \&
			R_p^3 \oplus R_p^3 \ar r {-(\chi^\dagger \oplus \psi)} \&
			R_p^3  \oplus R_p \ar r {\varphi^\dagger \oplus 0} \&
			R_p \oplus 0 \ar r \&
			0
			\\
		\end{tikzcd},
	\]
	where
	\[
		\varphi = 
		\begin{pmatrix}
			\Delta_x \\ \Delta_y \\ \Delta_z
		\end{pmatrix},
		\qquad
		\chi 
		=
		\begin{pmatrix}
			0 & - \Delta_z & \Delta_y
			\\
			\Delta_z & 0 & - \Delta_x
			\\
			- \Delta_y & \Delta_x & 0
		\end{pmatrix},
		\qquad
		\psi
		=
		\begin{pmatrix}
			\Delta_x & \Delta_y & \Delta_z
		\end{pmatrix},
	\]
	and $\Delta_\mu = 1 - \mu$. 
	Note that the underlying fusion complex is of the form $F \oplus D(F)$, where
	\[
		F = 0 \to R_p \xrightarrow {\varphi} R_p^3 \xrightarrow {\chi} R_p^3 \xrightarrow {\psi} R_p \to 0.
	\]
	This is in fact the Koszul complex for the module $S = R_p/(\Delta_x, \Delta_y, \Delta_z)$, which as an abelian group
	is just $\Z_p$. 
	Physically, $Q_0 = H_{-1}(F \oplus D(F))= H_{-1}(F)$ describes a single fully mobile point
	excitation. 
	The dual complex $D(F)$ computes the Ext modules of $\overline S$ against $R_p$ (Lemma~\ref{v2:lemma:Rn-coefficients} and Lemma~\ref{v2:lemma:real-coefficients}(b)); since $(\Delta_x,\Delta_y,\Delta_z)$ is a regular sequence, only $\Ext^3_{R_p}(\overline S, R_p)$ is nonvanishing.  
	It can also be seen that $\Ext^3_{R_p}(\overline S, R_p)$ is the only nonzero homology of the dual complex
	because, up to signs and antipodes, $F$ is manifestly symmetric under duality: $\chi^\dagger = - \bar \chi$, $
	\varphi^\dagger = \bar \psi$. 

	Our duality conventions put the nonzero homology of $D(F)$ in degree $-2$, and thus the module of
	loop excitations is $Q_1 = H_{-2}(F \oplus D(F)) = H_{-2}(D(F)) = R_p / (\bar \Delta_x, \bar \Delta_y, \bar
	\Delta_z)$, and the theory contains a single fully mobile $\Z_p$ loop excitation, as expected.
	The braiding map $\lambda : F \oplus D(F) \to D(F \oplus D(F))$ induces the maps on
	homology:
	\[
		\beta_{-1} : Q_0 \to \Ext^3_{R_p}(\overline{Q_1}, R_p), \qquad \beta_{-2} : Q_1 \to \Ext^3_{R_p}(\overline{Q_0}, R_p).
	\]
	The displayed matrices give $\lambda(x,u)=(u,-\can_F(x))$, so $\beta_{-1}=-H_{-1}(\can_F)$ and $\beta_{-2}=\id_{Q_1}$ after identifying the target homology modules with the Ext modules above. Here $\can_{F,-1}=-\can_{F_{-1}}$, so the two minus signs in the expression for $\beta_{-1}$ cancel.
	By the Rees theorem, using the ordered regular sequence $(\Delta_x,\Delta_y,\Delta_z)$ and the displayed Koszul bases, there is moreover an identification
	\[
		\rho : \Ext^3_{R_p}(\overline{Q_1}, R_p)
		\xrightarrow{\;\cong\;}
		\Hom_{R_p}(\overline{Q_1}, R_p/(\Delta_x, \Delta_y, \Delta_z)).
	\]
	Thus $\rho \circ \beta_{-1}$ assigns to each point excitation a linear functional on $\overline{Q_1}$.
	Evaluating this functional on a loop excitation and using the coefficient embedding
	$a \mapsto a/p+\Z$ gives the pairing
	\[
		b : Q_0 \times Q_1 \to \Q/\Z, \qquad
		b(e,m)=\frac{\rho(\beta_{-1}(e))(\overline m)}{p}+\Z,
	\]
	where the right-hand side is understood to carry a trivial action of translation.
	For $e=[1]\in Q_0$ and the oriented loop generator $m=[1]\in Q_1$, the degree $-1$ vertical matrix gives
	$\rho(\beta_{-1}(e))(\overline m)=1$, and hence $b(e,m)=1/p$.
	This pairing is nondegenerate because $\beta_{-1}$ descends from an isomorphism, and
	expresses the 3D toric code point-loop braiding rule $b(e,m)=1/p$.
\end{example}

Now we address the question of when two BFCs should be considered equivalent, in the sense that they describe the same universal fusion, mobility and statistics data.  Quasi-isomorphisms of fusion complexes should induce equivalences of BFCs; that is, if $\lambda : F \to D(F)$ is a BFC and $\alpha : G \to F$ is a quasi-isomorphism of tame fusion complexes, then we want $D(\alpha) \lambda \alpha : G \to D(G)$ to be an equivalent BFC, because the underlying fusion complexes are equivalent.   It is straightforward to check that such conjugation by a quasi-isomorphism preserves antisymmetry of $\lambda$ and constant-alternation in degree zero (Lemma~\ref{v2:lemma:pullback-braiding}).  Moreover, conjugation also preserves nondegeneracy of a BFC; if $\lambda$ is a quasi-isomorphism, then so is $D(\alpha) \lambda \alpha$, because $D(\alpha)$ is a quasi-isomorphism.

\begin{defn} \label{def:quasi-isometry}
If $\lambda : F \to D(F)$ and $\mu : G \to D(G)$ are tame BFCs, and $\alpha : G \to F$ is a quasi-isomorphism such that $\mu = D(\alpha) \lambda \alpha$, then $\alpha$ is a {\bf quasi-isometry} from $(G,\mu)$ to $(F,\lambda)$, and $\lambda$ and $\mu$ are {\bf quasi-isometric}. \end{defn}

While conjugating by $\alpha : G \to F$ maps a BFC over $F$ to one over $G$, it is important to note that this map only goes in one direction.  That is, $\alpha$ cannot in general be used to map a BFC over $G$ to one over $F$.  Looking ahead, it is helpful to discuss an example that illustrates this point.  The double semion model in $d=2$ can be realized by a $\Z_4$ Pauli code \cite{Ellison_2022}, and thus by a $\Z_4$-tame BFC over a complex $G_0$.  The fusion module is $S = \Z_2^2$. Choose a free $R_4$-resolution $G$ of $S$ in degrees $\geq-1$; Lemma~\ref{v2:lem:truncated-gamma}(b) gives a quasi-isomorphism $G\to G_0$, and pulling back the braiding gives $\lambda:G\to D(G)$. We choose a free resolution $F$ of $S$ by $R_2$-modules.  Viewing $F$ as an $R_4$-complex via $R_4\to R_2$, the same lemma gives a quasi-isomorphism $\alpha : G \to F$, but, on physical grounds, a chain map realizing the inverse quasi-isomorphism cannot  exist.  If such a chain map existed, we would be able to find a $\Z_2$-tame BFC description of the double semion model.  But this is impossible because such a BFC cannot produce semions; only bosonic or fermionic self-statistics are possible since the commutator $\omega$ takes values in $\Z_2$.

We now argue that an apparently different kind of equivalence of BFCs results from deforming $\lambda$ by certain chain homotopies.  It will actually turn out that such deformations are superfluous, in the sense that they do not change the equivalence relation on BFCs, because they can be modeled by conjugation through a suitable zigzag of quasi-isomorphisms (Proposition~\ref{v2:prop:homotopy-conjugation}).  Nonetheless, we find it useful to discuss them to provide some physical insight, and to make some of the available moves between equivalent BFCs more transparent.

Recall that a chain homotopy between two chain maps $\lambda, \lambda' : F \to D(F)$ is a collection of maps $h_i : F_i \to D(F)_{i+1} = F^*_{-(i+1)}$ such that
\begin{equation}
\lambda'_i - \lambda_i = d^{D(F)}_{i+1}  h_i + h_{i-1}  d_i = (-1)^i d^*_{-i} h_i + h_{i-1}  d_i \text{.} \label{eqn:chain-htpy}
\end{equation}
Schematically, we denote a chain homotopy by $h$ and write $\lambda' - \lambda = d^{D(F)} h + h d$. Deforming $\lambda$ by a chain homotopy means passing from $\lambda$ to $\lambda'$.  However, if $\lambda$ is a BFC, then $\lambda'$ might not be; the right-hand side of Equation~\ref{eqn:chain-htpy} need not be an antisymmetric chain map.  Therefore, for such deformations of $\lambda$ to make sense, we need to put some conditions on the chain homotopy.

\begin{defn}\label{def:antisymmetric-homotopy}
A chain homotopy $h$ is symmetric if $h_i =  h^*_{-(i+1)} \circ \can_{F_i}$ for all $i$, and antisymmetric if $h_i = - h^*_{-(i+1)} \circ \can_{F_i}$ for all $i$.
\end{defn}

\begin{proposition}
Let $h$ be an antisymmetric homotopy between chain maps $\lambda, \lambda' : F \to D(F)$, and let $\delta \lambda  = \lambda' - \lambda = d^{D(F)} h + h d$.  Then $\delta \lambda$ is an antisymmetric chain map that is constant-alternating in degree zero.
\end{proposition}

\begin{proof}
Showing antisymmetry is a straightforward computation (see Lemma~\ref{v2:lemma:constant-alternating}).  We observe that $\delta \lambda_0 = h_{-1}
d_0 - (h_{-1} d_0)^* \circ \can_{F_0}$, from which it follows that $\delta\lambda$ is constant-alternating in degree zero.
\end{proof}

\begin{figure}
	\includegraphics[width=0.6\textwidth]{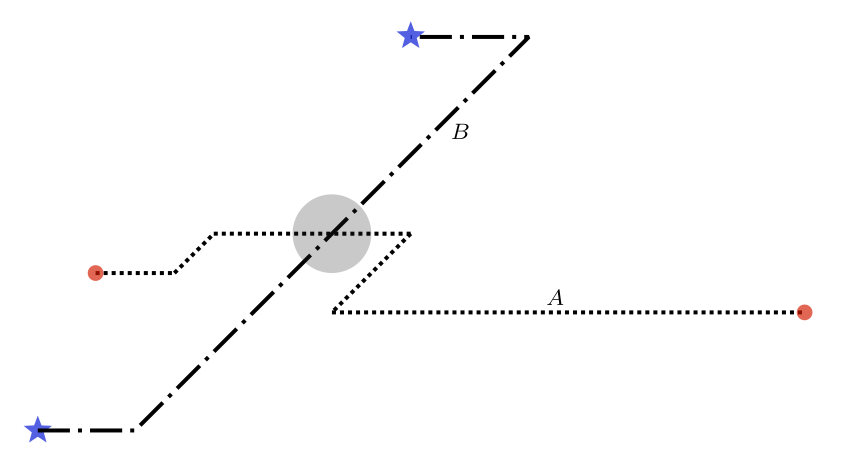}
	\caption{Two long but finite string operators $A$ (dotted line) and $B$ (dash-dot line) in a
		theory of 2d anyons, which create anyons at their endpoints (blue stars and red dots).  Each
		string operator is a product of local operators of small support, and the commutator of the
		string operators is determined by commutators of these small-support operators, within a region
		near the intersection point.}
	\label{fig:string-commutator}
\end{figure}

Thus we see that it makes sense mathematically to demand that BFCs are equivalent when their chain maps are antisymmetrically homotopic.  It  remains to explain why this is desirable from a physical point of view.  For concreteness, we work in $d=2$ and consider the commutator of two long but finite string operators $A,B \in F_0$, which intersect transversely and create mobile excitations at their endpoints, as illustrated in Fig.~\ref{fig:string-commutator}.  Mutual statistics in $d=2$ is completely determined by commutators of such string operators, far away from any endpoints where excitations are created (red dots and blue stars in Fig.~\ref{fig:string-commutator}).  More generally, based on the discussion at the beginning of this subsection, all statistical phases are determined by commutators of movers far away from locations where excitations are created.

Now suppose we deform the commutator on movers by $\lambda_0 \to \lambda_0 + \delta \lambda_0 =
\lambda_0 + h_{-1} d_0 - (h_{-1} d_0)^* \circ \can_{F_0}$, for some antisymmetric chain homotopy $h$.  The map $h_{-1} d_0$ acts on $A \in F_0$ by first obtaining the excitations $d_0(A)$ created by $A$, and then acting on these excitations with a map $h_{-1}$ that sends excitations (elements of $F_{-1}$) to functionals on movers (elements of $F^*_0$).  In the present example, $d_0(A)$ has support at the two endpoints of the $A$-string.  Because the modules are finitely generated and the pairings take values in finite-support Laurent polynomials, $h_{-1}$ has finite spatial range.  Suppose each string stays farther than this range from the endpoints of the other string.  Then the constant coefficient of $(h_{-1}d_0)(A)(\overline B)$ vanishes.  The same is true of $(h_{-1}d_0)(B)(\overline A)$ and its involution, so the constant coefficient of $\delta\lambda_0(B)(\overline A)$ vanishes.  Thus the physical commutator $\omega(A,B)$ is unaffected, since $\delta\Omega(A,B)=-\delta\lambda_0(B)(\overline A)$.  The full Laurent polynomial $\delta\lambda_0(B)(\overline A)$ need not vanish: its other coefficients compare one string with translates of the other.  Therefore we conclude that deformation by an antisymmetric homotopy is a kind of local modification of the commutator that leaves statistics unaffected.

The two types of BFC equivalence discussed above are easily seen to generate the following equivalence relation:
\begin{defn}[Equivalence of tame BFCs]  \label{def:tame-bfc-equivalence}
Let $\lambda : F \to D(F)$ and $\mu : G \to D(G)$ be two tame BFCs.  If $\lambda$ and $\mu$ are related by a zigzag of quasi-isometries up to antisymmetric homotopy, then the two BFCs are equivalent.  In more detail, this means that there exists a diagram of the form
\begin{equation}
\begin{tikzcd}
F \arrow[r, "f_1"] \arrow[d, "\lambda"] & C^1 \arrow[d, "\tau_1"] & C^2 \arrow[l, "s_1"] \arrow[d, "\tau_2"] \arrow[r, "f_2"] & \cdots & \arrow[l, "s_k"] G \arrow[d, "\mu"] \\
D(F) & D(C^1) \arrow[l, "D(f_1)"]  \arrow[r, "D(s_1)"] &    D(C^2)  & \arrow[l, "D(f_2)"]  \cdots \arrow[r, "D(s_k)"]  & D(G)
\end{tikzcd} \text{,} \label{eqn:bfc-equiv-diagram}
\end{equation}
where each $\tau_i : C^i \to D(C^i)$ is a tame BFC, each $f_i$ and $s_i$ is a quasi-isomorphism, and each square commutes up to antisymmetric homotopy.  We describe such a diagram as a {\bf zigzag of homotopy quasi-isometries}.  Equivalence classes of BFCs are referred to as {\bf braided fusion theories} (BFTs). 
\end{defn}

We recover the equivalences from our motivating discussion above as special cases of Definition~\ref{def:tame-bfc-equivalence}.  If $\lambda$ and $\mu$ are equivalent through a single square, which commutes exactly (not only up to homotopy), then we have conjugation by a quasi-isomorphism.  If we insist that each square in Equation~\ref{eqn:bfc-equiv-diagram} commute exactly, then Definition~\ref{def:tame-bfc-equivalence} reduces to the equivalence relation generated by quasi-isometries.  On the other hand, if we consider a diagram again with a single square, but where the horizontal chain maps are identities, then we recover equivalence of antisymmetrically homotopic BFCs over the same fusion complex.

So far we have defined tame BFCs and their equivalence.  This is not fully satisfactory because it forces us to work over certain representatives of a fusion theory, namely the tame fusion complexes.  We address this issue in Appendix~\ref{v2:app:tame-BFC} for the case of odd fusion order, where we give definitions of BFC and BFC equivalence that are more natural from the perspective of the derived category.  There we show that two tame BFCs are equivalent under the more abstract definition if and only if they are equivalent in the sense of Definition~\ref{def:tame-bfc-equivalence}, so there is only a single notion of equivalence (Theorem~\ref{v2:thm:odd-tame-derived-BFC} and Lemma~\ref{v2:lemma:isometric-equivalent}).  Moreover, every braided fusion theory under the abstract definition has a tame BFC representative.
There are considerable subtleties at even fusion order, closely related to well-known difficulties in the abstract theory of quadratic forms at even characteristic (see Remark~\ref{rem:2rem}).  As a result, it is left to future work to give an abstract definition of BFC and BFC equivalence that encompasses even fusion order.

At odd fusion order, we can upgrade Remark~\ref{rmk:alternating-superfluous} into the following result.  This tells us that the requirement that the braiding be constant-alternating in degree zero is superfluous, in the sense that it could be removed from the definition of BFC with no effect on the BFC equivalence classes we obtain.
\begin{proposition}
Define a tame pseudo-BFC to be a pair $(F, \lambda)$, where $F$ is a tame fusion complex, and $\lambda : F \to D(F)$ is an antisymmetric chain map.  Equivalence of tame pseudo-BFCs is defined as in Definition~\ref{def:tame-bfc-equivalence}.  Restricting to the case of odd fusion order, there is a natural bijection between the sets of tame BFC and tame pseudo-BFC equivalence classes.
\end{proposition}

\begin{proof}
Every tame BFC is a tame pseudo-BFC, and every equivalence between
BFCs is also an equivalence between pseudo-BFCs. Thus inclusion induces
a map $\Phi$ on equivalence classes. We show that $\Phi$ is surjective
and injective.

First, let $(F,\lambda)$ be a tame pseudo-BFC of odd fusion
order. Multiplication by $2$ on $F$ is a quasi-isomorphism,
since it induces an automorphism of each homology module. We have $D(2\id_F)\lambda(2\id_{F})=4\lambda$, so $(F, \lambda)$ and $(F, 4 \lambda)$ are equivalent pseudo-BFCs.  Moreover, $(F,4\lambda)$ is a BFC. Indeed, antisymmetry of
$\lambda$ gives $4 \lambda = \gamma - D(\gamma) \circ \can_F$, where $\gamma = 2 \lambda$.  For $x \in F_0$, 
we thus have $4\lambda_0(x)(\overline x) = \gamma_0(x)(\overline x)-\overline{\gamma_0(x)(\overline x)}$, whose constant coefficient
vanishes, so $4\lambda$ is constant-alternating in degree zero.  

For injectivity, suppose that two tame BFCs $(F,\lambda)$ and
$(G,\mu)$ are equivalent as pseudo-BFCs. Choose a zigzag of
homotopy quasi-isometries between them through tame pseudo-BFCs.
Multiply every braiding in this zigzag by $4$. By the preceding
calculation, every resulting object is a BFC. Keep all the
quasi-isomorphisms unchanged and multiply every antisymmetric homotopy
by $4$; these are still antisymmetric homotopies.  Thus the scaled zigzag is an equivalence of BFCs between
$(F,4\lambda)$ and $(G,4\mu)$.
Finally, multiplication by $2$ gives quasi-isometries of BFCs $(F,4\lambda) \to (F,\lambda)$ and $(G,4\mu) \to (G,\mu)$.  Appending these to the scaled zigzag proves that the original BFCs are
equivalent as BFCs, and hence $\Phi$ is injective.
\end{proof}

\begin{remark}[Rational and real BFCs] \label{rmk:rational-real-BFCs}
In the definition of BFC, the commutator $\omega$ takes values in $\R / \Z$, as appropriate in quantum mechanics.  However, because we only consider systems with $\Z$-torsion fusion modules, statistical phase factors should lie in $\Q / \Z \subset \R / \Z$, and therefore we believe it is unlikely that restricting the commutator to take values in $\Q / \Z$ would change the set of quantum phases that can be described by BFCs.  Indeed, the commutator of a $\Z_n$ Pauli code takes values in $\Z_n \subset \Q / \Z$, and examples discussed in this paper that are beyond Pauli codes also have the same property.  This leads us to contemplate the notion of {\bf rational BFCs}, as opposed to the {\bf real BFCs} we have already defined, where we repeat the definitions of this Section, but replace $\R / \Z$ with $\Q / \Z$, and accordingly we also replace $\RRZ$ with $\RQZ$.

It turns out that it is a matter of taste whether we work with rational or real BFCs, because there is a natural bijection between the sets of rational and real tame BFC equivalence classes, induced by the inclusion $\RQZ \hookrightarrow \RRZ$. This is shown in Proposition~\ref{v2:thm:real-rational-BFC-equivalence}.  We prefer to work with rational BFCs, because these encompass all our explicit examples, and because we find this aesthetically more natural given the restriction of $\Z$-torsion homology.  Therefore, {\bf after Section~\ref{sec:first} all BFCs will be rational in the main text, unless otherwise stated}. In subsequent sections of the main text, unless stated otherwise, the notation $M^*$ or $D(M)$ means $D_{\RQZ}(M)$, and the dual complex $D(F)$ means $D_{\RQZ}(F)$, with the same scalar action and differential signs as above.  For $R_n$-modules this makes no change to the dual: both $\RQZ$ and $\RRZ$ have the same $n$-torsion submodule, canonically identified with $R_n$.\end{remark}

As mentioned above, we do not need to include deformation by antisymmetric homotopies in the definition of equivalence of tame BFCs.  More precisely, deformation by antisymmetric homotopies does not enlarge the equivalence relation generated by quasi-isometries. Proposition~\ref{v2:prop:homotopy-conjugation} proves this by constructing a zigzag of quasi-isometries joining tame BFCs whose braidings are related by an antisymmetric homotopy.


\section{Braided fusion complexes from Pauli codes}
\label{sec:Pauli}

\subsection{Review of Pauli codes}
\label{subsec:Pauli-review}

Pauli codes provide a large class of examples of gapped quantum matter, including realizations of many fracton phases.  Indeed, Chamon discovered fractons by constructing and studying a Pauli code \cite{Chamon_2005}.  We first review Pauli codes, following the approach of Ruba and Yang \cite{Ruba_2024, Ruba_2025}, which in turn builds on earlier work of Haah \cite{Haah_2013}. Then in Section~\ref{subsec:Pauli-BFC} we extract a BFC from a Pauli code.  This further motivates the definitions of fusion complex and BFC, bolsters our physical interpretation, and provides a large class of examples.

We place $k$ $n$-dimensional qudits on each site of the $d$-dimensional hypercubic lattice, and assume translation symmetry ${\cal T} = \Z^d$. 
On each lattice site $\br$ we have Pauli operators $Z_{\br a}$ and $X_{\br a}$, where $a = 1,\dots,k$, satisfying the relations $Z^n_{\br a} = X^n_{\br a} = 1$ and
\begin{equation}
Z_{\br a} X_{\br' b} = \exp \Big( \frac{2 \pi i}{n} \delta_{a b} \delta_{\br \br'} \Big)  X_{\br' b} Z_{\br a} \text{.} \label{eqn:clock-shift}
\end{equation}
These are sometimes also referred to as clock and shift operators.  The Pauli group ${\cal P}$ consists of all finite products of $X$ and $Z$ Pauli operators multiplied by arbitrary phase factors.  We denote by ${\rm U}(1) \subset {\cal P}$ the subgroup consisting of phase factors times the identity operator.  For any $A,B \in {\cal P}$, the commutator $[A,B] = A B A^{-1} B^{-1} = e^{2 \pi i m / n}$ for some $m =0,\dots,n-1$, \emph{i.e.} the commutator takes values in the subgroup $\Z_n \subset {\rm U}(1)$ consisting of $n$th roots of unity.

The {\bf projective Pauli group} is defined as the quotient $P = {\cal P} / {\rm U}(1)$.  The group $P$ is an abelian group whose elements are projective operators as discussed in Section~\ref{subsec:fc1}.  We use additive notation for $P$, even though the addition operation represents operator multiplication.  The action of translation symmetry makes $P$ into an $R_n$-module, and it will be the module of movers $F_0 = P$ in a fusion complex.  Because the commutator takes values in $\Z_n$, it can be expressed as a spatially local $\Z_n$-valued commutator $\omega : P \times P \to \Z_n$ as in Section~\ref{subsec:bfc1}, from which we can also construct the $R_n$-valued commutator $\Omega : P \times P \to R_n$.

We identify $P$ with the free module $P = R_n^{2k}$, with basis $e_1, \dots, e_{2k}$, where $e_1, \dots, e_k$ represent $Z_{0 1}, \dots, Z_{0 k}$, and $e_{k+1}, \dots, e_{2k}$ represent $X_{0 1}, \dots, X_{0 k}$.  It is convenient to view elements of $P$ as $2k$-component column vectors with entries in $R_n$, and then $e_i$ is the vector with $1$ in the $i$th position and zeros elsewhere. With this convention, the commutator can be expressed by
\begin{equation}
\Omega(p, p') = p^\dagger \begin{pmatrix}
0 & \mathbbm{1} \\
- \mathbbm{1} & 0
\end{pmatrix} p' \text{,}
\end{equation}
where $\mathbbm{1}$ is the $k \times k$ identity matrix, and where the row vector $p^\dagger$ is the transpose of $p$ with the involution applied to each entry.  This is nothing but the standard symplectic form on $R_n^{2k}$. Here the Laurent-valued form $\Omega$ is sesquilinear and constant-alternating; its constant component is the alternating $\Z_n$-bilinear commutator $\omega$.  Throughout this section, we use $M^*=D_{R_n}(M)=\Hom_{R_n}(\overline{M},R_n)$ with ordinary scalar action, and $f^*$ for the induced dual map.  The embedding $a\bmod n\mapsto a/n+\Z$ identifies this dual with $D_{\RQZ}(M)$.  For complexes we write $D(F)=D_{R_n}(F)$ and $D(f)$, retaining the degree signs of Definition~\ref{def:dual-complex}.  We note that the associated map $\Omega : P \to P^*$ is an isomorphism, with $\Omega(p')(\bar p)=\Omega(p,p')$.

Informally, a Pauli code is a Hamiltonian that is a sum of commuting terms, each of which is a product of Pauli operators.  A ground state is a common unit-eigenvalue eigenstate of a set of stabilizers, which consists of the terms in the Hamiltonian and  products thereof.  The essential features of such a system are preserved upon passing to the projective Pauli group.  First we introduce some standard terminology:
\begin{defn}
Given a submodule $L \subset P$, the {\bf orthogonal} of $L$ is
\begin{equation}
L^\perp = \{ p \in P \, \vert \, \Omega(p, \ell) = 0 \text{ for all } \ell \in L \} \text{.} \nonumber
\end{equation}
We say that $L$ is {\bf isotropic} if $L \subset L^\perp$, and $L$ is {\bf Lagrangian} if $L = L^\perp$.
\end{defn}
\noindent The orthogonal $L^\perp$ is clearly a submodule.  
\begin{defn}[Ruba and Yang]
A translation-invariant {\bf projective Pauli code} consists of a free $R_n$-module $P = R_n^{2k}$ equipped with the standard symplectic form, and an isotropic submodule $L \subset P$.  A projective Pauli code is Lagrangian if $L$ is a Lagrangian submodule.
\end{defn}
Each element of $L$ corresponds to the image of a stabilizer in the projective Pauli group, so giving $L \subset P$ is an efficient way to specify the stabilizers.  This does not completely determine the stabilizers as elements of ${\cal P}$, \emph{i.e.} we still need to make a phase choice for each element of $L$, hence the appearance of the qualifier ``projective.''  One way to do this is to specify a Pauli code Hamiltonian of the form $H = - \sum_i (\hat{L}_i + \hat{L}_i^\dagger)$, where the stabilizers with phase choice are $\hat{L}_i \in {\cal P}$.  One also needs to assume that the $\hat{L}_i$ are mutually commuting, and are frustration-free, which means that there exists a simultaneous unit-eigenvalue eigenstate of all the $\hat{L}_i$.  Given such a Pauli code Hamiltonian, there is a natural projective Pauli code given by forgetting phase choices, \emph{i.e.} by sending $\hat{L}_i \mapsto L_i$ using the quotient map ${\cal P} \to P$ and taking $L \subset P$ as the submodule generated by the $L_i$.  Henceforth, all Pauli codes are understood to be projective Pauli codes, and we omit the qualifier ``projective.''

Translation-invariance is built into the definition via the $R_n$-module structure, and we only consider translation-invariant Pauli codes.  The Lagrangian condition is equivalent to the property that a Pauli code has a unique infinite-system ground state for a fixed consistent phase choice \cite[Appendix~A.3]{Ruba_2025}; in this context, sometimes this property is referred to as ``local topological order.''  We do not use that term in this paper, because ``topological order'' has been used with many different meanings.  We note that the data required to specify a Pauli code is the tuple $(n,k,L)$.

\subsection{Pauli fusion complexes and BFCs}
\label{subsec:Pauli-BFC}

Every Pauli code gives rise to a BFC, and, moreover, from this BFC we can reconstruct the Pauli code.  In fact, the underlying fusion complex plays a more important role than the BFC.  This is because we will be able to construct a special kind of fusion complex from a Pauli code, dubbed a Pauli fusion complex, which contains all the information needed to reconstruct the Pauli code from which we began.  Given a Pauli fusion complex, the BFC is given canonically and contains no additional data.

 To begin, we choose a free resolution over $R_n$ of the module $L$:
\begin{equation}
\begin{tikzcd}
\cdots \arrow[r] & F_3 \arrow[r, "d_3"] & F_2 \arrow[r, "d_2"] & F_1 \arrow[r, "\rho"] & L \arrow[r] & 0
\end{tikzcd} \text{,}
\end{equation}
where each $F_i$ is a finitely generated free $R_n$-module.
The module $F_1$ will be the module of stabilizers in the fusion complex; the images under $\rho$ of a basis for $F_1$ generate $L$. Defining $F_0 = P$, and letting $i : L \hookrightarrow F_0$ be the inclusion, we obtain a free resolution of the quotient $P/L$ by
\begin{equation}
\begin{tikzcd}
\cdots \arrow[r] & F_3 \arrow[r, "d_3"] & F_2 \arrow[r, "d_2"] & F_1 \arrow[r, "d_1"] & F_0 \arrow[r] & P / L \arrow[r] & 0
\end{tikzcd} \text{,}
\end{equation}
where we have defined $d_1 = i \rho$.  Here, exactness at $F_1$ holds because $\operatorname{ker} d_1 = \operatorname{ker} \rho$.

\begin{lemma}  \label{lem:PL-tf}
In a Lagrangian Pauli code, $P/L$ is a torsion-free $R_n$-module.
\end{lemma}

\begin{proof}
Take $p \in P$ and suppose $r p \in L$ for some regular $r \in R_n$.  Then $r p \in L^\perp$, so for all $\ell \in L$ we have $0 = \Omega(r p, \ell) = \bar{r} \Omega(p, \ell)$, which implies $\Omega(p, \ell) = 0$.  Thus $p \in L^\perp$ and so $p\in L$.
\end{proof}

Deleting $P/L$, viewing its free resolution as a complex with zeros in negative degree, and applying duality as in Definition~\ref{def:dual-complex} gives
\begin{equation}
\begin{tikzcd}
0 \arrow[r] & F^*_0 \arrow[r, "-d^*_1"] & F^*_1 \arrow[r, "d^*_2"] & F^*_2 \arrow[r, "-d^*_3"] & \cdots
\end{tikzcd} \text{.}
\end{equation}
We stitch this together with the free resolution of $P/L$ by defining $d_0 : F_0 \to F^*_1$ to be $d_0 =  d^*_1  \Omega = \rho^* i^* \Omega$.
We thus have the complex
 \begin{equation}
 \begin{tikzcd}
\cdots \arrow[r] & F_3 \arrow[r, "d_3"] & F_2 \arrow[r, "d_2"] & F_1 \arrow[r, "d_1"] & F_0 \arrow[r, "d_0"] & F^*_1 \arrow[r, "d^*_2"] & F^*_2 \arrow[r, "-d^*_3"] & \cdots
\end{tikzcd} \text{.}
\end{equation}
This is indeed a complex because $d_0 d_1 = \rho^* i^* \Omega i \rho = \rho^* ( i^* \Omega i) \rho = 0$ and $d^*_2 d_0 = d^*_2 d^*_1 \Omega
= (d_1 d_2)^* \Omega = 0$.  Here, $ i^* \Omega i=0$ because $L$ is isotropic.  We  define $F_{-i} = F^*_i$ for $i > 0$ and denote the complex by $F$. 

We already know that $F$ is exact in positive degree.  We have $\ker d_0 = \ker i^* \Omega = L^\perp$ (using injectivity of $\rho^*$), and $\im d_1 = L$, so exactness at $F_0$ is the same as $L = L^\perp$.  Moving on to negative degree, we have $H_{-i}(F) \cong \ExtRn^{i} (\overline{P/ L}, R_n)$ for $i > 0$.  This is immediate for $i > 1$, and holds when $i = 1$ because $\im d_0 = \im d_1^*$.  It follows from Proposition~7 of \cite{Ruba_2025} (see also Proposition~\ref{v2:prop:resolution-dual-ext}) that $H_{-i}(F) = 0$ for $i > d$, and therefore $F$ is a fusion complex: its homology modules are finitely generated and annihilated by $n$.  
Our discussion so far is summarized by the following definition and proposition:

\begin{defn}
A {\bf Pauli fusion complex} is a $\Z_n$-tame fusion complex $F$ such that the following conditions hold:
\begin{itemize}
\item $F_0 = R_n^{2k}$.  We denote the standard symplectic form on $F_0$ by $\Omega : F_0 \to F^*_0$.
\item  $F_{-i} = F^*_i$ for all $i > 0$.
\item  $d_0 = d^*_1 \Omega$, and $d_{-i} = (-1)^{i+1} d^*_{i+1}$ for all $i > 0$.
\end{itemize}
We say a Pauli fusion complex is Lagrangian if $H_0(F) = 0$.
\end{defn}

The fusion complex constructed above from a Pauli code is clearly a Pauli fusion complex.  From any Pauli fusion complex we obtain an associated Pauli code given by $P = F_0$ and $L = \operatorname{im} d_1$.  Moreover, if we start with a Pauli code, construct a Pauli fusion complex as above, and then extract the associated Pauli code, we recover the original Pauli code.  We can thus view a Pauli fusion complex as an alternative means of presenting a Pauli code, which of course involves additional data beyond the tuple $(n,k, L)$, given the need to choose a free resolution of $L$.  This goes along with the fact that in general there are many different Pauli fusion complexes with the same associated Pauli code.

In a Pauli fusion complex, the physical interpretation of elements of $F_i$ becomes simpler and more concrete (see Section~\ref{subsec:fc1}).  The movers (elements of $F_0$) are realized concretely as projective Pauli operators, referred to as Paulis for short.  Elements of $F_1$ represent Pauli stabilizers; more precisely, they are labels that become projective stabilizers upon mapping to $F_0$, but we simply refer to them as stabilizers when this does not cause confusion.  Elements of $F_2$ are relations among stabilizers, and so on.  Aside from the concreteness of Pauli operators, the main simplification occurs in negative degree.  For instance $F_{-1} = F^*_1$, so the elements of $F_{-1}$ (configurations) are literally spatially local functionals on stabilizers.  Elements of $\ker d_{-1}$ (excitations) are those functionals that vanish on relations and are thus states of the system.  Moreover these are excitations of bounded spatial support.  Elements of $F_{-2} = F^*_2$ (violations) are functionals on relations, and so on for elements of $F_{-3}$ and beyond (higher violations).

\begin{proposition}
A Pauli fusion complex has $H_0(F) = 0$ if and only if its associated Pauli code is Lagrangian. In fact,
$ H_0(F) = L^\perp/L$.
\end{proposition}

\begin{proof}
Since $d_0=d_1^*\Omega$ and $\operatorname{im}d_1=L$,
we have $\ker d_0=L^\perp$. Thus
$H_0(F)=\ker d_0/\operatorname{im}d_1=L^\perp/L$,
which vanishes precisely when the associated code is Lagrangian.
\end{proof}

\begin{proposition}  \label{prop:Pauli-fc-associated-code}
If two Pauli fusion complexes have the same associated Pauli code, the complexes are quasi-isomorphic; in fact, they are chain homotopy equivalent.
\end{proposition}

The proof is given in Proposition~\ref{v2:prop:Pauli-fc-associated-code}
of Appendix~\ref{v2:app:tame-BFC}.
Now we can write down a BFC describing a Pauli code:
\begin{defn}
A {\bf Pauli BFC} $(F, \lambda)$ is a $\Z_n$-tame BFC where $F$ is a Pauli fusion complex, and the braiding $\lambda$ is the chain map with components $\lambda_0 = -\Omega$, $\lambda_i = (-1)^{i+1} \can_{F_i}$ for $i > 0$, and $\lambda_i = \operatorname{id}$ for $i < 0$.  That is, we have the diagram
\begin{equation}
 \begin{tikzcd}
\cdots \arrow[r] & F_3 \arrow[r, "d_3"] \arrow[d, "\can_{F_3}"]& F_2 \arrow[r, "d_2"] \arrow[d, "-\can_{F_2}"] & F_1 \arrow[r, "d_1"] \arrow[d, "\can_{F_1}"] & F_0 \arrow[r, "d^*_1 \Omega"] \arrow[d, "-\Omega"] & F^*_1 \arrow[r, "d^*_2"] \arrow[d, "\operatorname{id}"] & F^*_2 \arrow[r, "-d^*_3"] \arrow[d, "\operatorname{id}"] & F^*_3 \arrow[r]  \arrow[d, "\operatorname{id}"]  &\cdots \\
\cdots \arrow[r] & F^{**}_3 \arrow[r, "-d^{**}_3"] & F^{**}_2 \arrow[r, "-d^{**}_2"] & F^{**}_1 \arrow[r, "d^*_0"] & F^*_0 \arrow[r, "-d^*_1"] & F^*_1 \arrow[r, "d^*_2"] & F^*_2 \arrow[r, "-d^*_3"] & F^*_3 \arrow[r] & \cdots
\end{tikzcd} \text{.} \nonumber
\end{equation}
\end{defn}
\noindent Here, $\lambda$ is a chain isomorphism, so if $F$ is Lagrangian, the Pauli BFC is nondegenerate.  

\begin{remark}
Given a Pauli fusion complex, there is a canonical Pauli BFC that involves no additional data.  This is not the case for a general fusion complex; physically, $F$  contains information about fusion and mobility, while $\lambda$ encodes statistics of excitations.  A Pauli fusion complex already contains this statistical information, because $F_0=R_n^{2k}$ comes with a commutator, namely the standard symplectic form $\Omega:F_0\to F_0^*$.  This is why we can recover a Pauli code from a Pauli fusion complex; in contrast, a general fusion complex does not provide enough information to reconstruct a physical system realizing it.

Put another way, in a BFC there is a clean separation between fusion and braiding information, while this is not the case in Pauli codes.  This is to be expected, because a Pauli code is a physical system that has both fusion and braiding properties.  Translating Definition~22 of \cite{Ruba_2024} into our language, Ruba and Yang's charge modules are $Q_{\mathrm{RY}}^i = H_{-(i+1)}(D(F))$ ($i \geq 0$).  In a general BFC, the fusion-complex data alone does not supply
a natural identification between
$Q_i=H_{-(i+1)}(F)$ and
$H_{-(i+1)}(D(F))$.
The braiding induces a map between these modules, which is an
isomorphism for a nondegenerate braiding.  In a Pauli BFC, this isomorphism is the identity, and there is no distinction between $Q_i$ and $Q^i_{\mathrm{RY}}$.  However, in general, the elements of  $H_{-(i+1)}(D(F))$ should not be interpreted
as excitation types.
\end{remark}

Ruba and Yang established the following Krull dimension bounds for the fusion modules of Pauli codes \cite[Proposition~25]{Ruba_2024}, which we state here in terms of Pauli fusion complexes:
\begin{proposition}
	\label{prop:pauli-krull-dim}
Let $F$ be a Pauli fusion complex.  Its homology modules satisfy the {\bf weak dimension bound}
\begin{equation}
\operatorname{dim} H_{-i}(F) \leq d-i , \qquad i = 0,\dots,d \text{.}
\end{equation}
If moreover $F$ is Lagrangian, then $H_{-d}(F) = 0$, and we have the {\bf strong dimension bound}
\begin{equation}
\operatorname{dim} H_{-i}(F) \leq d-i-1, \qquad i = 1,\dots,d-1 \text{.}
\end{equation}
\end{proposition}
\begin{proof}
For $i>0$, the charge module $Q_{\mathrm{RY}}^{i-1}$ of
\cite{Ruba_2024} is isomorphic to $H_{-i}(F)$.
The weak bound therefore follows from Proposition~25(3)
of that reference. Lagrangian codes are saturated, so
Proposition~25(2),(4) gives $H_{-d}(F)=0$
and the strong bound. For $i=0$, the weak bound follows
from $\dim R_n=d$.
\end{proof}
\noindent  In Section~\ref{sec:dual_complex}, we show these dimension bounds are satisfied for more general classes of fusion complexes.  Indeed, the strong dimension bounds hold for any fusion complex that participates in a nondegenerate BFC.

\begin{corollary}
If $F$ is a Pauli fusion complex, then $H_{-i}(F)$ is $R_n$-torsion for $i > 0$.  Moreover, if $F$ is Lagrangian, then $H_{i}(F)$ is $R_n$-torsion for all $i$.
  \label{cor:torsion}
\end{corollary}

\begin{proof}
Let $K_n = S^{-1}R_n$ be the localization of $R_n$ at the multiplicative subset $S$ of regular elements. This is the total ring of fractions of $R_n$.
We let $\iota  : R_n \to K_n$ be the localization map. 
For any $R_n$-module $M$, the torsion submodule of $M$ consists of elements $m\in M$ such that there exists a regular element $r\in R_n$ such that $rm=0$. 
This is precisely the kernel of the localization $\id_M\otimes \iota :  M=M\otimes_{R_n} R_n \to M\otimes_{R_n} K_n$. 
The total ring of fractions $K_n$ of $R_n$ is self-injective (see, e.g., \cite[Lem. 7]{Ruba_2024}), and so for $i>0$ we have
\[H_{-i}(F)\otimes_{R_n} K_n \cong  \Ext^{i}_{R_n}(\overline{P/L},R_n)\otimes_{R_n} K_n \cong \Ext^{i}_{K_n}(\overline{P/L}\otimes_{R_n} K_n,K_n)=0. \]
Therefore, $H_{-i}(F)$ is $R_n$-torsion for $i>0$. 

The last claim is obvious because $H_i(F) = 0$ for $i \geq 0$ when $F$ is Lagrangian.
\end{proof}

An alternative proof follows from Lemma~\ref{lemma:Ext-torsion}.

As we should expect given the discussion of Section~\ref{sec:first}, the homology $H_{-1}(F)$ of a Pauli fusion complex is the module of superselection sectors $S$ \cite{Ruba_2024}.  To see this, we observe that $\ker d^*_2 \subset F^*_1$ consists of spatially local functionals on stabilizers that vanish on the relations among stabilizers.  That is, we can identify $\ker d^*_2$ with $L^*$, and this is precisely the module of excitations of bounded support, which are eigenstates of all the stabilizers.  Taking the quotient by $\operatorname{im}  d_0$ is nothing but the quotient by locally creatable excitations, so indeed $S = H_{-1}(F)$.  Below in Sec.~\ref{subsec:hfm-interpretation}, we give a physical interpretation in terms of extended excitations for the higher fusion modules $H_{-i}(F)$ with $i > 1$.

There is a very close relationship between Pauli fusion complexes and an earlier algebraic description of projective Pauli codes due to Haah \cite{Haah_2013}.  In that approach, a Pauli code is described as a complex
\begin{equation}
\begin{tikzcd}
G \arrow[r, "\sigma"] & P \arrow[r, "\epsilon"] & E \text{,}
\end{tikzcd} \label{eqn:Haah-sequence} 
\end{equation}
where $P = R^{2k}_n$ and $G = E = R_n^t$ for some $k, t \in \N$.  Here, $P$ is the module of projective Pauli operators as in our treatment, and $\Omega : P \to P^*$ is the standard symplectic form.  We can identify $P^* = P$, and the maps $\sigma$, $\epsilon$ and $\Omega$ can be viewed as matrices with entries in $R_n$.  The map $\epsilon$ is given by $\epsilon = \sigma^\dagger \Omega$, where ``$\dagger$'' denotes matrix transpose combined with the involution applied to each entry.\footnote{Here we choose the overall sign of $\epsilon$ to agree with our conventions.  The opposite sign can also be chosen; this is merely a convention and not an important difference.}  This specifies a Pauli code by setting $L = \im \sigma$; the fact that Equation~\ref{eqn:Haah-sequence} is a complex implies that $L$ is isotropic.  Moreover, the Pauli code is Lagrangian if and only if the sequence is exact at $P$.  While Ref.~\cite{Haah_2013} does not interpret $E$ as the dual $E = G^*$, this identification is natural given the choice $G = E = R_n^t$.

Given a Pauli fusion complex, the diagram $F_1 \to F_0 \to F^*_1$ is a complex as in Equation~\ref{eqn:Haah-sequence}.  Strictly speaking, we must also choose a basis for $F_1$ in order to identify $F_1 = F^*_1 = R_n^t$.  Moreover, given a complex of the form in Equation~\ref{eqn:Haah-sequence}, we can extend the complex on the left and right to obtain a Pauli fusion complex, by recognizing that $G$ is nothing but the first term in a free resolution of $L$, and following our construction of a Pauli fusion complex from a Pauli code.

Reference~\cite{Haah_2013} showed that the module of superselection sectors is $S = T( \coker \epsilon)$, where $T(M)$ denotes the $R_n$-torsion submodule of $M$ (see also Proposition~24 of \cite{Ruba_2024}). The following result shows this is the same as $S = H_{-1}(F)$.

\begin{proposition}
Let $F$ be a Pauli fusion complex.  Then $H_{-1}(F) = T( \coker d_0)$.
\end{proposition}

\begin{proof}
By Corollary~\ref{cor:torsion}, $H_{-1}(F)$ is torsion, so $H_{-1}(F) \subset T (\coker d_0)$.  Choose $[\varphi] \in T( \coker d_0)$, with a representative $\varphi \in F^*_1$.  Then there is a regular $r \in R_n$ such that $r \varphi \in \im d_0$.  We have $d^*_2 (r \varphi) = 0$, so for any $f_2 \in F_2$ it follows that $(d^*_2 (r \varphi))(\overline{f_2}) = r \varphi (\overline{d_2 ( f_2 )}) = 0$, which implies $d^*_2 \varphi = 0$ and thus $[\varphi] \in 
H_{-1}(F)$.
\end{proof}

\subsection{Higher fusion modules and extended excitations}
\label{subsec:hfm-interpretation}

We have already seen that $Q_0$ has the physical interpretation of superselection sectors of pointlike excitations.  
More generally, for a subset of Pauli codes referred to as fully mobile models,
defined as those where $\dim Q_i \leq 0$ for all $i$, it was shown for
Lagrangian codes that $Q_i$ represents the excitation types of extended
$i$-dimensional excitations \cite{Ruba_2024}. It is expected that the fully mobile models are precisely
those Pauli codes with a TQFT continuum limit.
A braiding pairing was also defined
between the excitation groups $Q_i$ and $Q_j$ in complementary dimensions,
$i+j=d-2$, and was conjectured to be a perfect pairing. In even spatial
dimension $d=2m+2$, the middle-dimensional group $Q_m$ therefore carries a
self-pairing, which is symmetric when $m$ is even and alternating when $m$ is
odd. When $m$ is even, it was further conjectured that this pairing admits a
natural quadratic refinement, which physically plays the role of a spin; for
$d=2$, this refinement should agree with the one defined by the T-junction
construction \cite{Levin_2003}. In \cite{Ruba_2025}, these conjectures were proved for general
Pauli stabilizer codes in $d=2$, while in \cite{Ruba_2026} they were established
for fully mobile models in all dimensions.

Here, we sketch a physical interpretation of the higher fusion modules ($Q_i$ with $i > 0$) that holds in general Pauli codes.  In a certain sense, they continue to describe types of $i$-dimensional extended excitations, but this interpretation is more subtle than simply looking at the support of the excitation in space.  Our interpretation is based on a prescription to start with a state of the system, and to extract homology classes lying in $Q_i$ from it.  We cannot extract a class in $Q_i$ from an arbitrary state, but instead we will obtain a hierarchy of selection criteria that determine when this is possible.  We emphasize that the discussion here is a sketch; it may be useful to make it fully rigorous in future work.  Moreover, we expect it is straightforward to extend this discussion to general BFCs, but it seems unlikely that this would provide additional physical insight.

Throughout this subsection, all modules are $R_n$-modules unless stated otherwise, and we write $\widehat{M} = \Hom_{\Z_n}(M, \Z_n)$.  That is, we drop the explicit ``$\Z_n$'' decoration from the notation for the $\Z_n$ character dual of Appendix~\ref{subsec:notation}, both for modules and for maps.  For $f\in M^*$ the corresponding character is $\chi_f(x)=(f(\bar x))_1$, where $(\cdot)_1$ denotes the constant component; for finitely generated free $M$, this identifies $M^*$ with the bounded-support submodule of $\widehat M$.  Consider an element $\varphi \in \widehat{F_1}$.  This is a $\Z_n$-valued functional on stabilizers with no restriction on spatial support, and we thus refer to $\widehat{F_1}$ as the module of {\bf extended configurations}.  We have a map $\widehat{d_2} : \widehat{F_1} \to \widehat{F_2}$ induced by precomposition with $d_2$, and we define $\cS = \ker \widehat{d_2} \subset \widehat{F_1}$.  Elements of $\cS$ are functionals on stabilizers that vanish on relations, and we refer to these as {\bf states} because they describe physical stabilizer-eigenvalue assignments, with no restriction on spatial support. Elements of $\widehat{F_2}$ are functionals on relations (with no spatial support restriction), 
and will be called {\bf extended violations}, while those of $\widehat{F_i}$ with $i > 2$ are {\bf extended higher violations}.

For this discussion, we extend the notation for fusion modules by
defining $Q_i=H_{-(i+1)}(F)=0$ for $i\geq d$.
When we are interested in pointlike excitations, it is familiar that asking for the excitation type of a general state in $\cS$ is not a meaningful question.  Instead, we have to restrict to the $R_n$-submodule $\cS_0 \subset \cS$ of functionals of bounded spatial support, which is naturally identified with $\ker d_{-1} \subset F_{-1} = F^*_1$, \emph{i.e.} with the module of bounded excitations.  Similarly, if we want to assign a one-dimensional extended excitation type to a state, this will not be possible for an arbitrary state.  We will proceed by constructing the {\bf selection filtration}
\begin{equation}
\cS_0 \subset \cS_1 \subset \cdots \subset \cS_d \subset \cS \text{,}  \label{eqn:selection-filtration}
\end{equation}
where states within $\cS_i$ can be associated to a class in $Q_i$.  It is important to emphasize that $\cS_i$ is in general not an $R_n$-module for $i \neq 0,d$; the $\cS_i$ are all $\Z_n$-modules, and the selection filtration is a filtration of $\Z_n$-modules.  The definition of $\cS_i$ as a $\Z_n$-submodule of $\cS$ is a \emph{selection criterion} that specifies those states for which it makes sense to assign an $i$-dimensional excitation type.  States belonging to $\cS_i$ are referred to as $i${\bf -admissible}.  We note that a geometric superselection criterion for infinite flux-string excitations in the 3D toric code was proposed in \cite{Vadnerkar_2023} within a $C^*$-algebraic framework.

The selection filtration depends on additional geometrical data that we need to specify.  First we choose an $R_n$-basis
$f^i_1, \dots, f^i_{k_i}$ for each of the modules $F_i$ with $i>0$.  This will allow us to view elements of $\widehat{F_i}$ as physical objects lying on the spatial lattice $\Z^d$.    Focusing on $F_1$, we have a $\Z_n$-basis with elements of the form $t f^1_i$ for $i = 1,\dots,k_1$, and where $t \in \T = \Z^d$.  We introduce a support function $\supp_1 : \widehat{F_1} \to {\mathscr P}(\Z^d)$ where ${\mathscr P}(\Z^d)$ is the power set of $\Z^d$.
For any $\varphi \in \widehat{F_1}$, we define
\begin{equation}
\supp_1(\varphi) := \Big\{ t \in \Z^d \vert \varphi(t f^1_i) \neq 0 \text{ for some } i = 1,\dots,k_1 \Big\}
\nonumber \text{.}
\end{equation}
More intuitively, the subset $\supp_1(\varphi) \subset \Z^d$ is the collection of lattice points where $\varphi$ takes a nonzero value on one of the $\Z_n$-basis elements located at that point.  Similarly, we define functions $\supp_i : \widehat{F_i} \to {\mathscr P}(\Z^d)$ for all $i > 0$.

Elements of $\widehat{F_i}$ can be truncated to any subset $X \subset \Z^d$, by simply setting their values to zero ``outside $X$.''
More precisely, for any subset $X \subset \Z^d$, and any element $\varphi \in \widehat{F_i}$, we can define its truncation to $X$ by
\begin{equation}
\trunc^i_X(\varphi)( t f^i_j ) := \left\{ \begin{array}{ll} \varphi(t f^i_j), & \qquad t \in X \\
0, & \qquad t \notin X
\end{array}\right.  \text{,}
\end{equation}
and extending $\Z_n$-linearly.  Here $f^i_1, \dots, f^i_{k_i}$ is the chosen $R_n$-basis for $F_i$.  This gives $\Z_n$-linear maps $\trunc^i_X : \widehat{F_i} \to \widehat{F_i}$; we suppress the $i$ superscript when it is clear from context.

Now we describe $\cS_1$, the $\Z_n$-submodule of $1$-admissible states.  We choose a $(d-1)$-dimensional hyperplane $\Pi_1 \subset \R^d$, which divides $\Z^d$ into two $d$-dimensional half spaces.  We select one of the half spaces and denote it by $\Omega_1 \subset \Z^d$.  For definiteness, we take $\Omega_1$ to contain any lattice points lying in $\Pi_1$, although this is an unimportant detail.  Now we take a state $\varphi \in \cS$, and consider its truncation $\trunc_{\Omega_1}(\varphi)$.  In general the truncation is no longer a state, because it may be nonzero on the image of $d_2$.  Indeed, $\widehat{d_2} \trunc_{\Omega_1}(\varphi)$ is an extended violation whose support lies within a bounded distance of the hyperplane $\Pi_1$.
\begin{defn}
A state $\varphi \in \cS$ is $1$-admissible if $\varphi_2 := \widehat{d_2} \trunc_{\Omega_1}(\varphi)$ has bounded support.
\end{defn}
\noindent It is straightforward to see that this definition makes $\cS_1$ a $\Z_n$-submodule, and it is important to emphasize that $\cS_1$ depends on the choice of $\Pi_1$ and $\Omega_1$, even though this is not written explicitly.

For any $\varphi \in \cS_1$, $\varphi_2$ lies in $F_{-2} = F^*_2 \subset \widehat{F_2}$.  In fact, $\varphi_2 \in \ker d_{-2}$, because $d_{-2}=-d_3^*$ and $\widehat{d_3}\widehat{d_2}=\widehat{d_2d_3}=0$.  Thus to each $1$-admissible state we can associate a cycle in $F_{-2}$, and a homology class in $Q_1 = H_{-2}(F)$.  We  note that $\cS_0 \subset \cS_1$, because $\varphi \in \cS_0$ is already of bounded spatial support before truncation.  While states $\varphi \in \cS_0$ may have a nontrivial class in $Q_0$, their $Q_1$ class is always trivial because $\trunc_{\Omega_1}(\varphi) \in F^*_1 = F_{-1}$, so $\varphi_2 = d^*_2 \trunc_{\Omega_1}(\varphi)$ is a boundary in $F_{-2}$.  This is consistent with the intuition that $Q_1$ measures the nontriviality of extended excitations.

\begin{figure}
	\centering\includegraphics[width=0.86\textwidth]{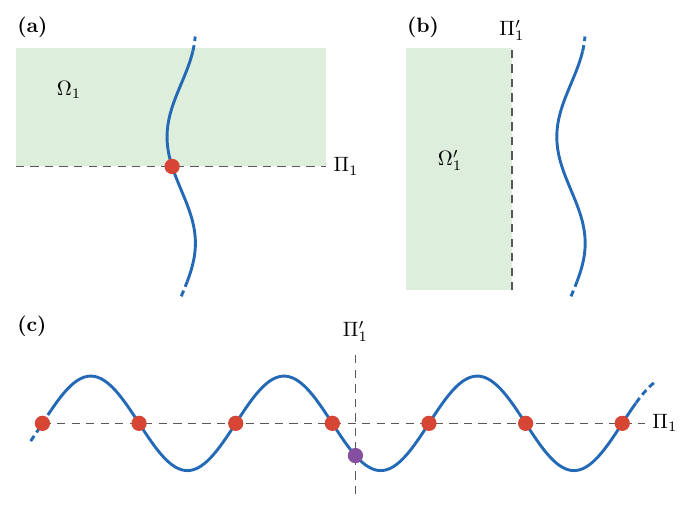}\par
	\caption{  Application of the $1$-admissibility selection criterion to flux-line excitations of the 3D toric code (blue lines), showing two-dimensional cross sections of three-dimensional space.  Dashed blue ends indicate continuations of the flux-line excitations to infinity.
Panels (a) and (b) show the same flux line excitation with different choices of the plane, $\Pi_1$ in (a) and $\Pi'_1$ in (b), shown as dashed lines.  The regions $\Omega_1$ and $\Omega'_1$ are denoted with green shading in (a) and (b).  In (a), the line excitation intersects the plane at one point, where upon truncation a single pointlike violation is exposed (red dot).  The homology class is the nontrivial element of $Q_1 \cong \Z_2$.  In (b), the flux line does not intersect the plane, and the $Q_1$-class is trivial.  Panel (c) shows a ``wavy'' flux line.  Choosing the plane $\Pi_1$,  the violation obtained on truncation (red dots) has infinite spatial support, and the state is not $1$-admissible.  However, there is only a single intersection with the plane $\Pi'_1$ (purple dot), and the state is $1$-admissible in that case.  These illustrations show that the $Q_1$ class, and even the property of $1$-admissibility itself, depend on the choice of the plane $\Pi_1$. (Images generated by ChatGPT-6 Astra based on author-provided sketches.)}
	\label{fig:Q1-3D-toric-code}
\end{figure}

Further intuition is provided by applying the $1$-admissibility criterion to extended flux-line excitations in the 3D toric code, as illustrated in Figure~\ref{fig:Q1-3D-toric-code}.  We see that $Q_1$ is measuring extended excitations via their intersection with the plane $\Pi_1$.  The excitations measured are naturally viewed as one-dimensional because $\Pi_1$ is of codimension one.  However, the spatial support of such excitations may be complicated and, in contrast to the 3D toric code, even simple representative excitations might not look like $1$-manifolds.  We do not have an example, but this may be so in some systems where extended one-dimensional excitations are themselves of restricted mobility, which we expect to occur in four spatial dimensions and above.  See \cite{Pai_2018_fractonic_line} for a discussion of restricted-mobility line objects in the context of a three-dimensional gapless tensor gauge theory.

\begin{figure}
	\centering\includegraphics[width=0.86\textwidth]{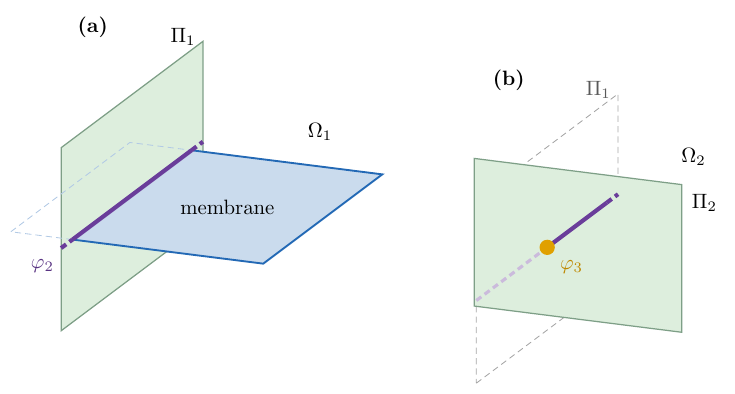}\par
	\caption{  Illustration of how the $Q_2$ class detects a nontrivial $2$-admissible membrane-like excitation; such excitations can occur in the 4D toric code with qudits on edges \cite[Section~7.2]{Ruba_2024}; each panel shows a three-dimensional cross section of four-dimensional space.  In panel (a), the blue membrane excitation is truncated to $\Omega_1$, exposing a purple line of violations along the green plane $\Pi_1$.  This line has infinite support and the state is thus not $1$-admissible.  Panel (b) shows the further truncation of the violation line to $\Omega_2$, bounded by the green plane $\Pi_2$, exposing a single pointlike higher violation (gold dot).  Faint dashed outlines indicate removed portions; only finite portions of the infinite objects are shown. (Images generated by ChatGPT-6 Astra based on author-provided descriptions.)}
	\label{fig:Q2-membrane}
\end{figure}

Now we proceed to define $\cS_2$, the $\Z_n$-submodule of $2$-admissible states.  Keeping $(\Pi_1,\Omega_1)$ fixed, we choose a second hyperplane and half-space pair $(\Pi_2,\Omega_2)$, requiring that $\Pi_1$ and $\Pi_2$ have linearly independent normal vectors.  Given a state $\varphi \in \cS$, as before we first truncate to $\Omega_1$ and pass to an extended violation $\varphi_2 = \widehat{d_2} \trunc_{\Omega_1}(\varphi)$, which has support near $\Pi_1$.  Then we proceed to truncate a second time.  While $\varphi_2 \in \ker \widehat{d_3} \subset \widehat{F_2}$, this need not hold for the truncated functional $\trunc_{\Omega_2}(\varphi_2)$.  We thus pass to the extended higher violation
$\varphi_3 = \widehat{d_3} \trunc_{\Omega_2}(\varphi_2)$, which has spatial support near the $(d-2)$-dimensional intersection $\Pi_1 \cap \Pi_2$.  This leads us to the following:
\begin{defn}
A state $\varphi \in \cS$ is $2$-admissible if $\varphi_3$ (as constructed above) has bounded support.
\end{defn}
In $d=2$, it is clear that every state is $2$-admissible, and so $\cS_2 = \cS$; in this case $Q_2=0$.  Moreover, as before, $2$-admissible states can be associated to a homology class in $Q_2 = H_{-3}(F)$.  We have $\cS_1 \subset \cS_2$, and the $Q_2$-class is trivial for any $1$-admissible state.  Figure~\ref{fig:Q2-membrane} illustrates how the $Q_2$ class can detect a nontrivial two-dimensional membrane-like excitation.

To construct the entire selection filtration, we choose pairs of hyperplanes and half-spaces $(\Pi_1, \Omega_1), \allowbreak (\Pi_2, \Omega_2), \allowbreak \dots, \allowbreak (\Pi_d, \Omega_d)$, with the requirement that the hyperplane normal vectors are linearly independent in $\R^d$.  Setting $\varphi_1=\varphi$, we recursively define
\[
\varphi_{i+1}=\widehat{d_{i+1}}\trunc_{\Omega_i}(\varphi_i), \qquad 1\leq i\leq d.
\]
The $\Z_n$-submodule $\cS_i$ consists of states for which $\varphi_{i+1}$ has bounded support.  States in $\cS_i$ can be associated to a homology class in $Q_i$, and for $i<d$ the $Q_{i+1}$-class of an $i$-admissible state is trivial.  Moreover we have $\cS_d = \cS$, because the final support lies within a bounded distance of the point $\Pi_1 \cap \cdots \cap \Pi_d$.


\section{Examples beyond Pauli BFCs and realization of abelian anyons}
\label{sec:examples}

As discussed in the previous section, every Pauli code gives rise to a BFC. In Sections~\ref{subsec:example1D} and~\ref{subsec:example2D}, we present $d$-dimensional examples of BFCs that cannot be realized by a Pauli code in the same spatial dimension.  Then in Section~\ref{subsec:anyon-realization}, we construct a  nondegenerate pointlike BFC describing an arbitrary theory of abelian anyons in 2D.  Throughout this section, translations act trivially on the fusion module $S = H_{-1}(F)$, and we use polynomial indeterminates  $x$ and $y$ to represent translations in the $x$- and $y$-directions, respectively.

\subsection{1D theories with \texorpdfstring{$S=\Z_2$}{S=Z2}}
\label{subsec:example1D}

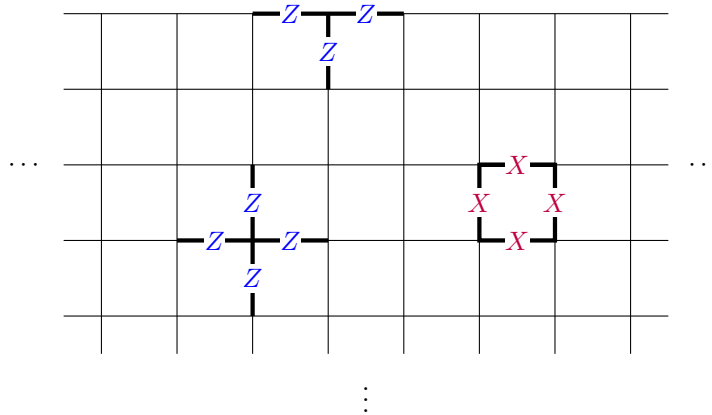
\begin{figure}[t]
    \centering
    \begin{tikzpicture}[scale=1]
        \draw[color=black] (-.5,0)--(7.5,0);
        \draw[color=black] (-.5,-1)--(7.5,-1);
        \draw[color=black] (-.5,-2)--(7.5,-2);
        \draw[color=black] (-.5,-3)--(7.5,-3);
        \draw[color=black] (-.5,-4)--(7.5,-4);
        \draw[color=black] (0,0)--(0,-4.5);
        \draw[color=black] (1,0)--(1,-4.5);
        \draw[color=black] (2,0)--(2,-4.5);
        \draw[color=black] (3,0)--(3,-4.5);
        \draw[color=black] (4,0)--(4,-4.5);
        \draw[color=black] (5,0)--(5,-4.5);
        \draw[color=black] (6,0)--(6,-4.5);
        \draw[color=black] (7,0)--(7,-4.5);
        \draw[color=black,ultra thick] (1,-3)--(3,-3);
        \draw[color=black,ultra thick] (2,-2)--(2,-4);
        \draw[color=black,ultra thick] (5,-2)--(6,-2)--(6,-3)--(5,-3)--(5,-2)--(6,-2);
        \draw[color=black,ultra thick] (2,0)--(4,0);
        \draw[color=black,ultra thick] (3,0)--(3,-1);
        \node[draw=none,fill=white,inner sep=0pt] at (2.5,0) {$\textcolor{blue}{Z}$};
        \node[draw=none,fill=white,inner sep=0pt] at (3.5,0) {$\textcolor{blue}{Z}$};
        \node[draw=none,fill=white,inner sep=2pt] at (3,-.5) {$\textcolor{blue}{Z}$};
        \node[draw=none,fill=white,inner sep=0pt] at (1.5,-3) {$\textcolor{blue}{Z}$};
        \node[draw=none,fill=white,inner sep=0pt] at (2.5,-3) {$\textcolor{blue}{Z}$};
        \node[draw=none,fill=white,inner sep=2pt] at (2,-3.5) {$\textcolor{blue}{Z}$};
        \node[draw=none,fill=white,inner sep=2pt] at (2,-2.5) {$\textcolor{blue}{Z}$};
        \node[draw=none,fill=white,inner sep=0pt] at (5.5,-2) {$\textcolor{purple}{X}$};
        \node[draw=none,fill=white,inner sep=0pt] at (5.5,-3) {$\textcolor{purple}{X}$};
        \node[draw=none,fill=white,inner sep=2pt] at (5,-2.5) {$\textcolor{purple}{X}$};
        \node[draw=none,fill=white,inner sep=2pt] at (6,-2.5) {$\textcolor{purple}{X}$};
        \node at (-1,-2) {$\cdots$};
        \node at (8,-2) {$\cdots$};
        \node at (3.5,-5) {$\vdots$};
    \end{tikzpicture}
    \caption{The 2D toric code with smooth boundary. The bulk and boundary terms of (\ref{eq:HTC}).}
    \label{fig:2Dtoriccodeboundary}
\end{figure}

We begin with a particularly simple example of a nontrivial BFC in $d=1$, which arises as a description of the boundary excitations of the 2D toric code with smooth boundary \cite{Bravyi_1998}. First we briefly review this model. The 2D toric code is composed of a qubit on each edge $\ell$ of a square lattice, governed by the Hamiltonian
\begin{equation}
    H_\mathrm{TC}=-\sum_v A_v -\sum_p B_p.
    \label{eq:HTC}
\end{equation}
Here, $v$, $\ell$, and $p$ denote the vertices, edges, and plaquettes of the lattice, respectively.  The Hamiltonian is written in terms of vertex stabilizers $A_v = \prod_{\ell \ni v}Z_\ell$ and plaquette stabilizers $B_p = \prod_{\ell \in p}X_\ell$.
We identify excitations of the vertex terms as $m$ anyons and excitations of the plaquette terms as $e$ anyons. To introduce a smooth boundary, we consider the half-plane lattice geometry shown in Fig.~\ref{fig:2Dtoriccodeboundary}. The Hamiltonian retains the form of Eq.~(\ref{eq:HTC}), with the sums restricted to the truncated lattice geometry (see Fig.~\ref{fig:2Dtoriccodeboundary}). This realizes an $e$-condensed gapped boundary, in which individual $e$ anyons can be created or annihilated by local boundary operators, but $m$ anyons cannot. Thus, the boundary theory has a single nontrivial superselection sector, corresponding to the $m$ anyon.

We now construct a BFC describing this boundary theory. We will consider states in which a finite number of boundary vertices are occupied by $m$ anyons, and there are no excitations within the bulk. Let $V$ denote the set of boundary vertices. In a given excited state, each boundary vertex term has eigenvalue $\pm1$, with the value $-1$ indicating the presence of an $m$ anyon. Since only finitely many vertices are excited, a boundary excitation configuration is naturally represented by an element of the module $C_0=\bigoplus_{v\in V}\Z_2$. Neighboring pairs of boundary excitations can be created or annihilated by applying a Pauli $X$ operator to the intervening edge. Accordingly, movers are represented by elements of the module $C_1=\bigoplus_{\ell\in E}\Z_2$, where $E$ denotes the set of boundary edges. Altogether, the boundary theory is modeled by the fusion complex
\begin{equation}
    \begin{tikzcd}[ampersand replacement=\&]
        0\&C_1\&C_0\&0
        \arrow[from=1-1, to=1-2]
        \arrow[from=1-2, to=1-3,"\partial"]
        \arrow[from=1-3, to=1-4]
    \end{tikzcd}
\end{equation}
where $\partial$ maps each edge to the sum of its two endpoints. This is precisely the cellular chain complex of the 1D lattice with $\mathbb{Z}_2$ coefficients, with $C_j$ placed in degree $j-1$.

Note that $C_0$ and $C_1$ are both isomorphic to $R_2$; hence this fusion complex is a free resolution of $S=\Z_2$ over $R_2$. To identify these modules with $R_2$, we must choose a basis for each one. We take the vertex at the origin as the generator of $C_0$, and the edge immediately to its right as the generator of $C_1$. With these choices of basis, $\partial=x+1$, since the generator of $C_1$ is mapped to the sum of the origin and its neighbor in the positive $x$ direction. Moreover, every mover corresponds to a Pauli $X$ string, so any two movers commute. The resulting braiding is therefore trivial. Altogether, we obtain the BFC
\begin{equation}
    \begin{tikzcd}[ampersand replacement=\&]
        0\&0\&R_2\&R_2\&0\\
        0\&R_2\&R_2\&0\&0
        \arrow[from=1-1, to=1-2]
        \arrow[from=1-2, to=1-3]
        \arrow[from=1-3, to=1-4,"x+1"]
        \arrow[from=1-4, to=1-5]
        \arrow[from=2-1, to=2-2]
        \arrow[from=2-2, to=2-3,"\ol x+1"]
        \arrow[from=2-3, to=2-4]
        \arrow[from=2-4, to=2-5]
        \arrow[from=1-2, to=2-2,"0"]
        \arrow[from=1-3, to=2-3,"0"]
        \arrow[from=1-4, to=2-4,"0"]
    \end{tikzcd} \text{.}
    \label{eq:2DtoriccodeboundaryBFC}
\end{equation}

We now generalize this construction to describe the boundary excitations of the double semion model \cite{LevinGu_2012,Ellison_2022}.  We denote the nontrivial bosonic anyon by $e$ and consider an $e$-condensed boundary.  The nontrivial boundary sector is denoted by $m$.  Rather than deriving the theory from a microscopic lattice model, we present a representative BFC directly. In contrast to the toric code, $\mathbb{Z}_2$ coefficients are no longer sufficient. Instead, the fusion complex is taken to be a free resolution $B$ of the fusion group $S=\mathbb{Z}_2$ over $R$.

To construct this resolution, let $\tilde C$ denote the cellular chain complex of the 1D lattice with $\mathbb{Z}$ coefficients. Note that $\tilde C_0$ and $\tilde C_1$ are both isomorphic to $R$. In particular, $\tilde C_0=\bigoplus_{v\in V}\Z$, and $\tilde C_1=\bigoplus_{\ell\in E}\Z$. The resolution begins with the excitation module $B_{-1}=\tilde C_0$, whose basis elements represent $m$ excitations localized at the vertices of the 1D lattice. Since these excitations are mobile, there is a mover associated with each edge, which is mapped by the differential to the difference of the two endpoints. Furthermore, because the fusion group is $\mathbb{Z}_2$, fusing an excitation with itself yields the vacuum sector. Accordingly, the mover module is $B_0=\tilde C_1\oplus\tilde C_0$, with differential $d_0=\begin{pmatrix}\partial&2\end{pmatrix}$. We refer to the two types of movers as ``hoppers'' and ``doublers,'' respectively. To obtain a pointlike fusion complex, we require $H_0(B)=0$. However, $\ker d_0$ is nontrivial: for each edge, twice the corresponding hopper is equivalent to a dipole of doublers. These relations are generated by the stabilizer module $B_1=\tilde C_1$, with differential $d_1=\begin{psmallmatrix}2\\-\partial\end{psmallmatrix}$. Altogether, this yields the fusion complex
\begin{equation}
    \begin{tikzcd}[ampersand replacement=\&]
        0\&\tilde C_1\&\tilde C_1\oplus\tilde C_0\&\tilde C_0\&0
        \arrow[from=1-1, to=1-2]
        \arrow[from=1-2, to=1-3,"{\begin{psmallmatrix}2\\-\partial\end{psmallmatrix}}"]
        \arrow[from=1-3, to=1-4,"{\begin{psmallmatrix}\partial&2\end{psmallmatrix}}"]
        \arrow[from=1-4, to=1-5]
    \end{tikzcd} \text{.}
\end{equation}
In order to identify $\tilde C_0$ and $\tilde C_1$ with $R$, we must choose bases for these modules. As before, we take the vertex at the origin as the generator of $\tilde C_0$, and the edge immediately to its right as the generator of $\tilde C_1$. With these choices of basis, the boundary operator is $\partial=x-1$.

We now equip the fusion complex $B$ with a braiding $\lambda:B\to D_{\RQZ}(B)$:
\begin{equation}
        \begin{tikzcd}[ampersand replacement=\&]
            0\&R\&[2em]R^2\&[2em]R\&0\\[2ex]
            0\&R_{\Q/\Z}\&R_{\Q/\Z}^2\&R_{\Q/\Z}\&0
            \arrow[from=1-1, to=1-2]
            \arrow[from=1-2, to=1-3,"{\begin{psmallmatrix}2\\1-x\end{psmallmatrix}}"]
            \arrow[from=1-3, to=1-4,"{\begin{psmallmatrix}x-1&2\end{psmallmatrix}}"]
            \arrow[from=1-4, to=1-5]
            \arrow[from=2-1, to=2-2]
            \arrow[from=2-2, to=2-3,"{\begin{psmallmatrix}\ol x-1\\2\end{psmallmatrix}}"']
            \arrow[from=2-3, to=2-4,"{\begin{psmallmatrix}-2&\ol x-1\end{psmallmatrix}}"']
            \arrow[from=2-4, to=2-5]
            \arrow[from=1-2, to=2-2,"\frac{1}{2}(x+1)"]
            \arrow[from=1-3, to=2-3,"{\frac{1}{4}\begin{psmallmatrix}\ol x-x&0\\0&0\end{psmallmatrix}}"]
            \arrow[from=1-4, to=2-4,"{\frac{1}{2}(\ol x+1)}"]
        \end{tikzcd}
\end{equation}
This braiding captures the \emph{one-dimensional fusion statistics} of the boundary excitations. The physical comparison is between two ways of splitting one excitation into three, or, in reverse, fusing three excitations into one. Such processes measure the associativity of fusion even when particles cannot be exchanged in space. The nontrivial sign in this example is the same fusion-associator sign found at the boundary of the nontrivial $\Z_2$ symmetry-protected topological phase~\cite{LevinGu_2012,KawagoeLevin_2021}; the semionic domain-wall operator algebra was also discussed in Ref.~\cite{ChenLuVishwanath_2014}.

To see the sign directly from the BFC, choose four well-separated boundary vertices $r_0<r_1<r_2<r_3$. Let $U_j$ be a product of hoppers moving one excitation from $r_j$ to $r_{j+1}$, and let $D$ be the doubler at $r_1$. The operator $W=DU_1$ creates a pair at $r_1,r_2$. Starting with one excitation at $r_1$, compare
\begin{equation}
    P_{\mathrm L}=U_1^{-1}U_0^{-1}WU_2U_1,\qquad
    P_{\mathrm R}=U_2WU_0^{-1}.
    \label{eq:1d-fusion-histories}
\end{equation}
The first process moves the original excitation to $r_3$, creates a pair at $r_1,r_2$, and moves that pair to $r_0,r_1$. The second moves the original excitation to $r_0$, creates the same pair, and moves its right member to $r_3$. Both end with excitations at $r_0,r_1,r_3$, and arbitrary overall phases of the chosen operators cancel in the comparison.

For movers $u,v$, let $\omega(u,v)\in\Q/\Z$ be the constant component of $\lambda_0(u)(\bar v)$. It determines the operator commutator $[u,v]=e^{2\pi i\omega(u,v)}\mathbbm{1}$, as in Section~\ref{subsec:bfc1}. For the displayed braiding, the only nonzero contributions between these long hopping strings come from neighboring edges at their common endpoint. Thus $U_0U_1=cU_1U_0$ and $U_1U_2=cU_2U_1$, with $c=\pm i$ according to the orientation convention; $U_0$ and $U_2$ commute, and $D$ commutes with all the hoppers. Substituting $W=DU_1$ and reordering gives
\begin{equation}
    P_{\mathrm L}=c^{-2}P_{\mathrm R}=-P_{\mathrm R}.
    \label{eq:1d-fusion-sign}
\end{equation}
With vacuum fusion normalized to the identity, the fusion associator has the value $F(m,m,m)=-1$~\cite{KawagoeLevin_2021}. The invariant is the fusion-process sign, not the individual $\pm i$ commutator in a chosen representative. For the toric-code boundary, the corresponding sign is $+1$. A more general mixed statistic constructed from one-dimensional fusion statistics is discussed in Refs.~\cite{HsinChen_2026, Xue_2026}; it should be distinguished from the self-statistic evaluated here.

\subsection{\texorpdfstring{$\Z_2$}{Z2} anyons}
\label{subsec:example2D}

Now we go to two dimensions and derive a BFC describing the $m$ anyon sector of the 2D $\Z_2$ toric code. The construction closely parallels that of the smooth boundary discussed in the previous subsection, and we follow the notation and conventions introduced there.  Letting $V$ denote the set of vertices of the square lattice, we are interested in  configurations of $m$ anyons represented by elements of $C_0=\bigoplus_{v\in V}\Z_2$. Movers correspond to Pauli $X$ operators and are represented by elements of $C_1=\bigoplus_{\ell\in E}\Z_2$, where $E$ denotes the set of edges of the square lattice. Finally, moving an $m$ anyon around a plaquette and returning it to its original position gives a mover that creates no excitations. The corresponding stabilizers are therefore represented by $C_2=\bigoplus_{p\in P}\Z_2$, where $P$ denotes the set of plaquettes. Altogether, the theory is described by the fusion complex
\begin{equation}
    \begin{tikzcd}[ampersand replacement=\&]
        0\&C_2\&C_1\&C_0\&0
        \arrow[from=1-1, to=1-2]
        \arrow[from=1-2, to=1-3,"\partial_2"]
        \arrow[from=1-3, to=1-4,"\partial_1"]
        \arrow[from=1-4, to=1-5]
    \end{tikzcd}
\end{equation}
where $\partial_2$ maps each plaquette to the sum of its four edges, and $\partial_1$ maps each edge to the sum of its two endpoints. This is precisely the cellular chain complex of the 2D lattice with $\mathbb{Z}_2$ coefficients, with $C_j$ placed in degree $j-1$.

Note that $C_0$, $C_1$, and $C_2$ are respectively isomorphic to $R_2$, $R_2^2$, and $R_2$. Hence this fusion complex is a free resolution of $S=\mathbb{Z}_2$ over $R_2$. To make these identifications, we choose bases for each. We take the vertex at the origin as the generator of $C_0$, the edges immediately to the east and north as the generators of $C_1$, and the plaquette to the northeast as the generator of $C_2$. With these choices of basis, the differentials take the form shown below. Since every mover corresponds to a Pauli $X$ string, all movers commute, and the braiding is trivial. Thus, we obtain the BFC
\begin{equation}
        \begin{tikzcd}[ampersand replacement=\&]
            0\&R_2\&[2em]R_2^2\&[2em]R_2\&0\\[2ex]0\&R_2\&R_2^2\&R_2\&0
            \arrow[from=1-1,to=1-2]
            \arrow[from=1-2,to=1-3,"\begin{psmallmatrix}y+1\\x+1\end{psmallmatrix}"]
            \arrow[from=1-3,to=1-4,"{\begin{psmallmatrix}x+1&y+1\end{psmallmatrix}}"]
            \arrow[from=1-4,to=1-5]
            \arrow[from=2-1,to=2-2]
            \arrow[from=2-2,to=2-3,"{\begin{psmallmatrix}\ol x+1\\\ol y+1\end{psmallmatrix}}"']
            \arrow[from=2-3,to=2-4,"{\begin{psmallmatrix}\ol y+1&\ol x+1\end{psmallmatrix}}"']
            \arrow[from=2-4,to=2-5]
            \arrow[from=1-2,to=2-2,"0"]
            \arrow[from=1-3,to=2-3,"0"]
            \arrow[from=1-4,to=2-4,"0"]
        \end{tikzcd} \text{.}
\end{equation}
This degenerate BFC describes a fully mobile $\Z_2$ boson.

Next, we generalize this construction to describe a $\mathbb{Z}_2$ fermion. Although such a BFC can be derived from the fermionic excitations of the 2D toric code, we present the construction directly:
\begin{equation}
        \begin{tikzcd}[ampersand replacement=\&]
            0\&R_2\&[2em]R_2^2\&[2em]R_2\&0\\[2ex]0\&R_2\&R_2^2\&R_2\&0
            \arrow[from=1-1,to=1-2]
            \arrow[from=1-2,to=1-3,"\begin{psmallmatrix}y+1\\x+1\end{psmallmatrix}"]
            \arrow[from=1-3,to=1-4,"{\begin{psmallmatrix}x+1&y+1\end{psmallmatrix}}"]
            \arrow[from=1-4,to=1-5]
            \arrow[from=2-1,to=2-2]
            \arrow[from=2-2,to=2-3,"{\begin{psmallmatrix}\ol x+1\\\ol y+1\end{psmallmatrix}}"']
            \arrow[from=2-3,to=2-4,"{\begin{psmallmatrix}\ol y+1&\ol x+1\end{psmallmatrix}}"']
            \arrow[from=2-4,to=2-5]
            \arrow[from=1-2,to=2-2,"(1+xy)"]
            \arrow[from=1-3,to=2-3,"{\begin{psmallmatrix}0&\ol x+y\\x+\ol y&0\end{psmallmatrix}}"]
            \arrow[from=1-4,to=2-4,"(1+\ol{xy})"]
        \end{tikzcd}
\end{equation}
To verify the fermionic statistics, we use the T-junction hopping process of Ref.~\cite{Levin_2003}. Denote the east and north basis movers by $h_x,h_y$, and take outgoing movers along the east, north, and west arms to be $h_x,h_y,\bar x h_x$, respectively. The displayed $\lambda_0$ says that the east mover commutes with the other two, while the north and west movers anticommute. For operator representatives $U_{\mathrm E},U_{\mathrm N},U_{\mathrm W}$, it follows that
\begin{equation}
    U_{\mathrm E}U_{\mathrm N}^{-1}U_{\mathrm W}
    =-U_{\mathrm W}U_{\mathrm N}^{-1}U_{\mathrm E}.
    \label{eq:fermion-hopping-statistics}
\end{equation}
Acting on an initial state with two particles, one at the origin and one to its north, the two operator orderings correspond to processes that differ by an exchange of the two particles.
Their relative phase is $-1$, so the excitation is a fermion.  This $\Z_2$ fermion BFC is degenerate: the map $\lambda_{-1}=1+\bar x\bar y$ induces the zero map on $H_{-1}\cong\Z_2$, so $\lambda$ is not a quasi-isomorphism.  This is expected, because the fermionic self statistics implies trivial mutual statistics between two fermions.

Finally, we generalize this construction to describe a $\mathbb{Z}_2$ chiral semion. Unlike the bosonic and fermionic cases, this requires going beyond $\mathbb{Z}_2$ coefficients, since a free $\mathbb{Z}_2$ complex cannot capture the $\pm i$ phases associated with semionic statistics. We therefore consider a fusion complex given by a free resolution $A$ of $S=\mathbb{Z}_2$ over $R$.

The construction of $A$ closely parallels that of $B$ in the previous subsection. Let $\tilde C$ denote the cellular chain complex of the square lattice with $\mathbb{Z}$ coefficients. In particular,
\begin{equation}
\tilde C_0=\bigoplus_{v\in V}\mathbb{Z},\qquad
\tilde C_1=\bigoplus_{\ell\in E}\mathbb{Z},\qquad
\tilde C_2=\bigoplus_{p\in P}\mathbb{Z}.
\end{equation}
The resolution begins with the excitation module $A_{-1}=\tilde C_0$, whose basis elements represent semions localized at the vertices of the lattice. Since these excitations are mobile, there is a mover associated with each edge, and the differential maps each mover to the difference of its two endpoints. Moreover, because the fusion group is $\mathbb{Z}_2$, fusing an excitation with itself yields the vacuum sector. Accordingly, the mover module is $A_0=\tilde C_1\oplus\tilde C_0$,
with differential $d_0=\begin{pmatrix}\partial_1&2\end{pmatrix}$.
We refer to the two types of movers as ``hoppers'' and ``doublers,'' respectively. 
There are two types of stabilizers.  The first arises from moving a semion around a plaquette. The second encodes a relation among movers:  for each edge, twice the corresponding hopper is equivalent to a dipole of doublers. Thus, the stabilizer module is $A_1=\tilde C_2\oplus\tilde C_1$,
with differential $d_1=\begin{psmallmatrix}
\partial_2&2\\
0&-\partial_1
\end{psmallmatrix}$.
There is also a relation among stabilizers: moving a semion around a plaquette twice is equivalent to a sum of dipole stabilizers over the edges of that plaquette. Therefore, to enforce $H_1(A)=0$, we introduce $A_2=\tilde C_2$,
with differential
$d_2=\begin{psmallmatrix}
2\\
-\partial_2
\end{psmallmatrix}$.
Altogether, this yields the fusion complex
\begin{equation}
    \begin{tikzcd}[ampersand replacement=\&]
        0\&\tilde C_2\&[2em]\tilde C_2\oplus\tilde C_1\&[2em]\tilde C_1\oplus\tilde C_0\&[2em]\tilde C_0\&0
        \arrow[from=1-1, to=1-2]
        \arrow[from=1-2, to=1-3,"{\begin{psmallmatrix}2\\-\partial_2\end{psmallmatrix}}"]
        \arrow[from=1-3, to=1-4,"{\begin{psmallmatrix}\partial_2&2\\0&-\partial_1\end{psmallmatrix}}"]
        \arrow[from=1-4, to=1-5,"{\begin{psmallmatrix}\partial_1&2\end{psmallmatrix}}"]
        \arrow[from=1-5, to=1-6]
    \end{tikzcd} \text{.}
\end{equation}
We may identify $\tilde C_0$, $\tilde C_1$, and $\tilde C_2$ with $R$, $R^2$, and $R$, respectively, by choosing bases. As in the case of $\Z_2$ coefficients, we take the vertex at the origin as the generator of $\tilde C_0$, the edges immediately to the east and north as the generators of $\tilde C_1$, and the plaquette to the northeast as the generator of $\tilde C_2$. We orient the edges in the positive coordinate directions and the plaquette counterclockwise.

We then equip $A$ with a braiding $\lambda:A\to D_{\RQZ}(A)$:
\begin{equation}
\begin{tikzcd}[ampersand replacement=\&]
    0\&R\&[3.5em]R^3\&[3.5em]R^3\&[3.5em]R\&[3.5em]0\&\\[2ex]
    \&0\&R_{\Q/\Z}\&[1em]R_{\Q/\Z}^3\&[4em]R_{\Q/\Z}^3\&[4em]R_{\Q/\Z}\&0
    \arrow[from=1-1, to=1-2]
    \arrow[from=1-2, to=1-3,"{\begin{psmallmatrix}2\\y-1\\1-x\end{psmallmatrix}}"]
    \arrow[from=1-3, to=1-4,"{\begin{psmallmatrix}1-y&2&0\\x-1&0&2\\0&1-x&1-y\end{psmallmatrix}}"]
    \arrow[from=1-4, to=1-5,"{\begin{psmallmatrix}x-1&y-1&2\end{psmallmatrix}}"]
    \arrow[from=1-5, to=1-6]
    \arrow[from=2-2, to=2-3]
    \arrow[from=2-3, to=2-4,"{\begin{psmallmatrix}\ol x-1\\\ol y-1\\2\end{psmallmatrix}}"']
    \arrow[from=2-4, to=2-5,"{\begin{psmallmatrix}\ol y-1&1-\ol x&0\\-2&0&\ol x-1\\0&-2&\ol y-1\end{psmallmatrix}}"']
    \arrow[from=2-5, to=2-6,"{\begin{psmallmatrix}2&\ol y-1&1-\ol x\end{psmallmatrix}}"']
    \arrow[from=2-6, to=2-7]
    \arrow[from=1-2, to=2-2]
    \arrow[from=1-3, to=2-3,"\lambda_1"]
    \arrow[from=1-4, to=2-4,"\lambda_0"]
    \arrow[from=1-5, to=2-5,"\lambda_{-1}"]
    \arrow[from=1-6, to=2-6]
\end{tikzcd}
\end{equation}
where
\begin{equation}
    \lambda_1=\frac{1}{2}\begin{pmatrix}
        x&1+x&1+y
        \end{pmatrix},\quad
    \lambda_0=\frac{1}{4}\begin{pmatrix}
        \ol{x}-x&-1+\ol{x}-y-\ol{x}y&0\\1-x+\ol{y}+x\ol{y}&\ol{y}-y&0\\0&0&0
        \end{pmatrix},\quad
    \lambda_{-1}=\frac{1}{2}\begin{pmatrix}
        \ol x\\1+\ol x\\1+\ol y
        \end{pmatrix}.
\end{equation}
This nondegenerate braiding captures the statistics of a $\Z_2$ chiral semion.

\subsection{BFC realization of abelian anyons}
\label{subsec:anyon-realization}

Here we construct a nondegenerate pointlike BFC realizing an arbitrary theory of abelian anyons in two spatial dimensions.  As discussed in Section~\ref{subsec:background-motivation}, such a theory is a pair $({\cal A}, \vartheta)$, where ${\cal A}$ is a finite abelian group and $\vartheta : {\cal A} \to \Q/\Z$ is a nondegenerate quadratic form encoding topological spins.  The associated symmetric bilinear form expressing mutual statistics is denoted $b : {\cal A} \times {\cal A} \to \Q/\Z$.  Without loss of generality, we assume ${\cal A} =\bigoplus_{i=1}^K\Z_{n_i}$, and let $a_i$ denote the generator of the $i$th cyclic summand.  The quadratic form $\vartheta$ is completely specified by the values of $\vartheta$ and $b$ on generators, \emph{i.e.} by $\vartheta_i := \vartheta(a_i)$ and $b_{i j} := b(a_i, a_j)$.  

We define the $\Z$-tame fusion complex $F^{(n)}$, which is a free resolution of $\Z_n$:
\begin{equation}
\begin{tikzcd}[ampersand replacement=\&]
    0\&R\&[1em]R^3\&[4em]R^3\&[3.25em]R\&0
    \arrow[from=1-1, to=1-2]
    \arrow[from=1-2, to=1-3,"{\begin{psmallmatrix}n\\y-1\\1-x\end{psmallmatrix}}"]
    \arrow[from=1-3, to=1-4,"{\begin{psmallmatrix}1-y&n&0\\x-1&0&n\\0&1-x&1-y\end{psmallmatrix}}"]
    \arrow[from=1-4, to=1-5,"{\begin{psmallmatrix}x-1&y-1&n\end{psmallmatrix}}"]
    \arrow[from=1-5, to=1-6]
\end{tikzcd}.
\end{equation}
Here the rightmost $R$ term lies in degree $-1$, and translations act trivially on $H_{-1}(F^{(n)}) \cong \Z_n$.  We let $F=\bigoplus_{i=1}^KF^{(n_i)}$, and equip $F_0$ with the commutator
\begin{equation}
    \Omega=- Q\otimes L .
\end{equation}
Here $\otimes$ denotes the Kronecker product of matrices $Q$ and $L$, with
\begin{equation}
    L=\begin{pmatrix}
    \bar{x}-x&-1+\bar{x}-y-\bar{x}y&0\\1-x+\bar{y}+x\bar{y}&\bar{y}-y&0\\0&0&0
    \end{pmatrix},
\end{equation}
and $Q$ is a symmetric $K\times K$ matrix with entries in $\Q/\Z$ satisfying
\begin{equation}
    Q_{ii}=\vartheta_i,\qquad 2Q_{ij}=b_{ij},\qquad n_in_jQ_{ij}=0.
\end{equation}
It is straightforward to verify that such a matrix exists and that $\Omega$ is a commutator.  
Up to a reversal of one lattice coordinate and a change of generator
basis, the upper-left $2\times 2$ block of $L$ is the commutation
matrix used in Haah's constructions of Clifford quantum cellular
automata and invertible subalgebras
\cite{Haah_2021_QCA,Haah_2023_invertible}.

Using the matrix BFC conventions of Remark~\ref{rmk:matrix-bfcs},  this commutator lifts to a BFC braiding $\lambda : F \to D_{\RQZ}(F)$ with nonzero components 
\[
\lambda_{-1}=J,\qquad \lambda_0=-\Omega,\qquad
\lambda_1=J^\dagger \text{,}
\]
where $J$ is a $3K \times K$ matrix viewed as a block matrix of $3 \times 1$ columns.  The $i j$ block, with $i,j = 1,\dots,K$, is given by
\[
J_{ij}=Q_{ij}
\begin{pmatrix}2\bar x\\n_i(1+\bar x)\\n_i(1+\bar y)\end{pmatrix}.
\]

Both $F$ and $D_{\RQZ}(F)$ have homology concentrated in degree $-1$.
Elements of $D_{\RQZ}(F)_{-1}$ are $3K$-component column vectors with entries
in $\RQZ$.  Denote the $i$th 3-component block of a cycle $\varphi \in D_{\RQZ}(F)_{-1}$
by $(u_i,v_i,w_i)^t$, where $t$ denotes the transpose.  
The cycle condition implies $n_i u_i(1,1)=0$,
so we define a map from the cycles in degree $-1$ to
$\Hom_{\Z}({\cal A},\Q/\Z)$ by
\[
\varphi\mapsto\Bigl(a_i\mapsto u_i(1,1)\Bigr)
\]
for the generators $a_i \in {\cal A}$.
This map vanishes on boundaries and induces an isomorphism
$H_{-1}(D_{\RQZ}(F))\cong\Hom_{\Z}({\cal A},\Q/\Z)$.
Under this identification, the map induced by $\lambda$ on homology is
$a \mapsto b(-, a)$, since the first entry of $J_{ij}$ evaluates to
$2Q_{ij}=b_{ij}$ at $x=y=1$.  Thus $\lambda$ is a quasi-isomorphism if and
only if $b$ is nondegenerate, equivalently if and only if
$\vartheta$ is nondegenerate.

So far we have constructed a nondegenerate pointlike BFC.  It remains to verify that it encodes the input theory of abelian anyons.  Denote the first two generators of the $i$th summand of $F_0$ by $W_{xi}$ and $W_{yi}$.  Their boundaries are
\begin{equation}\label{eq:movers}
d_0W_{xi}=(x-1)e_i,\qquad d_0W_{yi}=(y-1)e_i ,
\end{equation}
where $e_i$ generates the $i$th summand of $F_{-1}$, and moreover $a_i = [e_i] \in S$.  Thus
$W_{xi}$ and $W_{y i}$ move a generating $i$-type excitation one lattice spacing rightward and
upward, respectively.  From these elementary movers, we can construct arbitrary string operators for the generating $a_i$ anyons.  By evaluating commutators of these strings, we recover the mutual statistics $b_{i j}$ among the generating anyons.  Moreover, a counterclockwise T-junction exchange of two $a_i$ anyons
\cite{Levin_2003} gives the phase $e^{2\pi i\vartheta_i}$,
recovering the input topological spin.


\section {Proper and complete fusion complexes}
\label{sec:dual_complex}

The dual complex provides a convenient means to characterize certain properties of fusion complexes, without the need to specify a BFC in which the fusion complex participates. Throughout this section, we write $D=D_{\RQZ}$.

\begin{defn}\label{defn:propercomplete}
	A fusion theory is {\bf proper} (resp. {\bf complete}) if, upon choosing a tame
	representative $F$, both the complex $F$ and its dual $D(F)$ have
	homology concentrated in nonpositive (resp. negative) degrees. 
\end{defn}\noindent These properties do not depend on the tame representative: a quasi-isomorphism between tame complexes induces a quasi-isomorphism on their duals (Corollary~\ref{v2:cor:tame-dual}(a),(c)). We call a tame representative proper or complete when its fusion theory has that property.

If $\lambda : F \to D(F)$ is a nondegenerate BFC, then $F$ is complete.  Informally, we can thus think of complete fusion complexes as those that have a chance to participate in a nondegenerate BFC.  The term complete comes from Conjecture~\ref{conj:nondeg-BFC}, namely that nondegenerate BFCs are precisely those realized as gapped phases of essentially $d$-dimensional physical systems.  A complete fusion complex thus has a chance to be realized ``on its own'' in an essentially $d$-dimensional system, without the need for a higher-dimensional bulk that supports fractional excitations.

To give some evidence for Conjecture~\ref{conj:nondeg-BFC}, and also to motivate the definition of proper fusion complexes, we discuss the physical interpretation of  $H_0(D(F))$.  We argue that for systems with a unique ground state, $H_0(D(F))$ measures the presence of fractional excitations of a higher-dimensional bulk that are condensed within the $d$-dimensional system of interest.  We consider the following piece of the dual complex:  $F^*_{-1} \to F^*_0 \to F^*_1$.  Elements of $F^*_0$ are spatially local functionals on movers, and to compute the homology we are interested in $\operatorname{ker} d^*_1 \subset F^*_0$, which consists of those functionals that vanish on stabilizers.  We assume that the physical system of interest has a tensor product Hilbert space as a $d$-dimensional system, or that it is the boundary of a higher-dimensional tensor product bulk. We also assume the ground state is unique.  In either case, we expect that every  $\varphi \in F^*_0$ can be represented as the commutator with some projective operator $A$, \emph{i.e.} $\varphi$ corresponds to $\Omega(A, -)$, where we are slightly abusing notation because $A$ need not be a mover.  For instance, $A$ may be an operator that extends a finite distance into the bulk.  The condition $\varphi \in \operatorname{ker} d^*_1$ then becomes the condition that $A$ commutes with all stabilizers.

Now we consider the quotient $H_0(D(F)) =  \operatorname{ker} d^*_1 / \operatorname{im} d^*_0$.  Using the assumption of a unique ground state, the functional $\Omega(A, -)$ lies in $\operatorname{im} d^*_0$ precisely when the operator $A$ creates no excitations.   To see this, let $B \in F_0$, denote the ground state by $| \psi_0 \rangle$, and consider $\hat{A} \hat{B} | \psi_0 \rangle = e^{2 \pi i \omega(A, B) } \hat{B} \hat{A} | \psi_0 \rangle$, where $\hat{A}, \hat{B}$ are unitary operators representing the corresponding projective operators.  If $A$ creates no excitations and the ground state is unique, then by an appropriate choice of the overall phase of $\hat{A}$, we have $\hat{A} | \psi_0 \rangle = | \psi_0 \rangle$, implying  $\hat{A} \hat{B} | \psi_0 \rangle = e^{2 \pi i \omega(A, B) } \hat{B} | \psi_0 \rangle$ .  Thus the commutator $\omega(A, -)$ (and hence $\Omega(A, -)$) arises from a functional given by acting with $A$ on excitations.

 Because $A$ commutes with stabilizers, it will never create any of the excitations we keep track of in $F_{-1}$, but it could create other excitations.  In particular, $A$ could create excitations that lie a finite distance into the bulk.  However, in an essentially $d$-dimensional system, any bulk excitations are of trivial type and can be destroyed by acting with a local operator supported within the bulk.  So in that case, we can modify the operator $A$ so that it creates no excitations at all, without changing its commutator with the movers that lie on the boundary.  However, if the system is not essentially $d$-dimensional, then $A$ may create a nontrivial bulk excitation that is condensed at the boundary.  In such a case, we  expect there is no way to modify $A$ so that it creates no excitations without affecting its commutator with movers, and the homology $H_0(D(F))$ may be nontrivial.  Therefore we expect that $H_0(D(F)) = 0$ for essentially $d$-dimensional systems.  Since a complete fusion complex always has $H_0(D(F)) = 0$, this supports Conjecture~\ref{conj:nondeg-BFC}.

The above general arguments are illustrated by the 1D examples of Section~\ref{subsec:example1D}. For both the toric-code and double-semion boundaries, $H_{-1}(F)\cong\Z_2$ describes the nontrivial boundary sector $m$, while $H_0(D(F))\cong\Z_2$ and all other dual homologies vanish; these fusion complexes are thus proper but not complete.  The nonzero degree-zero dual homology is consistent with the condensed bulk boson $e$ in each example.

The results of Section~\ref{subsec:ed-bfc} provide further evidence for Conjecture~\ref{conj:nondeg-BFC} and motivation for Definition~\ref{defn:propercomplete}.   Theorem~\ref{thm:extended-evaluation-comparison}, together with Proposition~\ref{prop:irc-coinduction}, shows that $\Hom_{\Z}(H_{-i}(D_{\RQZ}(F)), \Q/\Z) \cong H_i(\fD)$, where $\fD$ is the detection complex (Definition~\ref{def:detection-modules}). Remark~\ref{rmk:dual-complex-vanishing-homologies} uses this fact to make a physical argument that $D_{\RQZ}(F)$ has vanishing non-negative-degree homology for a fusion complex describing an essentially $d$-dimensional system.  In degree zero, the argument is distinct from that given above.  

Theorem~\ref{thm:extended-evaluation-comparison} also provides a physical interpretation of the dual homologies $H_{-(i+1)}(D_{\RQZ}(F))$ when $i \geq 0$.  We have isomorphisms
\begin{equation}  \label{eqn:detector-detectee-pairing}
D_i \cong \Hom_{\Z}(H_{-(i+1)}(D_{\RQZ}(F)), \Q/\Z) ,
\end{equation}
where $D_i$ is the $i$th detection module (see Definition~\ref{def:detection-modules}). Its elements are called detectors, which can be understood physically as ``projective operators supported at infinity'' that serve to detect $i$-dimensional excitations; this requires a braiding.  Without a braiding, we can think of $H_{-(i+1)}(D_{\RQZ}(F))$ as a module of objects that can be detected via the pairing corresponding to Equation~\ref{eqn:detector-detectee-pairing}.  We thus refer to $H_{-(i+1)}(D_{\RQZ}(F))$ as the module of $i$-dimensional {\bf detectees}.  We also refer to elements of $H_{-1}(D_{\RQZ}(F))$ as {\bf pointlike detectees}.

A commutator on a complete tame fusion complex extends to a braiding, uniquely up to antisymmetric chain homotopy.  This substantiates the physical expectation, discussed in Section~\ref{subsec:bfc1}, that the commutator alone should determine the outcome of all statistical processes in essentially $d$-dimensional systems.  For a proper tame fusion complex, the same conclusion holds once a compatible stabilizer--excitation pairing supplies the first component of the extension. Lemmas~\ref{v2:lemma:commutator-extension} and \ref{v2:lemma:braiding-uniqueness} and Corollary~\ref{v2:cor:complete-commutator} in Appendix~\ref{v2:app:tame-BFC} give the precise statements and proofs. Example~\ref{v2:ex:nonextendible-commutator} shows that a commutator on an incomplete fusion complex need not extend.

Complete and proper fusion complexes obey constraints on the mobility of excitations that can be expressed as bounds on the Krull dimensions of the fusion modules. Ruba and Yang established these bounds for Pauli codes, with the stronger bounds holding for saturated codes and hence for Lagrangian codes \cite[Proposition~25]{Ruba_2024}; see Proposition~\ref{prop:pauli-krull-dim}. Here we characterize properness and completeness of general tame fusion complexes by these bounds, and show that the bounds are preserved by duality. The proof is deferred to the appendix.

The proof uses the Postnikov tower (Definition~\ref{def:postnikov}), which expresses the quasi-isomorphism type of a complex as successive extensions of its homology modules. Dualizing relates the homologies of the dual complex to Ext modules of the original homologies. The dimension bounds constrain which of these Ext modules can be nonzero, as illustrated by Proposition~\ref{example:proper-d2}, Theorem~\ref{example:proper-d3}, and Theorem~\ref{example:proper-d4} below.

\begin{defn}
	Let us say that a complex $F \in \Ch^p(R)$ satisfies the \textbf{weak dimension bound}
	if it has homology concentrated in nonpositive degrees and satisfies
	\[
		\dim H_{-i}(F) \leq d - i \quad \text{for all $i$}. 
	\]
	Let us say it satisfies the \textbf{strong dimension bound} if it has homology concentrated in
	negative degrees and satisfies 
	\[
		\dim H_{-i}(F) < d - i \quad \text{for all $i$}. 
	\]
\end{defn}

\begin{remark}
	Note that these are the same Krull dimension bounds as those that appear in Proposition~\ref{prop:pauli-krull-dim}.
\end{remark}

\begin{theorem}
	\label{theorem:dimension-bounds}
	Let $F \in \Ch^p(R)$ be a tame complex with $\Z$-torsion homology. 
	The following are equivalent:
	\begin{itemize}
		\item $F$ satisfies the weak (resp.\ strong) dimension bound. 
		\item $D(F)$ satisfies the weak (resp.\ strong) dimension bound. 
		\item $F$ is a proper (resp.\ complete) fusion complex.  
	\end{itemize}
\end{theorem}

\begin{proof}
	This is essentially a restatement of Theorem~\ref{theorem:dimension-bounds-R} (applied to $F$ or its derived dual, using Corollary~\ref{cor:QZ-biduality}), using that
	when $F$ is a tame complex, $D(F)$ is quasi-isomorphic to the derived dual (Theorem~\ref{thm:dualityisderived} and Lemmas~\ref{v2:lemma:RQZ}(a),(b) and \ref{lemma:QZ-duality}). 
\end{proof}

Based on this result, we make the tentative conjecture that gapped boundaries of gapped essentially $(d+1)$-dimensional systems realize proper fusion complexes.  To argue for this, we consider a  $(d+1)$-dimensional physical system with boundary, where the bulk is described by a nondegenerate BFC.  There is thus a complete $(d+1)$-dimensional fusion complex $F^{\mathrm{bulk}}$ describing the bulk excitations. Write $R_{\mathrm{bulk}}$ and $R_{\mathrm{bdy}}$ for the bulk and boundary translation rings, respectively. Let 
$x_{\mathrm{bulk}} \in H_{-i}(F^{\mathrm{bulk}})$ be a bulk excitation for which $\dim R_{\mathrm{bulk}}x_{\mathrm{bulk}}$ saturates the strong dimension bound, and suppose that $x_{\mathrm{bulk}}$ can be transported to the boundary where it is represented by $x_{\mathrm{bdy}} \in H_{-i}(F^{\mathrm{bdy}})$.  The mobility cannot decrease when passing from $x_{\mathrm{bulk}}$ to $x_{\mathrm{bdy}}$, because the boundary excitation can always move back into the bulk and take advantage of the mobility there.  Mathematically we thus expect that $\dim R_{\mathrm{bdy}}x_{\mathrm{bdy}} \leq \dim R_{\mathrm{bulk}}x_{\mathrm{bulk}}$, \emph{i.e.} the Krull dimension cannot increase.  If $\dim R_{\mathrm{bdy}}x_{\mathrm{bdy}} = \dim R_{\mathrm{bulk}}x_{\mathrm{bulk}}$, then $H_{-i}(F^{\mathrm{bdy}})$ violates the strong dimension bound while it may still satisfy the weak dimension bound.  

This discussion is illustrated by the toric-code and double-semion boundaries of Section~\ref{subsec:example1D}. In both cases, the nontrivial boundary sector $m$ is fully mobile and has Krull dimension zero, as does its bulk representative. The boundary fusion module therefore satisfies the weak dimension bound in $d=1$, but violates the strong bound, which would require it to vanish.   While the aforementioned conjecture still needs further investigation, we note that proper fusion complexes, like complete fusion complexes, are an interesting class of fusion complexes that is closed under duality.

Now we present some results in specific spatial dimensions that describe how the dual complex homologies $H_{-(i+1)}(D(F))$  are related to the fusion modules of a complete fusion complex $F$.  The relationships become more complex with increasing dimension.  If we also have a nondegenerate braiding, these results can be understood as constraints on the fusion modules themselves, via the induced isomorphisms $H_{-(i+1)}(F) \cong H_{-(i+1)}(D(F))$.

\begin{proposition}
\label{example:proper-d2}
Let $F$ be a complete tame fusion complex in $d=2$. Its dual homology is concentrated in degree $-1$, with a natural isomorphism
\[
H_{-1}(D(F))\cong\Ext^2_R(\overline{H_{-1}(F)},\RQZ).
\]
\end{proposition}

\begin{proof}
The strong dimension bound leaves only $H_{-1}(F)$, of dimension at most zero. Thus $F$ is quasi-isomorphic to $H_{-1}(F)[-1]$. Its derived dual has homology $\Ext_R^{i+1}(\overline{H_{-1}(F)},\RQZ)$ in degree $-i$ by Proposition~\ref{prop:truncation-duals}(a),  which vanishes unless $i=1$ by
Corollary~\ref{cor:torsion-ext-degrees}(c), since a nonzero
zero-dimensional module is Cohen--Macaulay.
\end{proof}

The following results are obtained by dualizing the Postnikov tower and inspecting the resulting long exact sequences on homology.  Supporting background on Postnikov towers is given in
Appendix~\ref{app:derived_background};
Proposition~\ref{prop:truncation-duals} in
Appendix~\ref{app:derived-fusion} describes their dualization,
with the tame-dual comparison in Theorem~\ref{v2:thm:tame-dual}.

	\begin{theorem}
\label{example:proper-d3}
Let $F$ be a complete tame fusion complex in $d=3$, and write $Q_0=H_{-1}(F)$ and $Q_1=H_{-2}(F)$. There is a natural exact sequence
\[
0\longrightarrow\Ext^3_R(\overline{Q_1},\RQZ)
\longrightarrow H_{-1}(D(F))
\longrightarrow\Ext^2_R(\overline{Q_0},\RQZ)\longrightarrow0
\]
and a natural isomorphism
\[
H_{-2}(D(F))\cong\Ext^3_R(\overline{Q_0},\RQZ).
\]
Let $M_0=\{x\in Q_0:\dim_R(Rx)\leq0\}$ be the maximal zero-dimensional submodule, including zero, and let $I_0=Q_0/M_0$. Then $I_0$ is zero or pure one-dimensional, and the quotient and inclusion maps induce isomorphisms
\[
\Ext^2_R(\overline{I_0},\RQZ)\xrightarrow{\cong}\Ext^2_R(\overline{Q_0},\RQZ),
\qquad
\Ext^3_R(\overline{Q_0},\RQZ)\xrightarrow{\cong}\Ext^3_R(\overline{M_0},\RQZ).
\]
\end{theorem}
\begin{proof}
Completeness gives $\dim Q_0\leq1$, $\dim Q_1\leq0$, and no other nonzero homology. Consequently the only possibly nonzero Ext modules of $\overline{Q_0}$ against $\RQZ$ are in degrees $2,3$, and those of $\overline{Q_1}$ are in degree $3$ by Corollary~\ref{cor:torsion-ext-degrees}(a). 
Applying derived duality to the truncation triangle (Proposition~\ref{prop:truncation-duals}),
\[
Q_0[-1]\longrightarrow F\longrightarrow Q_1[-2]\longrightarrow Q_0[0],
\]
and taking the resulting long exact sequence on homology gives the first two assertions. See
Section~\ref{dra:new-postnikov} and Definition~\ref{v2:def:derived-dual} for the definitions of exact triangles and derived duality. 

The zero-dimensional cyclic submodules of $Q_0$ sum to a zero-dimensional submodule $M_0$, since $Q_0$ is Noetherian. Any zero-dimensional submodule of $I_0$ would have a zero-dimensional inverse image in $Q_0$, contradicting maximality of $M_0$. Thus $I_0$ is zero or pure one-dimensional. By Corollary~\ref{cor:torsion-ext-degrees}(c), 
$\Ext^i_R(\overline{I_0},\RQZ)=0$ for $i \neq 2$, while the dimension bound gives $\Ext^i_R(\overline{M_0},\RQZ)=0$ for $i
\neq 3$. The long exact Ext sequence of \[
0\longrightarrow M_0\longrightarrow Q_0\longrightarrow I_0\longrightarrow0
\]
now gives the two displayed isomorphisms.
\end{proof}
	
Here $M_0$ consists of the fully mobile point sectors, while $I_0$ retains the restricted-mobility point sectors modulo fully mobile ones. Also, $Q_1$ has dimension at most zero, as expected for fully mobile loop sectors. The theorem says that pointlike detectees receive contributions from the Ext-duals of loop sectors and restricted-mobility point sectors, whereas loop detectees depend only on the fully mobile point sectors. The first exact sequence need not split. When a nondegenerate braiding is present, it identifies excitation modules with the corresponding detectee modules and gives these statements their interpretation in terms of mutual detection.

\begin{corollary}  \label{cor:3d-no-loop-only}
Every nontrivial nondegenerate BFC in $d=3$ has at least one nontrivial pointlike excitation. In particular, this property holds for nontrivial Lagrangian Pauli codes in $d=3$.
\end{corollary}

We note that all complexes with bounded finitely generated homologies admit tame replacements, by Lemmas~\ref{lemma:bounded-replacement}(c) and \ref{v2:lemma:tame-in-FC}(c).

\begin{theorem}
\label{example:proper-d4}
Let $F$ be a complete tame fusion complex in $d=4$, and write $Q_i=H_{-(i+1)}(F)$ for $i=0,1,2$. 
Let $G$ be a tame replacement for the smart truncation $\tau_{\leq -2} F$, which
retains the homologies of $F$ only in degrees $\leq -2$, and let
\[
	A = H_{-1} D(G).
\]
Then the homologies of the dual of $F$ fit into natural exact sequences
\[
\begin{aligned}
0\longrightarrow A \longrightarrow H_{-1}(D(F))
&\longrightarrow\Ext^2_R(\overline{Q_0},\RQZ)
\longrightarrow\Ext^4_R(\overline{Q_1},\RQZ)\\
&\longrightarrow H_{-2}(D(F))
\longrightarrow\Ext^3_R(\overline{Q_0},\RQZ)\longrightarrow0,
\end{aligned}
\]
\begin{equation} \label{eqn:A-ses}
0\longrightarrow\Ext^4_R(\overline{Q_2},\RQZ)
\longrightarrow A
\longrightarrow\Ext^3_R(\overline{Q_1},\RQZ)\longrightarrow0,
\end{equation}
and a natural isomorphism
\[
H_{-3}(D(F))\cong\Ext^4_R(\overline{Q_0},\RQZ).
\]
The map $A \to H_{-1}(D(F))$ is induced by the truncation map $F \to G$
under derived duality.
\end{theorem}
\begin{proof}
The strong dimension bounds give $\dim Q_i\leq2-i$. By Corollary~\ref{cor:torsion-ext-degrees}(a), the only possibly nonzero Ext degrees against $\RQZ$ are $2,3,4$ for $\overline{Q_0}$, $3,4$ for $\overline{Q_1}$, and $4$ for $\overline{Q_2}$. All other homologies of $F$ vanish.

Take the derived dual of the truncation triangle (Proposition~\ref{prop:truncation-duals})
\[
Q_1[-2]\longrightarrow G\longrightarrow Q_2[-3]\longrightarrow Q_1[-1].
\]
The homology long exact sequence gives the short exact sequence of Equation~\ref{eqn:A-ses}, as well as
\[
H_{-2} D(G)
\cong\Ext^4_R(\overline{Q_1},\RQZ);
\]
all other homologies of $D(G)$ vanish except in degree $-1$. Taking the derived dual of the
triangle (Proposition~\ref{prop:truncation-duals})
\[
Q_0[-1]\longrightarrow F\longrightarrow G\longrightarrow Q_0[0]
\]
and substituting these calculations into the resulting long exact sequence gives the remaining
assertions.
\end{proof}

Here, the module $A$ collects the contribution of $Q_1$ and $Q_2$ to pointlike detectees,
generically as a non-split extension.
In particular, fully mobile membrane excitations and pure $1$-dimensional loops
contribute. Pure 2-dimensional pointlike excitations also contribute to the pointlike detectee module, via
$\Ext^2_R(\overline{Q_0},\RQZ)$; however, the connecting map $\Ext^2_R(\overline{Q_0},\RQZ) \to \Ext^4_R(\overline{Q_1},\RQZ)$
further couples the contributions of the different excitation modules.
These sequences illustrate the additional complexity of detection in four and higher dimensions. 
Indeed similar long exact sequences that relate homologies of the dual complex to Ext modules of the
direct homologies can be written in arbitrary dimension. 
We leave a fuller physical analysis to future work.


\section{Characterization of pointlike fusion modules}
\label{sec:pfm-char}

In this section we give a characterization of certain modules, which applies in particular to the fusion modules of nondegenerate pointlike BFCs.  This is a mathematical expression of the physical idea that spatial locality imposes nontrivial conditions on the possible patterns of fusion and mobility in gapped phases. Throughout this section, we write $D=D_{\RQZ}$.

\begin{defn}
	We say a fusion theory is \textbf{totally pointlike} if, given any tame representative
	fusion complex
	$F$, both $F$ and $D(F)$ have homology concentrated in degree $-1$.  A braided fusion theory is totally pointlike if its underlying fusion theory has that property.
\end{defn}

We note that a pointlike nondegenerate braided fusion theory is automatically totally pointlike (Remark~\ref{v2:rmk:self-dual-tp} and Theorem~\ref{v2:thm:tame-dual}).

\begin{defn}
	We say that a module $M$ is \textbf{pure-dimensional} if all its nonzero submodules have the same Krull dimension, and \textbf{pure $k$-dimensional} if $M$ is pure-dimensional and $\dim M = k$.
\end{defn}

See Lemma~\ref{v2:lemma:assoc-prime-pure-dimen} for an alternative characterization of pure-dimensionality.
By Corollary~\ref{cor:torsion-ext-degrees}(a),(b), a nonzero finitely generated pure-dimensional $R_n$-module $M$ is Cohen-Macaulay if and only if $\Ext^i_{R_n}(M, R_n)$ is nonzero for at most one value of $i$.

\begin{lemma}  \label{lem:ext-RQZ-Rn}
Let $M$ be a finitely generated $\Z$-torsion $R$-module of fusion order $n$.  Then $\Ext^i_R(\overline M, \RQZ) \cong \Ext^i_{R_n}(\overline M, R_n)$ for all $i \geq 0$.  
\end{lemma}

\begin{proof}
Let $0 \to \RQZ \to I$ be an injective resolution.  Taking the subcomplex of $n$-torsion elements, by Lemma~\ref{v2:lemma:Rn-coefficients}(a) we have an exact sequence $0 \to R_n \to {}_n I$, where ${}_n I \hookrightarrow I$ is the degreewise subcomplex of $n$-torsion elements in $I$.  It is straightforward to see that each term of ${}_n I$ is injective as an $R_n$-module, so ${}_n I$ is an injective resolution of $R_n$ as an $R_n$-module.  

Now let $A$ be any chain complex of $R$-modules where each term $A_i$ is $n$-torsion.  Then we have chain isomorphisms $\Hom_{R_n}(A, {}_n I) \cong \Hom_R(A, {}_n I) \cong \Hom_R(A, I)$, where the first isomorphism holds because $R$-linear maps of $n$-torsion modules are the same as $R_n$-linear maps, and the second isomorphism is induced by ${}_n I \hookrightarrow I$.  Setting $A = \overline M[0]$ and taking homology gives the desired result.
\end{proof}

\begin{theorem}
	\label{theorem:totally-pointlike}
	A pointlike fusion theory is totally pointlike if and only if its fusion module $S$ is zero or pure $(d-2)$-dimensional Cohen-Macaulay.
\end{theorem}

\begin{proof}
	We choose a tame representative fusion complex $F$, and we have $S = H_{-1}(F)$. The case $S=0$ is clear, so assume $S \neq 0$.
	Then because $F$ is pointlike, it is quasi-isomorphic to $S[-1]$, 
	and the homologies of $D(F)$ are isomorphic to Ext modules of $\overline S$ against $\RQZ$ by Theorem~\ref{v2:thm:tame-dual} and Proposition~\ref{prop:truncation-duals}(a). In
	particular, 
	\[
		H_{-i}(D(F)) \cong \Ext^{1+i}_{R}(\overline S, \RQZ)
		\cong \Ext^{1+i}_{R_m}(\overline S, R_m),
	\]
	where $m$ is the fusion order of $S$ and the second isomorphism follows from
	Lemma~\ref{lem:ext-RQZ-Rn}.
	The result is now a consequence of Lemma~\ref{v2:lemma:tp-modules}(a),(b) and Corollary~\ref{cor:torsion-ext-degrees}(a),(b). 
\end{proof}

\begin{defn}
	An $R$-module is \textbf{totally pointlike} if it is the fusion module of a
	totally pointlike fusion theory. 
\end{defn}

Note that every finitely generated $\Z$-torsion $R$-module is the fusion module of some pointlike fusion
complex.  Thus, as a corollary of the theorem:

\begin{corollary}
	A finitely generated, $\Z$-torsion $R$-module is totally pointlike if and only if it is zero or 
	pure $(d-2)$-dimensional Cohen-Macaulay. 
\end{corollary}

We thus arrive at a characterization of fusion modules of nondegenerate pointlike BFCs:

\begin{corollary}  \label{cor:pfm-characterization}
The fusion module $S$ of a nondegenerate pointlike BFC is a finitely generated, $\Z$-torsion $R$-module that is either zero or pure $(d-2)$-dimensional Cohen-Macaulay.  In particular, these properties hold for the fusion module of a pointlike Lagrangian Pauli code.
\end{corollary}

This is not a full characterization, in the sense that modules with these properties can fail to be the fusion module of any nondegenerate pointlike BFC.  Indeed, we discuss examples of this phenomenon in Section~\ref{subsec:statistical-duality-examples}.  However, Corollary~\ref{cor:full-fm-characterization} does provide a full characterization in the case of odd fusion order, with the additional requirement of an antisymmetric isomorphism to the statistical dual.  We also note that if a module is pure $(d-2)$-dimensional Cohen-Macaulay, then so are any of its nonzero direct summands.  Therefore Corollary~\ref{cor:pfm-characterization} gives useful necessary conditions on nonzero direct summands of fusion modules, such as the $X$-sector or $Z$-sector of a pointlike Lagrangian CSS Pauli code, even in cases when these direct summands cannot occur on their own.

 \begin{remark}
 	A finitely generated zero-dimensional module over a Noetherian ring is pure zero-dimensional Cohen-Macaulay.  Moreover, a finitely generated, pure $1$-dimensional $R$-module is
	 automatically Cohen-Macaulay (Corollary~\ref{cor:pure-one-dim-CM}).  This is not true for pure $n$-dimensional modules when $n > 1$.

	 The distinction is as follows. The Krull dimension provides information about mobility, which in some sense decreases with increasing dimension.\footnote{We emphasize that the Krull dimension does not distinguish between type I and type II pointlike fracton orders in $d=3$, which all have fusion modules with Krull dimension $d-2=1$, whether or not partially mobile excitations (\emph{i.e.} lineons and planons) are present.}
	On the other hand, the Cohen-Macaulay condition speaks to whether the dual fusion complex has homologies in
	 multiple degrees. Let $S$ be a pure $(d-2)$-dimensional $R$-module, and let $F$ be a tame pointlike fusion complex with $H_{-1}(F) = S$.
	 In dimensions $d \leq 3$, $S$ is automatically Cohen-Macaulay, so the dual complex $D(F)$
	 has homology in degree $-1$, which is to say the only nontrivial detectees are pointlike.
	 In higher dimensions, however, such pointlike tame fusion complexes can have dual homologies in multiple degrees, so they also contribute
	 extended detectees. We give an example of this phenomenon below.
\end{remark}

 \begin{example}
		Working in $d = 4$, with translation generators $x,y,z,w$, consider the ideals $\mathfrak p_1 =
		(\Delta_x, \Delta_y)$ and $\mathfrak p_2 = (\Delta_z, \Delta_w)$ in $R_2$, $\Delta_\mu = 1 - \mu$. Their intersection is $\mathfrak p_1 \cap \mathfrak
		p_2 = (\Delta_x \Delta_z, \Delta_x \Delta_w, \Delta_y \Delta_z, \Delta_y \Delta_w)$. 
		Now consider the module $S = R_2 / (\mathfrak p_1 \cap \mathfrak p_2)$. Its associated primes are
		precisely $\mathfrak p_1, \mathfrak p_2$; hence it is pure $2$-dimensional. However, the module is not
		Cohen-Macaulay. One way to see this is as follows. There is a general exact sequence
		\[
			0 \to \frac {R_2} {\mathfrak p_1 \cap \mathfrak p_2} 
			\to \frac {R_2} {\mathfrak p_1} \oplus \frac {R_2} {\mathfrak p_2} 
			\to \frac {R_2} {\mathfrak p_1 + \mathfrak p_2} 
			\to 0.
		\]
		For $i = 1, 2$, set $A_i = R_2 / \mathfrak p_i$, and set $B = R_2 / (\mathfrak p_1 + \mathfrak p_2)$. Note
		that $\mathfrak p_1 + \mathfrak p_2 = (\Delta_x, \Delta_y, \Delta_z, \Delta_w)$. 
		By the Koszul complex, the $A_i$ are both pure $2$-dimensional Cohen-Macaulay, while $B$ is zero
		dimensional, and moreover they are self-dual:
		\[
			A_i \cong \Ext^2_{R}(\overline{A_i}, R_{\Q/\Z}),
			\qquad
			B \cong \Ext^4_R(\overline{B}, R_{\Q/\Z}).
		\]
		Dualizing the sequence $0 \to S \to A_1 \oplus A_2 \to B \to 0$, we get isomorphisms
		\[
			\Ext^2_R(\overline{S}, R_{\Q/\Z}) \cong A_1 \oplus A_2,
			\qquad
			\Ext^3_R(\overline{S}, R_{\Q/\Z}) \cong B.
		\]
		Hence $S$ is not Cohen-Macaulay. Physically, any nondegenerate BFT featuring $S$ as its pointlike fusion module must also feature a fully mobile loop.
 \end{example}

The Krull dimension provides an effective numerical measure of the mobility of excitations. A useful observation is that zero Krull dimension and full mobility are equivalent conditions. Thus an apparently commutative-algebraic invariant exactly detects a physically meaningful notion of mobility. We reproduce the proof from \cite{Ruba_2024}, adding intuition and an explicit numerical scale on the string operators. First, we need a lemma on divisibility.

\begin{lemma} \label{lem:divisibility}
Let $n>1$ be an integer, and for each positive integer $r$ let $L_n(r)$ denote the largest integer $L$ such that
\[
    (t-1)^L \mid t^{n^r}-1
\]
in $\mathbb Z_n[t]$. Then
\[
    \limsup_{r\to\infty} L_n(r)=\infty.
\]
Equivalently, as $r$ varies, the polynomials $t^{n^r}-1$ are divisible by arbitrarily large powers of $t-1$ in $\mathbb Z_n[t]$.

For example, if $n=p$ is prime, then
\[
    t^{p^r}-1 \equiv (t-1)^{p^r} \pmod p,
\]
so $L_p(r)=p^r$.

This phenomenon is special to polynomial rings over $\mathbb Z_n$. It has no analogue over $\mathbb Z[t]$, where $t^m-1$ is divisible by $t-1$ but never by $(t-1)^2$.
\end{lemma}\begin{proof}
We first note that $L_n(1)\geq 2$. Indeed,
\[
t^n-1=(t-1)(1+t+\cdots+t^{n-1}),
\]
and, in $\mathbb Z_n[t]$,
\[
1+t+\cdots+t^{n-1}
=
(t-1)\sum_{j=0}^{n-2}(n-1-j)t^j.
\]
Thus
\[
t^n-1
=
(t-1)^2\sum_{j=0}^{n-2}(n-1-j)t^j,
\]
so $(t-1)^2$ divides $t^n-1$ in $\mathbb Z_n[t]$.

We next show that  $L_n(2r)\geq L_n(r)^2$.
Let $L=L_n(r)$, and write
$t^{n^r}-1=(t-1)^L f(t)$
for some $f(t)\in \mathbb Z_n[t]$. Substituting $t^{n^r}$ for $t$ gives
\[
t^{n^{2r}}-1
=
(t^{n^r}-1)^L f(t^{n^r}).
\]
Using the previous expression for $t^{n^r}-1$, we obtain
\[
t^{n^{2r}}-1
=
\bigl((t-1)^L f(t)\bigr)^L f(t^{n^r})
=
(t-1)^{L^2} f(t)^L f(t^{n^r}).
\]
Therefore $(t-1)^{L_n(r)^2}$ divides $t^{n^{2r}}-1$, and hence $L_n(2r)\geq L_n(r)^2$.

Since $L_n(1)\geq 2$, iterating this inequality gives $L_n(2^k)\geq 2^{2^k}$
for all $k\geq 0$. In particular, the sequence $L_n(r)$ is unbounded along the subsequence $r=2^k$. Thus
\[
\limsup_{r\to\infty} L_n(r)=\infty.
\]
\end{proof}

\begin{proposition} \label{prop:mob}
Let $\mathfrak a\subset R_n$ be {a proper ideal}. Then
    $\dim(R_n/\mathfrak a)=0$
if and only if there exists a positive integer $D$ such that
    $t_i^D-1\in \mathfrak a$
for every $i=1,\dots,d$.
\end{proposition}\begin{proof}
We first prove the reverse implication. Suppose that there exists a positive
integer $L$ such that
    $t_i^L-1\in \mathfrak a$
for every $i=1,\dots,d$. Then $R_n/\mathfrak a$ is a quotient of
\[
    \mathbb Z_n[t_1^{\pm 1},\dots,t_d^{\pm 1}]
    /
    (t_1^L-1,\dots,t_d^L-1).
\]
The latter is a finite ring, since it is generated as a $\mathbb Z_n$-module
by the finitely many monomials
\[
    t_1^{e_1}\cdots t_d^{e_d},
    \qquad 0\leq e_i<L.
\]
Hence $R_n/\mathfrak a$ is finite, and therefore Artinian. In particular,
    $\dim(R_n/\mathfrak a)=0$.

Conversely, suppose that
    $\dim(R_n/\mathfrak a)=0$.
Since $R_n$ is Noetherian, the quotient $R_n/\mathfrak a$ is a
zero-dimensional Noetherian ring, hence Artinian. Therefore there are only
finitely many maximal ideals of $R_n$ containing $\mathfrak a$; denote them by
    $\mathfrak m_1,\dots,\mathfrak m_k$.
For each $j$, the residue field $R_n/\mathfrak m_j$ is a finite field, say
    $R_n/\mathfrak m_j \cong \mathbb F_{q_j}$.
Since each $t_i$ is a unit in $R_n$, its image in $R_n/\mathfrak m_j$ lies in
$\mathbb F_{q_j}^{\times}$. Hence
    $t_i^s-1\in \mathfrak m_j$
whenever $q_j-1$ divides $s$.

Let
\[
    r=\operatorname{lcm}_{1\leq j\leq k}(q_j-1).
\]
Then, for every $i=1,\dots,d$,
    $t_i^r-1\in \mathfrak m_j$
for all $j$. Hence
\[
    (t_1^r-1,\dots,t_d^r-1)
    \subset
    \bigcap_{\mathfrak m\supset \mathfrak a}\mathfrak m
    =
    \sqrt{\mathfrak a}.
\]
Since $R_n$ is Noetherian, there exists $N\gg 0$ such that
    $\sqrt{\mathfrak a}^{\,N}\subset \mathfrak a$.
In particular,
    $(t_i^r-1)^N\in \mathfrak a$
for every $i$.

By Lemma~\ref{lem:divisibility}, we may choose $m$ such that
    $L_n(m)>N$.
{In particular},
    $(t-1)^N \mid t^{n^m}-1$
in $\mathbb Z_n[t]$. Substituting $t_i^r$ for $t$, we obtain
    $(t_i^r-1)^N \mid t_i^{rn^m}-1$
in $R_n$. Therefore
   $t_i^{rn^m}-1\in \mathfrak a$
for every $i=1,\dots,d$.

Thus, setting
$D=rn^m$,
we have
    $t_i^D-1\in \mathfrak a$
for all $i=1,\dots,d$, as desired.

For a given ideal $\mathfrak a$, the proof is constructive in principle:
one first determines the finitely many residue fields
    $R_n/\mathfrak m_j\cong \mathbb F_{q_j}$
and a nilpotence exponent $N$ satisfying
$\sqrt{\mathfrak a}^{\,N}\subset \mathfrak a$. Then one chooses $m$ such that
$L_n(m)>N$, and the resulting mobility length is
\[
    D=rn^m,
    \qquad
    r=\operatorname{lcm}_j(q_j-1).
\]
This bound is generally not optimal.
\end{proof}

\begin{corollary}[Mobility theorem]
A nonzero finitely generated $R_n$-module $S$ has $\dim S = 0$ if and only if there exists a positive integer $D$ such that $(t_i^D -1)x = 0$ for all $x \in S$ and all $i = 1,\dots,d$.  Viewing $S$ as the fusion module of a physical system, the latter condition is the statement that $S$ is fully mobile, \emph{i.e.} every excitation can be translated $D$ lattice spacings in each coordinate direction by an appropriate mover.
\end{corollary}

\begin{proof}
	Take $\mathfrak{a} = \mathrm{Ann}(S)$.
\end{proof}

\begin{corollary}  \label{cor:2d-mobile-pointlike}
	A complete fusion theory in $d=2$ has fully mobile pointlike excitations. 
\end{corollary}

\begin{proof}
	If $Q_0=0$, the statement is immediate. Otherwise, completeness implies $\dim Q_0 = 0$, and we apply the mobility theorem.
\end{proof}

\section{Algebraic theories of pointlike excitations}
\label{sec:homological-braiding}

We have seen above that a BFC encodes statistical information about excitations in the braiding $\lambda$.  If we are interested in the universal statistical properties of a phase, we can ask about the distinct braided fusion theories with the same underlying fusion theory.  Distinct such theories will correspond to translation-invariant phases with the same universal fusion and mobility properties, but different statistics.  While this is a useful point of view, it can be simplified and made more concrete if we focus on some simpler class of physical systems.  

To give an idea of the kind of simplification we are looking for, we recall that an algebraic theory of abelian anyons in $d=2$ is a pair $({\cal A}, \vartheta_a)$, where ${\cal A} = S$ is the finite abelian group of superselection sectors, and $\vartheta_a : {\cal A} \to \Q / \Z$ is a nondegenerate quadratic form giving the topological spin.  If we only consider mutual statistics, then the associated theory of anyon mutual statistics is the pair $({\cal A}, b_a)$, where $b_a : {\cal A} \times {\cal A} \to \Q / \Z$ is a symmetric bilinear form. It should be possible to extract $({\cal A}, \vartheta_a)$ and/or $({\cal A}, b_a)$ from a BFC in $d=2$, and moreover $({\cal A}, \vartheta_a)$ is expected to be a complete invariant for such BFCs.  At odd fusion order, this translates into the expectation that $({\cal A}, b_a)$ is a complete invariant, because in that case $b_a$ and $\vartheta_a$ contain the same information.

More generally, we can ask whether it is possible to obtain a similar theory of self or mutual statistics for gapped phases with only pointlike excitations.  Of course, apart from anyons in $d=2$, such phases are fracton phases with restricted mobility excitations.  Because the module $S$ is a complete fusion and mobility invariant, the desired theory -- to be extracted from a BFC --  should schematically consist of a pair $(S, \theta)$, where $\theta$ consists of a function (or functions) involving $S$ that encodes the statistical data.  Identifying such statistical data in a mathematically efficient and elegant form has been a longstanding problem in the theory of fractons.  In this section, we solve this problem in the case of odd fusion order, and give a partial solution for even fusion order.

We have already discussed a prototype for the desired result,
at the level of fusion, namely Proposition~\ref{prop:S-complete-invt},
which showed that the isomorphism class of the fusion module
$S = H_{-1}(F)$ is a complete invariant for the fusion theory
of a pointlike fusion complex $F$. The right perspective on
this result is that $F$ and $S$ are two different presentations
of the same underlying theory. In more formal language,
taking degree $-1$ homology gives an equivalence between the
full subcategory of the derived category on
pointlike fusion complexes and the category of finitely
generated $R$-modules of finite fusion order
(Lemma~\ref{v2:lemma:pointlike-H}).

Starting with a nondegenerate pointlike BFC $(F, \lambda)$, it is then natural to guess that taking degree $-1$ homology also captures the universal information about statistics, and produces another equivalent description of the same theory.  This guess turns out to be correct at odd fusion order. At even fusion order, the resulting description preserves some of the universal statistical data but loses other information.  The resulting theory is a pair $(S,b)$, where $b : S \to S^\vee$ is an isomorphism and $S^\vee$ is a module called the statistical dual of $S$.  We will see in Section~\ref{sec:rdms} that the pair $(S,b)$ encodes mutual statistics, and is thus referred to as a theory of pointlike mutual statistics.  (See the physical discussion of Section~\ref{subsec:ed-phys}, and Remark~\ref{rmk:hb-encodes-ms} in Section~\ref{subsec:ed-bfc}.)  The map $b$ is called the homological braiding, and it satisfies a certain antisymmetry property.  Given a BFC, its associated theory of mutual statistics is an invariant of the braided fusion theory.  Moreover, at odd fusion order, the theory of pointlike mutual statistics is a complete invariant, in the sense that there is a bijection between nondegenerate pointlike braided fusion theories and equivalence classes of pointlike theories of mutual statistics.  Therefore, at odd fusion order it is justified to call $(S,b)$ a theory of pointlike excitations, because it encodes all the universal fusion, mobility and statistical properties of a translation-invariant phase.  
The proof uses an equivalence of categories with duality,
together with the relation between derived braidings and
braided fusion theories at odd fusion order. These results
make precise the viewpoint that $(F,\lambda)$ and $(S,b)$
are two different presentations of the same theory.

\subsection{Derived duality and canonical evaluation}
\label{subsec:derived-duality}

To define statistical duality, we will need to work with more general chain complexes than tame fusion complexes.  The duality $D_{\RQZ}$ is not well-behaved in this context, so we shall replace the coefficient module $\RQZ$ by an injective resolution, which provides a derived model for the duality.  To proceed, we first introduce canonical evaluation for a general coefficient complex.  Let $\cR$ be a commutative ring with involution, and let $T$ be a chain complex with involution. A homogeneous element of degree $q$ in $D_T(A)$ is a collection of maps $\overline{A_p}\to T_{p+q}$, so its evaluation on $\overline a\in\overline{A_p}$ is defined.

\begin{defn}[Canonical evaluation pairing] \label{def:canonical-evaluation-pairing}
For any chain complex $A$, the {\bf canonical evaluation pairing} is the degree zero map
\begin{equation}
\ev^T_{A} : \overline{A}\otimes_\cR D_T(A) \longrightarrow T
\end{equation}
defined on homogeneous elements $a\in A_p$ and $\varphi\in D_T(A)_q$ by
\begin{equation}
\ev^T_{A}(\overline a\otimes\varphi)
=(-1)^{pq}\varphi(\overline a)\text{,}
\label{eqn:canonical-evaluation-pairing}
\end{equation}
and extended additively.  
\end{defn}

This formula is $\cR$-linear in $\overline a$ and $\varphi$, so it defines a map on the tensor product.  The sign $(-1)^{pq}$ is the Koszul sign for placing $a$ before $\varphi$ in the evaluation pairing.

\begin{defn}[Canonical evaluation map]
For any chain complex $A$, the {\bf canonical evaluation map}
$\can^T_{A}:A\to D_T^2(A)$ is obtained from the conjugated
evaluation pairing with its inputs interchanged. Explicitly,
\begin{equation}
\can^T_A(a)(\overline\varphi)
:= \overline{\ev^T_A(\overline a\otimes\varphi)}\text{.}
\end{equation}
In particular, for homogeneous $a\in A_p$ and
$\varphi\in D_T(A)_q$,
$\can^T_A(a)(\overline\varphi)
=(-1)^{pq}\overline{\varphi(\overline a)}$.
\end{defn}

Both the pairing and its associated map are chain maps. We omit the $A$ decoration when it is clear from context. Canonical evaluation is natural and satisfies
\[
D_T(\can^T_{A})\circ\can^T_{D_T(A)}=\id_{D_T(A)}.
\]
An involution-compatible chain map $T\to T'$ changes the duality by postcomposition. This change is compatible with canonical evaluation and preserves symmetry and antisymmetry.  The precise statements and proofs are in Corollary~\ref{v2:cor:can-identities} and Lemma~\ref{v2:lemma:chi-antisymmetric}.

To introduce the derived duality,  we choose once and for all an injective resolution $0 \to \RQZ \to I$, or in more detail
\begin{equation}
\begin{tikzcd}
0 \arrow[r] & \RQZ \arrow[r, "\chi"] & I_0 \arrow[r] & I_{-1} \arrow[r] & \cdots \arrow[r] & I_{-d} \arrow[r] & 0  \text{,}
\end{tikzcd}
\end{equation}
where we used the fact that $\RQZ$ has injective dimension $d$ to choose a resolution of length $d$.  Moreover, we can take $I$ to be a chain complex with involution:
\begin{proposition}
There exists a length-$d$ injective resolution $I$ of $\RQZ$ that is equipped with an involution $\inv : I \to I$, which is compatible with the involution on $\RQZ$ in the sense that the following diagram commutes:
\begin{equation}
\begin{tikzcd}
\RQZ \arrow[r, "\chi"] \arrow[d, "\inv_{\RQZ}"'] & I_0 \arrow[d, "\inv_0"] \\
\RQZ \arrow[r, "\chi"] & I_0 
\end{tikzcd} \nonumber \text{.}
\end{equation}
For a fixed injective resolution $I$, any two involutions on $I$ compatible with the involution on $\RQZ$ are chain homotopic.
\end{proposition}

\begin{proof}
Existence is Lemma~\ref{v2:lemma:involutive-resolution}. Homotopy uniqueness on the fixed injective resolution follows from Lemma~\ref{v2:lemma:involutive-model}.
\end{proof}

\begin{defn}
{\bf Derived duality} is the functor $D_I=\Hom_R(\overline{(-)}, I) : \Ch(R)^\op \to \Ch(R)$.  That is, given a chain complex $F$, the derived dual complex is $D_I(F)=\Hom_R(\overline F,I)$.  Given a chain map $f : F \to G$, its derived dual is denoted $D_I(f) : D_I(G) \to D_I(F)$.
\end{defn}

This duality is derived in the sense that if $\alpha : F \to G$ is a quasi-isomorphism, then so is the derived dual $D_I(\alpha) : D_I(G) \to D_I(F)$, because $I$ is a $K$-injective complex of $R$-modules (Lemma~\ref{v2:lemma:dual-invariance}(b)).  Unlike the duality of Definition~\ref{def:dual-complex}, this holds even when $F$ and $G$ are not tame fusion complexes.  It follows that the quasi-isomorphism type of $D_I(F)$ only depends on the quasi-isomorphism type of $F$.

\begin{remark}  \label{rmk:dd-Ext}
Given an $R$-module $M$, for $i \geq 0$ we have
\begin{equation}
H_{-i}(D_I(M[0])) = H_{-i} (\Hom_R(\overline{M}[0], I) ) = \Ext^i_R(\overline M, \RQZ) \text{.} \nonumber
\end{equation}
This is simply the standard definition of $\Ext^i_R(\overline M,\RQZ)$ in terms of an injective resolution for the second argument.
\end{remark}

Given a tame BFC with braiding $\lambda : F \to D_{\RQZ}(F)$, we can change duality to obtain a map $\lambda_I : F \to D_I(F)$, referred to as the derived braiding.  To do this we observe that the injective resolution $0 \to \RQZ \to I$ can be reinterpreted as a chain map $\chi : \RQZ[0] \to I$, by placing $\chi : \RQZ \to I_0$ in degree zero.  Moreover, $\chi$ is compatible with the involutions on $\RQZ[0]$ and $I$ (Lemma~\ref{v2:lemma:involutive-resolution}), 
so Lemma~\ref{v2:lemma:transposition-natural}(b) applies, 
and we have a natural transformation between the dualities with components $\chi_{* F} : D_{\RQZ}(F) \to D_I(F)$, for any chain complex $F$.  It follows from Theorem~\ref{v2:thm:tame-dual} that if $F$ is a tame fusion complex, then $\chi_{* F}$ is a quasi-isomorphism.

\begin{defn}
Given a tame BFC $(F, \lambda)$, the associated {\bf derived braiding} is the composite $\lambda_I = \chi_{* F} \circ \lambda$.  
\end{defn}

\begin{proposition}  \label{prop:derived-braiding-antisymmetric}
The derived braiding of a tame BFC $(F, \lambda)$ satisfies $\lambda_I = - D_I(\lambda_I) \circ \can^I_F$.  Moreover, if the BFC is nondegenerate, then $\lambda_I$ is a quasi-isomorphism.
\end{proposition}

\begin{proof}
 Since $\lambda$ is antisymmetric (Definition~\ref{def:chain-map-sym}), the first claim is Lemma~\ref{v2:lemma:chi-antisymmetric}.
The second claim holds because $\chi_{*F}$ is a quasi-isomorphism when $F$ is a tame fusion complex, as noted above and proved in Theorem~\ref{v2:thm:tame-dual}.
\end{proof}

We now extend the notion of antisymmetric homotopy to a general
coefficient complex.
Let $T$ be a complex with involution.
A homotopy $h$ between chain maps
$A\to\Hom_R(\overline A,T)$ is
{\bf graded antisymmetric} if, for every $x\in A_p$ and $y\in A_q$,
\[
h(x)(\overline y)=-(-1)^{pq}\overline{h(y)(\overline x)}.
\]
Here $h(x)(\overline y)\in T_{p+q+1}$, and the bar denotes the involution on
$T$.  
Equivalently, the degree-one form $\Theta(\overline{x}\otimes y)=(-1)^{pq}h(y)(\overline x)$ attached to $h$ is antisymmetric, that is, $t(\overline h)=-h$ (Remark~\ref{v2:rmk:forms}).

This extends Definition~\ref{def:antisymmetric-homotopy}. Indeed, when $T=\RQZ[0]$
(or $\RRZ[0]$), the evaluations vanish unless $q=-p-1$.
Then $pq=-p(p+1)$ is even, so the condition becomes
\[
h_p(x)(\overline y)=-\overline{h_{-p-1}(y)(\overline x)},
\quad\text{equivalently}\quad
h_p=-h_{-p-1}^*\circ\can_{A_p}.
\]
Here $\can_{A_p}$ is the module-level evaluation map, exactly as
in Definition~\ref{def:antisymmetric-homotopy}. We use this extended convention with
$T=I$ in the following proposition.
\par

\begin{proposition}
If two tame BFCs $(F, \lambda)$ and $(G, \mu)$ are equivalent and thus joined by a zigzag of homotopy quasi-isometries as written in Equation~\ref{eqn:bfc-equiv-diagram}, then  the derived braidings $\lambda_I : F \to D_I(F)$ and $\mu_I : G \to D_I(G)$ are joined by the following zigzag of homotopy quasi-isometries:
\begin{equation}
\begin{tikzcd}
F \arrow[r, "f_1"] \arrow[d, "\lambda_I"] & C^1 \arrow[d, "\tau_1"] & C^2 \arrow[l, "s_1"] \arrow[d, "\tau_2"] \arrow[r, "f_2"] & \cdots & \arrow[l, "s_k"] G \arrow[d, "\mu_I"] \\
D_I(F) & D_I(C^1) \arrow[l, "D_I(f_1)"]  \arrow[r, "D_I(s_1)"] &    D_I(C^2)  & \arrow[l, "D_I(f_2)"]  \cdots \arrow[r, "D_I(s_k)"]  & D_I(G)
\end{tikzcd} \text{.} \nonumber
\end{equation}
Here, each square commutes up to a graded antisymmetric homotopy.
\end{proposition}

\begin{proof}
For each homotopy quasi-isometry in the zigzag, postcompose its antisymmetric homotopy $h$ with $\chi$ to obtain $h_I=\chi_{*F}h$. Naturality of $\chi_*$ and its chain-map property show that $h_I$ is a homotopy between the derived braidings. It remains graded antisymmetric: since $h$ is graded antisymmetric, $t(\overline h)=-h$ (Remark~\ref{v2:rmk:forms}), and Lemma~\ref{v2:lemma:transposition-natural}(b) for $\psi=\chi$, applied to the degree-one element $h$, gives $t^I(\overline{h_I})=\chi_*t(\overline h)=-h_I$. Applying this to each arrow proves the proposition.
\end{proof}

\begin{proposition}  \label{prop:can-dd-qi}
Let $F$ be any perfect complex with $\Z$-torsion homology.  Then the canonical evaluation map of the derived duality is a quasi-isomorphism; that is, $\can^I_{F} : F \to D_I^2(F)$ is a quasi-isomorphism.
\end{proposition}

\begin{proof}
Choose a $K$-projective resolution $P\to F$. Corollary~\ref{cor:QZ-biduality}, together with the chain-map description in Lemma~\ref{v2:lemma:derived-models}(b), implies that $\can^I_P:P\to D_I^2(P)$ is a quasi-isomorphism. Naturality of $\can^I$ and the fact that $D_I$ preserves quasi-isomorphisms then imply that $\can^I_F:F\to D_I^2(F)$ is a quasi-isomorphism.
\end{proof}

\subsection{Statistical duality and theories of pointlike excitations}
\label{subsec:tpe}

Now we define statistical duality, and proceed to prove the main result of this section.  While we are mainly interested in applying statistical duality to totally pointlike modules, it is convenient to define it for arbitrary $R$-modules:

\begin{defn}
Let $S$ be an $R$-module.  The {\bf statistical dual} of $S$ is $S^\vee = H_{-1} ( D_I(S[-1]) )$.  That is, we place $S$ in degree $-1$ to obtain the chain complex $S[-1]$, take the derived dual, and then take homology in degree $-1$.  The statistical dual of an $R$-linear map $f : S_1 \to S_2$ is $f^\vee = H_{-1}(D_I(f[-1]) )$, where $f[-1]$ denotes the chain map obtained by placing $f : S_1 \to S_2$ in degree $-1$.
\end{defn}

\begin{remark}  \label{rmk:sd-Ext}
Referring to Remark~\ref{rmk:dd-Ext}, for any $i \geq - 1$ we have
\begin{equation}
H_{-i}(D_I(S[-1])) = \Ext^{i+1}_R(\overline S, \RQZ) \cong \Ext^{i+1}_{R_n}(\overline S, R_n) \text{,}
\end{equation}
where the last isomorphism assumes the fusion order of $S$ is $n$ (see Lemma~\ref{v2:lemma:Rn-coefficients}(c)).  In particular, statistical duality is the same as $\Ext^2_R(\overline{(-)}, \RQZ)$.  That is, $S^\vee = \Ext^2_R(\overline S, \RQZ)$. For totally pointlike modules, this agrees with the statistical dual of Appendix~\ref{v2:app:pointlike}, by Proposition~\ref{v2:prop:sD-model}(a).\end{remark}

For any pointlike fusion complex $F$, setting $S = H_{-1}(F)$, we have a canonical zigzag of quasi-isomorphisms
\begin{equation} \label{eqn:SF-zigzag}
\begin{tikzcd}
S[-1]
& \tau_{\geq -1}F
  \arrow[l, "p_F"', "\simeq"]
  \arrow[r, "i_F", "\simeq"']
& F
\end{tikzcd}\text{.}
\end{equation}
Here $\tau_{\geq -1}F$ is the smart truncation defined in
Appendix~\ref{dra:new-postnikov}.  Both maps induce the identity on degree $-1$ homology (Remark~\ref{v2:lemma:pointlike-zigzag}). Applying $D_I$ and taking homology gives the comparison isomorphism
\[
\alpha_F=H_{-1}(D_I(p_F))^{-1}\circ H_{-1}(D_I(i_F)):
H_{-1}(D_I(F))\longrightarrow S^\vee.
\]
The truncation zigzag is natural in $F$, so for a chain map $f:F\to G$ between pointlike complexes,
\[
\alpha_F\circ H_{-1}(D_I(f))=H_{-1}(f)^\vee\circ\alpha_G.
\]
For totally pointlike $F$, this is the comparison of Proposition~\ref{v2:prop:sD-model}(b); the construction above also applies when $F$ is only pointlike.

\begin{proposition}
If $S$ is a totally pointlike fusion module, then so is $S^\vee$.
\end{proposition}

\begin{proof}
Write $S=H_{-1}(F)$ with $F$ tame and totally pointlike. Then $S^\vee\cong H_{-1}(D_I(F))$ by Remark~\ref{v2:lemma:pointlike-zigzag}, and $D_I(F)$ is totally pointlike by Theorem~\ref{v2:thm:tame-dual} and Corollary~\ref{cor:QZ-biduality}.
\end{proof}

\begin{corollary}
	\label{corollary:tp-cat-duality}
Let ${\cal M}_{tp}$ denote the full subcategory of the category of $R$-modules, whose objects are totally pointlike fusion modules.  Statistical duality is a functor $\vee : {\cal M}^\op_{tp} \to {\cal M}_{tp}$.
\end{corollary}

\begin{proof}
This follows from the proposition because statistical duality is a composition of contravariant and covariant functors, and hence is a contravariant functor on the category of $R$-modules.
\end{proof}

\begin{defn}\label{def:homological-braiding}
Let $(F, \lambda)$ be a tame pointlike BFC, with $S = H_{-1}(F)$ and $\lambda_I$  the associated derived braiding.  The {\bf homological braiding} $b:S\to S^\vee$ is
\[
b=\alpha_F\circ H_{-1}(\lambda_I)=a_F\circ H_{-1}(\lambda),
\qquad
a_F=\alpha_F\circ H_{-1}(\chi_{*F}):H_{-1}(D_{\RQZ}(F))\xrightarrow{\cong}S^\vee.
\]
Here $\alpha_F$ is the comparison induced by Equation~\ref{eqn:SF-zigzag}; $a_F$ also includes the change from ordinary to derived duality. The map $a_F$ is an isomorphism by Theorem~\ref{v2:thm:tame-dual}.
\end{defn}
If the BFC is nondegenerate, then clearly $b$ is an isomorphism.  We will study the pair $(S,b)$; the first step is to understand how the homological braiding transforms under quasi-isometry.

\begin{proposition}  \label{prop:bfc-tms-equivalence}
Let $(F, \lambda)$ and $(G, \mu)$ be two tame pointlike BFCs, with $(S,b)$ and $(T,c)$ the associated pairs of fusion module and homological braiding, respectively.  If the two BFCs are related by the quasi-isometry
\begin{equation}
\begin{tikzcd}
F \arrow[d, "\alpha"] \arrow[r, "\lambda"] & D_{\RQZ}(F) \\
G \arrow[r, "\mu"] & D_{\RQZ}(G) \arrow[u, "D_{\RQZ}(\alpha)"'] 
\end{tikzcd} \text{,} \nonumber
\end{equation}
then $b = H_{-1}(\alpha)^\vee \circ c \circ H_{-1}(\alpha)$.
\end{proposition}

\begin{proof}
Naturality of the comparison gives
\[
a_F\circ H_{-1}(D_{\RQZ}(\alpha))=H_{-1}(\alpha)^\vee\circ a_G.
\]
Combining this with $\lambda=D_{\RQZ}(\alpha)\circ\mu\circ\alpha$ and taking degree $-1$ homology gives the result.
\end{proof}

Our next task is to show that the homological braiding satisfies a certain antisymmetry property.  The canonical evaluation map descends to statistical duality by taking homology:
\begin{defn}
Let $S$ be a totally pointlike fusion module.   The {\bf statistical canonical map} $\scan_S : S \to S^{\vee \vee}$ is the map induced on degree $-1$ homology from the zigzag
\begin{equation} \label{eqn:scan-defining-zigzag}
\begin{tikzcd}
S[-1] \arrow[r, "\can^I_{S[-1]}"] & D_I^2(S[-1]) \arrow[r, "\simeq"]  & D_I(\tau_{\geq -1}D_I(S[-1])) & D_I(S^\vee[-1])  \arrow[l, "\simeq"']
\end{tikzcd}  \text{.}
\end{equation}
The zigzag after the first map is obtained by applying Equation~\ref{eqn:SF-zigzag} to $D_I(S[-1])$ and then taking the derived dual. This is the construction in Proposition~\ref{v2:prop:sD-model}(c).
\end{defn}

\begin{proposition}
	\label{proposition:scan-identity}
The statistical canonical map is a natural isomorphism from the identity to the statistical double dual functor on the category ${\cal M}_{tp}$.  Moreover, for any $S \in {\cal M}_{tp}$ we have
\begin{equation} \label{eqn:scan-duality-property}
(\scan_S)^\vee \circ \scan_{S^\vee} = \id_{S^\vee} \text{.} 
\end{equation}
\end{proposition}

These properties follow from Theorem~\ref{v2:thm:pointlike-duality}. 

Now we can state and establish the desired antisymmetry property, which mirrors the antisymmetry of bilinear forms (see Equation~\ref{eqn:form-sym}) and chain maps (see Definition~\ref{def:chain-map-sym} and Proposition~\ref{prop:derived-braiding-antisymmetric}).  However, it is important to emphasize that the homological braiding $b$ is not literally the associated map of a bilinear form.

\begin{proposition}  \label{prop:b-antisym}
Let $(F, \lambda)$ be a tame totally pointlike BFC, with fusion module $S$ and homological braiding $b$.  Then
\begin{equation}
b = - b^\vee \circ \scan_S \text{.} \label{eqn:b-antisym}
\end{equation}
\end{proposition}

The map $D_I(\lambda_I)\circ\can^I_F$ induces
$b^\vee\circ\scan_S$ on homological braidings. Since
$\lambda_I=-D_I(\lambda_I)\circ\can^I_F$, we obtain
$b=-b^\vee\circ\scan_S$.
This is Corollary~\ref{v2:cor:tame-tms}(a),(b).

We have arrived at a characterization of the pair $(S,b)$, which motivates the following definition:
\begin{defn} \label{def:tms}
A (pointlike) {\bf theory of mutual statistics} is a pair $(S,b)$, where $S$ is a totally pointlike $R$-module, and $b : S \to S^\vee$ is an antisymmetric isomorphism in the sense that $b = - b^\vee \circ \scan_S$.  A theory of mutual statistics has fusion order $n$ if $S$ is of fusion order $n$.  An isometry between theories of mutual statistics $(S,b)$ and $(S',b')$ is an isomorphism $h : S \to S'$ such that $b = h^\vee \circ b' \circ h$.  Isometric theories of mutual statistics are considered equivalent.
\end{defn}

\begin{remark}
If $b'$ is antisymmetric, then so is $h^\vee \circ b' \circ h$, by naturality of $\scan$.
\end{remark}

The terminology in this definition differs from that used for BFCs in the usage of the term ``theory,'' which here refers to a specific pair $(S,b)$, while a braided fusion theory is an equivalence class of BFCs.  The reason for this differing usage is to emphasize the analogy between theories of mutual statistics and theories of anyon mutual statistics as discussed at the beginning of Sec.~\ref{sec:homological-braiding}.  Recall that a theory of anyon mutual statistics is a pair $({\cal A}, b_a)$, where ${\cal A}$ is a finite abelian group and $b_a : {\cal A} \times {\cal A} \to \Q / \Z$ is a symmetric bilinear form.  Equivalence of two such theories is also given by isometry.  

We will explain how $(S,b)$ encodes mutual statistics in Section~\ref{sec:rdms}; in particular, see the physical discussion of Section~\ref{subsec:ed-phys}, and Remark~\ref{rmk:hb-encodes-ms} in Section~\ref{subsec:ed-bfc}.  At odd fusion order, it follows from Theorem~\ref{thm:Sb-invariant} that the theory of mutual statistics actually encodes \emph{all} the statistics (and fusion) data of a BFC, and in that case we are justified in referring to $(S,b)$ as a {\bf theory of pointlike excitations}.  This does not hold at even fusion order.  For instance, consider the chiral semion and chiral anti-semion theories, where we have $S \cong S^\vee \cong \Z_2$.  There is thus only a single possible isomorphism $b : S \to S^\vee$, so both semion and anti-semion theories must have the same theory of mutual statistics $(S,b)$.  However, these theories differ in the self-statistics of the nontrivial particle, which is evidently not encoded in $(S,b)$.  Of course, for anyons, we know how to encode the self-statistics as a quadratic form, but more generally defining a theory of pointlike excitations remains an open problem at even fusion order.

Now we come to the main result of this section.  First of all, the theory of mutual statistics, or more precisely its equivalence class, is an invariant of the braided fusion theory, and hence, under Conjecture~\ref{conj:nondeg-BFC}, a phase invariant.  This result upgrades the classification of pointlike fusion complexes by their fusion modules established in Proposition~\ref{prop:S-complete-invt}. The appendix shows that passing to homology is compatible with duality and canonical evaluation, a statement formalized as an equivalence of categories with duality for derived totally pointlike complexes and totally pointlike modules (Theorem~\ref{v2:thm:pointlike-duality}). Consequently, antisymmetric derived braidings correspond to antisymmetric maps $b:S\to S^\vee$ (Corollary~\ref{v2:cor:tms}). At odd fusion order, this correspondence gives mutually reconstructible descriptions of the same theory: a nondegenerate pointlike BFC up to equivalence, or a theory of mutual statistics $(S,b)$ up to isometry (Corollary~\ref{v2:cor:odd-tms}). The reconstruction below makes this correspondence concrete. Precisely, we have

\begin{theorem} \label{thm:Sb-invariant}
\leavevmode
\begin{enumerate}
\item If two pointlike nondegenerate BFCs are equivalent, their associated theories of mutual statistics are also equivalent.  

\item Let $n$ be an odd integer.  Two pointlike nondegenerate BFCs of fusion order $n$ are equivalent if and only if their associated theories of mutual statistics are equivalent.  Moreover, if $(S,b)$ is a theory of mutual statistics of fusion order $n$, then there exists a nondegenerate pointlike BFC for which $(S,b)$ is the associated theory of mutual statistics.
\end{enumerate}

In other words, at fixed fusion order $n$, there is a function from the set of pointlike nondegenerate braided fusion theories to the set of equivalence classes of theories of mutual statistics.  If $n$ is odd, this function is a bijection.
\end{theorem}

\begin{proof}[Proof sketch]
The appendix proves the result using an equivalence of categories with duality. Taking degree $-1$ homology gives an equivalence between derived totally pointlike complexes and totally pointlike modules, compatible with duality and canonical evaluation (Theorem~\ref{v2:thm:pointlike-duality}). This compatibility ensures that antisymmetric forms correspond under the equivalence; moreover, their isometries correspond. Consequently, nondegenerate antisymmetric derived braidings,
up to isometry in the derived category, are classified by
theories of mutual statistics $(S,b)$, up to isometry
(Corollary~\ref{v2:cor:tms}).  For a nondegenerate pointlike BFC, this equivalence sends its derived braiding to its homological braiding $b$.

Equivalent tame BFCs have isometric derived braidings, and at odd fusion order the converse holds (Lemmas~\ref{v2:lemma:equivalent-isometric} and~\ref{v2:lemma:isometric-equivalent}). Combining these facts with the preceding correspondence proves invariance and, at odd fusion order, completeness of $(S,b)$.

For existence, a theory of mutual statistics $(S,b)$ determines a nondegenerate antisymmetric derived braiding on $S[-1]$. At odd fusion order, Lemma~\ref{v2:lemma:odd-realization} realizes this by a tame BFC. Its underlying complex is isomorphic to $S[-1]$ in the derived category, and its derived braiding is an isomorphism, so the BFC is pointlike and nondegenerate. Its theory of mutual statistics is isometric to $(S,b)$. This is the argument of Corollary~\ref{v2:cor:odd-tms}.
\end{proof}

\begin{corollary}
Every nondegenerate pointlike braided fusion theory of odd fusion order $n$ has a representative BFC with coefficient order $n$.
\end{corollary}
\noindent We note that this result is not expected to hold at even fusion order.  As discussed in Section~\ref{subsec:bfc1}, the double semion model has fusion order $2$, but a coefficient order $2$ BFC should not exist on physical grounds.  

\begin{proof}
Let $(S,b)$ be the theory of mutual statistics of a given
nondegenerate pointlike braided fusion theory, of odd fusion
order $n$.
Choose a resolution $F$ of $S$ by finitely generated free $R_n$-modules, placed so that $F_i=0$ for $i<-1$. Lemma~\ref{v2:lem:truncated-gamma}(b) lifts $a_F^{-1}\circ(b/2)$ to a chain map $\gamma:F\to D_{\RQZ}(F)$. Antisymmetrizing gives
\[
\lambda=\gamma- D_{\RQZ}(\gamma) \circ\can_F^{\RQZ}.
\]
By Lemma~\ref{v2:lemma:transposition-can} and Corollary~\ref{v2:cor:tame-tms}(a), its homological braiding is
\[
b/2-(b/2)^\vee\circ\scan_S=b.
\] This braiding is constant-alternating in degree zero, and it is nondegenerate because $b$ is an isomorphism and both complexes are pointlike.  By Theorem~\ref{thm:Sb-invariant}, the reconstructed BFC represents the original
braided fusion theory, since their theories of mutual
statistics are isometric.
\end{proof}

Combined with the results of Section~\ref{sec:pfm-char}, we have also arrived at a useful characterization of the fusion modules of nondegenerate pointlike BFCs.  It is a full characterization at odd fusion order.
\begin{corollary} \label{cor:full-fm-characterization}
The fusion module $S$ of a nondegenerate pointlike BFC is totally pointlike and admits an antisymmetric isomorphism $b : S \to S^\vee$ in the sense of Equation~\ref{eqn:b-antisym}.  Conversely, if $S$ satisfies these properties and is also of odd fusion order, then it is the fusion module of some nondegenerate pointlike BFC.
\end{corollary}


\subsection{Constraints on fusion modules from statistical duality}
\label{subsec:statistical-duality-examples}

Being totally pointlike is a necessary condition for a module to arise
as the fusion module of a nondegenerate pointlike BFC, but it is not sufficient. By
Proposition~\ref{prop:b-antisym}, the homological braiding must be an
antisymmetric isomorphism $b:S\to S^\vee$, satisfying  $b=-b^\vee\circ\scan_S$.
This imposes further conditions on $S$. In any pointlike CSS Pauli code, the fusion module
is a direct sum $S = S_X \oplus S_Z$, where the summands are $X$ and $Z$ sector fusion
modules.  We show that the sector fusion modules of the X-cube and anisotropic
models cannot arise on their own, as the fusion module of some nondegenerate
pointlike BFC.  In the X-cube case, $S_X$ is not isomorphic to its statistical dual,
and similarly for $S_Z$, while $S_Z \cong (S_X)^\vee$.  In the anisotropic model,
$S_X \cong S_Z$ and moreover $S_X \cong (S_X)^\vee$, but no antisymmetric
isomorphism exists. Throughout this subsection, we write $D(F) = D_{\RQZ}(F)$.
Moreover,
$R_n=\Z_n[x^{\pm1},y^{\pm1},z^{\pm1}]$, $n\geq2$, and $\Delta_\mu=1-\mu$. 

\subsubsection{X-cube model}

The fusion complex of the $\Z_n$ X-cube model~\cite{Vijay_2016, Vijay_2017} is
$F\oplus D(F)$, where
\[
 F:
 0\longrightarrow R_n^2
 \xrightarrow{\begin{pmatrix}\Delta_x&0\\0&\Delta_y\\-\Delta_z&-\Delta_z\end{pmatrix}}
 R_n^3
 \xrightarrow{\begin{pmatrix}\Delta_y\Delta_z&\Delta_z\Delta_x&\Delta_x\Delta_y\end{pmatrix}}
 R_n\longrightarrow0,
\]
with $R_n^3$ in degree zero. The fracton and lineon sectors have fusion modules
$S_X=H_{-1}(F)$ and $S_Z=H_{-1}(D(F))$, respectively.
Both are totally pointlike as direct summands of $S$.  The
braiding takes the form
\[
 \lambda=
 \begin{pmatrix}
  0&\id_{D(F)}\\
  -\can_F^{\RQZ}&0
 \end{pmatrix}
 :F\oplus D(F)\longrightarrow D(F)\oplus D^2(F).
\]
Using standard bases on the terms of $F$ and induced dual bases on those of $D(F)$,
one can recover the canonical Pauli form of the braiding by
 interchanging the two
summands in degrees $0$ and $-1$, leaving degree $1$ unchanged.  The braiding then has
$\lambda_{-1}=\lambda_1=I_3$ and $\lambda_0=-\Omega$.

The braiding as displayed pairs the two sectors with each other.
In particular, its restriction to $D(F)$ induces the isomorphism
\[
 S_Z=H_{-1}(D(F))
 \xrightarrow[\cong]{a_F} S_X^\vee,
\]
where $a_F$ is the comparison map of
Definition~\ref{def:homological-braiding}.
Restriction to $F$ gives an isomorphism $S_X \cong S_Z^\vee$.
The displayed complex and its dual give presentations
\begin{align*}
 S_X&\cong R_n/(\Delta_y\Delta_z,\Delta_z\Delta_x,\Delta_x\Delta_y),\\
 S_Z&\cong
 \frac{R_n w_1\oplus R_n w_2}
 {\langle \Delta_xw_1,\ \Delta_yw_2,\ \Delta_z(w_1+w_2)\rangle}.
\end{align*}
The numerator in the presentation of $S_Z$ is the free $R_n$-module with basis $\{w_1, w_2\}$.
The single generator of $S_X$ is a fracton (cube excitation), while the two generators
of $S_Z$ are lineons (vertex excitations) each free to move along one of two perpendicular directions.

\begin{proposition}\label{prop:xc-sector-obstruction}
There is no nondegenerate totally pointlike BFC whose fusion module is
isomorphic to only the $X$-sector or only the $Z$-sector of the $\Z_n$ X-cube
model, for $n\geq2$.
\end{proposition}

\begin{proof}
It is enough to show that $S_X$ and $S_Z$ are not isomorphic, because
a nondegenerate BFC with fusion module $S_X$ would provide an isomorphism
$S_X \cong S^\vee_X \cong S_Z$, and similarly for one with fusion module $S_Z$.

Let $\mathfrak a=(\Delta_x,\Delta_y,\Delta_z)$. Passing to the quotient by
$\mathfrak a$ sets $x=y=z=1$, so all the  relations in the
displayed presentations of $S_X$ and $S_Z$
become trivial. The single generator of $S_X$ and the two
generators of $S_Z$ therefore give
\[
 S_X/\mathfrak aS_X \cong R_n / \mathfrak a R_n \cong \Z_n,
 \qquad S_Z/\mathfrak aS_Z  \cong R_n^2 / \mathfrak a R_n^2 \cong   \Z_n^2.
\]
Thus $S_X$ and $S_Z$ are not isomorphic.
\end{proof}

It is important to verify that this result continues to hold upon lowering the translation
symmetry; this is done in Proposition~\ref{prop:xc-sector-restructuring} below.
There is thus an intrinsic obstruction  to realizing the $X$
and $Z$ sector fusion modules on their own, as opposed to a mere ``symmetry-enriched'' obstruction
that occurs for some fixed action of translation symmetry.  We recall that coarsening of translation symmetry
means passing to a finite-index subgroup  $\T' \hookrightarrow \T$ \cite{Haah_2013}.  This induces
an injective ring homomorphism $R'_n = \Z_n[ \T'] \hookrightarrow R_n$, which allows us to view $R_n$-modules
as $R'_n$-modules; this procedure is referred to as restriction of scalars, and is a functor $\Mod(R_n) \to \Mod(R'_n)$.
For an $R_n$-module $M$, the corresponding $R'_n$-module is written $M \vert_{R'_n}$.

\begin{proposition}\label{prop:xc-sector-restructuring}
The conclusion of Proposition~\ref{prop:xc-sector-obstruction}
remains true after arbitrary coarsening of translation symmetry.
\end{proposition}

\begin{proof}
Lemma~\ref{v2:lemma:coarsening-statistical} shows that coarsening commutes with statistical duality.  So if a non-degenerate pointlike BFC has fusion module $S_X \vert_{R'_n}$ for some coarsening $\T' \hookrightarrow \T$, then its braiding induces an isomorphism
\[
S_X \vert_{R'_n} \cong ( S_X \vert_{R'_n} )^\vee \cong S_Z \vert_{R'_n} .
\]
It is thus enough to show that $S_X \vert_{R'_n}$ and $S_Z \vert_{R'_n}$ are not isomorphic for any choice of coarsening.  The quotient superselection sectors form an abelian group $\mathrm{QSS}(S)$ constructed from an $R_n$-module $S$ that is invariant under coarsening \cite{Shirley_2019, Wickenden_2025}, and
\[
\mathrm{QSS}(S_X) \cong \Z_n , \qquad \mathrm{QSS}(S_Z) \cong \Z^2_n ,
\]
as shown in \cite{Shirley_2019}.  This establishes the desired result.
\end{proof}

\subsubsection{Anisotropic model}

The $\Z_n$ anisotropic model~\cite{Shirley_2019} also has a fusion complex of the form
$F\oplus D(F)$, now with
\begin{equation}\label{eq:aniso-sector-resolution}
 F:
 0\longrightarrow R_n
 \xrightarrow{\begin{pmatrix}\Delta_x\Delta_y\\\Delta_z\end{pmatrix}}
 R_n^2
 \xrightarrow{\begin{pmatrix}\Delta_z&-\Delta_x\Delta_y\end{pmatrix}}
 R_n\longrightarrow0,
\end{equation}
where $R_n^2$ is in degree zero. Set
\[
 A=R_n/(\Delta_z,\Delta_x\Delta_y).
\]
The $X$-sector has fusion module $S_X = H_{-1}(F) \cong A$, and the $Z$-sector has
$S_Z = H_{-1}(D(F)) \cong A^\vee$. In this case the two sectors are isomorphic. To see this, use the matrix conventions of
Remark~\ref{rmk:matrix-bfcs} and define $J:F\to D(F)$ by
\begin{equation}\label{eq:aniso-chain-self-duality}
 J_1=-1,\qquad
 J_0=\begin{pmatrix}0&z^{-1}\\(xy)^{-1}&0\end{pmatrix},\qquad
 J_{-1}=(xyz)^{-1}.
\end{equation}
Using $\bar \Delta_\mu=-\mu^{-1}\Delta_\mu$ and
$d_i^{D(F)}=(-1)^{i-1}d_{1-i}^\dagger$, one checks that $J$ is a chain
isomorphism. It induces an isomorphism $j:A\to A^\vee$, using the
comparison $a_F$ of Definition~\ref{def:homological-braiding}.

However, this self-duality does not suffice to realize $A$ as the fusion module
of a nondegenerate BFC. We first determine which maps $A\to A^\vee$ are
antisymmetric. The components of $\can_F^{\RQZ}$ in the chosen bases 
are $(-1)^i\id_{F_i}$, and the matrices in
Equation~\ref{eq:aniso-chain-self-duality} satisfy
\[
 -D(J)\circ\can_F^{\RQZ}=-xyz\,J, 
\]
where $D(J)_i = J^\dagger_{-i}$.
Since $J$ is an isomorphism, $F$ is totally pointlike (Remark~\ref{v2:rmk:self-dual-tp}), so Corollary~\ref{v2:cor:tame-tms}(a) applies with $\gamma=J$ so that $a_F\circ H_{-1}(J)=j$. Since $a_F$ and $H_{-1}$ are $R$-linear, this gives
\begin{equation}\label{eq:aniso-self-duality-adjoint}
 -j^\vee\circ\scan_A=-xyz\,j.
\end{equation}
Every $R_n$-linear map $b:A\to A^\vee$ can be written uniquely as
$b=j\circ m_\beta$ for $\beta\in A$, where $m_\beta$ is multiplication
by $\beta$. The statistical dual of $m_\beta$ is multiplication by
$\bar\beta$, as follows by dualizing scalar multiplication on a free
resolution of $A$. Thus Equation~\ref{eq:aniso-self-duality-adjoint} implies
\begin{equation}\label{eq:aniso-antisymmetric-scalar}
 b=-b^\vee\circ\scan_A
 \quad\Longleftrightarrow\quad
 \beta=-xyz\,\bar\beta.
\end{equation}
In particular, an antisymmetric isomorphism would require a unit $\beta$
satisfying this equation.

\begin{proposition}\label{prop:aniso-sector-obstruction}
There is no nondegenerate totally pointlike BFC whose fusion module is
isomorphic to only the $X$-sector of the $\Z_n$ anisotropic model,
for $n\geq2$. The same holds for the $Z$-sector, which is isomorphic to it.
\end{proposition}

\begin{proof}
Suppose $\beta\in A$ satisfies
Equation~\ref{eq:aniso-antisymmetric-scalar}. Under the projection
\[
 \pi:A\longrightarrow A/(\Delta_x)\cong\Z_n[y^{\pm1}],
\]
write $\pi(\beta)=\sum_{i\in\Z}c_i y^i$. Since $x=z=1$ in the target,
Equation~\ref{eq:aniso-antisymmetric-scalar} becomes
\[
 \sum_i c_i y^i=-y\sum_i c_i y^{-i},
 \qquad\text{so}\qquad c_i=-c_{1-i}.
\]
The function $i\mapsto1-i$ has no fixed integer. The finitely many
nonzero coefficients therefore cancel in pairs, giving
\begin{equation}\label{eq:aniso-zero-augmentation}
 \varepsilon(\beta)=\sum_i c_i=0,
\end{equation}
where $\varepsilon:A\to A/(\Delta_x,\Delta_y,\Delta_z)A
\cong\Z_n$ is the quotient ring homomorphism, which sets $x=y=z=1$.
A unit must map to a unit
under $\varepsilon$, so $\beta$ cannot be a unit. Hence there is no
antisymmetric isomorphism $A\to A^\vee$, which proves the claim.
\end{proof}

The obstruction also survives adding any number of the following two types of planons. This is the key to treating coarsening, which adds precisely these summands.

\begin{lemma}\label{lem:aniso-planon-stability}
For any $p,q\geq0$, the module
\[
 M=A\oplus Q_{xz}^{\oplus p}\oplus Q_{yz}^{\oplus q},
 \qquad Q_{xz}=R_n/(\Delta_x,\Delta_z),\quad Q_{yz}=R_n/(\Delta_y,\Delta_z),
\]
admits no antisymmetric isomorphism $M\to M^\vee$.
\end{lemma}

\begin{proof}
Let $i:A\to M$ be the inclusion sending $A$ to the first summand. If $b$ were such an isomorphism,
then
\[
 h=j^{-1}\circ i^\vee\circ b:M\longrightarrow A
\]
would be surjective.  Indeed, the projection $\pi:M\to A$
satisfies $\pi i=\mathrm{id}_A$, so
$i^\vee\pi^\vee=\mathrm{id}_{A^\vee}$, showing that $i^\vee$
is surjective.  The pullback $i^\vee b i:A\to A^\vee$ is antisymmetric by Lemma~\ref{v2:lemma:transpose}.
Therefore $hi=m_\beta$ for $\beta \in A$ satisfying
Equation~\ref{eq:aniso-antisymmetric-scalar}. By Equation~\ref{eq:aniso-zero-augmentation},
$\varepsilon(\beta)=0$, so
\[
 h(i(A))=\beta A\subset\ker\varepsilon.
\]

Under $h$, the planon summands also map into $\ker\varepsilon$. Any $R_n$-linear map
$Q_{xz}\to A$ has image in $\Ann_A(\Delta_x)=(\Delta_y)A$, and similarly any map
$Q_{yz}\to A$ has image in $\Ann_A(\Delta_y)=(\Delta_x)A$. These annihilator
identities follow by cancelling $\Delta_x$, respectively $\Delta_y$, in
$\Z_n[x^{\pm1},y^{\pm1}]$, where both are non-zero-divisors.
Both ideals lie in $\ker \varepsilon$.  Thus
$h(M)\subset\ker\varepsilon\subsetneq A$, where the last
proper inclusion follows from $\varepsilon(1)=1$.
This contradicts surjectivity of $h$.
\end{proof}

\begin{proposition}\label{prop:aniso-sector-restructuring}
The conclusion of Proposition~\ref{prop:aniso-sector-obstruction} remains
true after arbitrary coarsening of translation symmetry.
\end{proposition}

\begin{proof}
Every finite-index translation subgroup contains the subgroup generated by $x^r,y^r,z^r$ for some positive integer $r \geq 2$. By Corollary~\ref{v2:cor:coarsening-BFC} and Lemma~\ref{v2:lemma:coarsening}, further coarsening preserves nondegeneracy. It therefore suffices to treat these uniform coarsenings. Write
\[
 R_n'=\Z_n[X^{\pm1},Y^{\pm1},Z^{\pm1}],
 \qquad X=x^r,\quad Y=y^r,\quad Z=z^r.
\]
Let $A'=R_n'/(1-Z,(1-X)(1-Y))$, and let
$Q_{XZ}'=R_n'/(1-X,1-Z)$ and $Q_{YZ}'=R_n'/(1-Y,1-Z)$.
Restriction of scalars gives a decomposition
\begin{equation}\label{eq:aniso-coarsened-module}
 \left.A\right|_{R_n'}\cong
 A'\oplus(Q_{XZ}')^{\oplus(r-1)}\oplus(Q_{YZ}')^{\oplus(r-1)}.
\end{equation}
This decomposition follows from the foliated renormalization group (RG) of the
$\Z_n$ anisotropic model established in
Ref.~\cite[Sec.~V.C and Proposition~B.3]{Wickenden_2025}.
The construction respects the $X$ and $Z$ sectors separately,
and each of the $r-1$ toric-code layers extracted in each
orientation contributes one planon summand to either sector.
The corresponding foliated RG transformation was obtained
previously for the $\Z_2$ model in
Ref.~\cite[Appendix~A]{Shirley_2019}.

After relabelling the coarse variables as $x,y,z$, the right-hand side has the form in Lemma~\ref{lem:aniso-planon-stability}, with $p=q=r-1$. It therefore admits no antisymmetric statistical self-duality and cannot be the fusion module of a nondegenerate pointlike BFC. The same holds for the $Z$ sector, since its fusion module is isomorphic to $A$ before and after restriction of scalars.
\end{proof}


\section{Remote detection and mutual statistics}
\label{sec:rdms}

In \cite{Wickenden_2026}, some of the present authors proposed the following:
\begin{framed}
\noindent
\textbf{Excitation-detector principle.}  Consider a bosonic, gapped, essentially $d$-dimensional system with a unique ground state in infinite Euclidean space. Every nontrivial excitation can be detected by some detector, where detectors are operators whose support lies at spatial infinity.  In addition, every nontrivial detector is effective, meaning that it detects at least one nontrivial excitation.
\end{framed}
\noindent 
The first half of this statement is the well-known {\bf principle of remote detectability} \cite{Kitaev_2006, Lan_2018}, which is a necessary and sufficient condition for the physical realizability of a theory of abelian anyons $({\cal A}, \vartheta_a)$, where it becomes nondegeneracy of $\vartheta_a$.  In fact, Reference~\cite{Wickenden_2026} introduced the above principle for strictly $d$-dimensional systems, but it should hold equally well for essentially $d$-dimensional systems, because any remote detection processes do not involve a higher-dimensional bulk in that case.  The excitation-detector principle need not hold for $d$-dimensional boundaries of essentially $(d+1)$-dimensional bulk systems.  For instance, consider the 2D toric code boundary discussed in Section~\ref{subsec:example1D}.  Because the $m$ anyons can move into the $d=2$ bulk, they cannot be remotely detected by any operator whose support is restricted to the $d=1$ boundary.

For algebraic theories of anyons in 2D, the principle of remote detectability implies the excitation-detector principle.  In contrast, \cite{Wickenden_2026} showed that there exist algebraic theories of planon-only fracton orders that satisfy the principle of remote detectability, do not satisfy the excitation-detector principle, and are not physically realizable.  Moreover, within the planon-only setting, detectors were given a precise definition, and it was proved that the excitation-detector principle is necessary for physical realizability.  It was conjectured that the excitation-detector principle holds in general gapped quantum phases of matter.  To establish or disprove the conjecture, a more precise statement of the principle is required, including a precise definition of detectors.
Reference~\cite{Ruba_2025} later defined pointlike and higher detectors in Pauli codes and identified these as certain $\Tor$ modules.  

In Section~\ref{subsec:ed-phys}, we give a physical discussion of the excitation-detector principle in the context of Pauli codes, illustrated by the 2D toric code.  We argue that remote detection, as realized by the excitation-detector pairing, is the correct analog of mutual statistics in fracton phases.  Then in Section~\ref{subsec:ed-bfc}, we state the excitation-detector principle for BFCs and relate it to earlier formulations.  We prove that the principle is satisfied by nondegenerate BFCs.  Moreover, in pointlike BFCs we explain how to obtain the excitation-detector pairing directly from the homological braiding $b$, justifying the claim that $b$ encodes mutual statistics.  Finally, in Section~\ref{subsec:2d-detector-particles}, we identify
detectors with particle types in two dimensions and use this to relate the
excitation-detector pairing to the familiar mutual braiding statistics of anyons.  

\subsection{Physical discussion of the excitation-detector principle}
\label{subsec:ed-phys}

To introduce the construction of detectors in Pauli codes, we begin with a physical discussion.  For illustration, we first consider the 2D toric code model, which was reviewed in Section~\ref{subsec:example1D}.    Unlike in that section, we are interested in the bulk 2D phase of matter realized by the toric code, and accordingly we consider the model on an infinite square lattice with no spatial boundaries.  After discussing this example, we make some of the ideas discussed precise in a simple special case, before developing the more general theory in the following subsection.

\begin{figure}[t]
\centering
\includegraphics[width=\textwidth]{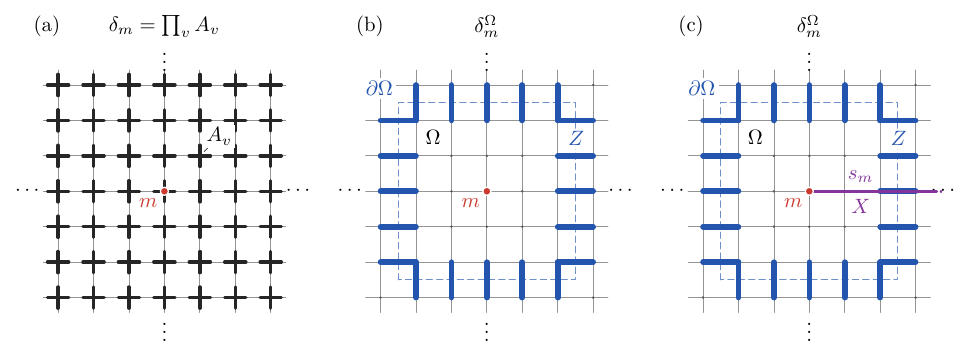}
\caption{Remote detection in the 2D toric code. (a) The detector $\delta_m$ is an infinite product of vertex stabilizers $A_v$. Each cross represents one stabilizer, with its arms shortened to distinguish neighboring factors. A vertex excitation $m$ is shown in red. (b) Truncating the product to vertices in a large square region $\Omega$ leaves Pauli $Z$ operators (thickened blue edges) on the links crossing $\partial\Omega$ (dashed blue line). These form an $e$-string along the dual-lattice loop $\partial \Omega$. (c) The action of the truncated detector $\delta^\Omega_m$ on the central $m$ excitation can be understood in terms of the commutator of $\delta^\Omega_m$ with a semi-infinite Pauli $X$ string $s_m$ (purple) that creates the $m$ excitation.  The two operators overlap on one link and anticommute, so the truncated detector has eigenvalue $-1$ on this excitation.  Equivalently, this eigenvalue is the product of the vertex-stabilizer eigenvalues in $\Omega$: $-1$ at the excited vertex and $+1$ at every other vertex.  (Images generated by ChatGPT-6 Astra based on author-provided descriptions.)}
\label{fig:toric-code-detector}
\end{figure}

Given an $m$-particle (vertex excitation in our conventions), we would like to construct an ``operator supported at infinity'' $\delta_m$ that is capable of detecting $m$, in the sense that it should return eigenvalue $-1$ when acting on any state in the same superselection sector as a single $m$-particle, and return unity when acting on any trivial excitation, including the ground state.  This job is accomplished by
\begin{equation}
\delta_m = \prod_{v}  A_v \text{,}
\end{equation}
an infinite product of the vertex stabilizers (Figure~\ref{fig:toric-code-detector}a).  The reader might be taken aback by the preceding statement, because $\delta_m$ is naively equal to the identity operator.  The stabilizers cancel locally, because the Pauli operator $Z_\ell$ on each edge $\ell$ appears twice in the product, once for each of the two vertices $v$ adjacent to $\ell$.  However, making sense of such an infinite product is subtle, and a better prescription produces a more interesting result.

Suppose we have a state $| \psi \rangle$ that differs locally from the ground state only within a finite simply connected region $\Omega'$.  We then choose a much larger finite simply connected region $\Omega$ that contains $\Omega'$, and truncate the product in $\delta_m$ by selecting only those vertices lying within $\Omega$.  The result is a well-defined truncated operator $\delta^\Omega_m$, which is supported only on the boundary $\partial \Omega$, far away from any excitations contained in $\Omega'$.  Indeed, $\delta^\Omega_m$ is an $e$-string operator running along $\partial \Omega$; that is, it is an operator that moves an $e$-particle (plaquette excitation) in a closed loop along the boundary (Figure~\ref{fig:toric-code-detector}b).  Acting on $| \psi \rangle$ with $\delta^\Omega_m$ thus braids an $e$-particle around whatever excitation is contained within $\Omega'$, which detects $m$ via its nontrivial mutual statistics with $e$.  This can equivalently be understood in terms of the commutator between $\delta^\Omega_m$ and a semi-infinite string operator creating the $m$ excitation (Figure~\ref{fig:toric-code-detector}c).

Similarly, we can construct a detector for $e$-particles as the infinite product $\delta_e = \prod_p B_p$ of plaquette stabilizers, which reduces to an $m$-string on the boundary upon truncation.  More generally, detectors in Pauli codes should be thought of as locally canceling infinite products of stabilizers.  Our formal approach will make this precise, and in fact will avoid the need to consider explicit truncations at all.

\begin{figure}[t]
\centering
\includegraphics[width=.65\textwidth]{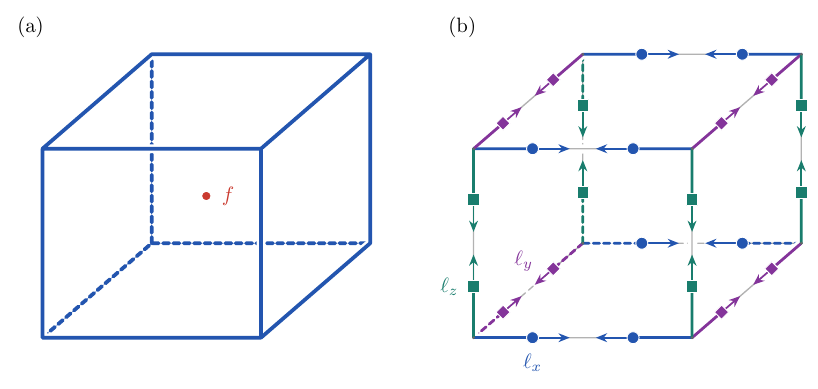}
\caption{Remote detection in the X-cube model. (a) A truncated product of cube stabilizers gives a cage operator supported on the twelve blue edges, detecting the enclosed fracton $f$. (b) The corresponding lineon process. A triple of perpendicular lineons is created at each corner, and the lineons move along the edges in the directions indicated by the arrows. Blue circles, purple diamonds and green squares denote $\ell_x$, $\ell_y$ and $\ell_z$, respectively. The lineons are shown before reaching the edge midpoints, where pairs of the same type annihilate to complete the cage operator. Dashed segments indicate rear edges; the microscopic cubic lattice is omitted. (Images generated by ChatGPT-6 Astra based on author-provided descriptions.)}
\label{fig:xcube-cage}
\end{figure}

In the toric code, the connection between remote detection and mutual statistics is obvious.  The truncated detectors are literally anyon string operators.  In fracton phases the situation is more subtle.  For instance, consider the X-cube model detector formed as an infinite product of all cube stabilizers, a truncated version of which is shown in Figure~\ref{fig:xcube-cage}a.  This is a ``cage operator,'' which detects any single fracton enclosed within.  The operator can be understood as a process where, for instance, we create triples of perpendicular lineons at each vertex of the cube, and then we move these along their corresponding edges and annihilate them at the edge midpoints (Figure~\ref{fig:xcube-cage}b).  This is a process where particles move around ``at infinity,'' but unlike in the case of an anyon string operator, we cannot associate this motion with a single particle type.  Nonetheless, it is the analog of mutual statistics in a fracton system, in the sense that it describes a collection of particles moving around ``at infinity'' to detect another particle.  This leads us to the following slogan:
\par\nopagebreak[4]
\vspace{0.4\baselineskip}
{\centering
\begingroup
\setlength{\fboxsep}{6pt}
\fbox{\textbf{Remote detection = mutual statistics}}
\par
\endgroup
}

Now we will make these ideas precise in a simple special case. We use the module $\widehat{R_n}^{\Q/\Z}\cong\widehat{R_n}^{\Z_n}$ of formal Laurent series with $\Z_n$ coefficients (see Appendix~\ref{subsec:notation}).
We consider a Lagrangian $\Z_n$ Pauli code, described by a Pauli fusion complex $F$. We assume the code has only pointlike excitations, so that $F$ is totally pointlike, as is its fusion module $S=H_{-1}(F)$. We also assume the fusion order is $n$. Finally, we assume that stabilizers can be chosen with no local relations, so we can take $F_i=0$ for $i>1$. The fusion complex is then simply
$0\to F_1\xrightarrow{d_1}F_0\xrightarrow{d_0}F_1^*\to0$.
These assumptions encompass many systems of interest, including the 2D toric code and 3D fracton models such as the X-cube model and the cubic code. The 3D toric code is excluded because of its local relations among plaquette stabilizers. We note that $F$ is a free resolution of $S$ over $R_n$, with $S$ placed in degree $-1$ as usual.

The next step is to consider $F \otimes_R \widehat{R_n}^{\Q/\Z}$.  Since each $F_i$ is a finitely generated free $R_n$-module, a choice of isomorphisms $F_i \cong (R_n)^{k_i}$ (equivalently, a choice of bases) gives
\[
F_i\otimes_R\widehat{R_n}^{\Q/\Z}
\cong\bigl(\widehat{R_n}^{\Q/\Z}\bigr)^{k_i}.
\]
Thus elements in degree one represent possibly infinite products of stabilizers ({\bf infinite stabilizers}, for short). Similarly, elements in degree zero represent possibly infinite products of Pauli operators, and in degree $-1$ we have excitations of possibly infinite support. We call these {\bf infinite Paulis} and {\bf infinite excitations}, respectively, even though these modules also contain finite products of Pauli operators and excitations of finite support.

We define the detection module (see Definition~\ref{def:detection-modules}) as
\[
D=H_1\bigl(F\otimes_R\widehat{R_n}^{\Q/\Z}\bigr)
=\ker\bigl(d_1\otimes\id\bigr).
\]
Elements of $D$, called detectors, are precisely the infinite stabilizers that map to zero in the module of infinite Paulis: the locally canceling products sought in the physical discussion above.  Since $F$ resolves $S$ in degree $-1$, we have
$D\cong\Tor^{R_n}_2(S,\widehat{R_n}^{\Q/\Z})$.
This shows that the detection module depends only on the fusion and mobility data encoded in $S$. It also agrees with the definition in \cite{Ruba_2025}, expressed in our notation as $D\cong\Tor^{R_n}_1(P/L,\widehat{R_n}^{\Q/\Z})$. Indeed, the construction in Section~\ref{subsec:Pauli-BFC} gives a free resolution $0\to F_1\xrightarrow{d_1}F_0\to P/L\to0$ over $R_n$, so this Tor module is the same kernel $\ker(d_1\otimes\id)$.

Because elements of $D$ are (projective) stabilizers, they naturally act on excitations of bounded support, as represented by elements of $F_{-1} = F^*_1$.  One way to think of this action is via the truncation procedure described above.  Because a detector is built from local stabilizers, its eigenvalue does not depend on the truncation as long as we truncate to a sufficiently large region.  Moreover, the eigenvalue only depends on the superselection sector of an excitation; the support of a truncated detector lies far away from the excitation in question, so acting with a local operator near the excitation to modify it without changing its superselection sector cannot change the eigenvalue.  Equivalently, the detector eigenvalue relative to the vacuum is the commutator phase of the
truncated detector with a creation operator for the enclosed excitation, as illustrated in Figure~\ref{fig:toric-code-detector}c for the toric code. Such a creation operator is infinite for a nontrivial excitation, but it too can be cut off far away from the truncated
detector. 

Therefore we obtain the excitation-detector pairing
\begin{equation} \label{eqn:physical-ED-pairing}
\theta : D \times S \to \Z_n \subset \Q/ \Z   \text{,}
\end{equation}
which is $\Z$-bilinear in its arguments and translation-invariant, in the sense that $\theta(t \delta, t x) = \theta(\delta, x)$ for all $t \in \T$, $\delta \in D$ and $x \in S$.  
Section~\ref{subsec:ed-bfc} describes how to obtain this pairing directly from a BFC.

In this setting, the excitation-detector principle says that $\theta$ is two-sided nondegenerate: both associated maps
\[
\theta_L:D\to\Hom_{\Z_n}(S,\Z_n),\qquad
\theta_R:S\to\Hom_{\Z_n}(D,\Z_n)
\]
are injective. Injectivity of $\theta_L$ says that every nonzero detector is effective, while injectivity of $\theta_R$ is the principle of remote detectability.

\subsection{Excitation-detector principle for BFCs}
\label{subsec:ed-bfc}

Here we develop the general theory of the excitation-detector principle for BFCs. We use the character duals and their formal Laurent-series realizations introduced in Appendix~\ref{subsec:notation}, together with Proposition~\ref{prop:irc-coinduction}.

The first step is to introduce detectors.  It will be useful to formulate the excitation-detector pairing at the level of chain complexes, and the following definition is designed with this in mind:
\begin{defn}\label{def:detection-modules}
Given a tame fusion complex $F$, its associated {\bf detection complex} is $\fD := F \otimes_R \hR$.  The $i$th {\bf detection module} is $D_i:=H_{i+1}(\fD)$, where $i\geq0$.  We refer to $D_0$ as the {\bf pointlike detection module}, with elements called pointlike detectors.  The modules $D_i$ with $i>0$ are the {\bf higher detection modules}, and their elements are called higher detectors.  If $F$ is totally pointlike, then $D_0$ is the only possibly nonzero detection module (see Corollary~\ref{cor:detection-module-vanishing}), so in that case we often write $D = D_0$ and refer to $D$ as the detection module.
\end{defn}

\begin{remark} \label{rmk:detection-complex-is-derived}
The detection complex represents the derived tensor product $F \otimes_R^L \hR$, so, in particular, the quasi-isomorphism type of  $\fD$ only depends on that of $F$.  This result relies on the fact that $F$ is a tame fusion complex, and is shown in Proposition~\ref{v2:prop:tame-derived-tensor}.
\end{remark}

Our next task is to construct the excitation-detector pairing given a BFC.  First we need to introduce a key ingredient, the extended evaluation pairing of a tame fusion complex $F$.  It is similar to, and in fact can be obtained from, the canonical evaluation pairing of Definition~\ref{def:canonical-evaluation-pairing}.  It is important to emphasize that there is no braiding information in the extended evaluation pairing; it can be thought of as a ``pre-excitation-detector pairing'' that lets us obtain the actual excitation-detector pairing from the BFC braiding $\lambda$.

\begin{lemma}\label{lem:mu-map}
There is a canonical injective $R$-linear map, compatible with involutions,
\[
\mu:\RQZ\otimes_R\hR\longrightarrow\hR^{\Q/\Z},
\qquad \mu(q\otimes\widehat r)=q\widehat r,
\]
where the right-hand side is multiplication of the finite Laurent polynomial $q$ by the formal Laurent series $\widehat r$.  Its image is the $\Z$-torsion submodule of $\hR^{\Q/\Z}$, consisting of series annihilated by a single positive integer.  The map $\mu$ is not surjective.
\end{lemma}

\begin{proof}
Finite-support multiplication defines the stated $R$-linear map and commutes with involution.  
Any $x\in\RQZ\otimes_R\hR$ is a finite sum $x=\sum_i q_i\otimes\widehat r_i$. Choose a common denominator $m\geq1$, so that $q_i=(1/m+\Z)s_i$ for some $s_i\in R$. Then
\[
x=(1/m+\Z)\otimes\widehat r',\qquad
\widehat r'=\sum_i s_i\widehat r_i\in\hR.
\]
If $\mu(x)=0$, every coefficient of $\widehat r'$ is divisible by $m$, so $\widehat r'=m\widehat s$ for some $\widehat s\in\hR$. Hence
\[
x=(1/m+\Z)\otimes m\widehat s
=m(1/m+\Z)\otimes\widehat s=0,
\]
proving injectivity.

We also have $mx=0$, so every element in the image is annihilated by a positive integer. Conversely, if a series in $\hR^{\Q/\Z}$ is annihilated by $m$, each coefficient can be written as $a_t/m+\Z$ with $a_t\in\Z$. It is therefore the image of $(1/m+\Z)\otimes\sum_{t\in\T}a_t t$, proving the claim about the image. Finally, the series $\sum_{k\geq1}(1/k+\Z)t_1^k$ is not annihilated by any positive integer, so $\mu$ is not surjective.
\end{proof}

\begin{defn}\label{def:extended-evaluation}
Given a chain complex $A$ of $R$-modules, the {\bf extended evaluation pairing} is the degree zero map
\begin{equation}
\widehat{\ev}_A:\overline{A\otimes_R\hR}\otimes_R D_{\RQZ}(A)
\longrightarrow\hR^{\Q/\Z}[0]
\end{equation}
given, for homogeneous $a\in A_p$, $\varphi\in D_{\RQZ}(A)_q$ and $\hat r\in\hR$, by
\begin{equation}\label{eqn:extended-evaluation}
\widehat{\ev}_A(\overline{a\otimes\hat r}\otimes\varphi)
=
\begin{cases}
(-1)^{p}\mu(\varphi(\bar a)\otimes\overline{\hat r}) = (-1)^{p}\varphi(\bar a)\overline{\hat r}   & \qquad p = - q,
\\
0  & \qquad \text{otherwise},
\end{cases}
\end{equation}
and extended additively.  
Here $\hR$ is concentrated in degree zero, and the bar on $\hat r$ denotes the involution on $\hR$.
\end{defn}

This pairing is a chain map.  Indeed, using compatibility of $\mu$ with involution, its defining formula can be written as
\[
\widehat{\ev}_A(\overline{a\otimes\hat r}\otimes\varphi)
=\mu\bigl(\ev_A^{\RQZ}(\bar a\otimes\varphi)\otimes\overline{\hat r}\bigr).
\]
We thus obtain $\widehat{\ev}_A$ from the canonical evaluation chain map $\ev_A^{\RQZ}$ of Definition~\ref{def:canonical-evaluation-pairing} by making the natural identification
$\overline{A\otimes_R\hR}
\cong\overline{A}\otimes_R\overline{\hR}$,
permuting a degree zero factor,
applying $\ev_A^{\RQZ}\otimes\id$, applying the involution
$\overline{\hR}\to\hR$ to the remaining factor, and
composing with $\mu$.  This expresses $\widehat{\ev}_A$ as a composite of chain maps.
Writing $J=\hR^{\Q/\Z}$, the left and right chain adjoints of this
pairing are
\[
\begin{aligned}
\widehat{\ev}_{A,L}:A\otimes_R\hR&\longrightarrow D_J(D_{\RQZ}(A)),&
\widehat{\ev}_{A,L}(u)(\overline\varphi)
&=\overline{\widehat{\ev}_A(\bar u\otimes\varphi)},\\
\widehat{\ev}_{A,R}:D_{\RQZ}(A)&\longrightarrow D_J(A\otimes_R\hR),&
\widehat{\ev}_{A,R}(\varphi)(\bar u)
&=(-1)^{|u||\varphi|}\widehat{\ev}_A(\bar u\otimes\varphi),
\end{aligned}
\]
for homogeneous $u$ and $\varphi$. Both are chain maps.

\begin{theorem}\label{thm:extended-evaluation-comparison}
If $F$ is a tame fusion complex, then the left associated map of the extended evaluation pairing,
\begin{equation}
\widehat{\ev}_{F,L}: \fD
\longrightarrow\Hom_R(\overline{D_{\RQZ}(F)},\hR^{\Q/\Z}),\nonumber
\end{equation}
is a quasi-isomorphism.  In the $\Z_n$-tame case, it is a chain isomorphism.  For every tame $F$, it induces natural $R$-linear isomorphisms
\begin{equation}\label{eqn:detection-complex-homology}
	\frak e_{k,L} : 
H_k(\fD)\xrightarrow{\cong}
\Hom_R\bigl(\overline{H_{-k}(D_{\RQZ}(F))},\hR^{\Q/\Z}\bigr),
\qquad k\in\Z.
\end{equation}
\end{theorem}

\begin{proof}
See Appendix~\ref{app:extended-evaluation-proof}.
\end{proof}

\begin{corollary}\label{cor:detection-module-vanishing}
Let $F$ be a tame fusion complex in spatial dimension $d$.  If $F$ is complete, then the homology of the detection complex $\fD$ is concentrated in degrees $i = 1,\dots,d-1$.  In other words, the detection modules $D_0, D_1, \dots, D_{d-2}$ are the only possibly nonzero homologies.  If $F$ is proper, then the homology of $\fD$ is concentrated in degrees $i = 0, \dots, d$. If $F$ is totally pointlike, then the homology of $\fD$ is concentrated in degree $1$, with $D_0$ its only possibly nonzero homology module.
\end{corollary}

\begin{proof}
	By Equation~\ref{eqn:detection-complex-homology}, this follows from the opposite-degree bounds 
	for $D_{\RQZ}(F)$, which hold by Theorem~\ref{theorem:dimension-bounds}.
\end{proof}

\begin{defn}\label{def:homological-evaluation}
Let $F$ be a tame fusion complex. The {\bf homological evaluation pairings} are the $R$-sesquilinear pairings 
\[
	{\mathfrak e}_i : \overline {H_{i}(\frak D)} \otimes_R H_{-i}(D_{\RQZ}(F)) \to \widehat{R}^{\Q/\Z}
\text{,}\qquad i \in \Z
\]
induced by the extended evaluation pairing. Explicitly, if $\delta \in H_{i}\frak D$ and $x\in H_{-i}(D_{\RQZ}(F))$ are represented by cycles $u\in\fD_{i}$ and $\varphi\in D_{\RQZ}(F)_{-i}$, then
\[
\mathfrak e_i(\bar\delta\otimes x)=\widehat{\ev}_F(\bar u\otimes\varphi).
\]
The chain-map property makes this independent of the representatives. Equivalently, its left associated map is the isomorphism $\mathfrak e_{i,L}$ introduced in Theorem~\ref{thm:extended-evaluation-comparison}.
\end{defn}

The right associated maps of the homological evaluation pairings are
\[
\mathfrak e_{i,R}:H_{-i}(D_{\RQZ}(F))
\longrightarrow
\Hom_R\bigl(\overline{H_i(\frak D)},\hR^{\Q/\Z}\bigr),
\qquad
\mathfrak e_{i,R}(x)(\bar\delta)
=\mathfrak e_i(\bar\delta\otimes x).
\]
Under the canonical identification $H_{-i}\bigl(D_J(\frak D)\bigr) \cong
\Hom_R\bigl(\overline{H_i(\frak D)},J\bigr)$, these satisfy
$H_{-i}(\widehat{\ev}_{F,R})=(-1)^i\mathfrak e_{i,R}$.
While the maps $\mathfrak e_{i,L}$ are isomorphisms, this is often
not the case for $\mathfrak e_{i,R}$.  Although $H_{-i}(D_{\RQZ}(F))$ is always countable, $H_i (\frak D)$ can be uncountable. For example, consider a stack of 2D $\Z_2$ toric codes in $d=3$, with layers parallel to the $xy$-plane and regularly spaced along $z$. Here $H_{-1}(D_{\RQZ}(F)) \cong  S \cong(\Z_2[t^{\pm1}])^2$, where $t$ is the unit translation along $z$, and
$D_0 = H_1 (\frak D) \cong\bigl(\Hom_{\Z_2}(\Z_2[t^{\pm1}],\Z_2)\bigr)^2$.
Thus pointlike detectors are pairs of unrestricted formal Laurent series with $\Z_2$ coefficients, and $D_0$ is uncountable. Its algebraic dual $\Hom_{\Z_2}(D_0,\Z_2)$ is also uncountable, so the map $\mathfrak e_{1, R}$ from the countable module $H_{-1}(D_{\RQZ}(F))$ cannot be surjective.

In \cite{Ruba_2025}, it was emphasized that $\widehat{R_n}^{\Q/\Z}$ has the useful property of being an injective cogenerator.  We recall the definition:
\begin{defn}
Let $\cR$ be a commutative ring.  An {\bf injective cogenerator} $E$ in the category of $\cR$-modules is an injective module with the property that for every nonzero $\cR$-module $M$, there is a nonzero $\cR$-linear map $M \to E$.
\end{defn}

\begin{proposition}\label{prop:character-injective-cogenerator}
The module $\hR^{\Q/\Z}$ is an injective cogenerator in the category of $R$-modules.  Similarly, $\widehat{R_n}^{\Q/\Z}$ is an injective cogenerator in the category of $R_n$-modules.
\end{proposition}

\begin{proof}
Proposition~\ref{prop:irc-coinduction} showed injectivity of $\hR^{\Q/\Z}$ and $\widehat{R_n}^{\Q/\Z}$.
Let $M$ be a nonzero $R$-module and choose nonzero $x \in M$.  Denote by $\langle x \rangle \subset M$ the subgroup generated by $x$.  There is a nonzero group homomorphism $\langle x \rangle \to \Q / \Z$; if $x$ has finite order $n$, then choose the homomorphism defined by $x \mapsto 1/n + \Z$.  Otherwise, we can choose a homomorphism by sending $x$ to any nonzero element of $\Q/ \Z$.  Injectivity of $\Q/\Z$ implies this homomorphism can be extended to $M$, giving a nonzero element of $\Hom_\Z(M,\Q/\Z)$.  This shows $\hR^{\Q/\Z}$ is an injective cogenerator.

The argument for $\widehat{R_n}^{\Q/\Z}$ is very similar; we let $M$ be a nonzero $R_n$-module and replace $\Q / \Z$ with $\Z_n$.
\end{proof}

\begin{remark}  \label{rmk:dual-complex-vanishing-homologies}
Equation~\ref{eqn:detection-complex-homology} and the injective cogenerator property imply that $H_k(\fD)$ is nonzero if and only if $H_{-k}(D_{\RQZ}(F))$ is nonzero, for all $k \in \Z$ and for $F$ a tame fusion complex.  This allows us to make physical arguments for the vanishing of certain homologies of the dual fusion complex based on the homology of $\fD$.    Throughout this remark we assume a unique ground state, and we focus primarily on essentially $d$-dimensional systems.

First we consider $H_0(\fD)$, which has the physical interpretation
\begin{equation}
H_0(\fD) = \frac{\text{Infinite movers that create no excitations}}{\text{Infinite stabilizers}} \text{.} \label{eqn:H0-dc-interpretation}
\end{equation}
In an essentially $d$-dimensional system, this quotient should be zero; any operator that creates no excitations should be a product of stabilizers.  Otherwise, the Euclidean-space ground state would not be unique.  This implies $H_0(D_{\RQZ}(F)) = 0$, and in Section~\ref{sec:dual_complex} we gave a direct physical argument -- in the sense that it did not rely on Equation~\ref{eqn:detection-complex-homology} -- for the vanishing of this homology.  It was also noted that $H_0(D_{\RQZ}(F))$ can be nonzero at the boundary of an essentially $(d+1)$-dimensional bulk.  Indeed, this is consistent with the interpretation above. At the 2D toric code boundary discussed in Section~\ref{subsec:example1D}, an infinite product of Pauli $X$ operators over the boundary edges cannot be written as a product of stabilizers with support near the boundary, so the quotient of Equation~\ref{eqn:H0-dc-interpretation} is nonzero.  However, this infinite boundary operator can be written as an infinite product of bulk stabilizers, as expected given the essentially $2$-dimensional nature of the bulk.

Next we consider
\begin{equation}
H_{-1}(\fD) = \frac{\text{Infinite excitations}}{\text{Excitations created by possibly infinite movers}} \text{.} \label{eqn:Hn1-dc-interpretation}
\end{equation}
In an essentially $d$-dimensional system, this should be zero; if a configuration cannot be created by a possibly infinite mover, it is not sensible to think of it as an excitation of a physical system.  We also have
\begin{equation}
H_{-2}(\fD) = \frac{\text{Infinite violations that do not induce higher violations}}{\text{Violations that come from infinite configurations}} \text{.} \label{eqn:Hn2-dc-interpretation}
\end{equation}
Because the role of violations is to measure the failure of configurations to be excitations, this quotient should again be zero in an essentially $d$-dimensional system; if there is a violation that cannot be produced by any configuration, this is a flaw in our description.  Summarizing and proceeding along similar lines for the homologies in larger negative degree, we conclude that $H_{-k}(\fD) = 0$ for all $k \geq 0$ in an essentially $d$-dimensional system, and thus the dual fusion complex has its homology concentrated in negative degree.  We thus reach the conclusion that fusion complexes describing essentially $d$-dimensional systems should be complete.
\end{remark}

\begin{lemma}  \label{lem:right-associated-injective}
Let $M$ and $N$ be $R$-modules, and let $\mu : \overline{M} \otimes_R N \to \hR^{\Q/\Z}$ be an $R$-sesquilinear pairing with left and right associated maps $\mu_L : M \to \widehat{N}^{\Q/\Z}$ and $\mu_R : N \to \widehat{M}^{\Q/\Z}$.  Here we use the identifications of Proposition~\ref{prop:irc-coinduction}. If one of the associated maps is an isomorphism, then the other is injective.
\end{lemma}

\begin{proof}
Write $J=\hR^{\Q/\Z}$ and first suppose that $\mu_L$ is an isomorphism. For $0\ne x\in N$, the cogenerator property gives a nonzero map $R\bar x\to J$, which is necessarily nonzero on its generator $\bar x$. Injectivity of $J$ extends it to an $R$-linear map $f:\overline N\to J$ with $f(\bar x)\ne0$. Surjectivity of $\mu_L$ supplies $m\in M$ with $\mu_L(m)=f$. Thus
\[
\mu_R(x)(\bar m)=\mu(\bar m\otimes x)
=\overline{f(\bar x)}\ne0,
\]
where we view the associated maps as Hom-valued maps using Proposition~\ref{prop:irc-coinduction}. Hence $\mu_R$ is injective. If instead $\mu_R$ is an isomorphism, apply the same argument to the conjugate pairing $\bar n\otimes m\mapsto\overline{\mu(\bar m\otimes n)}$, which interchanges the two associated maps.
\end{proof}

\begin{corollary}\label{prop:homological-evaluation-right-injective}
Let $F$ be a tame fusion complex.  Then the right associated maps of the homological evaluation pairings,
\[
	\mathfrak e_{i, R} : H_{-i}(D_{\RQZ}(F)) \to \Hom_R(\overline{H_{i}(\frak D)}, \hR^{\Q/\Z} ), \qquad
	i\in \Z,
\]
are injective.
\end{corollary}

\begin{proof}
By Theorem~\ref{thm:extended-evaluation-comparison}, $\mathfrak e_{i,L}$ is an isomorphism. Lemma~\ref{lem:right-associated-injective} therefore implies that $\mathfrak e_{i,R}$ is injective.
\end{proof}

\begin{defn}
Let $(F, \lambda)$ be a tame BFC.  The {\bf excitation-detector pairing} is the chain map
\begin{equation}
\Theta : \overline{\fD} \otimes_R F \to \hR^{\Q/\Z}[0]
\end{equation}
given by $\Theta = -\widehat{\ev}_F \circ (\id_{\overline{\fD}} \otimes \lambda)$.  The {\bf homological
excitation-detector pairings},
\[\theta_i : \overline{D_i} \otimes_R Q_i \to \hR^{\Q/\Z},\] with $Q_i=H_{-(i+1)}(F)$ for all $i\geq0$, are the pairings induced on homology by $\Theta$; that is,
$\theta_i = -{\mathfrak e}_{i+1} \circ (\id \otimes H_{-(i+1)}(\lambda) )$, for $i \geq 0$.  We refer to $\theta_0$ as the pointlike excitation-detector pairing, and $\theta_i$ with $i > 0$ as the higher excitation-detector pairings.  If $F$ is totally pointlike, when the distinction with $\Theta$ is clear, we often refer to $\theta_0$ as the excitation-detector pairing.
\end{defn}

For a direct description, represent $\delta\in D_i$ and $x\in Q_i$ by cycles $u\in\fD_{i+1}$ and $v\in F_{-(i+1)}$. Then
\[
\theta_i(\bar\delta\otimes x)=\Theta(\bar u\otimes v).
\]
Changing either cycle by a boundary does not change this value, because $\Theta$ is a chain map to a
complex concentrated in degree zero. 
This agrees with the definition above since $\lambda(v)$ represents $H_{-(i+1)}(\lambda)(x)$.
Note that the constant coefficient of the pointlike pairing $\theta_0$ agrees with the physical $\Q/\Z$-valued pairing
$\theta$ described in Section~\ref{subsec:ed-phys}.

\begin{remark} [Homological braiding encodes mutual statistics]  \label{rmk:hb-encodes-ms}
Let $(F, \lambda)$ be a pointlike BFC, and recall the isomorphism $a_F : H_{-1}(D_{\RQZ}(F)) \to S^\vee$ from Definition~\ref{def:homological-braiding}.  Defining $\tilde{\mathfrak e} : \overline{D_0} \otimes_R S^\vee \to \hR^{\Q/\Z}$ by  $\tilde{\mathfrak e} = \mathfrak e_1 \circ (\id_{\overline{D_0}} \otimes a^{-1}_F )$, we have
$\theta_0 = -\tilde{\mathfrak e} \circ (\id_{\overline{D_0}} \otimes b )$.  The homological braiding $b$ thus directly determines the excitation-detector pairing.  This shows that $b$ encodes mutual statistics, and justifies calling pairs $(S,b)$ theories of mutual statistics as in Definition~\ref{def:tms}. 
\end{remark}

\begin{remark}[Physical interpretation of the higher excitation-detector pairings]  \label{rmk:physical-higher-ed-pairings}
For concreteness and simplicity, we focus on Pauli BFCs.  The physical interpretation of the pointlike excitation-detector pairing was given in Section~\ref{subsec:ed-phys}.  Here we sketch an interpretation of the higher excitation-detector pairings, focusing on $\theta_1 : D_1 \times Q_1 \to \Q/ \Z$. Within this remark, $\theta_i$ denotes the constant component of the formal-series-valued pairing defined above.  We give an interpretation of the values of $\theta_1$ in terms of the action of operators on states.

Our discussion builds on Section~\ref{subsec:hfm-interpretation}, and we follow the setup and notational conventions of that section.  In particular, all modules are $R_n$-modules, except that the selection modules $\cS_i$ are only assumed to be $\Z_n$-modules. $\widehat{M}$ denotes the $\Z_n$ character dual.  In Section~\ref{subsec:hfm-interpretation}, we described how to associate a class in $Q_1$ to a state $\varphi \in \cS_1 \subset \widehat{F_1}$.  Using the chosen pair $(\Pi_1, \Omega_1)$ of a $(d-1)$-dimensional hyperplane $\Pi_1$ and a half-space $\Omega_1$, we consider $\varphi_2 = \widehat{d_2} \trunc_{\Omega_1}(\varphi)$, which is a cycle in $F_{-2} = F^*_2$.  We thus denote the associated $Q_1$ class by $[\varphi_2]$.

To proceed, we choose a detector $[\delta] \in D_1$, and a representative cycle $\delta \in \fD_2$.  The cycle $\delta$ is an infinite formal sum of local relations, but does not directly have the interpretation of an operator that can act on $\varphi$.  Similar to elements of $\widehat{F_i}$ as discussed in Section~\ref{subsec:hfm-interpretation}, elements of $\fD_i$ can be truncated to a region $X \subset \Z^d$.  Using the same bases $f^i_1,\dots,f^i_{k_i}$ as in that section, an element of $\fD_i$ is a formal sum $\sum_{t\in\T}\sum_{j=1}^{k_i}a_{tj}\,t f^i_j$ with $a_{tj}\in\Z_n$, with no restriction on support. Its truncation to $X$ retains precisely the terms with $t\in X$. These truncations are $\Z_n$-linear, and use the same lattice positions as the truncations of functionals in Section~\ref{subsec:hfm-interpretation}.  

Let $B_\ell$ be the ball of radius $\ell \in \N$ centered at a fixed point of $\Pi_1$. Truncation to $B_\ell$ means truncation to $B_\ell\cap\Z^d$.  We consider the sequence of truncated detectors $\delta_\ell = \trunc_{B_\ell}(\delta)$.  These are finite sums of local relations and thus $\delta_\ell \in F_2$.  We define
\begin{equation}
\gamma_\ell =  \trunc_{\Omega_1}(d_2 \delta_\ell) - d_2 \trunc_{\Omega_1}(\delta_\ell) \text{.}
\end{equation}
This is a finite formal sum of stabilizers, so $\gamma_\ell \in F_1$, and moreover $\gamma_\ell$ has support within a bounded distance of $\Pi_1$ as $\ell$ increases. The associated Pauli operator $d_1\gamma_\ell$ has support within a bounded distance of the intersection $\Pi_1\cap\partial B_\ell$, with the bound independent of $\ell$. Its support therefore recedes to infinity as $\ell$ increases.

This construction produces a family of operators $\gamma_\ell$ that can act on the state of interest. The resulting additive eigenvalue phase is $\varphi(\gamma_\ell)$, meaning that the eigenvalue is $e^{2\pi i\varphi(\gamma_\ell)}$. Because $\varphi$ is a state, it vanishes on the finite relation $d_2\trunc_{\Omega_1}(\delta_\ell)$, and hence
\begin{equation}
\begin{aligned}
\varphi(\gamma_\ell)
&=\varphi\bigl(\trunc_{\Omega_1}(d_2\delta_\ell)\bigr)
 =(\trunc_{\Omega_1}\varphi)(d_2\delta_\ell)\\
&=(\widehat{d_2}\trunc_{\Omega_1}\varphi)(\delta_\ell)
 =\varphi_2(\delta_\ell)\text{.}
\end{aligned}
\end{equation}
Because $\varphi_2$ has bounded support, the value $\varphi(\gamma_\ell)$ stabilizes to the value $\varphi_2(\delta)$ for sufficiently large $\ell$, once $B_\ell$ contains the support of $\varphi_2$.  Taking the constant component in the definition of $\theta_1$, we therefore obtain
\begin{equation}
-\theta_1([\delta], [\varphi_2]) = \varphi_2(\delta) = \varphi(\gamma_\ell) ,
\end{equation}
for $\ell$ sufficiently large.  This construction is illustrated for the remote detection of a flux line excitation in the 3D toric code in Figure~\ref{fig:higher-detector-flux-line}.
\begin{figure}
\centering\includegraphics[width=0.88\textwidth]{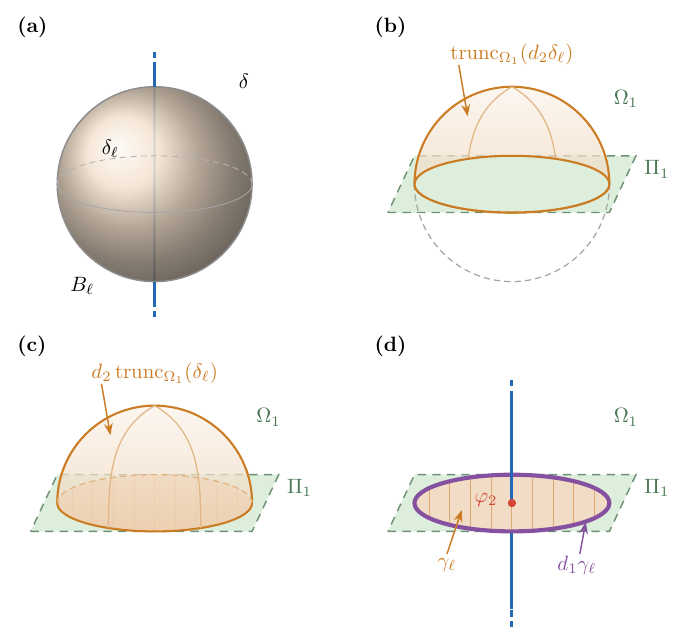}
\caption{Remote detection of a flux line (blue) in the 3D $\Z_2$ toric code. Each cube gives a local relation among its six plaquette stabilizers. {\bf (a)} The cycle $\delta$ is the formal sum of all cube relations; the truncation $\delta_\ell$ retains those in $B_\ell$.  {\bf (b)} The stabilizer $d_2\delta_\ell$ has support near the spherical shell $\partial B_\ell$; truncation to $\Omega_1$ retains its upper cap (orange). {\bf (c)} If instead we first truncate $\delta_\ell$ to $\Omega_1$, and then apply $d_2$, we obtain a stabilizer supported near a hollow half-ball (orange) consisting of the same cap in panel (b) plus the disk below it lying in $\Pi_1$.  {\bf (d)} Taking the difference of the stabilizers depicted in (b) and (c) gives $\gamma_\ell$.  The caps cancel, leaving a disk of plaquette stabilizers (orange).  The associated Pauli operator $d_1 \gamma_\ell$ has support only near the circle $\Pi_1\cap\partial B_\ell$ (purple). The circle links the blue flux line, resulting in an eigenvalue of $-1$ when the Pauli operator acts on the flux-line state. The red point marks the bounded violation $\varphi_2$ exposed by truncating the state at $\Pi_1$. In all panels, the plane $\Pi_1$ is shown in green and $\Omega_1$ is the upper half-space. Dashed blue ends indicate continuation to infinity. (Images generated by ChatGPT-6 Astra based on author-provided descriptions.)}
\label{fig:higher-detector-flux-line}
\end{figure}

For $1< i < d$, use the successive cuts of Section~\ref{subsec:hfm-interpretation}, so that $\varphi_{j+1}=\widehat{d_{j+1}}\trunc_{\Omega_j}(\varphi_j)$. We write $C_j=\trunc_{\Omega_j}d-d\trunc_{\Omega_j}$, with truncation acting in the appropriate degree. We choose balls $B_\ell$ with a common center lying in $\Pi_1 \cap \cdots \cap \Pi_i$.  Given a detector cycle $\delta\in\fD_{i+1}$, set $\gamma_\ell=C_1\cdots C_i\delta_\ell\in F_1$; note that the cuts act on the detector in the reverse order from the action on states. Since each $\varphi_j$ vanishes on the image of $d_{j+1}$, repeated evaluation gives $\varphi(\gamma_\ell)=\varphi_{i+1}(\delta_\ell)$. This stabilizes for $\varphi\in\cS_i$ at sufficiently large $\ell$.  The support of $d_1 \gamma_\ell$ lies within a bounded distance of the intersection $\partial B_\ell \cap \Pi_1 \cap \cdots \cap \Pi_i$, with the bound independent of $\ell$.  With the grading convention of Equation~\ref{eqn:extended-evaluation} and the minus sign in the definition of $\Theta$, the stabilized phase is $(-1)^i\theta_i([\delta],[\varphi_{i+1}])$.
\end{remark}

\begin{defn}[Excitation-detector principle for BFCs]
Let $(F, \lambda)$ be a tame BFC. We say $(F, \lambda)$ satisfies the excitation-detector principle if $\theta_i$ is two-sided nondegenerate for all $i \geq 0$; that is, both associated maps $\theta_{i,L}$ and $\theta_{i, R}$ are injective.
\end{defn}

The following theorem is the main result of this section.  Essentially, it tells us that the excitation-detector principle is equivalent to nondegeneracy of a BFC.  The theorem restricts to fusion complexes with $H_0(F) = 0$, which is natural from a physical point of view when discussing remote detection.  We recall the physical interpretation that $H_0(F) = 0$ corresponds to a system with a unique ground state in Euclidean space.  Remote detection has to do with properties of excitations, and should always be discussed for excitations \emph{above a particular ground state}.  The simplest way to ensure this is to assume the ground state is unique.  Physically speaking, we lose nothing by making this assumption; if a system has degenerate gapped ground states, we can always add a small perturbation to the Hamiltonian to select one of them while leaving the universal fusion and statistics properties of the excitations above it unchanged (ignoring domain wall excitations).

\begin{theorem}\label{thm:bfc-excitation-detector}
Let $F$ be a tame fusion complex with $H_0(F) = 0$, and let $(F, \lambda)$ be a BFC.  Then the following are equivalent:
\begin{enumerate}
\item The BFC $(F, \lambda)$ is nondegenerate.
\item The left associated map $\Theta_L$ of the excitation-detector pairing is a quasi-isomorphism.
\item The excitation-detector principle is satisfied by the BFC $(F, \lambda)$.
\end{enumerate}
\end{theorem}

\begin{proof}
	Write $J=\hR^{\Q/\Z}$. The definition of $\Theta$ and naturality of the Hom--tensor adjunction give
	\[
	\Theta_L=-D_J(\lambda)\circ\widehat{\ev}_{F,L},
	\]
	where $D_J(\lambda):\Hom_R(\overline{D_{\RQZ}(F)},J)\to\Hom_R(\overline F,J)$ is precomposition with $\overline\lambda$. By Theorem~\ref{thm:extended-evaluation-comparison}, $\widehat{\ev}_{F,L}$ is a quasi-isomorphism. Since $J$ is an injective cogenerator, the functor $\Hom_R(\overline{(-)},J)$ is exact and detects isomorphisms. In particular,
	\[
	H_k\Hom_R(\overline A,J)\cong\Hom_R(\overline{H_{-k}(A)},J)
	\]
	shows that $D_J(\lambda)$ is a quasi-isomorphism if and only if $\lambda$ is.
	This shows that (1) holds if and only if (2) does. 

	Next,
	for $i\geq0$, put $f_i=H_{-(i+1)}(\lambda)$. The homological pairings satisfy
	\[
	\theta_{i,L}=-D_J(f_i)\circ\mathfrak e_{i+1,L},\qquad
	\theta_{i,R}=-\mathfrak e_{i+1,R}\circ f_i.
	\]
	The map $\mathfrak e_{i+1,L}$ is an isomorphism, while $\mathfrak e_{i+1,R}$ is injective. Moreover,
	\[
	\ker D_J(f_i)\cong\Hom_R(\overline{\coker f_i},J),
	\]
	so the cogenerator property implies that $D_J(f_i)$ is injective if and only if $f_i$
	is surjective. 
	Hence, $\theta_{i,L}$ is injective if and only if $f_i$ is surjective, while $\theta_{i,R}$ is
	injective if and only if $f_i$ is injective.
	Thus, the excitation-detector principle holds if and only if $f_{i}$ is an isomorphism for
	all $i \geq 0$. If $F$ is complete, these negative degrees are all the potentially nonzero homology degrees of $F$ and $D_{\RQZ}(F)$, since $H_0(F)=0$. In general, Theorem~\ref{v2:thm:tame-dual} identifies $D_{\RQZ}(F)$ with $\mathbb D_{R[1]}(F)$ in $\D(R)$, and Lemma~\ref{lemma:antisymmetric-iso} with $s=1$ shows that this condition is equivalent to $\lambda$ being a quasi-isomorphism. This completes the proof.
\end{proof}


\subsection{Detectors and mutual statistics in two dimensions}
\label{subsec:2d-detector-particles}

For abelian anyons in two dimensions, a truncated detector is a closed anyon string operator. The excitation-detector pairing should therefore
express the usual braiding of anyons, reducing to a symmetric
bilinear form on $S$ once we identify a
detector with the particle carried by its associated string operator.
Here we construct this
identification directly from a fusion complex. Cutting a detector along
a line gives an infinite string; cutting that string in half exposes
its particle at the endpoint.

Throughout this subsection, we work in $d=2$ and
take $F$ to be a pointlike tame
fusion complex.  We write $D = H_1(\fD)$ for the detection module.
\begin{defn}
A pointlike fusion complex $F$ is {\bf maximally mobile}
if translations act trivially on $S = H_{-1}(F)$.
\end{defn}
\noindent We assume $F$ is maximally
mobile throughout this subsection.  Physically, maximal mobility means every particle type
can be moved by one lattice spacing in either direction without changing its type.

Since $S$ is finitely generated over $R$ and
has finite fusion order, maximal mobility implies that $S$ is a
finite abelian group.  We note that a totally pointlike fusion complex in $d=2$, without assuming
maximal mobility, has
a finite abelian fusion group. Since translations act by
automorphisms of this finite group, their actions have finite
order, so maximal mobility can always be achieved after
sufficient coarsening.

We choose bases in the terms of $F$, and regard elements of
$\fD=F\otimes_R\hR$ as unrestricted configurations on the spatial lattice.
In each degree $k$, define additive projections
$X,Y:\fD_k\to\fD_k$: $X$ retains the sites $(i,j)$ with $j\geq0$
and sets all other coefficients to zero, while $Y$ retains those
with $i<0$. Define
\begin{equation}\label{eq:ed2-cuts}
 K_x=dX-Xd,\qquad K_y=dY-Yd,\qquad C=K_yK_x.
\end{equation}
The values of $K_x$ are supported near the horizontal axis,
while those of $K_y$ have support near the vertical axis.
Moreover, $C$ takes any configuration to one of bounded support
near the spatial origin. We have identities
\begin{equation}\label{eq:ed2-boundaries}
 dK_r+K_rd=0,\qquad dC=Cd.
\end{equation}

If $\delta\in\fD_1$ is a detector cycle, then $K_x\delta=dX\delta$
is a string with no endpoint, while its left half has the finite
endpoint $dYK_x\delta=C\delta$. Thus the corner operator $C$ defines
the additive map
\begin{equation}\label{eq:ed2-Sigma}
 \Sigma:D\longrightarrow S,\qquad [\delta]\longmapsto[C\delta].
\end{equation}
The map $\Sigma$ is the desired assignment of a particle type to a detector,
and no braiding was used in its construction.
We note that the orientation matters: cutting the vertical string instead gives
\begin{equation} \label{eqn:corner-identity}
[K_xK_y\delta]= [dX K_y \delta] =-[C\delta] .
\end{equation}
Indeed,
\[
 K_yK_x+K_xK_y=dH+Hd,\qquad H=YK_x-K_xY,
\]
and the values of $H$ have bounded support.

\begin{proposition}\label{prop:ed2-Sigma}
The map $\Sigma$ is an $R$-module isomorphism, natural under
chain maps between maximally mobile pointlike fusion complexes.
It is independent of the chosen bases 
and finite translations of the cuts specified by $X$ and $Y$.
\end{proposition}

\begin{proof}
We first check naturality, before using maximal mobility. For a chain
map $f:F\to G$, also writing $f$ for its extension
$f\otimes_R\operatorname{id}_{\widehat R}$ to the detection complexes, put
\[
 \Delta_x=X_Gf-fX_F,\qquad \Delta_y=Y_Gf-fY_F,
 \qquad J=\Delta_yK_x^F-K_y^G\Delta_x.
\]
The values of $\Delta_x,\Delta_y$ are supported in strips, while those of $J$ have
bounded support. Direct expansion gives  $C_Gf-fC_F=dJ+Jd$, and therefore
$\Sigma_G \circ H_1(f) =
H_{-1}(f) \circ \Sigma_F$, which is the statement of naturality.
 This also treats
basis changes and translated cuts, by writing the identity chain map
with the two choices. Taking $f$ to be a lattice translation proves
$R$-linearity.

It remains to prove that $\Sigma$ is an isomorphism. First consider
$S=\Z_n$ with trivial translation action. We can represent this
fusion module by the vertex-excitation sector of the $\Z_n$ toric
code. Its fusion complex $P(n)$ is
\[
 R_n P \xrightarrow{d}
 R_n W_x\oplus R_n W_y \xrightarrow{d} R_n u,
\]
concentrated in degrees $1,0,-1$, with
\[
 dP=-(y-1)W_x+(x-1)W_y,\qquad
 dW_x=(x-1)u,\qquad dW_y=(y-1)u.
\]
Here $P$ represents a plaquette stabilizer, while $W_x,W_y$
are elementary movers of the vertex excitation $u$.
This is a tame fusion complex with homology $\Z_n$ in degree
$-1$ and zero in every other degree.

In the detection complex, a configuration of plaquette stabilizers
is a cycle precisely when its coefficients agree on neighboring
plaquettes. Thus every detector cycle is a multiple of $NP$, where
\[
 N=\sum_{i,j\in\Z}x^iy^j.
\]
Consequently, the detection module is $\Z_n$, generated by $[NP]$.
Cutting this uniform plaquette configuration first produces an
infinite horizontal string, and cutting that string in half exposes
its endpoint:
\[
 K_x(NP)=\left(\sum_{i\in\Z}x^i\right)W_x,
 \qquad C(NP)=u.
\]
Hence $\Sigma$ is an isomorphism for $P(n)$.

Now choose a cyclic decomposition $S\cong\bigoplus_i\Z_{n_i}$.
Since translations act trivially, this is also an $R$-module
decomposition. For each summand, choose a bounded resolution $Q_i$ of $\Z_{n_i}[-1]$
by finitely generated free $R$-modules and a quasi-isomorphism
$Q_i\to P(n_i)$. Both complexes are tame, so naturality of
$\Sigma$ and invariance of detector homology under
quasi-isomorphisms (Remark~\ref{rmk:detection-complex-is-derived})
show that $\Sigma$ is an isomorphism for $Q_i$.
The finite direct sum $Q=\bigoplus_iQ_i$ is $\Z$-tame, and
$\Sigma$ is an isomorphism for $Q$ because the corner construction
respects direct sums. Finally,
Lemma~\ref{v2:lem:truncated-gamma}(b) gives a quasi-isomorphism
$Q\to F$ inducing the chosen identification of fusion modules.
Naturality and invariance of detector homology then imply that
$\Sigma$ is an isomorphism for $F$.
\end{proof}

Now we equip $F$ with a braiding $\lambda$.  Let $\operatorname{ct}$ denote extraction of the constant coefficient.
The physical $\Q/\Z$-valued excitation-detector pairing is
\[
 \theta([\delta], [e])
 =\operatorname{ct}\bigl(\theta_0([ \bar\delta ] \otimes [e])\bigr) 
 \in\Q/\Z ,
\]
where $\delta \in \fD_1$ and $e \in F_{-1}$ are cycles. Let $\Omega$ and $\Xi$ be the sesquilinear pairings with associated
maps $\Omega_R=-\lambda_0$ and $\Xi_L=\lambda_1$, as in
Section~\ref{subsec:bfc1}. The physical commutator and
stabilizer-excitation pairing are
$\omega=\operatorname{ct}\circ\Omega$ and
$\xi=\operatorname{ct}\circ\Xi$, respectively. In the theorem and proof below, each expression
with infinite string operators is a finite sum: only the bounded intersection
of transverse strips contributes. Equivalently, such expressions can be
understood as commutators of truncated strings, which stabilize for
sufficiently large truncations.

\begin{theorem}\label{thm:ed2-mutual}
The excitation-detector pairing determines a symmetric bilinear form
on particle types,
\begin{equation}\label{eq:ed2-mutual}
 b_a(a,b)=\theta(\Sigma^{-1}a,b).
\end{equation}
For detector cycles $\delta,\delta'$, it is the crossing commutator
\begin{equation}\label{eq:ed2-crossing}
 b_a(\Sigma[\delta],\Sigma[\delta'])
   =\omega(K_x\delta,K_y\delta').
\end{equation}
This form is invariant under BFC equivalence. It is nondegenerate
if and only if the BFC is nondegenerate.
\end{theorem}

\begin{proof}
The form $b_a$ is bilinear and well-defined because $\theta$ is a pairing
on homology and $\Sigma$ is an isomorphism. We establish the crossing
formula and symmetry directly, using the chain-map identity
\[
 \xi(Z,dU)=\omega(dZ,U).
\]

First, moving either cut does not change the crossing commutator.
Indeed, replacing $X$ by a translated cut $X'$ changes $K_x\delta$
by $d((X'-X)\delta)$, since $d\delta=0$. Consequently,
\[
 \omega(d((X'-X)\delta),K_y\delta')
 =\xi((X'-X)\delta,dK_y\delta')=0.
\]
The same argument applies to the other cut. These expressions are
finite sums because the strips supporting the two arguments have
bounded intersection.

Let $\delta,\delta'$ be detector cycles, and truncate $\delta$ to
a large square centered at the intersection of the cuts, obtaining
$\delta_B$. Its boundary $d\delta_B$ is supported near the boundary
of the square. For either half-string $U$ considered below, $dU$
has bounded support near the origin. Once the square is sufficiently
large, locality and the chain-map identity give
\[
 \theta([\delta],[dU])
 =\xi(\delta_B,dU)
 =\omega(d\delta_B,U).
\]
Thus the excitation-detector pairing can be computed at the intersection of
the half-string with the boundary of the square.

By Equation~\ref{eqn:corner-identity} we have $[C\delta']=[-dXK_y\delta']$.
Take $U=-XK_y\delta'$, the upward half of the vertical string with
its orientation reversed; it satisfies $[dU] = \Sigma [\delta']$.
The half-string $U$ intersects only the top edge of the
square. Near this intersection, $d\delta_B$ is $-K_x\delta$, with
the horizontal cut translated to that edge, while $U$ agrees with
$-K_y\delta'$. By independence of the crossing commutator from the positions
of the cuts, we obtain
\[
 \theta([\delta],\Sigma[\delta'])
 =\omega(K_x\delta,K_y\delta').
\]
This proves Equation~\ref{eq:ed2-crossing}.

Alternatively, use the representative
$C\delta'=dYK_x\delta'$ and take $U=YK_x\delta'$; again we have
$[dU] = \Sigma [\delta']$. The 
horizontal half-string $U$ intersects only the left edge of the square,
where $d\delta_B$ is $-K_y\delta$, with the vertical cut translated
to that edge. Hence
\[
 \theta([\delta],\Sigma[\delta'])
 =-\omega(K_y\delta,K_x\delta')
 =\omega(K_x\delta',K_y\delta),
\]
using antisymmetry of $\omega$. Comparing the two
expressions and using surjectivity of $\Sigma$ proves symmetry
of $b_a$.

Naturality of $\Sigma$ and the homological excitation-detector
pairing shows that quasi-isometries of BFCs preserve $b_a$.
An antisymmetric homotopy of the braiding does not change the
homological pairing: its boundary terms vanish when evaluated on
cycles. Thus $b_a$ is invariant under BFC equivalence.

Finally, $\Sigma$ identifies the two source modules of $\theta$ with $S$.
Since translations act trivially, the Laurent-series-valued pairing
$\theta_0$ has equal coefficients at every lattice site, so its
constant component contains its entire value. Thus $b_a$ is
nondegenerate precisely when $\theta_0$ is two-sided nondegenerate.
The higher fusion and detection modules vanish in this setting,
so this is precisely the excitation-detector principle.
Theorem~\ref{thm:bfc-excitation-detector} now implies that $b_a$
is nondegenerate if and only if the BFC is nondegenerate.
\end{proof}

The isomorphism $\Sigma$ makes the usual physical picture precise:
a detector carries the particle exposed by cutting its string, and its
action on another particle equals the commutator of their crossing
strings. Symmetry expresses the fact that this mutual statistic depends
on the two particle types, with either one serving as the detector.
The construction therefore identifies the excitation-detector pairing
with familiar mutual braiding statistics for maximally mobile pointlike theories in two
dimensions.

At odd fusion order, this identification also gives a complete
invariant, by the classification of theories of mutual statistics
in Section~\ref{subsec:tpe}.

\begin{corollary}\label{cor:ed2-odd-completeness}
If two maximally mobile, pointlike nondegenerate BFCs in two dimensions are equivalent,
then their associated symmetric bilinear forms $(S,b_a)$ and $(S',b_a')$ are
isometric.  The converse holds at odd fusion order.  Moreover, every nondegenerate symmetric bilinear
 form on a finite abelian group is realized by such a BFC.
\end{corollary}

\begin{proof}

First, every finite abelian group $S$ with trivial translation action
is totally pointlike. For a cyclic summand $\Z_n$, use the
toric-code fusion complex $P(n)$ from the proof of
Proposition~\ref{prop:ed2-Sigma}. Its dual is quasi-isomorphic to the plaquette excitation
sector of the toric code, which has homology only in degree $-1$, where it is
isomorphic to $\Z_n$.
Each cyclic summand is therefore totally
pointlike. Total pointlikeness is preserved under finite direct sums, so $S$
is totally pointlike. Moreover, $S^\vee$ is a finite direct sum of
groups isomorphic to the cyclic summands of $S$, and is therefore finite.

We use the homological evaluation pairing $\tilde{\mathfrak e}$ of
Remark~\ref{rmk:hb-encodes-ms}.  After taking constant coefficients,
its left associated map gives an isomorphism
 $D \cong \Hom_{\Z}(S^\vee, \Q / \Z)$.  Because the groups are finite,
we thus have $|D| = |\Hom_{\Z}(S^\vee, \Q / \Z)| = | S^\vee |$.  The right associated
map gives an injection $S^\vee \to \Hom_{\Z}(D, \Q / \Z)$; this is an isomorphism
because the finite groups have the same order. We define
\begin{equation}\label{eq:ed2-kappa}
 \begin{split}
 \kappa_S:S^\vee&\longrightarrow\Hom_\Z(S,\Q/\Z),\\
 \kappa_S(z)(a)&=-\operatorname{ct}\bigl(
 \tilde{\mathfrak e}(\overline{\Sigma^{-1}a}\otimes z)
 \bigr).
 \end{split}
\end{equation}
This map requires no choice of braiding, and the remarks above show it is
an isomorphism.  Since translations act trivially on $S$ and its character dual,
postcomposition with $\kappa_S$ induces an isomorphism
\[
 \kappa_{S *} : \Hom_R(S,S^\vee)
 \longrightarrow
 \Hom_{\Z}\bigl(S,\Hom_{\Z}(S,\Q/\Z)\bigr).
\]
That is, it sends maps $S\to S^\vee$ to right associated maps
of $\Q/\Z$-valued bilinear forms on $S$.
For the homological
braiding $b$ of a BFC, Remark~\ref{rmk:hb-encodes-ms} gives
\begin{equation}\label{eq:ed2-b-comparison}
 b_a(a,c)= \theta(\Sigma^{-1} a, c) = \kappa_S(b(c))(a).
\end{equation}
Equivalently, for the right associated map $b_{a R}$ of $b_a$, we have $b_{a R} = \kappa_{S *} ( b )$.

Every group isomorphism $h:S\to S'$ is $R$-linear because translations
act trivially. Naturality of extended evaluation and $\Sigma$ gives
\[
 \kappa_S(h^\vee z')=\kappa_{S'}(z')\circ h.
\]
In particular, $\kappa_S$ is independent of the chosen fusion complex
representing $S$.  This implies $\kappa_{S*} (h^\vee b' h) = h^* b'_{a R} h$, where
$h^*$ denotes precomposition with $h$.  It follows that $h$ is an isometry of 
the scalar forms if and only if $b = h^\vee b' h$.  Invariance of the scalar form under BFC equivalence was established
above. At odd fusion order, the converse now follows from
Theorem~\ref{thm:Sb-invariant}.

For realization, let $\beta:S\times S\to\Q/\Z$ be a nondegenerate
symmetric bilinear form on a finite abelian group. Choose a cyclic
decomposition $S=\bigoplus_i\Z_{n_i}$ with generators $a_i$.
For each $i$, choose $t_i\in\Q/\Z$ such that
\[
2t_i=\beta(a_i,a_i),\qquad n_i^2t_i=0.
\]
If $n_i$ is even, any half of $\beta(a_i,a_i)$ satisfies the second
condition, since
$n_i^2t_i=(n_i/2)n_i\beta(a_i,a_i)=0$.
If $n_i$ is odd, choose its unique half in the subgroup annihilated
by $n_i$. Define
\[
\vartheta\left(\sum_i m_i a_i\right)
=\sum_i m_i^2t_i+
\sum_{i<j}m_i m_j\beta(a_i,a_j).
\]
The relations $n_i^2t_i=0$ and
$2n_it_i=n_i\beta(a_i,a_i)=0$, together with bilinearity of $\beta$,
make this independent of the integer representatives $m_i$.
It is a homogeneous quadratic form with associated bilinear form
$\beta$, and is therefore nondegenerate.
Section~\ref{subsec:anyon-realization} constructs a maximally mobile,
pointlike nondegenerate BFC realizing $(S,\vartheta)$. Its associated
mutual-statistics form is $\beta$, as required.
\end{proof}

\section{Discussion}
\label{sec:discussion}

We have developed an algebraic framework for abelian fusion, mobility, and statistics in bosonic systems with excitations of finite fusion order. Fusion complexes encode fusion and mobility up to quasi-isomorphism, while a braiding equips a fusion complex with statistical data.  The resulting BFCs, up to equivalence, define BFTs. The framework encompasses arbitrary translation-invariant Pauli codes and every theory of abelian anyons in two dimensions, including chiral theories. We characterized properness and completeness of fusion complexes by bounds on the Krull dimensions of their homology modules (Theorem~\ref{theorem:dimension-bounds}), extending the corresponding bounds for Pauli codes \cite{Ruba_2024}. These results constrain the excitation content of nondegenerate BFCs; in particular, in two dimensions all excitations are pointlike and fully mobile.

For pointlike nondegenerate BFCs, the fusion module $S$ is either zero or pure $(d-2)$-dimensional Cohen--Macaulay, and the homological braiding is an antisymmetric isomorphism $b:S\to S^\vee$ to the statistical dual. At odd fusion order, these conditions give a full characterization of the possible fusion modules (Corollary~\ref{cor:full-fm-characterization}). The theory of mutual statistics $(S,b)$ is an invariant of the BFT, and at odd fusion order it gives an equivalent description: pointlike nondegenerate BFTs correspond bijectively to isometry classes of theories of mutual statistics (Theorem~\ref{thm:Sb-invariant}). Statistical duality also yields obstructions to realizing certain totally pointlike fusion modules.  To wit, neither the $X$ nor the $Z$ sector of the X-cube and anisotropic models can occur alone in a nondegenerate BFT, even after coarsening translation symmetry. In two dimensions, after suitable coarsening, the odd-order correspondence recovers abelian anyon theories of odd fusion order.  

We interpreted remote detection as the generalization of mutual statistics to fracton orders and formulated excitation-detector pairings for both pointlike and extended excitations. For a BFC with $H_0(F)=0$, nondegeneracy is equivalent to the excitation-detector principle (Theorem~\ref{thm:bfc-excitation-detector}). This gives a physical interpretation of nondegeneracy and supports its role in Conjecture~\ref{conj:nondeg-BFC}. There we proposed that every nondegenerate BFT is realized by a gapped phase of an essentially $d$-dimensional bosonic system, and that the BFT describing such a phase is a phase invariant. Establishing these claims is an important direction for future work.

A separate question is which gapped phases admit a BFT description.  We expect that every gapped abelian phase can be described by a BFT, but defining what it means to be gapped abelian is subtle when extended excitations are present.  At least some of the subtleties are illustrated by models with nontrivial three-loop braiding \cite{WangLevin_2014, Jiang_2014, WangWen_2015} in $d=3$, in particular the nontrivial Lin-Levin model \cite{LinLevin_2015}.  This model has two elementary pointlike charge types ($e_a$ and $e_b$) and two corresponding elementary flux loops ($m_a$ and $m_b$).  Two identical $a$-type loops that are unlinked with other excitations satisfy the fusion rule $m_a^2 = 1$.  However, when both loops are linked to a base $b$-type loop, the fusion rule becomes $m_a^2 = e_b$, as follows from the three-loop braiding statistics computed in \cite{LinLevin_2015}.  Thus the fusion rules for the $a$-type flux loops depend on the presence of the ``spectator'' $b$-type loop. This illustrates the difficulty of assigning loop types and fusion rules independently of the surrounding excitation configuration.

We suggest that a definition of gapped abelian systems can be given by focusing on configurations of asymptotically well-separated excitations with a surrounding region assigned to each excitation.  We assume that a suitable excitation type can be assigned to each region in a manner that does not depend on the excitations in other regions.    
We propose calling such a gapped system abelian if, whenever two excitation regions are joined, the type of the resulting composite excitation is uniquely determined by the constituent types, independently of all spectator excitations.
The Lin-Levin model is expected not to satisfy this criterion.  Making these ideas precise is left as a problem for future work.

\noindent\textit{Disclosure of generative AI tool use.}
The authors used ChatGPT, Claude, and Gemini, with various models
and versions available over the course of this work,  to assist with drafting and revising exposition; literature searches; exploring, developing, and checking mathematical arguments;
and preparing \LaTeX{} and figures (see individual figure captions).
 These tools were used interactively and iteratively
under the authors' direction. The authors reviewed and revised the
resulting material, checked the mathematical arguments and references,
and take responsibility for the final manuscript.

\acknowledgments{We are grateful to Theo Johnson-Freyd, Michael Levin, Nathanan Tantivasadakarn, Noah Welikson, and Dominic Williamson for useful discussions.  EW and MH are also grateful to Arpit Dua for a collaboration on related work. BY is indebted to B\l{}a\.{z}ej Ruba for years of fruitful collaboration and stimulating discussions. The work of WS is supported by U.S. Department of Energy Grant 5-29073, and the Simons Collaboration on Ultra-Quantum Matter (UQM) (Simons Foundation grant no. 651444).  Prior to January 2026 (EW) and September 2024 (MH), the research of EW and MH was supported by the U.S. Department of Energy, Office of Science, Basic Energy Sciences (BES) under Award number DE-SC0014415.  Subsequent work of EW and MH was supported by the UQM Simons Collaboration (Simons Foundation grant no. 651440).  Work of AB is supported by a Travel Support for Mathematicians grant from Simons Foundation International (SFI). BY is supported by the AMS–Simons Travel Grant and acknowledges the support of the Simons Collaboration on Global Categorical Symmetries. This work benefited from meetings of the UQM Simons Collaboration (Simons Foundation grant no. 618615).}

\appendix

\section{Notation, conventions and definitions}
\label{subsec:notation}

Here we briefly establish some of our mathematical notation and conventions, and recall some key standard definitions.
For chain-complex conventions, see Appendix~\ref{dra:new-conventions}.
 Standard mathematical terminology and notation are used freely throughout the paper without definition; we refer the reader to \cite{Weibel_1994, Rotman_2009} for basic material on homological algebra, and \cite{Atiyah_1969, Reid_1995, Eisenbud_1995} for commutative algebra.

The set of natural numbers is $\N = \{ 1,2,3,\dots \}$.  All rings are commutative and have an identity element.  The spatial dimension is denoted by $d \geq 1$, and the group of lattice translations is ${\cal T} = \Z^d$, which we write as a multiplicative group.  The symbol $R$ always denotes the $d$-variable Laurent polynomial ring $R = \Z[ {\cal T} ] \cong \Z[ t^\pm_1, \dots, t^\pm_d ]$ with integer coefficients.  Here the indeterminate $t_i$ represents a lattice translation by one unit in the $i$-direction, and a general translation $t \in {\cal T}$ is a monomial $t = t_1^{m_1} \cdots t^{m_d}_d$ for $m_i \in \Z$.  Any $r \in R$ can be written $r = \sum_{t \in {\cal T}} a_t t$, where only a finite number of the coefficients $a_t \in \Z$ are nonzero.  The {\bf constant component} of $r$ is the coefficient $a_1$.  Similarly, $R_n$ denotes the ring of Laurent polynomials with $\Z_n$ coefficients $R_n = \Z_n[ {\cal T} ] \cong \Z_n[ t^\pm_1, \dots, t^\pm_d ]$, with coefficients $a_t\in\Z_n$ in the analogous expansion.  Any $R_n$-module can be viewed as an $R$-module because $R_n = R / (n)$, where $(n)$ is the ideal generated by $n$.  When we wish to refer to a general commutative ring, we use a symbol such as ${\cal R}$ to avoid confusion with the Laurent polynomial ring $R$.

 A {\bf ring with involution} is a ring $\cR$ with a ring automorphism $r\mapsto\bar r$ whose square is the identity. An additive map $f:M\to N$ of $\cR$-modules is {\bf antilinear} if $f(rm)=\bar r f(m)$. A {\bf module with involution} is a module $M$ equipped with an antilinear map $m\mapsto\overline{m}$ whose square is the identity.

For any $\cR$-module $M$, its {\bf conjugate module} $\overline M$ has the same underlying abelian group, with scalar multiplication twisted by the involution on $\cR$. We write the element corresponding to $m\in M$ as $\overline{m}\in\overline M$, so that $r\overline{m}=\overline{\bar r m}$. This notation does not require an involution on $M$: a bar here denotes an element of a different module, whereas a specified involution sends $M$ to itself. An antilinear map $f:M\to N$ can equivalently be viewed as a linear map $\overline M\to N$ or $M\to\overline N$, or as an antilinear map $\overline M\to\overline N$, using the same function on the underlying sets. We often pass between these descriptions without comment. Conjugation is an exact involutive functor on $\cR$-modules.

The ring $R$ (resp. $R_n$) has a natural involution induced as the $\Z$-linear (resp. $\Z_n$-linear) extension of the group automorphism ${\cal T} \to {\cal T}$ that sends $t \mapsto t^{-1}$.  We denote the involution by $r \mapsto \bar{r}$.  Explicitly, $\overline{ \sum_{t \in {\cal T}} a_t t } = \sum_{t \in {\cal T}} a_t  \bar{t} = \sum_{t \in {\cal T}} a_t  t^{-1}$.  For modules over a commutative ring $\cR$ with involution, $\Hom_\cR(M,N)$ consists of $\cR$-linear maps, with scalar multiplication given by
\begin{equation}
(rf)(m)=r f(m)=f(rm).
\label{eq:new-hom-action}
\end{equation}
We use $\mathrm{Mod}(\cR)$ for the category of $\cR$-modules and linear maps. We let $M\otimes_\cR N$ be the tensor product of $\cR$-modules. The category $\mathrm{Mod}(\cR)$ is closed symmetric monoidal, with adjunction isomorphism
\[ \Hom_\cR(M \otimes_\cR N, L) \cong \Hom_\cR( N, \Hom_\cR(M, L)).  \]
This isomorphism sends $f \colon M\otimes_\cR N \to L$ to the map $\Phi(f) \colon N \to \Hom_\cR(M, L)$ given for $x\in M$ and $y\in N$ by
\[\Phi(f)(y)(x)= f(x\otimes y).\]
In particular, this convention applies over $R$ and $R_n$.

We define $R$-modules $\RQZ$ and $\RRZ$ by $\RQZ = \Q/\Z [t^\pm_1, \dots, t^\pm_d]$ and 
$\RRZ =  \R/\Z [t^\pm_1, \dots, t^\pm_d]$; that is, $\RQZ$ and $\RRZ$ consist of Laurent polynomials with $\Q/\Z$ and $\R/\Z$ coefficients, respectively.  There is an obvious inclusion $\RQZ \hookrightarrow \RRZ$, induced by $\Q/\Z \hookrightarrow \R/\Z$.  We note that $\RQZ$ and $\RRZ$ are $R$-modules with involution, where the involution acts on polynomials as in $R$ and $R_n$.

For a $\cR$-module $T$, we write $D_T(M)=\Hom_\cR(\overline M,T)$ for the dual of $M$ with coefficients in $T$. For a linear map $f:M\to N$, the dual map $D_T(f):D_T(N)\to D_T(M)$ is defined by precomposition, $D_T(f)(\varphi)=\varphi\circ\overline f$. When the coefficient target is clear, we sometimes write $M^*=D_T(M)$ and $f^*=D_T(f)$, or omit the subscript and write $D(M)$ and $D(f)$. In Section~\ref{sec:first}, we mainly use $T=\RRZ$; in subsequent sections of the main text, we use $T=\RQZ$ unless stated otherwise. For an $R_n$-module $M$, both coefficient choices give duals canonically isomorphic to $D_{R_n}(M)=\Hom_{R_n}(\overline M,R_n)$, using the coefficient embedding $\Z_n\hookrightarrow\Q/\Z\hookrightarrow\R/\Z$ given by $a\bmod n\mapsto a/n+\Z$.

For $\cR$-modules $M,N,C$, a {\bf bilinear pairing} $\eta:M\times N\to C$ is additive and $\cR$-linear in each variable; equivalently, it is a linear map $M\otimes_\cR N\to C$. A {\bf sesquilinear pairing} is additive in both variables, antilinear in the first and linear in the second:
\[
\eta(rm,sn)=\bar r s\,\eta(m,n).
\]
Equivalently, it is a linear map $\overline M\otimes_\cR N\to C$, sending $\overline{m}\otimes n$ to $\eta(m,n)$. A pairing on $M\times M$ is called a {\bf form}. A bilinear form is {\bf symmetric} or {\bf antisymmetric} if $\eta(n,m)=\eta(m,n)$ or $\eta(n,m)=-\eta(m,n)$, respectively. When $C$ has an involution, a sesquilinear form is symmetric or antisymmetric if $\eta(n,m)=\overline{\eta(m,n)}$ or $\eta(n,m)=-\overline{\eta(m,n)}$, respectively. A bilinear form is {\bf alternating} if $\eta(m,m)=0$ for every $m$; this implies antisymmetry, but antisymmetry alone need not imply alternation when the target has nonzero $2$-torsion.

For a sesquilinear pairing with values in a module $C$ with involution, the {\bf left and right associated maps} are
\[
\begin{aligned}
\eta_L:M&\to\Hom_\cR(\overline N,C),& \eta_L(m)(\overline n)&=\overline{\eta(m,n)},\\
\eta_R:N&\to\Hom_\cR(\overline M,C),& \eta_R(n)(\overline m)&=\eta(m,n).
\end{aligned}
\]
Both are linear. For a symmetric or antisymmetric form, $\eta_R=\eta_L$ or $\eta_R=-\eta_L$, respectively, and we call $\eta_R$ the associated map, often also denoting it by $\eta$. For the coefficient duals defined above, we thus write $\eta:M\to M^*$.

\label{def:irc-duality}
Given an $R$-module $M$, we define its {\bf integral dual} as $\widehat{M} = \Hom_\Z(M, \Z)$, and its {\bf rational character dual} as $\widehat{M}^{\Q/\Z} = \Hom_\Z(M, \Q/\Z)$. The $R$-module structure is given by $(r\varphi)(x)=\varphi(\bar r x)$ for $r\in R$, $x\in M$, and $\varphi$ in either dual. If $M$ is equipped with an involution, we define involutions on both duals by $\overline\varphi(x)=\varphi(\bar x)$. For an $R$-linear map $f:M\to N$, its dual maps $\widehat f:\widehat N\to\widehat M$ and $\widehat f^{\Q/\Z}:\widehat N^{\Q/\Z}\to\widehat M^{\Q/\Z}$ are defined by precomposition, $\varphi\mapsto\varphi\circ f$.

Given an $R_n$-module $M$, we define its {\bf $\Z_n$ character dual} as $\widehat{M}^{\Z_n}=\Hom_{\Z_n}(M,\Z_n)$, with $R_n$-action $(r\varphi)(x)=\varphi(\bar r x)$. If $M$ is equipped with an involution, we define $\overline\varphi(x)=\varphi(\bar x)$.   For an $R_n$-linear map we define $\widehat f^{\Z_n}:\widehat N^{\Z_n}\to\widehat M^{\Z_n}$ by precomposition, as above.

The embedding $\Z_n\hookrightarrow\Q/\Z$ given by $a\bmod n\mapsto a/n+\Z$ induces a natural $R_n$-linear isomorphism $\widehat{M}^{\Z_n}\cong\widehat{M}^{\Q/\Z}$ for every $R_n$-module $M$.

Elements of $\hR$ and $\hR^{\Q/\Z}$ can be thought of as formal Laurent series with $\Z$ and $\Q/\Z$ coefficients, respectively, with no restriction on their support. For concreteness, consider $\hR$. An element $\varphi\in\hR$ is specified by its values $\varphi(t)$ on monomials $t\in\T$, and we associate it with the formal expression $\widehat r=\sum_{t\in\T}\varphi(t)t$. Here $t$ denotes a monomial $t_1^{k_1}\cdots t_d^{k_d}$, and infinitely many of the coefficients $\varphi(t)$ are allowed to be nonzero. The $R$-module structure and involution above give the usual operations on these series: multiplication by a Laurent polynomial and reversal of all monomial exponents. With this in mind, we often denote elements of $\hR$ by symbols like $\widehat r$. Multiplication by a Laurent polynomial is well-defined because it has finite support; we do not multiply pairs of formal Laurent series.  Similarly, $\widehat{R_n}^{\Z_n}\cong\widehat{R_n}^{\Q/\Z}$ can be viewed as the module of formal Laurent series with $\Z_n$ coefficients.

\begin{proposition}\label{prop:irc-coinduction}
Given an $R$-module $M$, there are natural $R$-linear isomorphisms
\[
\widehat M\cong\Hom_R(\overline M,\hR),\qquad
\widehat M^{\Q/\Z}\cong\Hom_R(\overline M,\hR^{\Q/\Z}).
\]
For an $R_n$-module $M$, there is likewise a natural $R_n$-linear isomorphism
\[
\widehat M^{\Z_n}\cong\Hom_{R_n}(\overline M,\widehat{R_n}^{\Z_n}).
\]
Moreover, $\hR^{\Q/\Z}$ is an injective $R$-module, and $\widehat{R_n}^{\Q/\Z}$ is an injective $R_n$-module.
\end{proposition}

\begin{proof}
For $B=\Z$, $B=\Q/\Z$, or $B=\Z_n$ (in the $R_n$-module case), send $\chi\in\Hom_\Z(M,B)$ to the map $f_\chi$ defined by
\[
f_\chi(\overline x) = \sum_{t \in \T}  \chi(t x) t ,\qquad x\in M \text{.}
\]
The inverse is given by taking the constant component. 

Because $\Q/\Z$ is injective as an abelian group, the functor $\Hom_\Z(-,\Q/\Z)$ is exact.  We thus have that $\Hom_R(\overline{-}, \hR^{\Q/\Z})$ is also exact; since conjugation is an exact involutive equivalence, $\Hom_R(-, \hR^{\Q/\Z})$ is exact, so $\hR^{\Q/\Z}$ is an injective $R$-module.   Similarly, injectivity of $\widehat{R_n}^{\Q/\Z}$ follows from 
the isomorphisms $\Hom_{\Z_n}(M, \Z_n) \cong \Hom_{\Z}(M, \Z_n) \cong \Hom_{R_n}(\overline M, \widehat{R_n}^{\Q/\Z})$ for any $R_n$-module $M$, the fact that $\Z_n$ is injective as a $\Z_n$-module, and the same exact conjugation equivalence.
\end{proof}


\section{Derived-category foundations}
\label{app:derived_background}

We briefly review the passage from chain complexes to the derived category.
Our purpose is not to give a comprehensive account, but to establish the
conventions and results used later in this appendix and the paper. Throughout this subsection,
$\cR$ is a Noetherian commutative ring with involution $c \colon \cR \to \cR$ written as $\bar a:=c(a)$.  We use homological grading conventions.

\subsection{Conventions in chain complexes}
\label{dra:new-conventions}

For module conventions and the Hom--tensor adjunction, see Appendix~\ref{subsec:notation}.

Let $\Ch(\cR)$ be the category of chain complexes of $\cR$-modules.  Thus a
complex $F$ is a graded $\cR$-module $\bigoplus_{i\in \Z} F_i$ with differentials 
\[d^F:F_i\to F_{i-1}\]
such that $d^Fd^F=0$. For $x\in F_i$, we define its degree by $|x|=i$.

For each $s\in \Z$, the $s$th homology of $F$ is the $\cR$-module
\[H_s(F) = \ker( d^F:F_s\to F_{s-1})/\im(d^F:F_{s+1}\to F_{s}).\]

A chain map $f\colon F \to G$ is a map of graded $\cR$-modules $\{f_i \colon F_i \to G_i\}_{i \in \Z}$, where each $f_i$ is an $\cR$-module map such that $f d^F = d^G f$. Let $f,g\colon F\to G$ be chain maps. A \emph{chain homotopy from $g$ to $f$} is a collection of $\cR$-linear maps $\{ h_i \colon F_i \to G_{i+1}\}_{i \in \Z}$ such that
\[
 f_i-g_i=d^G h_i+h_{i-1}d^F\colon F_i\longrightarrow G_i
 \qquad\text{for all }i\in\Z.
\]
The chain maps $f$ and $g$ are \emph{homotopic}, denoted $f\simeq g$, if there exists a chain homotopy from $g$ to $f$.
Chain maps induce maps on homology
\[H_*(f) \colon H_*(F) \to H_*(G)\]
and homotopic chain maps induce the same map on homology. That is, if $f\simeq g$ then $H_*(f)=H_*(g)$.

For complexes $F$ and $G$, the tensor product is given by
\begin{align}
 (F\otimes_\cR G)_k
   &=\bigoplus_{i+j=k}F_i\otimes_\cR G_j,
 &
 d(x\otimes y)
   &=d^Fx\otimes y+(-1)^{|x|} x\otimes d^Gy.
 \label{eq:new-tensor-complex}
 \end{align}
 The tensor product is symmetric, with symmetry isomorphism given by the chain map
\begin{align*}
\tau &\colon F\otimes_\cR G \to G\otimes_{\cR} F, \\
 \tau(x\otimes y)&=(-1)^{|x||y|}y\otimes x.
\end{align*}
 For complexes $F$ and $G$, the internal hom complex is given by
 \begin{align}
 \Hom_\cR(F,G)_k
   &=\prod_i\Hom_\cR(F_i,G_{i+k}),
 &
 d f
   &= d (f_i)_{i\in \Z}= (d^Gf_i-(-1)^k f_{i-1}d^F)_{i\in \Z}.
 \label{eq:new-hom-complex}
\end{align}
If $f\in\Hom_\cR(F,G)_k$ and $x\in F_i$, we let
\[
 f(x)=f_i(x)\in G_{i+k}.
\]
Conversely, $f$ is determined by the values $f(x)$ for $x\in F_i$ and $i\in\Z$, and every choice of $\cR$-linear maps $F_i\to G_{i+k}$ defines an element of $\Hom_\cR(F,G)_k$.

\begin{lemma}
\label{lemma:hom-cycles}
Let $F$ and $G$ be complexes.
\begin{enumerate}[(a)]
\item An element $f\in\Hom_\cR(F,G)_0$ is a cycle if and only if $f$ is a chain map.
\item An element $h\in\Hom_\cR(F,G)_1$ is a chain homotopy from a chain map $g$ to a chain map $f$ if and only if $d h=f-g$.
\end{enumerate}
So $H_0\Hom_\cR(F,G)$ is the $\cR$-module of chain maps $F\to G$ modulo chain homotopy.
\end{lemma}

\begin{proof}
By \eqref{eq:new-hom-complex}, the components of $d f$ and $d h$ are
\[
 (d f)_i=d^Gf_i-f_{i-1}d^F,
 \qquad
 (d h)_i=d^Gh_i+h_{i-1}d^F.
\]
The first vanishes for all $i$ exactly when $f$ is a chain map. The second is equal to $f_i-g_i$ for all $i$ exactly when $h$ is a chain homotopy from $g$ to $f$.
\end{proof}

For a chain map $f\colon F\to G$ and a complex $T$, let
\[
 \Hom_\cR(T,f)\colon\Hom_\cR(T,F)\to\Hom_\cR(T,G),
 \qquad
 \Hom_\cR(f,T)\colon\Hom_\cR(G,T)\to\Hom_\cR(F,T)
\]
be the chain maps $\psi\mapsto f\circ\psi$ and $\psi\mapsto\psi\circ f$. Let 
\[f\otimes_\cR\id_T\colon F\otimes_\cR T\to G\otimes_\cR T \qquad \id_T\otimes_\cR f\colon T\otimes_\cR F\to T\otimes_\cR G\] 
be the chain maps $x\otimes t\mapsto f(x)\otimes t$ and $t\otimes x\mapsto t\otimes f(x)$.

The category $\Ch(\cR)$ is closed symmetric monoidal, with tensor product and internal Hom forming an adjoint pair.
\begin{proposition}
\label{prop:tensor-hom}
Let $F$, $G$ and $T$ be complexes. There is an isomorphism of complexes
\begin{align*}
 \Phi\colon \Hom_\cR(F\otimes_\cR G,T)
 &\xlongrightarrow{\cong}
 \Hom_\cR\bigl(G,\Hom_\cR(F,T)\bigr)
\end{align*}
natural in $F$, $G$ and $T$,
given for $f\in  \Hom_\cR(F\otimes_\cR G,T)$ by
\begin{align*}
 \Phi(f)(y)(x)&=(-1)^{|x||y|}f(x\otimes y).
\end{align*}
\end{proposition}

\begin{proof}
By our convention, the formula reads as follows. Let $f$ have degree $k$, and let $x\in F_i$ and $y\in G_j$. Then $f(x\otimes y)=f_{i+j}(x\otimes y)\in T_{i+j+k}$. The element $\Phi(f)_j(y)\in\Hom_\cR(F,T)_{j+k}$ is the sequence of maps $F_i\to T_{i+j+k}$ given by
\[
 \Phi(f)_j(y)_i(x)=(-1)^{ij}f_{i+j}(x\otimes y).
\]
By \eqref{eq:new-tensor-complex} and \eqref{eq:new-hom-complex}, both sides in degree $k$ are products over pairs $(i,j)$:
\[
 \prod_{i,j}\Hom_\cR\bigl(F_i\otimes_\cR G_j,T_{i+j+k}\bigr)
 \qquad\text{and}\qquad
 \prod_{i,j}\Hom_\cR\bigl(G_j,\Hom_\cR(F_i,T_{i+j+k})\bigr).
\]
On each factor,  the formula for $\Phi$ is given in terms of the tensor--hom adjunction for $\cR$-modules, adjusted by the sign $(-1)^{ij}$. So $\Phi$ is bijective in each degree. Naturality is clear from the formula.

It remains to show that $\Phi$ commutes with the differentials. By \eqref{eq:new-hom-complex}, applied twice, and since $\Phi(f)(y)$ has degree $j+k$,
\begin{align*}
 \bigl(d\Phi(f)\bigr)(y)(x)
 &=d^T\bigl(\Phi(f)(y)(x)\bigr)
 -(-1)^{j+k}\,\Phi(f)(y)(d^Fx)
 -(-1)^k\,\Phi(f)(d^Gy)(x)\\
 &=(-1)^{ij}d^Tf(x\otimes y)
 -(-1)^{k+ij}f(d^Fx\otimes y)
 -(-1)^{k+ij+i}f(x\otimes d^Gy).
\end{align*}
On the other hand, by \eqref{eq:new-tensor-complex} and \eqref{eq:new-hom-complex},
\begin{align*}
 \Phi(d f)(y)(x)
 &=(-1)^{ij}(d f)(x\otimes y)\\
 &=(-1)^{ij}d^Tf(x\otimes y)
 -(-1)^{k+ij}f(d^Fx\otimes y)
 -(-1)^{k+ij+i}f(x\otimes d^Gy).
\end{align*}
The two expressions agree. \end{proof}

For  $s\in\Z$ and $M$ an $\cR$-module, $M[s]$ denotes the
complex with $M$ in degree $s$ and zero elsewhere. This gives a fully faithful embedding of $\Mod(\cR)$ into $\Ch(\cR)$ for every $s\in \Z$.  For $s=0$, we abuse notation and write $M \in \Ch(\cR)$ to denote $M[0]$.
More generally, we define the \emph{shift} $F[s]$ of $F$ by
\begin{equation}
 (F[s])_i=F_{i-s},\qquad d^{F[s]}=(-1)^sd^F.
 \label{eq:new-shift}
\end{equation}
That is,
\[F[s] \cong \cR[s] \otimes_\cR F .\]
Consequently $H_i(F[s])=H_{i-s}(F)$. For $f \colon F \to G$, $f[s]$ is just the map induced by $f$ in each degree.

\begin{remark}
\label{rem:new-shift-hom-tensor}
Let $F$ and $G$ be complexes, and let $s\in\Z$. We identify $(F[s])_{i+s}$ with $F_i$, so an element $x\in F_i$ is also an element of $F[s]$ of degree $i+s$. Then there are isomorphisms of complexes
\[
 \Hom_\cR(F,G)[s]\xrightarrow{\ \cong\ }\Hom_\cR(F,G[s])
 \qquad\text{and}\qquad
 \Hom_\cR(F,G)[s]\xrightarrow{\ \cong\ }\Hom_\cR(F[-s],G).
\]
For $f\in\Hom_\cR(F,G)_k$, viewed as an element of $\Hom_\cR(F,G)[s]$ of degree $k+s$, and for $x\in F_i$, they send $f$ to the maps given by
\[
 x\longmapsto f(x)
 \qquad\text{and}\qquad
 x\longmapsto(-1)^{sk}f(x).
\]
Similarly, there are isomorphisms of complexes
\[
 F[s]\otimes_\cR G\xrightarrow{\ \cong\ }(F\otimes_\cR G)[s]
 \qquad\text{and}\qquad
 F\otimes_\cR G[s]\xrightarrow{\ \cong\ }(F\otimes_\cR G)[s].
\]
For $x\in F_i$ and $y\in G_j$, they are given by
\[
 x\otimes y\longmapsto x\otimes y
 \qquad\text{and}\qquad
 x\otimes y\longmapsto(-1)^{si}x\otimes y.
\]
Each of these maps is bijective in each degree. That it commutes with the differentials is a direct check using \eqref{eq:new-tensor-complex}, \eqref{eq:new-hom-complex} and \eqref{eq:new-shift}. The signs follow the Koszul rule: a sign $(-1)^{s|z|}$ appears when the shift moves past an element $z$.
\end{remark}

\begin{defn}For a chain map $f\colon F\to G$, its mapping cone is
\begin{equation}
 \Cone(f)_i=G_i\oplus F_{i-1},\qquad
 d(y,x)=(d^Gy+f(x),-d^Fx).
 \label{eq:new-mapping-cone}
\end{equation}
\end{defn}

\begin{prop}[{\cite[\S1.5]{Weibel_1994}}]
There is a sequence of chain maps
\[
 \xymatrix{ F\ar[r]^-{f} & G \ar[r]^-i & \Cone(f)\ar[r]^-{\delta}  & F[1] }
\]
where $i(y)=(y,0)$ and $\delta(y,x)=x$,
which induce a long exact sequence on homology,
\[\xymatrix{
\cdots \ar[r] & H_s(F) \ar[r]^-{H_s(f)} & H_s(G) \ar[r] & H_s(\Cone(f)) \ar[r]^-{H_s(\delta)} &  H_s(F[1]) =H_{s-1}(F)\ar[r] & \cdots  }.\]  
\end{prop}

Hom and tensor commute with cones in $\Ch(\cR)$.

\begin{lemma}
\label{lem:new-cone-hom-tensor}
Let $f\colon F\to G$ be a chain map, and let $T$ be a complex. There are isomorphisms of complexes
\begin{align*}
 \Hom_\cR(T,\Cone(f))&\cong\Cone\bigl(\Hom_\cR(T,f)\bigr), & (\alpha,\beta)&\mapsto(\alpha,\beta),\\
 \Hom_\cR(\Cone(f),T)&\cong\Cone\bigl(\Hom_\cR(f,T)\bigr)[-1], & (\mu,\nu)&\mapsto\bigl((-1)^{k}\nu,-\mu\bigr),\\
 \Cone(f)\otimes_\cR T&\cong\Cone(f\otimes_\cR\id_T), & (y,x)\otimes t&\mapsto(y\otimes t,\ x\otimes t),\\
 T\otimes_\cR\Cone(f)&\cong\Cone(\id_T\otimes_\cR f), & t\otimes(y,x)&\mapsto\bigl(t\otimes y,\ (-1)^{|t|}\,t\otimes x\bigr).
\end{align*}
Here $t\in T$, $y\in G$, $x\in F$, $\alpha\in\Hom_\cR(T,G)_k$, $\beta\in\Hom_\cR(T,F)_{k-1}$, $\mu\in\Hom_\cR(G,T)_k$ and $\nu\in\Hom_\cR(F,T)_{k+1}$.
\end{lemma}

\begin{proof}
For the first isomorphism, see the proof of \cite[Tag 09LK]{stacks-project}, which uses cohomological grading. The others are checked in the same way, using \eqref{eq:new-tensor-complex}, \eqref{eq:new-hom-complex}, \eqref{eq:new-shift} and \eqref{eq:new-mapping-cone}.
\end{proof}

\begin{lemma}
\label{lem:new-cone-triangles}
Let $f\colon F\to G$ be a chain map, and let $T$ be a complex. Write $f_*=\Hom_\cR(T,f)$ and $f^*=\Hom_\cR(f,T)$. Then the following diagrams of complexes commute.
\[
\xymatrix@C=2.2pc{
 \Hom_\cR(T,F) \ar[r]^-{f_*} \ar@{=}[d] & \Hom_\cR(T,G) \ar[r]^-{i_*} \ar@{=}[d] & \Hom_\cR(T,\Cone(f)) \ar[r]^-{\delta_*} \ar[d]^-{\cong} & \Hom_\cR(T,F[1]) \ar[d]^-{\cong} \\
 \Hom_\cR(T,F) \ar[r]^-{f_*} & \Hom_\cR(T,G) \ar[r]^-{i} & \Cone(f_*) \ar[r]^-{\delta} & \Hom_\cR(T,F)[1]
}
\]
\[
\xymatrix@C=2.2pc{
 \Hom_\cR(F[1],T) \ar[r]^-{\delta^*} \ar[d]^-{\cong} & \Hom_\cR(\Cone(f),T) \ar[r]^-{i^*} \ar[d]^-{\cong} & \Hom_\cR(G,T) \ar[r]^-{f^*} \ar@{=}[d] & \Hom_\cR(F,T) \ar@{=}[d] \\
 \Hom_\cR(F,T)[-1] \ar[r]^-{-i[-1]} & \Cone(f^*)[-1] \ar[r]^-{-\delta[-1]} & \Hom_\cR(G,T) \ar[r]^-{f^*} & \Hom_\cR(F,T)
}
\]
\[
\xymatrix@C=2.2pc{
 F\otimes_\cR T \ar[r]^-{f\otimes\id} \ar@{=}[d] & G\otimes_\cR T \ar[r]^-{i\otimes\id} \ar@{=}[d] & \Cone(f)\otimes_\cR T \ar[r]^-{\delta\otimes\id} \ar[d]^-{\cong} & F[1]\otimes_\cR T \ar[d]^-{\cong} \\
 F\otimes_\cR T \ar[r]^-{f\otimes\id} & G\otimes_\cR T \ar[r]^-{i} & \Cone(f\otimes\id) \ar[r]^-{\delta} & (F\otimes_\cR T)[1]
}
\]
\[
\xymatrix@C=2.2pc{
 T\otimes_\cR F \ar[r]^-{\id\otimes f} \ar@{=}[d] & T\otimes_\cR G \ar[r]^-{\id\otimes i} \ar@{=}[d] & T\otimes_\cR\Cone(f) \ar[r]^-{\id\otimes\delta} \ar[d]^-{\cong} & T\otimes_\cR F[1] \ar[d]^-{\cong} \\
 T\otimes_\cR F \ar[r]^-{\id\otimes f} & T\otimes_\cR G \ar[r]^-{i} & \Cone(\id\otimes f) \ar[r]^-{\delta} & (T\otimes_\cR F)[1]
}
\]
Here the vertical isomorphisms are those of Lemma~\ref{lem:new-cone-hom-tensor} and Remark~\ref{rem:new-shift-hom-tensor}. In the first, third and fourth diagrams, the bottom row is a cone sequence. In the second, it is the cone sequence of $f^*$ rotated twice.
\end{lemma}

\begin{proof}
Each square commutes by a direct check, using the formulas of Lemma~\ref{lem:new-cone-hom-tensor} and Remark~\ref{rem:new-shift-hom-tensor}.
\end{proof}

\subsection{The derived category}
\label{dra:new-derived-category}

A chain map is a \emph{quasi-isomorphism} if it induces an isomorphism on
every homology group. 
A complex is \emph{acyclic} if all of its homology
groups vanish. Note that $f$ is a quasi-isomorphism if and only if $\Cone(f)$ is acyclic.

\begin{defn}
Define the \emph{homotopy category} $\K(\cR)$ to be the category whose objects are chain complexes and morphisms are homotopy classes of chain maps. That is,
\[\Hom_{\K(\cR)}(A,B) =\frac{\Hom_{\Ch(\cR)}(A,B)}{\simeq}  .\]
\end{defn}
\begin{remark}
By Lemma~\ref{lemma:hom-cycles},
\[\Hom_{\K(\cR)}(A,B) = H_0(\Hom_{\cR}(A,B)). \]
\end{remark}
Not all quasi-isomorphisms are chain homotopy equivalences. Let $\mathcal W$ be the class of quasi-isomorphisms in $\K(\cR)$. We want a category with the property that the morphisms in $\mathcal W$ are isomorphisms. 
\begin{defn}
\label{defn:new-localization}
A \emph{localization} of $\K(\cR)$ at $\mathcal W$ is a category $\D(\cR)$ together with a functor
\[\xymatrix{L \colon \K(\cR) \ar[r] & \D(\cR)  }\]
satisfying the following conditions. 
\begin{itemize}
\item For every $f\in\mathcal W$, the morphism $L(f)$ is an isomorphism.
\item Let $F\colon\K(\cR)\to\cC$ be a functor such that $F(f)$ is an isomorphism in $\cC$ for every $f\in\mathcal W$. Then there exist a functor $F'\colon\D(\cR)\to\cC$ and a natural isomorphism $\eta\colon F'\circ L\xrightarrow{\cong} F$. That is, the following diagram commutes up to natural isomorphism:
\[\xymatrix{\K(\cR)\ar[r]^-F \ar[d]_-{L} & \cC  \\
\D(\cR) \ar[ur]_-{F'} & }\]
\item The pair $(F',\eta)$ is unique up to unique isomorphism. That is, if $(F'',\eta')$ is another such pair, there is a unique natural isomorphism $\theta\colon F'\to F''$ with $\eta'\circ\theta L=\eta$.
\end{itemize}
We call $\D(\cR)$ the \emph{derived category} of $\cR$.
\end{defn}

\begin{remark}
If chain complexes $A$ and $B$ are isomorphic in $\D(\cR)$, we write $A\simeq B$. That is, $A\simeq B$ if and only if $L(A)\cong L(B)$.
\end{remark}

The description of $\D(\cR)$ is rather abstract. We now give a concrete construction.
\begin{defn}
Let $P$, $I$ and $Q$ be complexes.
\begin{itemize}
\item The complex $P$ is \emph{$K$-projective} if $\Hom_\cR(P,E)$ is acyclic for every acyclic complex $E$.
\item The complex $I$ is \emph{$K$-injective} if $\Hom_\cR(E,I)$ is acyclic for every acyclic complex $E$.
\item The complex $Q$ is \emph{$K$-flat} if $E\otimes_\cR Q$ is acyclic for every acyclic complex $E$.
\end{itemize}
\end{defn}

These complexes are important because of the following result.

\begin{lemma}
\label{lem:new-k-preserve-qi}
Let $f\colon A\to A'$ be a quasi-isomorphism.
\begin{enumerate}[(a)]
\item If $P$ is $K$-projective, then $\Hom_\cR(P,A)\to\Hom_\cR(P,A')$ is a quasi-isomorphism.
\item If $I$ is $K$-injective, then $\Hom_\cR(A',I)\to\Hom_\cR(A,I)$ is a quasi-isomorphism.
\item If $Q$ is $K$-flat, then $A\otimes_\cR Q\to A'\otimes_\cR Q$ and $Q\otimes_\cR A\to Q\otimes_\cR A'$ are quasi-isomorphisms.
\end{enumerate}
\end{lemma}

\begin{proof}
Since $f$ is a quasi-isomorphism, $\Cone(f)$ is acyclic. So $\Hom_\cR(P,\Cone(f))$, $\Hom_\cR(\Cone(f),I)$, $\Cone(f)\otimes_\cR Q$ and $Q\otimes_\cR\Cone(f)$ are acyclic, by the definitions of $K$-projective, $K$-injective and $K$-flat complexes, using the symmetry $\tau$ for the last one. By Lemma~\ref{lem:new-cone-hom-tensor}, these are, up to shift, the cones of the maps in (a), (b) and (c). So these maps are quasi-isomorphisms.
\end{proof}

\begin{defn}
For a complex $A$,
\begin{itemize}
\item a \emph{left resolution} of $A$ is a quasi-isomorphism $P \to A$.
\item a \emph{right resolution} of $A$ is a quasi-isomorphism $A \to I$.
\item a \emph{$K$-projective resolution} of $A$ is a left resolution $P\to A$ with $P$ $K$-projective.
\item a \emph{$K$-flat resolution} of $A$ is a left resolution $Q\to A$ with $Q$ $K$-flat.
\item a \emph{$K$-injective resolution} of $A$ is a right resolution $A\to I$ with $I$ $K$-injective.
\end{itemize}
\end{defn}
\begin{theorem}[{\cite[Thm.~C, Prop.~5.8]{Spaltenstein_1988}}]
\label{thm:new-k-resolutions}
Any complex $A$ has a $K$-injective resolution $A \to I$, a $K$-projective resolution $P\to A$ and a $K$-flat resolution $Q \to A$.
\end{theorem}

\begin{example}
In homological indexing, a bounded-below complex of projective modules is $K$-projective, a
bounded-above complex of injective modules is $K$-injective, and a bounded-below
complex of flat modules is $K$-flat.  Furthermore, $K$-projectives are $K$-flat \cite[Prop.~5.8]{Spaltenstein_1988}. However,  an unbounded complex whose terms are projective is not automatically $K$-projective.
\end{example}

The following proposition is essential for understanding the role of $K$-projective and $K$-injective complexes.
\begin{proposition}[{\cite[Props. 1.4, 1.5]{Spaltenstein_1988}}]
\label{thm:new-k-resolution-morphisms}
For every complex $P$ the following conditions are equivalent.
\begin{enumerate}[(a)]
\item $P$ is $K$-projective.
\item For every complex $S$, localization induces a bijection
 \[
 L\colon\Hom_{\K(\cR)}(P,S)\xrightarrow{\ \cong\ }\Hom_{\D(\cR)}(P,S).
 \]
\item For every diagram in $\K(\cR)$
 \[
 \xymatrix{ & X \ar[d]^-{s}_-{\simeq} \\ P \ar[r]^-{f} & Y }
 \]
 with $s$ a quasi-isomorphism, there exists a unique morphism
 $g\colon P\to X$ in $\K(\cR)$ such that $s\circ g=f$ in $\K(\cR)$.
\item For every quasi-isomorphism $u\colon S\to P$ there exists a morphism
 $v\colon P\to S$ in $\K(\cR)$ such that $u\circ v=\id_P$ in $\K(\cR)$.
\end{enumerate}
Similarly, for every complex $I$ the following conditions are equivalent.
\begin{enumerate}[(a)]
\item $I$ is $K$-injective.
\item For every complex $S$, localization induces a bijection
 \[
 L\colon\Hom_{\K(\cR)}(S,I)\xrightarrow{\ \cong\ }\Hom_{\D(\cR)}(S,I).
 \]
\item For every diagram in $\K(\cR)$
 \[
 \xymatrix{ Y \ar[r]^-{f} \ar[d]_-{s}^-{\simeq} & I \\ X & }
 \]
 with $s$ a quasi-isomorphism, there exists a unique morphism
 $g\colon X\to I$ in $\K(\cR)$ such that $g\circ s=f$ in $\K(\cR)$.
\item For every quasi-isomorphism $u\colon I\to S$ there exists a morphism
 $v\colon S\to I$ in $\K(\cR)$ such that $v\circ u=\id_I$ in $\K(\cR)$.
\end{enumerate}
\end{proposition}
This gives a concrete model for $\D(\cR)$. Let $\K_{\mathrm{proj}}(\cR)$ be the
full subcategory of $\K(\cR)$ on the $K$-projective complexes, and choose for each
complex $A$ a $K$-projective resolution $p_A\colon P_A\xrightarrow{\simeq}A$.
By Proposition~\ref{thm:new-k-resolution-morphisms}, every $f\colon A\to B$ lifts
uniquely up to homotopy to  a map $ P_A\to P_B$. This allows us to define a functor
\[
\K(\cR)\longrightarrow\K_{\mathrm{proj}}(\cR),\qquad A\longmapsto P_A,
\]
which sends quasi-isomorphisms to isomorphisms and satisfies the universal
property of $\D(\cR)$. Hence
\[
 \D(\cR)\simeq\K_{\mathrm{proj}}(\cR),
 \qquad
 \Hom_{\D(\cR)}(A,B)\cong\Hom_{\K(\cR)}(P_A,B)=H_0\bigl(\Hom_\cR(P_A,B)\bigr).
\]
Dually, using $K$-injective resolutions $B\xrightarrow{\simeq}I_B$, we can also model $\D(\cR)$ as
\[
 \D(\cR)\simeq\K_{\mathrm{inj}}(\cR),
 \qquad
 \Hom_{\D(\cR)}(A,B)\cong\Hom_{\K(\cR)}(A,I_B)=H_0\bigl(\Hom_\cR(A,I_B)\bigr).
\]

Since $\D(\cR)$ is obtained from $\K(\cR)$ by formally inverting
quasi-isomorphisms, a morphism $A\to B$ in $\D(\cR)$ is a priori a finite
zigzag in $\K(\cR)$
\[
 \xymatrix{A \ar[r] & X_1 & \ar[l]_-{\simeq} X_2 \ar[r] & \cdots & \ar[l]_-{\simeq} X_n \ar[r] & B}
\]
in which the backward arrows are quasi-isomorphisms. The following lemma shortens this to a single roof.

\begin{lemma}
\label{rem:roofs}
Let $A$ and $B$ be complexes, and let $p\colon P\to A$ be a $K$-projective resolution. Then every morphism $\varphi\colon A\to B$ in $\D(\cR)$ is represented by a roof
\[
 \xymatrix{A & \ar[l]_-{\simeq}^-{p} P \ar[r]^-{f} & B},
\]
that is, $\varphi=L(f)\circ L(p)^{-1}$ for a chain map $f\colon P\to B$. The chain map $f$ is unique up to chain homotopy.
\end{lemma}

\begin{proof}
Since $P$ is $K$-projective, condition (b) of Proposition~\ref{thm:new-k-resolution-morphisms} holds for $P$. So $\varphi\circ L(p)=L(f)$ for a chain map $f\colon P\to B$, unique up to chain homotopy.
\end{proof}

\begin{corollary}
\label{lemma:iso-zigzag}
Let $u\colon F\to G$ be an isomorphism in $\D(\cR)$, and let $p\colon P\to F$ be a $K$-projective resolution. Then the chain map $g\colon P\to G$ with $u=L(g)\circ L(p)^{-1}$ of Lemma~\ref{rem:roofs} is a quasi-isomorphism. In particular, two complexes are isomorphic in $\D(\cR)$ if and only if they are joined by a zigzag of quasi-isomorphisms.
\end{corollary}

\begin{proof}
The morphism $L(g)=u\circ L(p)$ is an isomorphism. Homology sends quasi-isomorphisms to isomorphisms, so by the universal property of $L$ it factors through $\D(\cR)$. Hence $H_*(g)$ is an isomorphism.

For the last statement, the roof $F\xleftarrow{\,p\,}P\xrightarrow{\,g\,}G$ is a zigzag of quasi-isomorphisms. Conversely, $L$ sends quasi-isomorphisms to isomorphisms.
\end{proof}

The category $\D(\cR)$ also has a tensor product and internal hom, with an associated hom-tensor adjunction. However, one has to take care in constructing these since the functors $\Hom_\cR$ and $\otimes_\cR$ do not preserve quasi-isomorphisms.
\begin{defn}
The \emph{derived hom-complex} $\RHom_\cR(A,B)\in\D(\cR)$ is modeled by
\[
 \RHom_\cR(A,B)=\Hom_\cR(A,I_B)\simeq\Hom_\cR(P_A,B),
\]
where $B\to I_B$ is a $K$-injective resolution and $P_A\to A$ is a $K$-projective
resolution. The \emph{derived tensor product} is modeled by
\[
 A\otimes_\cR^{\mathrm L}B=A\otimes_\cR Q_B,
\]
where $Q_B\to B$ is a $K$-flat resolution.
\end{defn}
It can be shown that these models do not depend on the choice of resolutions, up to unique isomorphism, and that they define functors on $\D(\cR)$ in both variables \cite[Props.~6.1, 6.5]{Spaltenstein_1988}.

\begin{remark}
If $A$ is $K$-projective, then $\Hom_\cR(A,-)\simeq\RHom_\cR(A,-)$. Similarly, if $A$ is
$K$-flat, then
$A\otimes_\cR(-)\simeq A\otimes_\cR^{\mathrm L}(-)$. Both of these hold, for example, if $A$ is a bounded-below complex of projective
$\cR$-modules.
\end{remark}

We can now state the hom-tensor adjunction in the derived category.
\begin{theorem}
\label{thm:new-derived-tensor-hom}
Let $A$, $B$ and $C$ be complexes. There is a natural isomorphism in $\D(\cR)$
\[
 \RHom_\cR(A\otimes_\cR^{\mathrm L}B,C)\simeq\RHom_\cR\bigl(B,\RHom_\cR(A,C)\bigr).
\]
\end{theorem}
\begin{proof}
Let $P_A\to A$ and $P_B\to B$ be $K$-projective resolutions, and let $C\to I_C$ be a $K$-injective resolution. Since $P_B$ is $K$-flat, the map $P_A\otimes_\cR P_B\to A\otimes_\cR P_B$ is a quasi-isomorphism by Lemma~\ref{lem:new-k-preserve-qi}(c). So $A\otimes^{\mathrm L}_\cR B$ is modeled by $P_A\otimes_\cR P_B$. Since $I_C$ is $K$-injective, the map $\Hom_\cR(A,I_C)\to\Hom_\cR(P_A,I_C)$ is a quasi-isomorphism by Lemma~\ref{lem:new-k-preserve-qi}(b). So $\RHom_\cR(A,C)$ is modeled by $\Hom_\cR(P_A,I_C)$. So the two sides are modeled by
\[
 \Hom_\cR(P_A\otimes_\cR P_B,I_C)
 \qquad\text{and}\qquad
 \Hom_\cR\bigl(P_B,\Hom_\cR(P_A,I_C)\bigr).
\]
By Proposition~\ref{prop:tensor-hom}, $\Phi$ is an isomorphism of complexes between them, natural in all three variables.
\end{proof}

\begin{remark}
\label{rem:new-derived-tensor-hom-rings}
More generally, for rings $\cR$ and $\cS$, an $(\cR,\cS)$-bimodule $N$, an
$\cS$-module $L$ and an $\cR$-module $M$, there is a natural isomorphism in
$\D(\Z)$
\[
 \RHom_\cR(N\otimes_\cS^{\mathrm L}L,M)\simeq\RHom_\cS\bigl(L,\RHom_\cR(N,M)\bigr).
\]
\end{remark}

\begin{remark}
The derived hom and tensor are related to Ext and Tor groups as follows. Let $M$ and $N$ be $\cR$-modules. Then
\[
 \Ext_\cR^q(M,N)=H_{-q}\RHom_\cR(M,N),
 \qquad
 \Tor_q^\cR(M,N)=H_q(M\otimes_\cR^{\mathrm L}N),\]
 where here we have used the convention $M=M[0]$ and $N=N[0]$ to view $M,N$ as objects of $\D(\cR)$.
 \end{remark}

\begin{theorem}\label{thm:dualityisderived}
Let $\cR$ be a commutative Noetherian ring, and let $T$ be an $\cR$-module of finite injective
dimension $n$. Let $F$ be a possibly unbounded chain complex of
$\cR$-modules, and suppose that
\[
\Ext_\cR^i(F_k,T)=0
\]
for every $k\in \Z$ and every $i>0$. If
\[
T\longrightarrow I
\]
is a $K$-injective resolution of $T$, then the induced map
\[
\Hom_\cR(F,T)\longrightarrow \Hom_\cR(F,I)
\]
is a quasi-isomorphism. In particular,
\[
	\Hom_{\cR}(F, T) \simeq \RHom_{\cR}(F, T).
\]
\end{theorem}

\begin{proof}
Since $T$ has injective dimension $n$, it admits a bounded injective
resolution
\[
T\longrightarrow J,
\qquad
J_q=0 \quad \text{for } q\notin\{-n,\ldots,0\}.
\]
For each $k$, the hypothesis implies that
\[
\Hom_\cR(F_k,T)[0] \longrightarrow \Hom_\cR(F_k,J)
\]
is a quasi-isomorphism, since the homology of the complex on the right is
\[
H_{-i}\Hom_\cR(F_k,J)\cong \Ext_\cR^i(F_k,T) = \begin{cases} \Hom_\cR(F_k,T) & i=0 \\
0 & \text{otherwise}.
\end{cases}
\]

Since $\Hom_\cR(F,\Cone(g))=\Cone(\Hom_\cR(F,g))$ for any chain map $g$, and a chain map is a quasi-isomorphism if and only if its cone is acyclic, it suffices to show that $\Hom_\cR(F,C)$ is acyclic, where $C=\Cone(T\to J)$. To do so, we compute its homology using the double complex spectral sequence constructed from the double complex $E^0_{k,j}:=\Hom_\cR(F_k,C_j)$. For each fixed $k$, the column is acyclic by the preceding observation.
Moreover, it has only finitely many nonzero rows, because $J$ is bounded, so
in each total degree only finitely many entries are nonzero. Filtering the total
complex by the $F$-degree $k$ therefore gives a filtration that is finite in each
degree, and the associated spectral sequence has zero $E_1$-page. Hence the total complex is acyclic,
so
\[
\Hom_\cR(F,T)\longrightarrow \Hom_\cR(F,J)
\]
is a quasi-isomorphism.

Finally, $J$ and $I$ are both $K$-injective resolutions of $T$, so they are
homotopy equivalent, compatibly with their maps from $T$ up to homotopy.
Consequently,
\[
\Hom_\cR(F,J)\longrightarrow \Hom_\cR(F,I)
\]
is a homotopy equivalence. The result follows. In particular, $\Hom_\cR(F,T)\simeq\RHom_\cR(F,T)$.
\end{proof}

 \subsection{Triangulated structure}

\begin{defn}
A sequence 
\[ \xymatrix{
A \ar[r]^-f &  B \ar[r]^-g & C \ar[r]^-{h} & A[1]
} \]
in $\D(\cR)$ is called a \emph{distinguished triangle} if there is a commutative diagram
\[ \xymatrix{A' \ar[d]^-\simeq \ar[r]^-{f'} &  B' \ar[r] \ar[d]^-\simeq &  \Cone(f') \ar[d]^-{\simeq} \ar[r] & A'[1] \ar[d]^-\simeq  \\
A \ar[r]^-f &  B \ar[r]^-g & C \ar[r]^-h & A[1]
} \]
where the vertical morphisms are isomorphisms in $\D(\cR)$, and the top row is the cone sequence of a chain map $f'\colon A' \to B'$.
\end{defn}

A distinguished triangle can be \emph{rotated} to give new distinguished triangles
\[ \xymatrix@C=3pc{
B \ar[r]^-g & C \ar[r]^-h & A[1] \ar[r]^-{-f[1]} & B[1]
} \]
and
\[ \xymatrix@C=3pc{
C[-1] \ar[r]^-{-h[-1]} & A \ar[r]^-f &  B \ar[r]^-g & C.
} \]

Applying Lemma~\ref{lem:new-cone-triangles} to $K$-projective, $K$-injective or $K$-flat models shows the following. The functors $(-) \otimes_\cR^{\mathrm L} F$, $F \otimes_\cR^{\mathrm L} (-)$ and $\RHom_\cR(F,-)$ send distinguished triangles to distinguished triangles. The functor $\RHom_\cR(-,F)$ sends a distinguished triangle $A\to B\to C\to A[1]$ to a distinguished triangle
\[
 \xymatrix{\RHom_\cR(A[1],F)\ar[r] & \RHom_\cR(C,F)\ar[r] & \RHom_\cR(B,F)\ar[r] & \RHom_\cR(A,F),}
\]
where $\RHom_\cR(A,F)\cong\RHom_\cR(A[1],F)[1]$ by Remark~\ref{rem:new-shift-hom-tensor}.

\begin{remark}
The shift functor $[1] \colon \D(\cR) \to \D(\cR)$ together with the collection of distinguished triangles gives $\D(\cR)$ the structure of a \emph{tensor-triangulated category} \cite{Balmer_2005}.
\end{remark}

\subsection{Postnikov and Whitehead towers}
\label{dra:new-postnikov}

For $F \in \Ch(\cR)$, define its truncations by
\begin{align*}
 (\tau_{\geq a}F)_i
 &=
 \begin{cases}
 F_i,&i>a,\\
 \ker(d\colon F_a\to F_{a-1}),&i=a,\\
 0,&i<a,
 \end{cases}
&
 (\tau_{\leq a}F)_i
 &=
 \begin{cases}
 0,&i>a,\\
 \coker(d\colon F_{a+1}\to F_a),&i=a,\\
 F_i,&i<a.
 \end{cases}
\end{align*}
We also use the notation $\tau_{>a}F = \tau_{\geq a+1}F$ and  $\tau_{<a}F = \tau_{\leq a-1}F$.
For $a\leq b$, we write $\tau_{[a,b]}F=\tau_{\leq b}\tau_{\geq a}F\simeq \tau_{\geq a}\tau_{\leq b}F$; its homology is $H_i(F)$ for $a\leq i\leq b$ and zero otherwise.
The natural maps fit into a distinguished triangle in $\D(\cR)$,
\[
 \xymatrix{\tau_{\geq a+1}F\ar[r] &  F\ar[r] & \tau_{\leq a}F
 \ar[r] & (\tau_{\geq a+1}F)[1]}.
\]
The map $ \tau_{\geq a+1}F\to F$  is a homology isomorphism in degrees strictly greater than $a$ and the map $F \to \tau_{\leq a}F$ 
is a homology isomorphism in degrees $\leq a$.

\begin{defn}
\label{def:postnikov}
The \emph{Postnikov
tower} of a complex $F$ is the sequence of truncations $\tau_{\leq i}F$ and
distinguished triangles
\[
\xymatrix{ H_i(F)[i] \ar[r] & \tau_{\leq i} F \ar[r] & \tau_{\leq i-1} F
 \ar[r] & H_i(F)[i+1]. }\]
Dually, the \emph{Whitehead tower} consists of the truncations $\tau_{\geq i}F$
and distinguished triangles
\[
\xymatrix{ \tau_{\geq i+1} F \ar[r] & \tau_{\geq i} F \ar[r] & H_i(F)[i]
 \ar[r] & \tau_{\geq i+1} F[1]. }\]
 \end{defn}
 
If $F$ has homology in only finitely many degrees, the towers are finite, so $F$
is obtained from its shifted homology modules in a finite number of exact triangles.
\begin{lemma}
\label{lemma:single-degree-zigzag}
Let $F$ be a complex with $H_i(F)=0$ for $i\neq a$. Then the natural maps
\[
 \xymatrix{F\ar[r]^-\simeq & \tau_{\leq a}F  & \ar[l]_-\simeq  \tau_{\geq a}(\tau_{\leq a}F)  =H_a(F)[a]}
\]
are quasi-isomorphisms. So $F\simeq H_a(F)[a]$ in $\D(\cR)$, by an isomorphism inducing the identity on $H_a$.
\end{lemma}

\begin{proof}
The map $F\to\tau_{\leq a}F$ is a quasi-isomorphism since $H_i(F)=0$ for $i>a$. The complex $\tau_{\geq a}(\tau_{\leq a}F)$ is zero outside degree $a$. In degree $a$, it is the kernel of
\[
 d\colon\coker(d\colon F_{a+1}\to F_a)\longrightarrow F_{a-1},
\]
which is $H_a(F)$. So its inclusion into $\tau_{\leq a}F$ is a quasi-isomorphism since $H_i(F)=0$ for $i<a$. Both maps induce the identity on $H_a$.
\end{proof}

\begin{lemma}
\label{lemma:single-degree}
Let $F,G\in\D(\cR)$ have homology concentrated in degree $a$. Then
\[
 H_a\colon\Hom_{\D(\cR)}(F,G)\longrightarrow\Hom_\cR\bigl(H_a(F),H_a(G)\bigr)
\]
is a bijection.
\end{lemma}

\begin{proof}
By Lemma~\ref{lemma:single-degree-zigzag}, $F\simeq H_a(F)[a]$ and $G\simeq H_a(G)[a]$, by zigzags inducing the identity on $H_a$. So, after shifting, we may assume that $a=0$ and that $F=M$ and $G=N$ are modules. We then need to show that
\[
 H_0\colon\Hom_{\D(\cR)}(M,N)\longrightarrow\Hom_\cR\bigl(M,N\bigr)
\]
is a bijection.
Let $P\to M$ be a projective resolution with $P_i=0$ for $i<0$. Since $P$ is $K$-projective, $\Hom_{\D(\cR)}(M,N)=\Hom_{\K(\cR)}(P,N)$ by Proposition~\ref{thm:new-k-resolution-morphisms}. Since $N$ is concentrated in degree $0$, a chain map $P\to N$ is a homomorphism $P_0\to N$ vanishing on $d(P_1)$, that is, a homomorphism $M\to N$, so the map is surjective. Furthermore, every $h\in\Hom_\cR(P,N)_1$ is zero, since $h_{-1}\colon P_{-1}\to N_0$ and $P_{-1}=0$. So no two distinct chain maps $P\to N$ are homotopic, and the map is injective.
\end{proof}

Recall that a ring $\cR$ has global dimension $\le d$ if every $\cR$-module $M$ has a projective resolution of length at most $d$:
\[0\to P_d\to\cdots\to P_1\to P_0\to M\to 0.\]
Equivalently, $\Ext^{n}_\cR(M,N)=0$ for all modules $M,N$ and all $n>d$. A ring has finite global dimension if it has global dimension at most $d$ for some $d<\infty$.

The following lemma is standard, and follows by combining several results of \cite{Yekutieli_2019}.
\begin{lemma}
\label{lemma:bounded-replacement}
Let $F\in\Ch(\cR)$ have bounded-below homology. Then there is a
quasi-isomorphism $P\to F$, where $P$ is a bounded-below complex of projective
$\cR$-modules. Moreover:
\begin{enumerate}[(a)]
\item if $\cR$ is Noetherian and every $H_n(F)$ is finitely generated, then
 each $P_n$ may be chosen finitely generated;
\item if $F$ has homology in only finitely many degrees and $\cR$ has finite
 global dimension, then $P$ may be chosen bounded;
\item if the hypotheses of (a) and (b) both hold, then $P$ may be chosen
 bounded and degreewise finitely generated. In particular, $F$ is perfect.
\end{enumerate}
\end{lemma}

 \subsection{Subcategories of the derived category}
 We define a few important subcategories of $\D(\cR)$. Recall a chain complex $F$ is bounded if $F_i=0$ for all but finitely many $i\in \Z$. A complex $F$ is \emph{bounded below} (resp. \emph{bounded above}) if there exists $N\in\Z$ such that $F_i=0$ for all $i<N$ (resp. all $i>N$).
 \begin{defn}
Define the full subcategories:
\begin{enumerate}[(a)]
\item $\D_f(\cR)$ on objects whose homology groups are all finitely generated $\cR$-modules.
\item $\D_f^b(\cR)$  on objects whose homology groups are finitely generated and non-zero in finitely many degrees.
\item $\D^p(\cR)$, the category of \emph{perfect complexes}, on objects which are quasi-isomorphic to a bounded chain complex of finitely generated projective $\cR$-modules.
\end{enumerate}
 \end{defn}
We write $\Ch^p(\cR)$ for the full subcategory of $\Ch(\cR)$ on perfect complexes.
 \begin{remark}
It follows from Lemma~\ref{lemma:bounded-replacement} that if $\cR$ is Noetherian of finite global dimension, then $\D_f^b(\cR) \simeq \D^p(\cR)$.
\end{remark}

Perfect complexes play a particularly important role in the derived category. Perfect complexes are to the derived category as finite-dimensional vector spaces are to the category of all vector spaces. To explain this, we recall a few more definitions.
\begin{defn}
Let $A \in \D(\cR)$.
\begin{enumerate}[(a)]
\item $A$ is \emph{compact} if, for every small family $\{C_i\}$ of objects
of $\D(\cR)$, the natural map
\[
 \bigoplus_i \RHom_\cR(A,C_i) \xrightarrow{\ \simeq\ } \RHom_\cR\Bigl(A , \bigoplus_i C_i\Bigr)
\]
is an isomorphism.
\item $A$ is \emph{strongly dualizable} if there exist an object $B$ and maps
\[
 \varepsilon \colon B \otimes_\cR^{\mathrm L} A \to \cR,
 \qquad
\eta \colon \cR \to A\otimes_\cR^{\mathrm L} B
\]
such that the composites
\[
 \xymatrix@C=4pc{A \ar[r]^-{\eta\otimes 1} & A \otimes_\cR^{\mathrm L} B \otimes_\cR^{\mathrm L} A \ar[r]^-{1\otimes \varepsilon} & A }
\]
and
\[
 \xymatrix@C=4pc{B \ar[r]^-{1\otimes \eta} & B \otimes_\cR^{\mathrm L} A \otimes_\cR^{\mathrm L} B \ar[r]^-{\varepsilon \otimes 1} & B }
\]
are both the identity in $\D(\cR)$.
\end{enumerate}
\end{defn}

The following characterizes perfect complexes. 
\begin{theorem}[{\cite[Tags 07LT, 0FPV]{stacks-project}}]
\label{thm:new-perfect-compact-dualizable}
For an object $P\in\D(\cR)$, the following are equivalent:
\begin{enumerate}[(a)]
\item $P$ is perfect;
\item $P$ is compact;
\item $P$ is strongly dualizable.
\end{enumerate}
\end{theorem}

\subsection{Duals, evaluation and canonical maps}
\label{v2:dra:duals}

In this section, we start introducing duality in $\Ch(\cR)$ and the derived category. As many examples come to us as explicit models in $\Ch(\cR)$, we first cover the material at the chain level before we consider duality in the derived context. We also start going into a lot more detail than before in order to make explicit some of the maps and isomorphisms involved so that they are concrete when working with specific models.

\begin{remark}
Conjugation extends degreewise to an exact involutive functor on $\Ch(\cR)$. So $\overline F$ is the complex with terms $(\overline F)_i=\overline{F_i}$ and differential $d^{\overline F}(\overline x)=\overline{d^F x}$. For a chain map $f\colon F\to G$, the chain map $\overline f\colon\overline F\to\overline G$ is given by $\overline f(\overline x)=\overline{f(x)}$.
\end{remark}

\begin{defn}
An \emph{anti-linear map} from $F$ to $G$ is equivalently a linear chain map $\overline F\to G$. An \emph{anti-linear involution} on a complex $T$ is represented by a linear chain map $c_T\colon\overline T\to T$ such that
\[
 c_T\circ\overline{c_T}=\id_T.
\]
\end{defn}

\begin{remark}
\label{v2:rmk:conjugation}
Let $A$ and $B$ be complexes. A map $A\to B$ is $\cR$-linear if and only if it is $\cR$-linear as a map $\overline A\to\overline B$. This gives an equality of complexes
\[
 \overline{\Hom_\cR(A,B)}=\Hom_\cR(\overline A,\overline B).
\]
Similarly, $\overline{A\otimes_\cR B}=\overline A\otimes_\cR\overline B$. Conjugation preserves quasi-isomorphisms and acyclic complexes, and it sends $K$-projective, $K$-flat and $K$-injective complexes to complexes of the same kind.
\end{remark}

Throughout, we let $T$ be a complex with a linear chain map $c_T\colon\overline T\to T$. In
the main case of interest, $c_T$ is an involution; we only use this where we say so.
\begin{defn}
\label{v2:def:dual}
The \emph{$T$-dual} of a complex $F$ is the complex
\[
 D_T(F)=\Hom_\cR(\overline F,T).
\]
For a chain map $f\colon F\to G$, the chain map $D_T(f)\colon D_T(G)\to D_T(F)$ is precomposition with $\overline f$.
\end{defn}

By Proposition~\ref{prop:tensor-hom}, applied to $\overline F$, $G$ and $T$, there is an isomorphism of complexes
\begin{equation}
 \Phi\colon\Hom_\cR(\overline F\otimes_\cR G,T)\xrightarrow{\ \cong\ }\Hom_\cR\bigl(G,D_T(F)\bigr),
 \label{v2:eq:chain-Phi}
\end{equation}
natural in $F$, $G$ and $T$. For $x\in G_i$ and $y\in F_j$, it is given by
\begin{equation}
 \Phi(\Lambda)(x)(\overline y)=(-1)^{ij}\Lambda(\overline y\otimes x).
 \label{v2:eq:sesquilinear}
\end{equation}
\begin{defn}
We call an element of $\Hom_\cR(\overline F\otimes_\cR G,T)$ a \emph{sesquilinear pairing}; when $F=G$, we call it a \emph{sesquilinear form}. 
\end{defn}
Indeed, this is a $\Z$-bilinear pairing which is anti-linear in $\cR$ in the left variable and $\cR$-linear in the right variable.

The symmetry isomorphism $\tau$ of tensor products induces an isomorphism
\[
 \tau\colon\overline G\otimes_\cR F\longrightarrow F\otimes_\cR\overline G=\overline{\overline F\otimes_\cR G}.
\]

\begin{defn}
\label{v2:def:ev-can}
The \emph{evaluation} is the chain map
\[
 \ev^T_F=\Phi^{-1}\bigl(\id_{D_T(F)}\bigr)\colon\overline F\otimes_\cR D_T(F)\longrightarrow T.
\]
The \emph{canonical map} is the chain map
\[
 \can^T_F=\Phi\bigl(c_T\circ\overline{\ev^T_F}\circ\tau\bigr)\colon F\longrightarrow D_T^2(F).
\]
\end{defn}

\begin{lemma}
\label{v2:lemma:ev-can-formulas}
For $x\in F_i$ and $f\in D_T(F)_j$,
\[
 \ev^T_F(\overline x\otimes f)=(-1)^{ij}f(\overline x),
 \qquad
 \can^T_F(x)(\overline f)=(-1)^{ij}c_T\bigl(\overline{f(\overline x)}\bigr).
\]
\end{lemma}

\begin{proof}
By \eqref{v2:eq:sesquilinear},
\[\Phi(\ev^T_F)(f)(\overline x)=(-1)^{ij}\ev^T_F(\overline x\otimes f),\] and this equals $f(\overline x)$. For $\can^T_F$, we have 
\[\tau(\overline f\otimes x)=(-1)^{ij}\,\overline{\overline x\otimes f}.\] 
So
\[
 \bigl(c_T\circ\overline{\ev^T_F}\circ\tau\bigr)(\overline f\otimes x)=(-1)^{ij}c_T\Bigl(\overline{\ev^T_F(\overline x\otimes f)}\Bigr)=c_T\bigl(\overline{f(\overline x)}\bigr).
\]
Applying \eqref{v2:eq:sesquilinear} gives the formula for $\can^T_F$.
\end{proof}

\begin{defn}
\label{v2:def:transposition}
Let $F$ and $G$ be complexes. The \emph{transposition} is the map of chain complexes
\[
 t\colon\overline{\Hom_\cR\bigl(G,D_T(F)\bigr)}\longrightarrow\Hom_\cR\bigl(F,D_T(G)\bigr),
 \qquad
 t(\overline\varphi)=\Phi\Bigl(c_T\circ\overline{\Phi^{-1}(\varphi)}\circ\tau\Bigr).
\]
\end{defn}

That is, $t$ is the composite
\begin{align*}
 \overline{\Hom_\cR\bigl(G,D_T(F)\bigr)}
 &\xrightarrow[\cong]{\ \overline{\Phi}^{-1}\ }\overline{\Hom_\cR\bigl(\overline F\otimes_\cR G,T\bigr)}\\
 &\xrightarrow[=]{\ \ }\Hom_\cR\bigl(\overline{\overline F\otimes_\cR G},\overline T\bigr)\\
 &\xrightarrow{\ (c_T)_*\ }\Hom_\cR\bigl(\overline{\overline F\otimes_\cR G},T\bigr)\\
 &\xrightarrow[\cong]{\ \tau^*\ }\Hom_\cR\bigl(\overline G\otimes_\cR F,T\bigr)\\
 &\xrightarrow[\cong]{\ \Phi\ }\Hom_\cR\bigl(F,D_T(G)\bigr),
\end{align*}
where the second map is the equality of Remark~\ref{v2:rmk:conjugation}, $(c_T)_*$ is postcomposition with $c_T$, and $\tau^*$ is precomposition with $\tau$. Each of these maps is a map of complexes. So $t$ is a map of complexes.

\begin{lemma}
\label{v2:lemma:transposition-formula}
Let $\varphi\in\Hom_\cR\bigl(G,D_T(F)\bigr)_k$. For $x\in G_i$ and $y\in F_j$,
\[
 t(\overline\varphi)(y)(\overline x)=(-1)^{ij}c_T\bigl(\overline{\varphi(x)(\overline y)}\bigr).
\]
\end{lemma}

\begin{proof}
Let $\Lambda=\Phi^{-1}(\varphi)$, so that $\Lambda(\overline y\otimes x)=(-1)^{ij}\varphi(x)(\overline y)$ by \eqref{v2:eq:sesquilinear}. Since $\tau(\overline x\otimes y)=(-1)^{ij}\,\overline{\overline y\otimes x}$,
\[
 \bigl(c_T\circ\overline\Lambda\circ\tau\bigr)(\overline x\otimes y)=(-1)^{ij}c_T\bigl(\overline{\Lambda(\overline y\otimes x)}\bigr)=c_T\bigl(\overline{\varphi(x)(\overline y)}\bigr).
\]
Applying \eqref{v2:eq:sesquilinear} once more gives the formula.
\end{proof}

\begin{remark}
\label{v2:rmk:forms}
Let $F$ be a complex. By \eqref{v2:eq:sesquilinear}, an element $\varphi\in\Hom_\cR\bigl(F,D_T(F)\bigr)$ of degree $k$ is the same as a sesquilinear form $\Lambda=\Phi^{-1}(\varphi)\colon\overline F\otimes_\cR F\to T$ of degree $k$, with
\[
 \Lambda(\overline x\otimes y)=(-1)^{|x||y|}\varphi(y)(\overline x).
\]
By Definition~\ref{v2:def:transposition}, $t(\overline\varphi)$ corresponds to the form $c_T\circ\overline\Lambda\circ\tau$. We call $\varphi$, or $\Lambda$, \emph{antisymmetric} if $t(\overline\varphi)=-\varphi$, that is, if $c_T\circ\overline\Lambda\circ\tau=-\Lambda$. By Lemma~\ref{v2:lemma:transposition-formula}, this holds if and only if, for all homogeneous $x,y\in F$,
\[
 \varphi(y)(\overline x)=-(-1)^{|x||y|}c_T\bigl(\overline{\varphi(x)(\overline y)}\bigr).
\]
\end{remark}

\begin{lemma}
\label{v2:lemma:transposition-natural}
\begin{enumerate}[(a)]
\item For chain maps $\alpha\colon G'\to G$ and $\beta\colon F'\to F$, the diagram
\[
\xymatrix@C=4pc{
 \overline{\Hom_\cR\bigl(G,D_T(F)\bigr)} \ar[r]^-{t} \ar[d]_-{\overline{D_T(\beta)_*\alpha^*}} & \Hom_\cR\bigl(F,D_T(G)\bigr) \ar[d]^-{D_T(\alpha)_*\beta^*} \\
 \overline{\Hom_\cR\bigl(G',D_T(F')\bigr)} \ar[r]^-{t} & \Hom_\cR\bigl(F',D_T(G')\bigr)
}
\]
commutes, where the vertical maps are pre- and postcomposition, that is, $t\bigl(\overline{D_T(\beta)\circ\varphi\circ\alpha}\bigr)=D_T(\alpha)\circ t(\overline\varphi)\circ\beta$ for every $\varphi\in\Hom_\cR\bigl(G,D_T(F)\bigr)$.
\item Let $\psi\colon T\to T'$ be a chain map with $\psi\circ c_T=c_{T'}\circ\overline\psi$, and let $t'$ be the transposition for $T'$. Then the diagram
\[
\xymatrix@C=4pc{
 \overline{\Hom_\cR\bigl(G,D_T(F)\bigr)} \ar[r]^-{t} \ar[d]_-{\overline{\psi_*}} & \Hom_\cR\bigl(F,D_T(G)\bigr) \ar[d]^-{\psi_*} \\
 \overline{\Hom_\cR\bigl(G,D_{T'}(F)\bigr)} \ar[r]^-{t'} & \Hom_\cR\bigl(F,D_{T'}(G)\bigr)
}
\]
commutes, where $\psi_*$ is postcomposition with $\psi$, that is, $t'\bigl(\overline{\psi_*\circ\varphi}\bigr)=\psi_*\circ t(\overline\varphi)$ for every $\varphi\in\Hom_\cR\bigl(G,D_T(F)\bigr)$.
\end{enumerate}
\end{lemma}

\begin{proof}
For (a), every map in Definition~\ref{v2:def:transposition} is natural in $F$ and $G$. This holds for $\Phi$ by Proposition~\ref{prop:tensor-hom}, and for $\tau$ because it is the symmetry. For (b), $\Phi$ and $\tau$ are natural in $T$, and $c_{T'}\circ\overline\psi=\psi\circ c_T$ by assumption.
\end{proof}

\begin{lemma}
\label{v2:lemma:transposition-twice}
For every $\varphi\in\Hom_\cR\bigl(F,D_T(G)\bigr)$,
\[
 t\bigl(\overline{t(\overline\varphi)}\bigr)=\bigl(c_T\circ\overline{c_T}\bigr)_*\circ\varphi.
\]
In particular, if $c_T$ is an involution, then $t\bigl(\overline{t(\overline\varphi)}\bigr)=\varphi$, and $t$ is bijective.
\end{lemma}

\begin{proof}
Let $x\in G_i$ and $y\in F_j$. By Lemma~\ref{v2:lemma:transposition-formula}, applied twice,
\[
 t\bigl(\overline{t(\overline\varphi)}\bigr)(y)(\overline x)=(-1)^{ij}c_T\Bigl(\overline{t(\overline\varphi)(x)(\overline y)}\Bigr)=c_T\Bigl(\overline{c_T\bigl(\overline{\varphi(y)(\overline x)}\bigr)}\Bigr)=\bigl(c_T\circ\overline{c_T}\bigr)\bigl(\varphi(y)(\overline x)\bigr).\qedhere
\]
\end{proof}

\begin{lemma}
\label{v2:lemma:transposition-can}
The canonical map is a transposition:
\[
 \can^T_F=t\bigl(\overline{\id_{D_T(F)}}\bigr).
\]
More generally, for every chain map $\varphi\colon G\to D_T(F)$,
\[
 t(\overline\varphi)=D_T(\varphi)\circ\can^T_F.
\]
\end{lemma}

\begin{proof}
By Definition~\ref{v2:def:ev-can}, $\Phi^{-1}(\id_{D_T(F)})=\ev^T_F$, so Definition~\ref{v2:def:transposition} sends $\overline{\id_{D_T(F)}}$ to $\Phi\bigl(c_T\circ\overline{\ev^T_F}\circ\tau\bigr)=\can^T_F$. For the second statement, write $\varphi=\id_{D_T(F)}\circ\varphi$. By Lemma~\ref{v2:lemma:transposition-natural}(a), with $\alpha=\varphi$ and $\beta=\id_F$,
\[
 t(\overline\varphi)=D_T(\varphi)\circ t\bigl(\overline{\id_{D_T(F)}}\bigr)=D_T(\varphi)\circ\can^T_F.\qedhere
\]
\end{proof}

The canonical maps now satisfy the expected identities, by formal arguments.

\begin{corollary}
\label{v2:cor:can-identities}
\begin{enumerate}[(a)]
\item The canonical map satisfies
\[
 D_T(\can^T_F)\circ\can^T_{D_T(F)}=\bigl(c_T\circ\overline{c_T}\bigr)_*.
\]
In particular, it is the identity if $c_T$ is an involution.
\item For every chain map $f\colon F\to G$, the diagram
\[
\xymatrix@C=3pc{
 F \ar[r]^-{\can^T_F} \ar[d]_-{f} & D_T^2(F) \ar[d]^-{D_T^2(f)} \\
 G \ar[r]^-{\can^T_G} & D_T^2(G)
}
\]
commutes, that is, $\can^T_G\circ f=D_T^2(f)\circ\can^T_F$.
\item Let $\psi\colon T\to T'$ be a chain map with $\psi\circ c_T=c_{T'}\circ\overline\psi$. Then the diagram
\[
\xymatrix@C=3pc{
 F \ar[r]^-{\can^T_F} \ar[d]_-{\can^{T'}_F} & D_T^2(F) \ar[d]^-{\psi_*} \\
 D_{T'}^2(F) \ar[r]^-{D_{T'}(\psi_*)} & \Hom_\cR\bigl(\overline{D_T(F)},T'\bigr)
}
\]
commutes, that is, $D_{T'}(\psi_*)\circ\can^{T'}_F=\psi_*\circ\can^T_F$.
\item Let $\psi\colon T\to T'$ be as in (c). For every chain map $\varphi\colon G\to D_T(F)$,
\[
 D_{T'}(\psi_*\circ\varphi)\circ\can^{T'}_F=\psi_*\circ D_T(\varphi)\circ\can^T_F
\]
as chain maps $F\to D_{T'}(G)$.
\end{enumerate}
\end{corollary}

\begin{proof}
For (a), apply Lemma~\ref{v2:lemma:transposition-can} to $\varphi=\can^T_F$, and then Lemma~\ref{v2:lemma:transposition-twice}:
\[
 D_T(\can^T_F)\circ\can^T_{D_T(F)}=t\bigl(\overline{\can^T_F}\bigr)=t\bigl(\overline{t(\overline{\id_{D_T(F)}})}\bigr)=\bigl(c_T\circ\overline{c_T}\bigr)_*.
\]
For (b), compute $t\bigl(\overline{D_T(f)}\bigr)$ in two ways, using Lemma~\ref{v2:lemma:transposition-natural}(a) and Lemma~\ref{v2:lemma:transposition-can}:
\begin{align*}
 t\bigl(\overline{D_T(f)}\bigr)&=t\bigl(\overline{\id_{D_T(G)}}\bigr)\circ f=\can^T_G\circ f,\\
 t\bigl(\overline{D_T(f)}\bigr)&=D_T^2(f)\circ t\bigl(\overline{\id_{D_T(F)}}\bigr)=D_T^2(f)\circ\can^T_F.
\end{align*}
For (c), compute $t'\bigl(\overline{\psi_*}\bigr)$ for the chain map $\psi_*\colon D_T(F)\to D_{T'}(F)$ in two ways, using Lemma~\ref{v2:lemma:transposition-natural}(b) and Lemma~\ref{v2:lemma:transposition-can}:
\begin{align*}
 t'\bigl(\overline{\psi_*}\bigr)&=\psi_*\circ t\bigl(\overline{\id_{D_T(F)}}\bigr)=\psi_*\circ\can^T_F,\\
 t'\bigl(\overline{\psi_*}\bigr)&=D_{T'}(\psi_*)\circ\can^{T'}_F.
\end{align*}
For (d), since $D_{T'}$ is a contravariant functor, $D_{T'}(\psi_*\circ\varphi)=D_{T'}(\varphi)\circ D_{T'}(\psi_*)$, where $D_{T'}(\varphi)$ is precomposition with $\overline\varphi$. Precomposition with $\overline\varphi$ commutes with postcomposition with $\psi$, so $D_{T'}(\varphi)\circ\psi_*=\psi_*\circ D_T(\varphi)$. Using (c),
\[
 D_{T'}(\psi_*\circ\varphi)\circ\can^{T'}_F=D_{T'}(\varphi)\circ D_{T'}(\psi_*)\circ\can^{T'}_F=D_{T'}(\varphi)\circ\psi_*\circ\can^T_F=\psi_*\circ D_T(\varphi)\circ\can^T_F.\qedhere
\]
\end{proof}

\begin{lemma}
\label{v2:lemma:dual-invariance}
\begin{enumerate}[(a)]
\item The functor $D_T$ sends chain homotopic maps to chain homotopic maps.
\item If $T$ is $K$-injective, then $D_T$ sends quasi-isomorphisms to quasi-isomorphisms.
\item Let $F$ be $K$-projective, and let $T\to T'$ be a quasi-isomorphism. Then postcomposition with it is a quasi-isomorphism $D_T(F)\to D_{T'}(F)$.
\item If $T$ is $K$-injective and $F$ is $K$-flat, then $D_T(F)$ is $K$-injective.
\end{enumerate}
\end{lemma}

\begin{proof}
Part (a) holds because conjugation and precomposition preserve chain homotopy. By Remark~\ref{v2:rmk:conjugation}, conjugation preserves quasi-isomorphisms and $K$-projective complexes, so (b) and (c) follow from Lemma~\ref{lem:new-k-preserve-qi}(b) and (a). For (d), let $E$ be acyclic. By Proposition~\ref{prop:tensor-hom},
\[
 \Hom_\cR\bigl(E,D_T(F)\bigr)\cong\Hom_\cR(\overline F\otimes_\cR E,T).
\]
Since $\overline F$ is $K$-flat, $\overline F\otimes_\cR E$ is acyclic. Since $T$ is $K$-injective, the right-hand side is acyclic.
\end{proof}

\subsection{Derived duals}
Now let $T$ be a complex with a morphism $c_T\colon\overline T\to T$ in $\D(\cR)$. We say that $c_T$ is an \emph{involution} if $c_T\circ\overline{c_T}=\id_T$ in $\D(\cR)$; we only assume this where we say so. This part mirrors the chain-level one, with $\RHom_\cR$ and $\otimes^{\mathrm L}_\cR$ in place of $\Hom_\cR$ and $\otimes_\cR$.

\begin{defn}
\label{v2:def:derived-dual}
The \emph{derived $T$-dual} of a complex $F$ is
\[
 \mathbb D_T(F)=\RHom_\cR(\overline F,T).
\]
For a morphism $f\colon F\to G$ in $\D(\cR)$, the morphism $\mathbb D_T(f)\colon\mathbb D_T(G)\to\mathbb D_T(F)$ is precomposition with $\overline f$.
\end{defn}

By Theorem~\ref{thm:new-derived-tensor-hom}, applied to $\overline F$, $G$ and $T$, there is an isomorphism in $\D(\cR)$
\[
 \Phi\colon\RHom_\cR\bigl(\overline F\otimes^{\mathrm L}_\cR G,T\bigr)\xrightarrow{\ \cong\ }\RHom_\cR\bigl(G,\mathbb D_T(F)\bigr),
\]
natural in $F$, $G$ and $T$. On $H_0$, it is a bijection between the corresponding Hom sets of $\D(\cR)$, which we also call $\Phi$.

\begin{defn}
\label{v2:def:derived-ev-can}
The \emph{evaluation} is the morphism
\[
 \ev^T_F=\Phi^{-1}\bigl(\id_{\mathbb D_T(F)}\bigr)\colon\overline F\otimes^{\mathrm L}_\cR\mathbb D_T(F)\longrightarrow T.
\]
The \emph{canonical map} is the morphism
\[
 \can^T_F=\Phi\bigl(c_T\circ\overline{\ev^T_F}\circ\tau\bigr)\colon F\longrightarrow\mathbb D_T^2(F).
\]
\end{defn}

\begin{defn}
\label{v2:def:derived-transposition}
Let $F$ and $G$ be complexes. The \emph{transposition} is the morphism
\[
 t\colon\overline{\RHom_\cR\bigl(G,\mathbb D_T(F)\bigr)}\longrightarrow\RHom_\cR\bigl(F,\mathbb D_T(G)\bigr)
\]
in $\D(\cR)$ given by the composite
\[
 t=\Phi\circ\tau^*\circ(c_T)_*\circ\overline{\Phi}^{-1},
\]
as in Definition~\ref{v2:def:transposition}. 
\end{defn}

On $H_0$, the transposition gives a map
\[
 t\colon\overline{\Hom_{\D(\cR)}\bigl(G,\mathbb D_T(F)\bigr)}\longrightarrow\Hom_{\D(\cR)}\bigl(F,\mathbb D_T(G)\bigr),
 \qquad
 t(\overline\varphi)=\Phi\Bigl(c_T\circ\overline{\Phi^{-1}(\varphi)}\circ\tau\Bigr),
\]
for morphisms $\varphi\colon G\to\mathbb D_T(F)$ in $\D(\cR)$, which we also call $t$.

The chain-level formulas of Lemma~\ref{v2:lemma:ev-can-formulas} and Lemma~\ref{v2:lemma:transposition-formula} are replaced by the following computation on models.

\begin{lemma}
\label{v2:lemma:involutive-model}
Let $\iota\colon T\to I$ be a $K$-injective resolution. There is a linear chain map $c_I\colon\overline I\to I$ such that the diagram
\[
\xymatrix@C=3pc{
 \overline T \ar[r]^-{\overline\iota} \ar[d]_-{c_T} & \overline I \ar[d]^-{c_I} \\
 T \ar[r]^-{\iota} & I
}
\]
commutes in $\D(\cR)$, that is, $c_I\circ\overline\iota=\iota\circ c_T$, and it is unique up to chain homotopy.
\end{lemma}

\begin{proof}
Since $\overline\iota$ is a quasi-isomorphism, the morphism $\iota\circ c_T\circ\overline\iota^{-1}\colon\overline I\to I$ in $\D(\cR)$ is defined. Since $I$ is $K$-injective, Proposition~\ref{thm:new-k-resolution-morphisms}(b) shows that it is represented by a chain map $c_I$, unique up to chain homotopy.
\end{proof}

\begin{lemma}
\label{v2:lemma:derived-models}
Let $\iota\colon T\to I$ and $c_I$ be as in Lemma~\ref{v2:lemma:involutive-model}, and let $P\to F$ and $Q\to G$ be $K$-projective resolutions.
\begin{enumerate}[(a)]
\item The complex $D_I(P)$ is $K$-injective, and it is a model for $\mathbb D_T(F)$. The complex $\overline P\otimes_\cR D_I(P)$ is a model for $\overline F\otimes^{\mathrm L}_\cR\mathbb D_T(F)$. On these models, the derived $\Phi$ is the chain-level $\Phi$ of \eqref{v2:eq:chain-Phi} for $I$.
\item The morphisms $\ev^T_F$ and $\can^T_F$ are represented by the chain maps $\ev^I_P$ and $\can^I_P$ of Definition~\ref{v2:def:ev-can}.
\item The transposition is represented by the transposition of Definition~\ref{v2:def:transposition} for $I$,
\[
 t\colon\overline{\Hom_\cR\bigl(Q,D_I(P)\bigr)}\longrightarrow\Hom_\cR\bigl(P,D_I(Q)\bigr).
\]
\end{enumerate}
\end{lemma}

\begin{proof}
For (a), $D_I(P)$ is $K$-injective by Lemma~\ref{v2:lemma:dual-invariance}(d), and it computes $\RHom_\cR(\overline F,T)$ since $\overline P$ is a $K$-projective resolution of $\overline F$. Since $\overline P$ is $K$-flat, $\overline P\otimes_\cR D_I(P)$ computes the derived tensor product. The derived $\Phi$ is computed by the chain-level $\Phi$ on $K$-projective and $K$-injective models, by the proof of Theorem~\ref{thm:new-derived-tensor-hom}.

For (b) and (c), each ingredient of Definition~\ref{v2:def:derived-ev-can} and Definition~\ref{v2:def:derived-transposition} is computed on these models by the corresponding chain-level map. For $\Phi$, this is (a). Conjugation and $\tau$ are computed by conjugation and $\tau$ on the models, and $c_T$ is computed by $c_I$. So the derived $\ev$, $\can$ and $t$ are computed by the chain-level ones, which are given by the same formulas in Definition~\ref{v2:def:ev-can} and Definition~\ref{v2:def:transposition}.
\end{proof}

\begin{lemma}
\label{v2:lemma:derived-transposition-natural}
\begin{enumerate}[(a)]
\item For morphisms $\alpha\colon G'\to G$ and $\beta\colon F'\to F$ in $\D(\cR)$, the diagram
\[
\xymatrix@C=4pc{
 \overline{\RHom_\cR\bigl(G,\mathbb D_T(F)\bigr)} \ar[r]^-{t} \ar[d]_-{\overline{\mathbb D_T(\beta)_*\alpha^*}} & \RHom_\cR\bigl(F,\mathbb D_T(G)\bigr) \ar[d]^-{\mathbb D_T(\alpha)_*\beta^*} \\
 \overline{\RHom_\cR\bigl(G',\mathbb D_T(F')\bigr)} \ar[r]^-{t} & \RHom_\cR\bigl(F',\mathbb D_T(G')\bigr)
}
\]
commutes in $\D(\cR)$, where the vertical maps are pre- and postcomposition. In particular,
\[
t\bigl(\overline{\mathbb D_T(\beta)\circ\varphi\circ\alpha}\bigr)
=\mathbb D_T(\alpha)\circ t(\overline\varphi)\circ\beta
\]
for every morphism $\varphi\colon G\to\mathbb D_T(F)$ in $\D(\cR)$.
\item Let $\psi\colon T\to T'$ be a morphism in $\D(\cR)$ with $\psi\circ c_T=c_{T'}\circ\overline\psi$, and let $t'$ be the transposition for $T'$. Then the diagram
\[
\xymatrix@C=4pc{
 \overline{\RHom_\cR\bigl(G,\mathbb D_T(F)\bigr)} \ar[r]^-{t} \ar[d]_-{\overline{\psi_*}} & \RHom_\cR\bigl(F,\mathbb D_T(G)\bigr) \ar[d]^-{\psi_*} \\
 \overline{\RHom_\cR\bigl(G,\mathbb D_{T'}(F)\bigr)} \ar[r]^-{t'} & \RHom_\cR\bigl(F,\mathbb D_{T'}(G)\bigr)
}
\]
commutes in $\D(\cR)$, where $\psi_*$ is postcomposition with $\psi$. In particular, $t'\bigl(\overline{\psi_*\circ\varphi}\bigr)=\psi_*\circ t(\overline\varphi)$ for every morphism $\varphi\colon G\to\mathbb D_T(F)$ in $\D(\cR)$.
\end{enumerate}
\end{lemma}

\begin{proof}
As for Lemma~\ref{v2:lemma:transposition-natural}. For (a), every map in Definition~\ref{v2:def:derived-transposition} is natural in $F$ and $G$. This holds for $\Phi$ by Theorem~\ref{thm:new-derived-tensor-hom}, and for $\tau$ because it is the symmetry. For (b), $\Phi$ and $\tau$ are natural in $T$, and $c_{T'}\circ\overline\psi=\psi\circ c_T$ by assumption.
\end{proof}

\begin{lemma}
\label{v2:lemma:derived-transposition-twice}
The composite of $\overline t$ and $t$ is postcomposition with $c_T\circ\overline{c_T}$. That is, the diagram
\[
\xymatrix@C=4pc{
 \RHom_\cR\bigl(F,\mathbb D_T(G)\bigr) \ar[r]^-{\overline t} \ar[dr]_-{(c_T\circ\overline{c_T})_*} & \overline{\RHom_\cR\bigl(G,\mathbb D_T(F)\bigr)} \ar[d]^-{t} \\
 & \RHom_\cR\bigl(F,\mathbb D_T(G)\bigr)
}
\]
commutes in $\D(\cR)$, that is, $t\circ\overline t=(c_T\circ\overline{c_T})_*$; for a morphism $\varphi\colon F\to\mathbb D_T(G)$ in $\D(\cR)$, $t\bigl(\overline{t(\overline\varphi)}\bigr)=(c_T\circ\overline{c_T})_*\circ\varphi$. In particular, if $c_T$ is an involution, then $t\circ\overline t=\id$, and $t$ is an isomorphism in $\D(\cR)$.
\end{lemma}

\begin{proof}
By Lemma~\ref{v2:lemma:derived-models}(c), $t$ is computed by the chain-level transposition for $I$. By Lemma~\ref{v2:lemma:transposition-twice}, the chain-level composite is postcomposition with $c_I\circ\overline{c_I}$. This chain map represents $c_T\circ\overline{c_T}$, by Lemma~\ref{v2:lemma:involutive-model}.
\end{proof}

\begin{lemma}
\label{v2:lemma:derived-transposition-can}
The canonical map is a transposition:
\[
 \can^T_F=t\bigl(\overline{\id_{\mathbb D_T(F)}}\bigr).
\]
More generally, for every morphism $\varphi\colon G\to\mathbb D_T(F)$ in $\D(\cR)$,
\[
 t(\overline\varphi)=\mathbb D_T(\varphi)\circ\can^T_F.
\]
\end{lemma}

\begin{proof}
As for Lemma~\ref{v2:lemma:transposition-can}. By Definition~\ref{v2:def:derived-ev-can}, $\Phi^{-1}(\id_{\mathbb D_T(F)})=\ev^T_F$, so Definition~\ref{v2:def:derived-transposition} sends $\overline{\id_{\mathbb D_T(F)}}$ to $\Phi\bigl(c_T\circ\overline{\ev^T_F}\circ\tau\bigr)=\can^T_F$. For the second statement, write $\varphi=\id_{\mathbb D_T(F)}\circ\varphi$. By Lemma~\ref{v2:lemma:derived-transposition-natural}(a), with $\alpha=\varphi$ and $\beta=\id_F$,
\[
 t(\overline\varphi)=\mathbb D_T(\varphi)\circ t\bigl(\overline{\id_{\mathbb D_T(F)}}\bigr)=\mathbb D_T(\varphi)\circ\can^T_F.\qedhere
\]
\end{proof}

\begin{corollary}
\label{v2:cor:derived-can-identities}
\begin{enumerate}[(a)]
\item The canonical map satisfies
\[
 \mathbb D_T(\can^T_F)\circ\can^T_{\mathbb D_T(F)}=\bigl(c_T\circ\overline{c_T}\bigr)_*.
\]
In particular, it is the identity if $c_T$ is an involution.
\item For every morphism $f\colon F\to G$ in $\D(\cR)$, the diagram
\[
\xymatrix@C=3pc{
 F \ar[r]^-{\can^T_F} \ar[d]_-{f} & \mathbb D_T^2(F) \ar[d]^-{\mathbb D_T^2(f)} \\
 G \ar[r]^-{\can^T_G} & \mathbb D_T^2(G)
}
\]
commutes, that is, $\can^T_G\circ f=\mathbb D_T^2(f)\circ\can^T_F$.
\item Let $\psi\colon T\to T'$ be a morphism in $\D(\cR)$ with $\psi\circ c_T=c_{T'}\circ\overline\psi$. Then the diagram
\[
\xymatrix@C=3pc{
 F \ar[r]^-{\can^T_F} \ar[d]_-{\can^{T'}_F} & \mathbb D_T^2(F) \ar[d]^-{\psi_*} \\
 \mathbb D_{T'}^2(F) \ar[r]^-{\mathbb D_{T'}(\psi_*)} & \RHom_\cR\bigl(\overline{\mathbb D_T(F)},T'\bigr)
}
\]
commutes, that is, $\mathbb D_{T'}(\psi_*)\circ\can^{T'}_F=\psi_*\circ\can^T_F$.
\end{enumerate}
\end{corollary}

\begin{proof}
The proof of Corollary~\ref{v2:cor:can-identities} applies word for word, using Lemma~\ref{v2:lemma:derived-transposition-natural}, Lemma~\ref{v2:lemma:derived-transposition-twice} and Lemma~\ref{v2:lemma:derived-transposition-can} in place of their chain-level versions.
\end{proof}

\subsection{Dualizing complexes}
\label{sec:dd}

Continue to write $\D_f^b(\cR)$ for the full subcategory of complexes with bounded, finitely generated homology.

We say that a complex $\omega$ has \emph{finite injective dimension} if there
are integers $a\leq b$ such that, for every $\cR$-module $M$,
\[
 \Ext^i_\cR(M,\omega):=H_{-i}\RHom_\cR(M,\omega)=0
 \qquad\text{for } i\notin[a,b].
\]
See \cite[Definition 12.4.8]{Yekutieli_2019} for the general notion. In
particular, if $N$ is an $\cR$-module of finite injective dimension, then
$\omega=N[0]$ has finite injective dimension as a complex.

For any complex $\omega$ there is a natural \emph{homothety map}
\[
 \cR\longrightarrow\RHom_\cR(\omega,\omega),
\]
sending $r\in\cR$ to multiplication by $r$. Concretely, choose a $K$-injective
resolution $\omega\to I$; then $\RHom_\cR(\omega,\omega)\simeq\Hom_\cR(I,I)$, and
the map sends $r$ to the chain map $r\cdot\id_I$, which is $\cR$-linear because
$\cR$ is commutative. Equivalently, it is the map adjoint to the unit
isomorphism $\cR\otimes_\cR^{\mathrm L}\omega\simeq \omega$.

\begin{defn}[{\cite[Defn. 13.1.9]{Yekutieli_2019}}]
\label{def:dualizing-complex}
A \emph{dualizing complex} over a commutative Noetherian ring $\cR$ is a
complex $\omega$ such that:
\begin{enumerate}[(a)]
\item $\omega\in \D_f^b(\cR)$;
\item $\omega$ has finite injective dimension ; and
\item the homothety map
      $\cR\to\RHom_\cR(\omega,\omega)$ is an isomorphism in $\D(\cR)$.
\end{enumerate}
\end{defn}

The following result is often called \emph{Grothendieck duality};
see \cite[\S 0.2]{Yekutieli_2019}.

\begin{theorem}\label{thm:derivedduality}
Let $\cR$ be a commutative Noetherian ring and let $\omega$ be a dualizing
complex with an involution represented by a morphism
$c_\omega \colon \overline{\omega} \to \omega$ in $\D(\cR)$, that is, $c_\omega\circ\overline{c_\omega}=\id_\omega$ in $\D(\cR)$. Then
\[
 \mathbb D_\omega\colon\D_f^{b}(\cR)^{\op}\longrightarrow\D_f^{b}(\cR)
\]
is an equivalence of triangulated categories, and the canonical map
\[
 \can^\omega\colon\id\xrightarrow{\ \simeq\ }\mathbb D_\omega^2
\]
is a natural isomorphism on $\D_f^b(\cR)$ satisfying
\[
 \mathbb D_\omega(\can_X^\omega)\circ\can^\omega_{\mathbb D_\omega(X)}
 =\id_{\mathbb D_\omega(X)}.
\]
\end{theorem}

\begin{proof}
For the ordinary duality $\RHom_\cR(-,\omega)$, all statements except the last follow from
\cite[Theorem 13.1.18 and Corollary 13.1.19]{Yekutieli_2019}. Since
$\mathbb D_\omega(F)=\RHom_\cR(\overline F,\omega)$ and conjugation is an exact
involutive equivalence preserving $\D_f^b(\cR)$, the same holds for
$\mathbb D_\omega$.
The identity is Corollary~\ref{v2:cor:derived-can-identities}(a), since $c_\omega$ is an involution in $\D(\cR)$.
\end{proof}

\begin{remark}\label{cor:derivedduality}
The triple $(\D^b_f(\cR),\mathbb D_\omega,\can^\omega)$ is a triangulated
category with duality in the sense of \cite{Balmer_2000}.
\end{remark}

\section{Commutative algebra and duality}
\label{subsec:commutative-algebra-duality}

Throughout this section,
$\cR$ is a Noetherian commutative ring with involution $c \colon \cR \to \cR$ written as $\bar a:=c(a)$.  We use homological grading conventions.
\par

\subsection{Cohen-Macaulay rings and canonical modules}
\label{app:commutative_background}

\begin{defn}\label{def:module-depth}
Let $(\cR,\mathfrak m)$ be a Noetherian local ring and let $M\neq0$ be a finitely generated $\cR$-module.
The \textbf{depth} $\depth_\cR M$ is the maximum length of an $M$-regular sequence in $\mathfrak m$.
Here a sequence $x_1,\ldots,x_r$ is $M$-regular if multiplication by $x_i$ is injective on
$M/(x_1,\ldots,x_{i-1})M$ for each $i$, and $M/(x_1,\ldots,x_r)M\neq0$.
We set $\depth_\cR0=+\infty$ and $\dim0=-\infty$.
\end{defn}
\par

\begin{lemma}
	Let $\cR$ be a Noetherian local ring and $M$ a nonzero finitely generated $\cR$-module. Then $\depth M \leq \dim M$. 
\end{lemma}

\begin{defn}
	A local ring is \textbf{Cohen-Macaulay} if its depth equals its Krull dimension. A
	general ring is \textbf{Cohen-Macaulay} if it is so at every localization. 
\end{defn}

\begin{defn}\label{def:maximal-cm}
Let $(\cR,\mathfrak m)$ be a Noetherian local ring. A nonzero finitely generated $\cR$-module $M$
is \textbf{Cohen--Macaulay} if $\depth_\cR M=\dim_\cR M$, and is \textbf{maximal Cohen--Macaulay}
if $\depth_\cR M=\dim \cR$. The zero module is included in both classes by convention.
Over a general Noetherian ring, a finitely generated module has either property if its
localization at every prime has that property.
\end{defn}

\begin{defn}\label{def:canonical-module}
For a Cohen--Macaulay local ring $\cR$, a \textbf{canonical module} is a finitely generated
$\cR$-module $\omega_{\cR}$ for which $\omega_{\cR}[0]$ is a dualizing complex in the sense of
Definition~\ref{def:dualizing-complex}. For a general Cohen--Macaulay ring $\cR$, a canonical module is
a finitely generated module $\omega_{\cR}$ such that $(\omega_{\cR})_{\mathfrak m}$ is a canonical
module of $\cR_{\mathfrak m}$ for every maximal ideal $\mathfrak m$.
Such a module also localizes to a canonical module at every prime
\cite[Theorem 3.3.5 and Definition 3.3.16]{Bruns_1998}.
\end{defn}

\begin{theorem}[Local canonical-module duality]\label{theorem:ext-bounds-local}
Let $(\cR,\mathfrak m)$ be a Cohen--Macaulay local ring of dimension $D$ with canonical module
$\omega_{\cR}$, and let $M\neq0$ be a finitely generated $\cR$-module. Set $k=\dim_\cR M$ and
$l=\depth_\cR M$. Then:
\begin{enumerate}[(a)]
\item $\Ext_\cR^{D-i}(M,\omega_{\cR})=0$ unless $l\leq i\leq k$.
\item $\Ext_\cR^{D-i}(M,\omega_{\cR})\neq0$ for $i=l$ and $i=k$.
\item $\dim_\cR\Ext_\cR^{D-i}(M,\omega_{\cR})\leq i$ for all $i$, with equality for $i=k$.
\item $M$ is Cohen--Macaulay if and only if $\Ext_\cR^j(M,\omega_{\cR})=0$ for all $j\neq D-k$.
In particular, $M$ is maximal Cohen--Macaulay if and only if $\Ext_\cR^j(M,\omega_{\cR})=0$
for all $j>0$.
\end{enumerate}
\end{theorem}

\begin{proof}
The vanishing, nonvanishing, and dimension inequality are
\cite[Corollary 3.5.11]{Bruns_1998}, with the Ext index written as $D-i$.
For the equality in (c), choose a minimal prime $\mathfrak p$ of $\Supp M$ with
$\dim \cR/\mathfrak p=k$. Then $\dim \cR_{\mathfrak p}=D-k$ by Lemma~\ref{lemma:cm-height-formula},
and $M_{\mathfrak p}$ is nonzero of dimension and depth zero. Since $(\omega_{\cR})_{\mathfrak p}$
is canonical, (b) over $\cR_{\mathfrak p}$ gives
\[
\Ext_\cR^{D-k}(M,\omega_{\cR})_{\mathfrak p}
\cong\Ext_{\cR_{\mathfrak p}}^{D-k}(M_{\mathfrak p},(\omega_{\cR})_{\mathfrak p})\neq0.
\]
Thus its dimension is at least $k$, proving equality. Finally, the two nonvanishing endpoints
coincide exactly when $l=k$, which proves the first assertion of (d). The higher Ext groups
all vanish exactly when $D-l=0$, which proves the last assertion.
\end{proof}
\par

\begin{defn}
	A ring $\cR$ is called \textbf{equicodimensional} if all of its maximal ideals have the same
	codimension. 
\end{defn}

\begin{lemma}
\label{lemma:cm-height-formula}
	In an equicodimensional Cohen-Macaulay ring,
	\[
		\height \mathfrak p + \dim \cR / \mathfrak p = \dim \cR 
	\]
	for all $\mathfrak p \in \Spec(\cR)$. 
\end{lemma}

\begin{proof}
For each maximal ideal $\mathfrak m$ containing $\mathfrak p$, the local Cohen--Macaulay
height formula gives
\[
\dim \cR_{\mathfrak m}/\mathfrak p\cR_{\mathfrak m}=\dim \cR-\operatorname{ht}\mathfrak p;
\]
see \cite[\href{https://stacks.math.columbia.edu/tag/00NA}{Lemma 00NA}]{stacks-project}.
Taking the maximum over such $\mathfrak m$ gives the formula.
\end{proof}
\par

\begin{theorem}
	\label{theorem:ext-bounds}

Let $\cR$ be an equicodimensional Cohen--Macaulay ring of dimension $D$ with canonical module $\omega_{\cR}$.
	Let $M\neq0$ be a finitely generated $\cR$-module of dimension $k$. Let $l$ be the minimum of
	$\depth M_{\mathfrak m}$ over the maximal ideals $\mathfrak m\in\Supp M$.
	\begin{enumerate}[(a)]
		\item $\Ext^{D-i}_\cR(M,\omega_{\cR})=0$ unless $l\leq i\leq k$.
		\item For every maximal ideal $\mathfrak m\in\Supp M$, $\Ext^{D-i}_\cR(M,\omega_{\cR})\neq0$ for
			$i=\dim M_{\mathfrak m}$ and for $i=\depth M_{\mathfrak m}$. In particular, this holds for $i=k$
			and $i=l$.
		\item $\dim\Ext^{D-i}_\cR(M,\omega_{\cR})\leq i$ for all $i$, with equality for $i=k$.
		\item $M$ is Cohen--Macaulay of pure dimension $k$ if and only if $\Ext^{i}_\cR(M,\omega_{\cR})=0$ for all
			$i\neq D-k$.
	\end{enumerate}
\end{theorem}

\begin{proof}

Let $\mathfrak m$ be a maximal ideal. The ring $\cR_{\mathfrak m}$ is Cohen--Macaulay local of
dimension $D$, with canonical module $(\omega_{\cR})_{\mathfrak m}$. Since $M$ is finitely
generated and $\cR$ is Noetherian,
\par

	\[
		\Ext^j_\cR(M,\omega_{\cR})_{\mathfrak m}\cong\Ext^j_{\cR_{\mathfrak m}}(M_{\mathfrak m},(\omega_{\cR})_{\mathfrak m}).
	\]
	Let $\mathfrak m\in\Supp M$. By Theorem~\ref{theorem:ext-bounds-local}, the module
	$\Ext^{D-i}_{\cR_{\mathfrak m}}(M_{\mathfrak m},(\omega_{\cR})_{\mathfrak m})$ vanishes unless
	$\depth M_{\mathfrak m}\leq i\leq\dim M_{\mathfrak m}$. It is nonzero in both extreme cases, and it has
	dimension at most $i$. A finitely generated module vanishes if and only if its localizations at all
	maximal ideals vanish. Its dimension is the maximum of the dimensions of these localizations. In
	particular, $k$ is the maximum of $\dim M_{\mathfrak m}$ over the maximal ideals $\mathfrak m\in\Supp M$.
	This gives (a), (b) and the inequality in (c).

	For the equality in (c), let $\mathfrak p$ be a minimal prime of $\Supp M$ with
	$\dim \cR/\mathfrak p=k$. Then $\operatorname{ht}\mathfrak p=D-k$, and $M_{\mathfrak p}$ is nonzero of
	dimension $0$.

The ring $\cR_{\mathfrak p}$ is Cohen--Macaulay local of dimension $D-k$, with canonical module $(\omega_{\cR})_{\mathfrak p}$.
\par

So
	$\Ext^{D-k}_\cR(M,\omega_{\cR})_{\mathfrak p}\neq0$, again by Theorem~\ref{theorem:ext-bounds-local}. Hence
	$\dim\Ext^{D-k}_\cR(M,\omega_{\cR})\geq\dim \cR/\mathfrak p=k$.

	For (d), recall that every associated prime of a Cohen--Macaulay module over a local ring has the
	dimension of the module \cite[Theorem 2.1.2]{Bruns_1998}. For a prime $\mathfrak p$ contained in a maximal
	ideal $\mathfrak m$, we have
	\[
		\dim \cR_{\mathfrak m}/\mathfrak p\cR_{\mathfrak m}=D-\operatorname{ht}\mathfrak p=\dim \cR/\mathfrak p.
	\]
	Every associated prime of $M$ lies in a maximal ideal in $\Supp M$, and the minimal primes of
	$\Supp M_{\mathfrak m}$ are associated primes of $M_{\mathfrak m}$. So $M$ is Cohen--Macaulay of pure
	dimension $k$ if and only if $\depth M_{\mathfrak m}=\dim M_{\mathfrak m}=k$ for every maximal ideal
	$\mathfrak m\in\Supp M$. By the local statements above, this holds if and only if
	$\Ext^i_\cR(M,\omega_{\cR})=0$ for all $i\neq D-k$.
\end{proof}
\par

\begingroup 
\providecommand{\EW}[1]{#1}
\renewcommand{\EW}[1]{#1}

\begin{lemma}
	\label{lemma:assoc-local-depth}
	Let $\cR$ be a Noetherian ring and $M$ a finitely generated $\cR$-module.
	Then a prime $\frak p$ is associated to $M$ if and only if $\depth_{\cR_{\frak p}} M_{\frak p} =
	0$. 
\end{lemma}

\begin{proof}
	If $\frak p$ is associated to $M$, then consider the cyclic module $N \subset M$ whose annihilator
	ideal is $\frak p$. Then the maximal ideal $\frak p \cR_{\frak p}$,
	which coincides with the non-units in $\cR_{\frak p}$, annihilates $N_{\frak p} \subset M_{\frak
	p}$, which shows that $M_{\frak p}$ has no regular elements.

	Conversely, suppose $\depth_{\cR_{\frak p}} M_{\frak p} = 0$. Then every element of $\frak p
	\cR_{\frak p}$ is a zero-divisor, hence, by \cite[Theorem 3.1]{Eisenbud_1995}, $\frak p \cR_{\frak p}$ is
	contained in the union of the associated primes of $M_{\frak p}$. By prime avoidance,
	\cite[Lemma 3.3]{Eisenbud_1995}, $\frak p \cR_{\frak p}$ is contained in an associated
	prime; but it is already a maximal ideal, so it equals an associated prime.
	By \cite[\href{https://stacks.math.columbia.edu/tag/0310}{Tag 0310}]{stacks-project},
	it follows that $\frak p$ is associated to $M$.
\end{proof}

Recall that a module is \textbf{pure-dimensional}, or \textbf{pure} for short, if it has no
nonzero lower-dimensional submodules.

\begin{lemma}
	\label{v2:lemma:assoc-prime-pure-dimen}
	Let $A$ be a Noetherian ring and let $M$ be a finitely generated $A$-module of finite Krull dimension.
	Then $M$ is pure-dimensional if and only if all of its associated primes have the same dimension.
\end{lemma}

\begin{proof}
	For a nonzero finitely generated module $N$, the minimal primes of $\Supp N$ are associated, so
	\[
		\dim N=\max_{\mathfrak p\in\Ass N}\dim A/\mathfrak p.
	\]
	Suppose that every associated prime of $M$ has dimension $\dim M$, and let $N\subseteq M$ be nonzero. Then
	$\Ass N$ is nonempty and contained in $\Ass M$, so $\dim N=\dim M$. Conversely, suppose that
	$\mathfrak p=\Ann(x)$ is an associated prime of $M$ with $\dim A/\mathfrak p<\dim M$. Then
	$Ax\cong A/\mathfrak p$ is a nonzero submodule of lower dimension.
\end{proof}

\begin{theorem}
	\label{theorem:pure-condition}
	Let $\cR$ be an equicodimensional Cohen-Macaulay ring of dimension $D$
	with canonical module $\omega_\cR$.
	Let $M$ be a finitely generated $\cR$-module of dimension $k$. Then $M$ is pure if and only if
	\[
		\dim \Ext^{D-i}_\cR(M,\omega_\cR) < i
	\]
	for all $i < k$.
\end{theorem}

\begin{proof}
	By Theorem~\ref{theorem:ext-bounds} and Lemma~\ref{lemma:assoc-local-depth}, 
	a prime $\frak p$ is associated to
	$M$ if
	and only if 
	\[
		\Ext^{\operatorname{ht} \frak p}_{\cR_{\frak p}}(M_{\frak p}, (\omega_\cR)_{\frak p})
		\neq 0.
	\]
	Hence, $M$ has an associated prime of height $D - i$ if and only if $\Ext^{D-i}_{\cR}(M,\omega_\cR)$ has
	an associated prime of the same height, which happens if and only if $\dim \Ext^{D-i}_{\cR}(M,
	\omega_\cR) = i$. 
\end{proof}

\begin{corollary}
\label{cor:pure-one-dim-CM}
With $\cR$ as in the preceding theorem,
a pure 1-dimensional finitely generated $\cR$-module $M$ is Cohen--Macaulay.
\end{corollary}

\begin{proof}
The only possibly nonzero modules $\Ext^i_\cR(\overline M,\omega_\cR)$ are those with $i=D$ and $i=D-1$.
Purity implies that the dimension of the former is negative, which is to say that it vanishes. By
Theorem~\ref{theorem:ext-bounds}, $\overline M$, and hence $M$, is Cohen--Macaulay.
\end{proof}

\subsection{Duality over Gorenstein rings}
\par
\endgroup

Now we record a few results on Gorenstein rings. See also
\cite[\href{https://stacks.math.columbia.edu/tag/0DW6}{Section 0DW6}]{stacks-project} and Appendix A of
Ref.~\cite{Ruba_2024}.
\par

\begin{defn}
	A Noetherian local ring is \textbf{Gorenstein} if it has finite injective dimension as a module over itself. 
	A general Noetherian ring is \textbf{Gorenstein} if it is a Gorenstein local ring at every localization. 
\end{defn}

\begin{lemma}
	\label{lemma:gorenstein-inj-dim}
	If a Gorenstein ring has finite Krull dimension, then its injective dimension equals its Krull
	dimension.
\end{lemma}

The following gives an equivalent characterization of a Gorenstein ring. 

\begin{proposition}
	\label{thm:gorenstein-canonical}
A Noetherian ring $\cR$ is Gorenstein if and only if it is Cohen-Macaulay
and $\cR$ itself is a canonical module. 
\end{proposition}

\begin{proof}
The local statement is \cite[Theorem 3.3.7]{Bruns_1998}, and the general case follows by
localization. 
\end{proof}

We also have the following characterization in terms of dualizing complexes:

\begin{proposition}
	\label{thm:dualizing}
	A Noetherian ring $\cR$ of finite Krull dimension is Gorenstein if and only if 
	$\cR[0]$ is a dualizing complex. 
\end{proposition}

\begin{proof}
	Supposing $\cR$ is Gorenstein, we have:
	\begin{enumerate}[(a)]
	\item $\cR\in\D_f^b(\cR)$, since $\cR$ is finitely generated and concentrated
	in degree $0$;
\item $\cR$ has finite injective dimension by Lemma~\ref{lemma:gorenstein-inj-dim};
	\item $\RHom_\cR(\cR,\cR)=\Hom_\cR(\cR,\cR)=\cR$, and under this identification
	the homothety map is the identity.
	\end{enumerate}
	This makes $\cR[0]$ into a dualizing complex. 
	Conversely, if $\cR[0]$ is a dualizing complex, then $\cR_\frak p$ has finite injective dimension
	over itself for every prime $\frak p$, hence $\cR$ is Gorenstein. 
\end{proof}

Recall that the \emph{support} of a complex $F$ of $\cR$-modules, written
$\Supp_\cR(F)$, is the set of prime ideals $\mathfrak p\in\Spec(\cR)$ such that
the localized complex $F_{\mathfrak p}$ is not acyclic. Since localization is
exact,
\[
 \Supp_\cR(F)=\bigcup_{i\in\Z}\Supp_\cR\bigl(H_i(F)\bigr).
\]
For a subset $S\subseteq\Spec(\cR)$, let $\dim S$ be the supremum of the
lengths $k$ of chains of primes $\mathfrak p_0\subsetneq\cdots\subsetneq
\mathfrak p_k$ in $S$. The \emph{Krull dimension} of $F$ is
$\dim_\cR(F)=\dim\Supp_\cR(F)$; for a finitely generated module $M$ this is the
usual $\dim_\cR(M)=\dim\cR/\Ann_\cR(M)$. Supports are closed under passing to larger primes. So any chain in $\Supp_\cR(F)=\bigcup_i\Supp_\cR(H_i(F))$ lies
in a single $\Supp_\cR(H_i(F))$, namely one containing its smallest prime. Hence
\[
 \dim_\cR(F)=\sup_{i\in\Z}\dim_\cR\bigl(H_i(F)\bigr).
\]

\begin{lemma}
\label{lemma:dual-dim}
Let $\cR$ be a commutative Noetherian Gorenstein ring of finite
Krull dimension, equipped with an involution $c$, and let $F\in\D_f^b(\cR)$. Then
\[
 \Supp_\cR\mathbb{D}_\cR(F)=c\bigl(\Supp_\cR(F)\bigr),
\]
where $c$ denotes the automorphism of $\Spec(\cR)$ induced by the
involution. In particular, 
\[\dim_\cR\mathbb D_\cR(F)=\dim_\cR(F).\]
\end{lemma}

\begin{proof}
Since $\mathbb D_\cR(F)=\RHom_\cR(\overline F,\cR)$, it suffices to show:
\begin{enumerate}[(i)]
\item $\Supp_\cR(\overline G)=c\bigl(\Supp_\cR(G)\bigr)$ for every complex $G$;
\item $\Supp_\cR\RHom_\cR(G,\cR)=\Supp_\cR(G)$ for every $G\in\D_f^b(\cR)$.
\end{enumerate}
For (i), $s\in\cR$ acts on $\overline G$ as $\overline s$ acts on $G$, so
inverting $\cR\setminus\mathfrak p$ on $\overline G$ amounts to inverting
$\cR\setminus c(\mathfrak p)$ on $G$. Hence
$(\overline G)_{\mathfrak p}=\overline{G_{c(\mathfrak p)}}$, and
$\mathfrak p\in\Supp(\overline G)$ if and only if
$c(\mathfrak p)\in\Supp(G)$.

For (ii), let $\mathfrak p\in\Spec(\cR)$. Computing $\RHom_\cR(G,\cR)$ with a
degreewise finitely generated free resolution of $G$
(Lemma~\ref{lemma:bounded-replacement}(a)), we have
\[
 \RHom_\cR(G,\cR)_{\mathfrak p}\simeq\RHom_{\cR_{\mathfrak p}}(G_{\mathfrak p},\cR_{\mathfrak p}).
\]
The local ring $\cR_{\mathfrak p}$ is again Gorenstein of finite Krull
dimension, so by Proposition~\ref{thm:dualizing} and Theorem~\ref{thm:derivedduality}, the functor
$\RHom_{\cR_{\mathfrak p}}(-,\cR_{\mathfrak p})$ is an equivalence on
$\D_f^b(\cR_{\mathfrak p})$ whose square is isomorphic to the identity. Hence
$G_{\mathfrak p}\simeq 0$ if and only if
$\RHom_{\cR_{\mathfrak p}}(G_{\mathfrak p},\cR_{\mathfrak p})\simeq 0$.

Finally, $c$ preserves inclusions of primes, so it preserves dimension.
\end{proof}

In the paper, we deal with some examples that come to us as unbounded complexes of projective modules. These need not be $K$-projective. However, we can still easily identify their $\cR$-duals under some assumptions, all of which hold in our context.

\begin{proposition}
    \label{cor:fc-duality-is-derived}
    Let $\cR$ be a commutative Noetherian Gorenstein ring of finite Krull dimension.
    Let $F$ be a possibly unbounded complex of projective $\cR$-modules.
    Then
    \[
        \Hom_\cR(F, \cR) \simeq \RHom_\cR(F, \cR).
    \]
\end{proposition}

\begin{proof}
	We apply Theorem~\ref{thm:dualityisderived} with $T = \cR[0]$ and $I$ a $K$-injective resolution of $T$. 
\end{proof}

Relatedly, we briefly mention another equivalent characterization of a Noetherian Gorenstein ring.

\begin{defn}[{\cite[5.2]{Iyengar_2006}}]
    Let $\cR$ be a ring.
    An acyclic complex $A$ of projective $\cR$-modules is \emph{totally acyclic} if the hom-complex $\Hom_\cR(A, Q)$ is
    acyclic for any projective module $Q$.
\end{defn}

\begin{theorem}[{\cite[Cor.~5.5]{Iyengar_2006}}]
    \label{theorem:gorenstein-acyclic}
    Let $\cR$ be a Noetherian ring with a dualizing complex. Then $\cR$ is Gorenstein if and only if every
    acyclic complex of projective $\cR$-modules is totally acyclic.
\end{theorem}

\section{Dimension Bounds}

In this section we use Krull dimension bounds on the homology of a complex to constrain the homological
degrees of its dual. We first work over an arbitrary equicodimensional Gorenstein ring, with the
duality $\mathbb D_{\cR}$. We then deduce the version used in the paper, for the shifted duality
$\mathbb D_{\cR[1]}$ on $\Q$-acyclic complexes, in Section~\ref{subsubsec:shifted-duality}.

\begin{defn}
	\label{def:coconnective}
	An object $F\in\D(\cR)$ is \emph{co-connective} if $H_i(F)=0$ for $i>0$, and \emph{co-connected} if
	$H_i(F)=0$ for $i\geq0$.
\end{defn}

\begin{defn}
	\label{def:dimension-bounds}
	Let $\ell$ be an integer. An object $F \in \D_f^b(\cR)$ satisfies the \textbf{weak dimension bound}
	in dimension $\ell$ if it is co-connective and
	\[
		\dim H_{-i}(F) \leq \ell - i \quad \text{for all $i$}.
	\]
	It satisfies the \textbf{strong dimension bound} in dimension $\ell$ if it is co-connected and
	\[
		\dim H_{-i}(F) < \ell - i \quad \text{for all $i$}.
	\]
	By convention, the zero module has dimension $-\infty$.
\end{defn}

\subsection{Dimension bounds and duality}

In this subsection, $\cR$ is a commutative Noetherian Gorenstein ring which is equicodimensional of
Krull dimension $D$; that is, every maximal ideal has height $D$. Our duality is
\[
	\mathbb D_\cR(F)=\RHom_\cR(\overline F,\cR)
\]
for $F \in \D^b_f(\cR)$.

\providecommand{\EW}[1]{#1}
\renewcommand{\EW}[1]{#1}

For a prime $\frak p$, 
because localization is exact and because $\overline
{F_\frak p} \cong \overline F_{\overline {\frak p}}$,
\[
	\mathbb D_{\cR}(F)_{\bar {\frak p}} \cong
	\RHom_{\cR_{\bar{\frak p}}}\bigl(\overline{F_{\frak p}},\cR_{\bar{\frak p}}\bigr),
\]
where the right hand side refers to Grothendieck duality with respect to the localized ring
$\cR_{\overline {\frak p}}$, which indeed remains a Gorenstein Noetherian ring.
Here we are also notating $c(\frak p) = \bar {\frak p}$.
Conjugation here is transported along the isomorphism $c\colon\cR_{\frak p}\to\cR_{\bar{\frak p}}$.

\begin{theorem}
	\label{theorem:dimension-bounds-general}
	Let $F\in \D_f^b(\cR)$, and let $D=\dim\cR$. The following are equivalent:
	\begin{enumerate}[(1)]
		\item $F$ satisfies the weak (resp.\ strong) dimension bound in dimension $D$.
		\item $\mathbb D_\cR(F)$ satisfies the weak (resp.\ strong) dimension bound in dimension $D$.
		\item Both $F$ and $\mathbb D_\cR(F)$ are co-connective (resp.\ co-connected).
	\end{enumerate}
\end{theorem}

\begin{proof}
	To treat both bounds at once, we let $s=0$ for the weak bound and $s=1$ for the strong bound.
	We first show (1) implies (2). If $F$ is acyclic, this is trivial. 
	Otherwise, by the Whitehead tower, $F$ is
	an iterated extension of its homologies, and $\mathbb D_{\cR}(F)$ is an iterated extension of the
	duals of the shifted homologies. It suffices to show that each of these duals satisfies the
	relevant dimension bound. 
	To that end, take a nonzero homology $N = H_{-j}(F)$. By the dimension 
	bound, $j \geq s$ and $\dim N \leq D - j - s$. 
	Then
	\[
		h_{j-k} := H_{j-k} \mathbb D_{\cR}(N[-j])
		\cong
		\Ext_{\cR}^{k} (\overline N, \cR).
	\]
	By Theorem~\ref{theorem:ext-bounds}, $h_{j-k}$ can be nonzero only if
	$k\geq D-\dim N\geq j+s$, and hence only if $j-k\leq-s$. Thus $\mathbb D_\cR(N[-j])$ has homology in the appropriate range,
	and moreover $\dim h_{j-k} \leq D - k$, hence 
	\[
		\dim h_{-k} \leq D - j - k \leq D - k - s.
	\]
	Thus, the dual of $N[-j]$ satisfies the relevant dimension bound. This shows (1) implies (2).

	Then (2) implies (1) follows because double duality is canonically isomorphic to the identity.
	And because (1) implies (2), it also implies (3).

	It remains to show (3) implies (1). 
	Suppose $F$ does not satisfy the weak (resp. strong) dimension bound. 
	Then take the largest $m$ such that 
	\[
		\dim H_{-i} F > D - i - s \qquad \text{and} \qquad \dim H_{-i} F = m.
	\]
	And take the largest $a$ such that
	\[
		\dim H_{-a} F = m.
	\]
	Then $m>D-a-s$.
	Now take an associated prime $\frak p$ of $H_{-a} F$ such that
		$\dim \mathcal R / \frak p = m$.
	By equicodimensionality, we also have
		$\dim \mathcal R_{\frak p} = D - m$.
	Now consider localizing at $\frak p$. Since $H_{-a} F$ is the rightmost homology having
	dimension $\geq m$, we have
	\[
		H_{-b} F_{\frak p} = 0 \qquad \text{for all} \qquad b > a. 
	\]
	Meanwhile, for $b < a$, by the assumed maximality properties, either (i) $H_{-b} (F)[-b]$ satisfies
	the weak (resp. strong) dimension bound, 
	in which case
	$\RHom_{\cR_{\bar{\frak p}}}\bigl(\overline{H_{-b}(F_{\frak p})}[-b],\cR_{\bar{\frak p}}\bigr)$
	is co-connective (resp.\ co-connected)
	or (ii) $\dim_{\mathcal R} H_{-b} F \leq m$, hence $\dim_{\mathcal
		R_{\frak p}} H_{-b} F_{\frak p} \leq 0$,
		and only the top-degree Ext-module of $\overline{H_{-b}(F_{\frak p})}$ against $\mathcal R_{\bar {\frak p}}$ may be nonzero, which is to say
	\[
		H_{c-(D-m)} \RHom_{\cR_{\bar{\frak p}}}\bigl(\overline{H_{-b}(F_{\frak p})}[-b],\cR_{\bar{\frak p}}\bigr) = 0
		\qquad
		\text{for all}
		\qquad
		c > b,
	\]
	and in particular this holds for $c = a$. 
	The localized duals of all the direct homologies with $b<a$ have homology concentrated strictly below
	$a-(D-m)>-s$. The dualized Whitehead tower therefore gives
	\[
	\begin{aligned}
		&H_{a-(D-m)} \RHom_{\cR_{\bar{\frak p}}}\bigl(\overline{F_{\frak p}},\cR_{\bar{\frak p}}\bigr)\\
		&\qquad\cong H_{a-(D-m)}
		\RHom_{\cR_{\bar{\frak p}}}\bigl(\overline{H_{-a}(F_{\frak p})}[-a],\cR_{\bar{\frak p}}\bigr)\\
		&\qquad\neq 0.
	\end{aligned}
	\]
	This implies
	\[
		H_{a-(D-m)} \mathbb D_{\cR}(F)
		\neq 0.
	\]
	And $a - (D-m) > - s$, so $\mathbb D_{\cR}(F)$ is not co-connective (resp.\ co-connected).
\end{proof}

\subsection{Coefficients in \texorpdfstring{$\Q/\Z$}{Q/Z}}
\label{subsubsec:QZ-coefficients}

In this subsection, $\cR$ is a commutative Noetherian Gorenstein
ring of finite Krull dimension, equipped with an involution $c$,
and $\cR$ is flat over $\Z$. Let
\[
 \cR_\Q=\cR\otimes_\Z\Q,
 \qquad
 \cR_{\Q/\Z}=\cR\otimes_\Z\Q/\Z,
\]
with the involutions induced by $c$. Since $\cR$ is flat over $\Z$, tensoring the short exact sequence $0\to\Z\to\Q\to\Q/\Z\to0$ with $\cR$ gives a short exact sequence of $\cR$-modules
\[\xymatrix{ 0 \ar[r] & \cR \ar[r] & \cR_\Q \ar[r] & \cR_{\Q/\Z} \ar[r] & 0,}\]
whose maps commute with the involutions. As for any short exact sequence of complexes, the quotient map $\Cone(\cR\to\cR_\Q)\to\cR_{\Q/\Z}$ is a quasi-isomorphism, and we obtain a distinguished triangle
\begin{equation}
\label{eq:ZQZ-triangle}
\xymatrix{ \cR \ar[r] & \cR_\Q \ar[r] & \cR_{\Q/\Z} \ar[r]^-{\delta} & \cR[1]}
\end{equation}
in $\D(\cR)$. The connecting map $\delta$ also commutes with the involutions.

\begin{lemma}
\label{lemma:QZ-duality}
If $F\in\D(\cR)$ is $\Q$-acyclic, then
\[
 \delta_*\colon\mathbb D_{\cR_{\Q/\Z}}(F)\longrightarrow\mathbb D_{\cR[1]}(F)
\]
is an isomorphism in $\D(\cR)$.
\end{lemma}

\begin{proof}
Since $\delta$ commutes with the involutions, $\delta_*$ is postcomposition with $\delta$, as in Lemma~\ref{v2:lemma:derived-transposition-natural}(b). Applying $\RHom_\cR(\overline F,-)$ to \eqref{eq:ZQZ-triangle} gives a distinguished triangle
\[
\xymatrix{
 \mathbb D_\cR(F) \ar[r] & \mathbb D_{\cR_\Q}(F) \ar[r] & \mathbb D_{\cR_{\Q/\Z}}(F) \ar[r]^-{\delta_*} & \mathbb D_{\cR[1]}(F),
}
\]
so it suffices to show that $\mathbb D_{\cR_\Q}(F)\simeq 0$. By extension of scalars along $\cR\to\cR_\Q$,
\[
 \RHom_\cR(\overline F,\cR_\Q)\simeq\RHom_{\cR_\Q}\bigl(\cR_\Q\otimes_\cR^{\mathrm L}\overline F,\cR_\Q\bigr).
\]
Since $\cR_\Q$ is a localization of $\cR$, it is flat over $\cR$, so $\cR_\Q\otimes_\cR^{\mathrm L}\overline F\simeq\overline F\otimes_\Z\Q$. Its homology is $H_i(\overline F)\otimes_\Z\Q$, which vanishes because $\overline F$ has the same underlying abelian groups as $F$.
\end{proof}

\begin{corollary}
\label{cor:QZ-biduality}
On the full subcategory of $\Q$-acyclic objects of $\D^b_f(\cR)$, the canonical map
\[
 \can^{\cR_{\Q/\Z}}\colon\id\xrightarrow{\ \simeq\ }\mathbb D_{\cR_{\Q/\Z}}^2
\]
is a natural isomorphism satisfying
\[
 \mathbb D_{\cR_{\Q/\Z}}\bigl(\can^{\cR_{\Q/\Z}}_F\bigr)\circ\can^{\cR_{\Q/\Z}}_{\mathbb D_{\cR_{\Q/\Z}}(F)}=\id_{\mathbb D_{\cR_{\Q/\Z}}(F)}.
\]
\end{corollary}

\begin{proof}
The identity is Corollary~\ref{v2:cor:derived-can-identities}(a), since $c$ is an involution. Let $F\in\D^b_f(\cR)$ be $\Q$-acyclic. By Lemma~\ref{lemma:QZ-duality}, $\mathbb D_{\cR_{\Q/\Z}}(F)\simeq\mathbb D_{\cR[1]}(F)$. This lies in $\D^b_f(\cR)$ by Theorem~\ref{thm:derivedduality} and Proposition~\ref{thm:dualizing}, and it is $\Q$-acyclic by Lemma~\ref{lemma:dual-dim}. By Corollary~\ref{v2:cor:derived-can-identities}(c) for $\delta$, the diagram
\[
\xymatrix@C=4pc{
 F \ar[r]^-{\can^{\cR_{\Q/\Z}}_F} \ar[d]_-{\can^{\cR[1]}_F} & \mathbb D_{\cR_{\Q/\Z}}^2(F) \ar[d]^-{\delta_*} \\
 \mathbb D_{\cR[1]}^2(F) \ar[r]^-{\mathbb D_{\cR[1]}(\delta_*)} & \mathbb D_{\cR[1]}\bigl(\mathbb D_{\cR_{\Q/\Z}}(F)\bigr)
}
\]
commutes. The right map is an isomorphism by Lemma~\ref{lemma:QZ-duality} for $\mathbb D_{\cR_{\Q/\Z}}(F)$, the bottom map is an isomorphism by Lemma~\ref{lemma:QZ-duality} for $F$, and the left map is an isomorphism by Theorem~\ref{thm:derivedduality}. So $\can^{\cR_{\Q/\Z}}_F$ is an isomorphism.
\end{proof}

\begin{lemma}
\label{lemma:torsion-dual-ext}
Let $M$ be a finitely generated $\Z$-torsion $\cR$-module, and let $j\in\Z$. Then, naturally in $M$,
\[
 H_k\mathbb D_{\cR[1]}(M[j])\cong\Ext^{-j-k}_\cR(\overline M,\cR_{\Q/\Z})
\]
for all $k$.
\end{lemma}

\begin{proof}
The module $M$ is $\Q$-acyclic, so by Lemma~\ref{lemma:QZ-duality},
\[
 H_k\mathbb D_{\cR[1]}(M[j])\cong H_k\RHom_\cR(\overline M[j],\cR_{\Q/\Z})=H_{k+j}\RHom_\cR(\overline M,\cR_{\Q/\Z})=\Ext^{-j-k}_\cR(\overline M,\cR_{\Q/\Z}).\qedhere
\]
\end{proof}

\subsection{The shifted duality on \texorpdfstring{$\Q$}{Q}-acyclic complexes}
\label{subsubsec:shifted-duality}

The goal of this subsection is to study the duality
\[\mathbb D_{\cR[1]}\]
when $\cR$ satisfies the following:
\begin{itemize}
	\item $\cR$ is a commutative Noetherian Gorenstein ring of finite Krull dimension $D$;
	\item $\cR$ is $\Z$-torsion-free;
	\item $\cR/n$ is equicodimensional of Krull dimension $D-1$ for every integer $n>1$ which is not a
		unit in $\cR$.
\end{itemize}

So, we make these assumptions on $\cR$ for the remainder of the
subsection, and we set $d=D-1$. 

\begin{example}
This applies to $\cR=\Z[\Z^d]$, where $d$ is the spatial dimension. 
\end{example}

Throughout, the canonical map
\[\can^{\cR[1]} \colon \id \xrightarrow{\simeq} \mathbb D_{\cR[1]}^2\]
is a natural isomorphism by Proposition~\ref{thm:dualizing}.

\begin{defn}
An object $F\in \D(\cR)$ is \emph{$\Q$-acyclic} if $H_i(F)\otimes \Q =0$ for all $i$. Let $\D^b_f(\cR)^{\mathrm{tor}}$ be the full subcategory of $\D^b_f(\cR)$ on $\Q$-acyclic complexes.
Let $\D^b_f(\cR)^{\mathrm{tor}}_{\leq 0}$ be the full subcategory of complexes that are moreover co-connective. 
\end{defn}

We take the dimension bounds of Definition~\ref{def:dimension-bounds} in dimension $d=D-1$, one less than the
Krull dimension of $\cR$. The reason is that, in this section, we consider duality with respect to the shift $\cR[1]$. Indeed, a finitely generated
$\Q$-acyclic module $N$ satisfies $nN=0$ for some $n>1$. Since $n$ is a nonzerodivisor on $\cR$, we have
$\RHom_\cR(\cR/n,\cR)\simeq\cR/n[-1]$, and adjunction gives
\[
	\RHom_\cR(N,\cR[1])
	\simeq
	\RHom_{\cR/n}\bigl(N,\RHom_\cR(\cR/n,\cR)[1]\bigr)
	\simeq
	\RHom_{\cR/n}(N,\cR/n).
\]
This is Rees' lemma \cite[Lemma 3.1.16]{Bruns_1998}. The ring $\cR/n$ is Gorenstein, as a quotient of
$\cR$ by a nonzerodivisor, and equicodimensional of dimension $d$. Applying this to $\overline N$, on $\Q$-acyclic modules,
$\mathbb D_{\cR[1]}$ is computed by the duality of the previous subsection over $\cR/n$. In particular,
Theorem~\ref{theorem:pure-condition} and Corollary~\ref{cor:pure-one-dim-CM} apply to finitely generated $\Q$-acyclic
$\cR$-modules, with $\Ext^i_\cR(\overline M,\cR)$ replaced by
\[
	\Ext^{i+1}_\cR(\overline M,\cR)=H_{-i}\mathbb D_{\cR[1]}(M).
\]

\begin{theorem}
	\label{theorem:dimension-bounds-appendix}
	\label{theorem:dimension-bounds-R}
	Let $F\in \D^b_f(\cR)^{\mathrm{tor}}$, let $D=\dim\cR$, and let $d=D-1$. The following are equivalent:
	\begin{enumerate}[(1)]
		\item $F$ satisfies the weak (resp.\ strong) dimension bound in dimension $d$.
		\item $\mathbb D_{\cR[1]}(F)$ satisfies the weak (resp.\ strong) dimension bound in dimension $d$.
		\item Both $F$ and $\mathbb D_{\cR[1]}(F)$ are co-connective (resp.\ co-connected).
	\end{enumerate}
\end{theorem}

\begin{proof}
The acyclic case is immediate. Otherwise, choose an integer $n>1$ annihilating all homology
modules of $F$ and $\mathbb D_{\cR[1]}(F)$. Such an integer exists because both complexes belong
to $\D^b_f(\cR)$ and are $\Q$-acyclic, by Theorem~\ref{thm:derivedduality} and Lemma~\ref{lemma:dual-dim}.
Put $A=\cR/n$. Then $A$ is equicodimensional Gorenstein of dimension $d$.

For each of these homology modules $N$, the preceding adjunction identifies
\[
    \mathbb D_{\cR[1]}(N)\simeq\RHom_A(\overline N,A).
\]
Dimensions and localized dimensions of $N$ are the same over $A$ and $\cR$.
The same adjunction after localization identifies the localized duals with
$\RHom_{A_{\bar{\frak p}}}(\overline{N_{\frak p}},A_{\bar{\frak p}})$,
where $\frak p$ denotes a prime of $A$ and its inverse image in $\cR$.
Thus, the proof of Theorem~\ref{theorem:dimension-bounds-general} goes through. 
\end{proof}

\begin{corollary}
	\label{cor:torsion-ext-degrees}
	Let $M\neq0$ be a finitely generated $\cR$-module with $nM=0$ for some integer $n>1$, and let $k=\dim M$.
	\begin{enumerate}[(a)]
		\item $H_{-j}\mathbb D_{\cR[1]}(M)=0$ unless $d-k\leq j\leq d$, and $H_{-(d-k)}\mathbb D_{\cR[1]}(M)\neq0$.
		\item $M$ is Cohen--Macaulay of pure dimension $k$ if and only if $H_{-j}\mathbb D_{\cR[1]}(M)=0$ for all
			$j\neq d-k$.
		\item If $k=0$, or if $k=1$ and $M$ is pure, then $H_{-j}\mathbb D_{\cR[1]}(M)=0$ for all $j\neq d-k$.
	\end{enumerate}
\end{corollary}

\begin{proof}
	Since $M\neq0$, the integer $n$ is not a unit in $\cR$. By the isomorphism above,
	\[
		H_{-j}\mathbb D_{\cR[1]}(M)\cong\Ext^j_{\cR/n}(\overline M,\cR/n).
	\]
	The dimension, depths and associated primes of an $\cR/n$-module are the same over $\cR/n$ as over $\cR$.
	Those of $\overline M$ are those of $M$, up to replacing $\mathfrak p$ by $c(\mathfrak p)$. So (a) and (b)
	are Theorem~\ref{theorem:ext-bounds} over $\cR/n$, applied to $\overline M$. For (c), if $k=0$, then every
	associated prime of $M$ has dimension $0$, and every localization of $M$ at a maximal ideal in $\Supp M$ has finite length, hence depth $0$.
	So $M$ is Cohen--Macaulay of pure dimension $0$. If $k=1$ and $M$ is pure, then $M$ is Cohen--Macaulay by
	Corollary~\ref{cor:pure-one-dim-CM}, which applies to $M$ as explained above. Now apply (b).
\end{proof}

\begin{lemma}
	\label{lemma:Ext-torsion}
	Let $n>1$ be an integer which is not a unit in $\cR$, and let $M$ be a finitely generated $\cR/n$-module.
	For $i>0$, the module $\Ext^i_{\cR/n}(M,\cR/n)$ is annihilated by a nonzerodivisor of $\cR/n$. In
	particular, it is $\cR/n$-torsion.
\end{lemma}

\begin{proof}
	Let $A=\cR/n$, and let $N=\Ext^i_A(M,A)$ with $i>0$. We may assume that $M\neq0$. The ring $A$ is
	Gorenstein and equicodimensional of dimension $d$, so Theorem~\ref{theorem:ext-bounds} applies over $A$. By
	part (c), $\dim N\leq d-i<d$. By part (d) applied to $M=A$, every associated prime of $A$ has
	dimension $d$. So no associated prime of $A$ lies in $\Supp N$, that is, contains $\Ann N$. The
	zerodivisors of $A$ form the union of its finitely many associated primes. By prime avoidance,
	$\Ann N$ contains a nonzerodivisor.
\end{proof}

\section{Derived Fusion Complexes}
\label{app:derived-fusion}

In this section, $\cR$ satisfies the assumptions of Section~\ref{subsubsec:shifted-duality}, and has finite global dimension. We record these assumptions.

\begin{convention}
\label{conv:fusion-ring}
The ring $\cR$ satisfies the following.
\begin{itemize}
\item $\cR$ is a commutative Noetherian Gorenstein ring of finite Krull dimension $D$, with an involution $c$.
\item $\cR$ is $\Z$-torsion-free.
\item $\cR/n$ is equicodimensional of Krull dimension $D-1$ for every integer $n>1$ which is not a unit in $\cR$.
\item $\cR$ has finite global dimension.
\end{itemize}
We set $d=D-1$.
\end{convention}
 The main example is $\cR=R=\Z[\Z^d]$, where $d$ is the spatial dimension. It is a Laurent polynomial ring over $\Z$, so it is regular of Krull dimension $d+1$, and $R/n=\Z_n[\Z^d]$ is equicodimensional of dimension $d$.

Since $\cR$ has finite global dimension, the inclusion
\[\D^p(\cR) \xrightarrow{\simeq} \D_f^b(\cR)\]
is an equivalence. As there,
\[
 \mathbb D_{\cR[1]}(F)=\RHom_\cR(\overline F,\cR[1]).
\]
We collect the facts we need in the following result, which is a direct consequence of Theorem~\ref{thm:derivedduality}, Remark~\ref{cor:derivedduality} and Proposition~\ref{thm:dualizing}.

\begin{theorem}
\label{thm:fusion-duality}
The object $\cR[1]$, with the anti-linear involution induced by $c$, is a dualizing complex. Consequently,
\[
 \mathbb D_{\cR[1]}\colon\D^p(\cR)^{\op}\xlongrightarrow{\simeq} \D^p(\cR)
\]
is an equivalence of triangulated categories, and the canonical map
\[
 \can^{\cR[1]}\colon\id\xrightarrow{\ \simeq\ }\mathbb D_{\cR[1]}^2
\]
is a natural isomorphism satisfying
\[
 \mathbb D_{\cR[1]}(\can^{\cR[1]}_F)\circ\can^{\cR[1]}_{\mathbb D_{\cR[1]}(F)}=\id_{\mathbb D_{\cR[1]}(F)}.
\]
So the triple $(\D^p(\cR),\mathbb D_{\cR[1]},\can^{\cR[1]})$ is a triangulated category with duality in the sense of \cite{Balmer_2000}.
\end{theorem}

Our goal in this section is to restrict this duality to the subcategories of  $\D^p(\cR)$ of interest in this paper.
Inspired by Definition~\ref{def:FC} and Definition~\ref{defn:propercomplete}, we define the following.
\begin{defn}
\label{def:derived-FC}
Say a complex $F \in \D^p(\cR)^\mathrm{tor}$ is \textbf{proper} (resp.\ \textbf{complete}) if both $F$ and
$\mathbb D_{\cR[1]}(F)$ are co-connective (resp. co-connected). 
Define a sequence of full subcategories
\[\FC^{\mathrm{cp}}(\cR)\subset \FC^{\mathrm{pr}}(\cR) \subset \FC(\cR) \subset \D^p(\cR)^{\mathrm{tor}}_{\leq 0}\]
as follows.
\begin{enumerate}[(a)]
\item Let $\FC(\cR)$ be the full subcategory on the $\Q$-acyclic complexes $F$ whose homology is concentrated in degrees $[-d,0]$, that is, $F\simeq\tau_{[-d,0]}F$. This is the \textbf{derived category of fusion complexes}.
\item Let $\FC^{\mathrm{pr}}(\cR)\subset\FC(\cR)$ be the full subcategory on the proper fusion
	complexes.
	This is the \textbf{derived category of proper fusion complexes}.
\item Let $\FC^{\mathrm{cp}}(\cR)\subset\FC^{\mathrm{pr}}(\cR)$ be the full subcategory on the
	complete fusion complexes.
	This is the \textbf{derived category of complete fusion complexes}.
\end{enumerate}
\end{defn}

The following result is a restatement of  Theorem~\ref{theorem:dimension-bounds-appendix} in this context.
\begin{theorem}
\label{thm:FC-dimension-bounds}
Let $F\in \D^b_f(\cR)^\mathrm{tor}$. The following are equivalent:
\begin{enumerate}[(1)]
\item $F$ is proper (resp.\ complete);
\item $F$ satisfies the weak (resp.\ strong) dimension bound in dimension $d$;
\item $\mathbb D_{\cR[1]}(F)$ satisfies the weak (resp.\ strong) dimension bound in dimension $d$.
\end{enumerate}
\end{theorem}

\begin{proof}
Since $F\in\D^b_f(\cR)^{\mathrm{tor}} \simeq \D^p(\cR)^{\mathrm{tor}}$, this is Theorem~\ref{theorem:dimension-bounds-appendix}.
\end{proof}

In the paper, we study another duality, $\mathbb D_{\cR_{\Q/\Z}}$, where $\cR_{\Q/\Z}=\cR\otimes_\Z\Q/\Z$. By Lemma~\ref{lemma:QZ-duality} and Corollary~\ref{cor:QZ-biduality}, it agrees with $\mathbb D_{\cR[1]}$ on $\Q$-acyclic complexes, via the connecting map $\delta\colon\cR_{\Q/\Z}\to\cR[1]$ of \eqref{eq:ZQZ-triangle}. We now restrict it to fusion complexes.

\begin{theorem}
\label{thm:FC-duality}
Let $\star\in\{\mathrm{pr},\mathrm{cp}\}$. The duality $\mathbb D_{\cR_{\Q/\Z}}$ is equivalent to  $\mathbb D_{\cR[1]}$, and so restricts to an equivalence
\[
 \mathbb D_{\cR_{\Q/\Z}}\colon\FC^\star(\cR)^{\op}\xlongrightarrow{\simeq} \FC^\star(\cR).
\]
The canonical map
\[
 \can^{\cR_{\Q/\Z}}\colon\id\xrightarrow{\ \simeq\ }\mathbb D_{\cR_{\Q/\Z}}^2
\]
is a natural isomorphism on $\FC^\star(\cR)$ satisfying
\[
 \mathbb D_{\cR_{\Q/\Z}}\bigl(\can^{\cR_{\Q/\Z}}_F\bigr)\circ\can^{\cR_{\Q/\Z}}_{\mathbb D_{\cR_{\Q/\Z}}(F)}=\id_{\mathbb D_{\cR_{\Q/\Z}}(F)}.
\]
So $\bigl(\FC^\star(\cR),\mathbb D_{\cR_{\Q/\Z}},\can^{\cR_{\Q/\Z}}\bigr)$ is a category with duality.
\end{theorem}

\begin{proof}
We first show the statement for $\mathbb D_{\cR[1]}$. Let $F\in\FC^\star(\cR)$. Then $\mathbb D_{\cR[1]}(F)$ is perfect, and it is $\Q$-acyclic by Lemma~\ref{lemma:dual-dim}. By Theorem~\ref{thm:FC-dimension-bounds}, $\mathbb D_{\cR[1]}(F)$ satisfies the weak dimension bound, so its homology lies in degrees $[-d,0]$. Since $\can^{\cR[1]}\colon F\simeq\mathbb D_{\cR[1]}^2(F)$, both $\mathbb D_{\cR[1]}(F)$ and $\mathbb D_{\cR[1]}^2(F)$ are co-connective (resp.\ co-connected), so $\mathbb D_{\cR[1]}(F)\in\FC^\star(\cR)$. The remaining statements for $\mathbb D_{\cR[1]}$ follow from Theorem~\ref{thm:fusion-duality}, since $\FC^\star(\cR)$ is a full subcategory.

By Lemma~\ref{lemma:QZ-duality}, $\delta_*\colon\mathbb D_{\cR_{\Q/\Z}}\Rightarrow\mathbb D_{\cR[1]}$ is a natural isomorphism on $\Q$-acyclic objects, so $\mathbb D_{\cR_{\Q/\Z}}$ also restricts to an equivalence on $\FC^\star(\cR)$. The statements for $\can^{\cR_{\Q/\Z}}$ are Corollary~\ref{cor:QZ-biduality}.
\end{proof}

The homology of the dual of a fusion complex can be computed from the truncation triangles of Section~\ref{dra:new-postnikov}.

\begin{proposition}
\label{prop:truncation-duals}
\begin{enumerate}[(a)]
\item Let $M$ be a finitely generated $\Z$-torsion $\cR$-module, and let $j\in\Z$. Then, naturally in $M$,
\[
 H_k\mathbb D_{\cR[1]}(M[j])\cong\Ext^{-j-k}_\cR(\overline M,\cR_{\Q/\Z})
\]
for all $k$. If $nM=0$, this is also $\Ext^{-j-k}_{\cR/n}(\overline M,\cR/n)$, and Corollary~\ref{cor:torsion-ext-degrees} describes the degrees in which it is nonzero.
\item Let $F\in\D^p(\cR)$ be $\Q$-acyclic, and let $a\in\Z$. Then $\tau_{\geq a+1}F$ and $\tau_{\leq a}F$ are $\Q$-acyclic objects of $\D^p(\cR)$, and applying $\mathbb D_{\cR[1]}$ to the truncation triangle gives a distinguished triangle
\[
\xymatrix{
 \mathbb D_{\cR[1]}(\tau_{\leq a}F) \ar[r] & \mathbb D_{\cR[1]}(F) \ar[r] & \mathbb D_{\cR[1]}(\tau_{\geq a+1}F) \ar[r] & \mathbb D_{\cR[1]}(\tau_{\leq a}F)[1]
}
\]
and hence a long exact sequence
\[
 \cdots\to H_k\mathbb D_{\cR[1]}(\tau_{\leq a}F)\to H_k\mathbb D_{\cR[1]}(F)\to H_k\mathbb D_{\cR[1]}(\tau_{\geq a+1}F)\to H_{k-1}\mathbb D_{\cR[1]}(\tau_{\leq a}F)\to\cdots.
\]
\end{enumerate}
By Lemma~\ref{lemma:QZ-duality}, the same holds with $\mathbb D_{\cR_{\Q/\Z}}$ in place of $\mathbb D_{\cR[1]}$.
\end{proposition}

\begin{proof}
For (a), the first isomorphism is Lemma~\ref{lemma:torsion-dual-ext}. If $nM=0$, then also $n\overline M=0$, and the identification of $\mathbb D_{\cR[1]}$ on $\Q$-acyclic modules with the duality over $\cR/n$ in Section~\ref{subsubsec:shifted-duality} gives $H_k\mathbb D_{\cR[1]}(M[j])\cong\Ext^{-j-k}_{\cR/n}(\overline M,\cR/n)$.

For (b), the homology of the truncations is part of the homology of $F$, so it is finitely generated, $\Z$-torsion and bounded. Hence the truncations are $\Q$-acyclic and lie in $\D^b_f(\cR)\simeq\D^p(\cR)$. The functor $\mathbb D_{\cR[1]}$ is a duality of triangulated categories by Theorem~\ref{thm:fusion-duality}, so it sends the truncation triangle to the displayed distinguished triangle. The long exact sequence is its homology long exact sequence.
\end{proof}

For example, let $F\in\FC(\cR)$ have homology concentrated in degree $-1$, with $H_{-1}(F)$ finite. Then $F\simeq H_{-1}(F)[-1]$, so $\mathbb D_{\cR[1]}(F)\simeq\mathbb D_{\cR[1]}(H_{-1}(F))[1]$. By Corollary~\ref{cor:torsion-ext-degrees}(c), $\mathbb D_{\cR[1]}(F)$ has homology concentrated in degree $1-d$. This is degree $-1$ when $d=2$.

In the following lemmas, we let $s = 0$ in the proper case and $s = 1$ in the complete case, to
treat both cases simultaneously. 

\begin{lemma}
	Let $A \to B \to C \to A[1]$ be a triangle in
	$\D^p(\cR)^\mathrm{tor}$. If $\tau_{\leq -s}
	A$ and $\tau_{\leq -s}
	C$ are proper (resp.\ complete), then so is $\tau_{\leq -s} B$. 
\end{lemma}

\begin{proof}
	From the long exact sequence of homology, we have exact sequences
	\[
		H_{-i} A \to H_{-i} B \to H_{-i} C 
	\]
	for all $i$. This implies $\dim H_{-i} B \leq \max
	(\dim H_{-i} A, \dim H_{-i} C)$ for all $i$. So if $\tau_{\leq {-s}} A$
	and $\tau_{\leq -s} C$ satisfy the weak (resp.\ strong) dimension bound, then so does $\tau_{\leq -s}
	B$. 
\end{proof}

\begin{lemma}
	Let $F \in \D^p(\cR)^\mathrm{tor}$ be co-connective (resp. co-connected). Then
	$\tau_{\leq-s}\mathbb D_{\cR[1]}(F)$ is proper (resp. complete).
\end{lemma}

\begin{proof}
	Write $\mathbb D = \mathbb D_{\cR[1]}$.
	If $M$ is any finitely generated $\Z$-torsion $\cR$-module, then by Theorem~\ref{theorem:ext-bounds},
	for any $i \geq s$,
	$\tau_{\leq -s} \mathbb D(M[-i])$ is proper (resp. complete). 
	In particular, $\tau_{\leq -s} \mathbb D(\tau_{\geq -s} F)$ is proper (resp. complete). 
	Now consider the dual of the Whitehead tower:
	\[
		\mathbb D(H_{-i-1} (F)[-i-1])
		\to 
		\mathbb D(\tau_{\geq -i-1} F)
		\to 
		\mathbb D(\tau_{\geq -i} F)
		\to
		\mathbb D(H_{-i-1} (F)[-i-1])[1].
	\]
	By induction, using the previous lemma,
	\[
		\tau_{\leq -s} \mathbb D(\tau_{\geq -i-1} F)
	\]
	is proper (resp. complete) for all $i \geq s$. 
	Since the homology is bounded, this completes the proof.
\end{proof}

\begin{lemma}
	\label{lemma:antisymmetric-iso}
	Let $F \in \D^p(\cR)^{\mathrm{tor}}$ be co-connective (resp. co-connected),
	and $\lambda : F \to \mathbb D_{\cR[1]}(F)$ a morphism.
	If $H_{-i}(\lambda)$ is an
	isomorphism for all $i \geq s$, then $F$ is proper (resp.\ complete) and $\lambda$ is a quasi-isomorphism.
\end{lemma}

\begin{proof}
	The hypothesis implies that the composite $F\xrightarrow{\lambda}\mathbb D_{\cR[1]}(F)\to\tau_{\leq -s}\mathbb D_{\cR[1]}(F)$ is a quasi-isomorphism. The right-hand side is proper (resp. complete) by the previous lemma, and therefore
	the left-hand side is too. Then, by Theorem~\ref{thm:FC-dimension-bounds}, $\lambda$ is an
	isomorphism on all of the nonzero homologies of $F$ and $\mathbb D_{\cR[1]}(F)$; hence it is a
	quasi-isomorphism.
\end{proof}

\section{Tame fusion complexes}
\label{v2:app:tame-fusion}

From now on, $\cR=R=\Z[\Z^d]$, and we let $R_n=R/n$. The results of Section~\ref{app:derived-fusion} apply with $\cR=R$. The paper computes with explicit chain complexes, the tame fusion complexes, and with their explicit duals
\[
 D_{R_{\Q/\Z}}(F)=\Hom_R(\overline F,R_{\Q/\Z}).
\]
These are not derived objects a priori: $F$ need not be $K$-projective, and $R_{\Q/\Z}$ is not injective. In this section, we show that these models can nevertheless be used to compute the derived dual, canonical map and transposition.

The argument has one key step. The module $R_{\Q/\Z}$ has a short injective resolution $\chi\colon R_{\Q/\Z}\to I$ with an involution, and $\chi$ induces a quasi-isomorphism on duals of tame complexes (Theorem~\ref{v2:thm:tame-dual}). The comparison of the remaining structure then follows from the general results on models of Section~\ref{v2:dra:duals} (Proposition~\ref{v2:prop:tame-dictionary}).

\subsection{Tame complexes}

\begin{defn}
Let $A$ be a chain complex of $R$-modules, and let $n$ be a positive integer.
\begin{itemize}
\item $A$ is \emph{$\Z_n$-tame} if each $A_i$ is a finitely generated free $R_n$-module; no boundedness condition is imposed.
\item $A$ is \emph{$\Z$-tame} if it is bounded and each $A_i$ is a finitely generated free $R$-module.
\item $A$ is \emph{tame} if it is $\Z$-tame or $\Z_n$-tame for some $n$.
\end{itemize}
A \emph{tame fusion complex} is a tame complex which lies in $\FC(R)$ of Definition~\ref{def:derived-FC}. That is, its homology is finitely generated, $\Z$-torsion, and concentrated in degrees $[-d,0]$.
\end{defn}

\begin{lemma}
\label{v2:lemma:tame-in-FC}
Let $F$ be a tame fusion complex.
\begin{enumerate}[(a)]
\item The complex $\overline F$ is again a tame fusion complex.
\item If $F$ is $\Z$-tame, then $F$ and $\overline F$ are $K$-projective.
\item Every object of $\D^p(R)$ is isomorphic to a $\Z$-tame complex. In particular, every object of $\FC(R)$ is represented by a $\Z$-tame fusion complex.
\end{enumerate}
\end{lemma}

\begin{proof}
Part (a) follows from the fact that, as an $R$-module, $R\cong \overline{R}$, so if $P$ splits off of a free module (i.e., is projective), then so does $\overline{P}$. For (b), a bounded complex of projective modules is $K$-projective. For (c), every perfect complex is quasi-isomorphic to a bounded complex of finitely generated projective $R$-modules. These modules are free, by the Quillen--Suslin--Swan theorem.
\end{proof}

\subsection{The coefficient module $R_{\Q/\Z}$}

Our next goal is to study the  $R_{\Q/\Z}$-duality and to show that for our models it is not always necessary to make projective or injective replacements. We start with some results about  $R_{\Q/\Z}$.

\begin{lemma}
\label{v2:lemma:RQZ}
\begin{enumerate}[(a)]
\item The $R$-module $R_{\Q/\Z}$ has injective dimension $d$.
\item If $P$ is a finitely generated free $R$-module or $R_n$-module, then $\Ext^i_R(P,R_{\Q/\Z})=0$ for $i>0$.
\item The map $R_n\to R_{\Q/\Z}$, $a\mapsto a/n$, identifies $R_n$ with the $n$-torsion submodule of $R_{\Q/\Z}$, compatibly with the involutions. In particular, if $F$ is $\Z_n$-tame, then so is $D_{R_{\Q/\Z}}(F)$.
\end{enumerate}
\end{lemma}

\begin{proof}
For (a), by \cite[Lemma 4.1.8]{Weibel_1994}, it suffices to show that $\Ext^{d+1}_R(R/\mathfrak a,R_{\Q/\Z})=0$ for every ideal $\mathfrak a$. The short exact sequence of \eqref{eq:ZQZ-triangle} gives an exact sequence
\[
 \Ext^{d+1}_R(R/\mathfrak a,R_\Q)\longrightarrow\Ext^{d+1}_R(R/\mathfrak a,R_{\Q/\Z})\longrightarrow\Ext^{d+2}_R(R/\mathfrak a,R).
\]
The right-hand term vanishes because $R$ has global dimension $d+1$. Since $R_\Q$ is flat over $R$, the left-hand term is
\[
 \Ext^{d+1}_{R_\Q}(R_\Q/\mathfrak aR_\Q,R_\Q),
\]
which vanishes because $R_\Q=\Q[\Z^d]$ has global dimension $d$. So the injective dimension is at most $d$.

For the lower bound, let $k=R/(p,t_1-1,\dots,t_d-1)\cong\F_p$ for a prime $p$. By the same flat base change argument, $\Ext^i_R(k,R_\Q)=0$ for all $i$. So the long exact sequence gives
\[
 \Ext^d_R(k,R_{\Q/\Z})\cong\Ext^{d+1}_R(k,R).
\]
The elements $p,t_1-1,\dots,t_d-1$ form a regular sequence in $R$. So the Koszul complex is a free resolution of $k$, and $\Ext^{d+1}_R(k,R)\cong k\neq0$.

For (b), the case of $R$ is clear. For $R_n$, use the resolution
\[
 0\longrightarrow R\xrightarrow{\ n\ }R\longrightarrow R_n\longrightarrow0.
\]
It shows that $\Ext^i_R(R_n,-)=0$ for $i\geq2$, and that $\Ext^1_R(R_n,R_{\Q/\Z})=R_{\Q/\Z}/nR_{\Q/\Z}$. This is zero because $\Q/\Z$ is divisible.

For (c), the $n$-torsion subgroup of $\Q/\Z$ is $\tfrac1n\Z/\Z$. Hence
\[
 D_{R_{\Q/\Z}}(F)_k=\Hom_R(\overline{F_{-k}},R_{\Q/\Z})\cong\Hom_{R_n}(\overline{F_{-k}},R_n),
\]
which is a finitely generated free $R_n$-module.
\end{proof}

\begin{lemma}
\label{v2:lemma:involutive-resolution}
There is an injective resolution $\chi\colon R_{\Q/\Z}\to I$, with $I_j=0$ for $j\notin[-d,0]$, together with an anti-linear involution $c_I$ on $I$ such that $\chi$ commutes with the involutions. In particular, $I$ is $K$-injective.
\end{lemma}

\begin{proof}
Embed $R_{\Q/\Z}$ into an injective module $J_0$ by a map $\jmath$, and let $I_0=J_0\oplus\overline{J_0}$, with the involution swapping the two summands. Then
\[
 q\longmapsto\bigl(\jmath(q),\overline{\jmath(\bar q)}\bigr)
\]
is an embedding $R_{\Q/\Z}\to I_0$ which commutes with the involutions. Repeat with its cokernel,
and so on. By Lemma~\ref{v2:lemma:RQZ}(a) and \cite[Lemma 4.1.8]{Weibel_1994}, the cokernel after
$d$ steps is already injective, and we stop there. A bounded complex of injective modules is $K$-injective.
\end{proof}

We fix such a resolution. Since $I$ is $K$-injective, $D_I(F)=\Hom_R(\overline F,I)$ is a model for $\mathbb D_{R_{\Q/\Z}}(F)$ for every complex $F$. Postcomposition with $\chi$ gives a natural chain map
\[
 \chi_*\colon D_{R_{\Q/\Z}}(F)\longrightarrow D_I(F).
\]

\begin{lemma}
\label{v2:lemma:chi-antisymmetric}
Let $F$ be a complex, and let $\lambda\colon F\to D_{R_{\Q/\Z}}(F)$ be an antisymmetric chain map, that is,
\[
 \lambda=-D_{R_{\Q/\Z}}(\lambda)\circ\can^{R_{\Q/\Z}}_F.
\]
Then $\lambda_I=\chi_*\circ\lambda\colon F\to D_I(F)$ is antisymmetric:
\[
 \lambda_I=-D_I(\lambda_I)\circ\can^I_F.
\]
\end{lemma}

\begin{proof}
View $R_{\Q/\Z}$ as a complex concentrated in degree $0$. By Lemma~\ref{v2:lemma:involutive-resolution}, $\chi\colon R_{\Q/\Z}\to I$ is a chain map with $\chi\circ c_{R_{\Q/\Z}}=c_I\circ\overline\chi$. So Corollary~\ref{v2:cor:can-identities}(d) applies with $T=R_{\Q/\Z}$, $T'=I$, $\psi=\chi$, $G=F$ and $\varphi=\lambda$. Since $\chi_*\circ\lambda=\lambda_I$, it gives
\[
 D_I(\lambda_I)\circ\can^I_F=\chi_*\circ D_{R_{\Q/\Z}}(\lambda)\circ\can^{R_{\Q/\Z}}_F.
\]
By antisymmetry of $\lambda$, $D_{R_{\Q/\Z}}(\lambda)\circ\can^{R_{\Q/\Z}}_F=-\lambda$. So the right-hand side is $\chi_*\circ(-\lambda)=-\lambda_I$.
\end{proof}

For $\Z_n$-tame complexes, it is natural to dualize over $R_n$ instead. The next lemma says that this gives the same answer.

\begin{lemma}
\label{v2:lemma:Rn-coefficients}
Let ${}_n I\subseteq I$ be the subcomplex of $n$-torsion elements.
\begin{enumerate}[(a)]
\item The map $\chi$ restricts to an injective resolution $R_n\to {}_n I$ of $R_n$-modules, compatible with the involutions, where $R_n\subseteq R_{\Q/\Z}$ as in Lemma~\ref{v2:lemma:RQZ}(c).
\item For every complex $A$ of $R_n$-modules,
\[
 \Hom_{R_n}(\overline A,{}_n I)=\Hom_R(\overline A,I)=D_I(A).
\]
This complex is a model both for $\mathbb D_{R_{\Q/\Z}}(A)$ and for $\RHom_{R_n}(\overline A,R_n)$.
\item For every $R_n$-module $M$ and every $i$,
\[
 \Ext^i_R(\overline M,R_{\Q/\Z})\cong\Ext^i_{R_n}(\overline M,R_n).
\]
\end{enumerate}
\end{lemma}

\begin{proof}
For (a), taking $n$-torsion elements is the functor $\Hom_R(R_n,-)$, so ${}_n I=\Hom_R(R_n,I)$.
\begin{itemize}
\item For an $R_n$-module $N$, we have $\Hom_{R_n}(N,({}_n I)_j)=\Hom_R(N,I_j)$. This is exact in $N$ because $I_j$ is injective. So $({}_n I)_j$ is an injective $R_n$-module.
\item The homology of ${}_n I=\Hom_R(R_n,I)$ is $H_{-i}({}_n I)=\Ext^i_R(R_n,R_{\Q/\Z})$. For $i>0$, this vanishes by Lemma~\ref{v2:lemma:RQZ}(b). For $i=0$, it is the $n$-torsion submodule of $R_{\Q/\Z}$, which is $R_n$ by Lemma~\ref{v2:lemma:RQZ}(c).
\end{itemize}
So $R_n\to {}_n I$ is an injective resolution. It is compatible with the involutions because $c_I$ preserves $n$-torsion.

For (b), the equality of Hom complexes holds because every $R$-linear map out of an $n$-torsion module lands in the $n$-torsion. The complex $I$ is $K$-injective over $R$. The complex ${}_n I$ is a bounded complex of injective $R_n$-modules, hence $K$-injective over $R_n$. So the two sides compute the two derived duals. Part (c) is (b) for $A=M$, after taking homology.
\end{proof}

\begin{theorem}
\label{v2:thm:tame-dual}
Let $F$ be a tame fusion complex. Then
\[
 \chi_*\colon D_{R_{\Q/\Z}}(F)\longrightarrow D_I(F)
\]
is a quasi-isomorphism. Consequently, the composite
\[
 \nu_F\colon D_{R_{\Q/\Z}}(F)\xrightarrow{\ \chi_*\ }D_I(F)\simeq\mathbb D_{R_{\Q/\Z}}(F)\xrightarrow{\ \delta_*\ }\mathbb D_{R[1]}(F)
\]
is an isomorphism in $\D(R)$. It is natural for chain maps $f\colon F\to G$ between tame fusion complexes, that is, the diagram
\[
\xymatrix@C=4pc{
 D_{R_{\Q/\Z}}(G) \ar[r]^-{D_{R_{\Q/\Z}}(f)} \ar[d]_-{\nu_G} & D_{R_{\Q/\Z}}(F) \ar[d]^-{\nu_F} \\
 \mathbb D_{R[1]}(G) \ar[r]^-{\mathbb D_{R[1]}(f)} & \mathbb D_{R[1]}(F)
}
\]
commutes in $\D(R)$.
\end{theorem}

\begin{proof}
We distinguish the two kinds of tameness.
\begin{itemize}
\item If $F$ is $\Z$-tame, then $\overline F$ is $K$-projective by Lemma~\ref{v2:lemma:tame-in-FC}(b). So $\chi_*$ is a quasi-isomorphism by Lemma~\ref{v2:lemma:dual-invariance}(c).
\item If $F$ is $\Z_n$-tame, then so is $\overline F$. By Lemma~\ref{v2:lemma:RQZ}(a) and (b), Theorem~\ref{thm:dualityisderived} applies to $\overline F$ with $T=R_{\Q/\Z}$, and it gives the claim.
\end{itemize}
The map $\delta_*$ is an isomorphism by Lemma~\ref{lemma:QZ-duality}, since $F$ is $\Q$-acyclic. Naturality holds because $\chi_*$ and $\delta_*$ are natural.
\end{proof}

\begin{remark}
For a $\Z_n$-tame $F$, Lemma~\ref{v2:lemma:RQZ}(c) and Lemma~\ref{v2:lemma:Rn-coefficients}(b) identify $\chi_*$ with the chain map
\[
 \Hom_{R_n}(\overline F,R_n)\longrightarrow\Hom_{R_n}(\overline F,{}_n I).
\]
So the theorem also says that $\Hom_{R_n}(\overline F,R_n)$ computes $\RHom_{R_n}(\overline F,R_n)$.
\end{remark}

\begin{proposition}
\label{v2:prop:resolution-dual-ext}
Let $G$ be a resolution of a finitely generated $R_n$-module $M$ by finitely generated free $R_n$-modules, with $G_i=0$ for $i<0$. Then
\[
 H_{-i}D_{R_{\Q/\Z}}(G)\cong\Ext^i_{R_n}(\overline M,R_n).
\]
These modules vanish for $i>d$.
\end{proposition}

\begin{proof}
By Lemma~\ref{v2:lemma:RQZ}(c), $D_{R_{\Q/\Z}}(G)=\Hom_{R_n}(\overline G,R_n)$, so the displayed identification follows by computing Ext with the resolution $\overline G$.
By Lemma~\ref{v2:lemma:Rn-coefficients}(c) and Lemma~\ref{v2:lemma:RQZ}(a), these modules vanish for $i>d$.
\end{proof}

\begin{corollary}
\label{v2:cor:tame-dual}
Let $F$ and $G$ be tame fusion complexes.
\begin{enumerate}[(a)]
\item If $\alpha\colon G\to F$ is a quasi-isomorphism, then so is $D_{R_{\Q/\Z}}(\alpha)\colon D_{R_{\Q/\Z}}(F)\to D_{R_{\Q/\Z}}(G)$.
\item The complex $D_{R_{\Q/\Z}}(F)$ is perfect and $\Q$-acyclic, and $H_i D_{R_{\Q/\Z}}(F)\cong H_i\mathbb D_{R[1]}(F)$ for all $i$.
\item The fusion complex $F$ is proper (resp.\ complete) in the sense of Definition~\ref{defn:propercomplete} if and only if $F\in\FC^{\mathrm{pr}}(R)$ (resp.\ $F\in\FC^{\mathrm{cp}}(R)$), and then so is $D_{R_{\Q/\Z}}(F)$. In particular, these notions do not depend on the tame representative.
\end{enumerate}
\end{corollary}

\begin{proof}
For (a), by Theorem~\ref{v2:thm:tame-dual},
\[
 D_{R_{\Q/\Z}}(\alpha)=\nu_G^{-1}\circ\mathbb D_{R[1]}(\alpha)\circ\nu_F
\]
in $\D(R)$, and $\mathbb D_{R[1]}(\alpha)$ is an isomorphism. Part (b) follows from Theorem~\ref{v2:thm:tame-dual}, since $\mathbb D_{R[1]}$ preserves $\D^p(R)$ and $\Q$-acyclic objects (Lemma~\ref{lemma:dual-dim}). The first claim of (c) follows from (b) and Definition~\ref{def:derived-FC}, and the second from (b) and Theorem~\ref{thm:FC-duality}.
\end{proof}

\subsection{Canonical maps and transpositions of tame complexes}
\label{subsection:canonical-maps}

The paper considers chain maps into $D_{R_{\Q/\Z}}(F)$. To ensure these constructions are preserved under quasi-isomorphisms and thus well defined in $\D(R)$, we need to know that the chain-level canonical maps and transpositions represent the derived ones. For a chain map $\gamma\colon G\to D_{R_{\Q/\Z}}(F)$ between tame fusion complexes, write
\[
 \gamma_{\mathbb D}=\nu_F\circ\gamma\colon G\longrightarrow\mathbb D_{R[1]}(F),
\]
where $\nu_F$ was defined in Theorem~\ref{v2:thm:tame-dual}.
By Lemma~\ref{v2:lemma:transposition-can}, if $\varphi\colon F\to D_{R_{\Q/\Z}}(F)$ is a chain map, then
\[
 t(\overline\varphi)=D_{R_{\Q/\Z}}(\varphi)\circ\can^{R_{\Q/\Z}}_F.
\]
By Lemma~\ref{v2:lemma:derived-transposition-can}, the derived transposition is given by the same formula in $\D(R)$.

\begin{proposition}
\label{v2:prop:tame-dictionary}
Let $F$ and $G$ be tame fusion complexes.
\begin{enumerate}[(a)]
\item For a chain map $\gamma\colon G\to D_{R_{\Q/\Z}}(F)$, the morphism $\gamma_{\mathbb D}$ depends only on the chain homotopy class of $\gamma$. For chain maps $\alpha\colon G'\to G$ and $\beta\colon F'\to F$ between tame fusion complexes,
\[
 \bigl(D_{R_{\Q/\Z}}(\beta)\circ\gamma\circ\alpha\bigr)_{\mathbb D}=\mathbb D_{R[1]}(\beta)\circ\gamma_{\mathbb D}\circ\alpha.
\]
\item The composite of chain maps
\[
 F\xrightarrow{\ \can^{R_{\Q/\Z}}_F\ }D^2_{R_{\Q/\Z}}(F)\xrightarrow{\ \chi_*\ }D_I\bigl(D_{R_{\Q/\Z}}(F)\bigr) \simeq \mathbb D^2_{R_{\Q/\Z}}(F)
\]
represents the derived canonical map $\can^{R_{\Q/\Z}}_F\colon F\to\mathbb D^2_{R_{\Q/\Z}}(F)$. Here the target is identified with $\mathbb D^2_{R_{\Q/\Z}}(F)$ using Theorem~\ref{v2:thm:tame-dual}.
\item Let $\gamma\colon F\to D_{R_{\Q/\Z}}(F)$ be a chain map, and let $\gamma_I=\chi_*\circ\gamma$. Then $\chi_*\circ t(\overline\gamma)$ represents the derived transposition $t(\overline{\gamma_I})$ for $T=R_{\Q/\Z}$, and
\[
 t(\overline\gamma)_{\mathbb D}=t(\overline{\gamma_{\mathbb D}}),
\]
where on the right, $t$ is the derived transposition for $T=R[1]$. That is, the diagram
\[
\xymatrix@C=4pc{
 F \ar[r]^-{t(\overline\gamma)_{\mathbb D}} \ar[d]_-{\can^{R[1]}_F} & \mathbb D_{R[1]}(F) \ar@{=}[d] \\
 \mathbb D^2_{R[1]}(F) \ar[r]^-{\mathbb D_{R[1]}(\gamma_{\mathbb D})} & \mathbb D_{R[1]}(F)
}
\]
commutes in $\D(R)$.
\end{enumerate}
\end{proposition}

\begin{proof}
Part (a) holds because homotopic chain maps agree in $\D(R)$, and by the naturality in Theorem~\ref{v2:thm:tame-dual}.

For (b), $\chi$ commutes with the involutions. So Corollary~\ref{v2:cor:can-identities}(c) gives a commutative diagram of chain maps
\[
\xymatrix@C=4pc{
 F \ar[r]^-{\can^{R_{\Q/\Z}}_F} \ar[d]_-{\can^I_F} & D^2_{R_{\Q/\Z}}(F) \ar[d]^-{\chi_*} \\
 D^2_I(F) \ar[r]^-{D_I(\chi_*)} & D_I\bigl(D_{R_{\Q/\Z}}(F)\bigr).
}
\]
The left map represents the derived canonical map. Indeed, choose a $K$-projective resolution $P\to F$, and apply Lemma~\ref{v2:lemma:derived-models}(b), Corollary~\ref{v2:cor:can-identities}(b) and Lemma~\ref{v2:lemma:dual-invariance}(b). The bottom map is a quasi-isomorphism, 
because $\chi_* : D_{R_{\Q/\Z}}(F) \to D_I(F)$ is by Theorem~\ref{v2:thm:tame-dual}, and hence
$D_{I}(\chi_*)$ is by Lemma~\ref{v2:lemma:dual-invariance}(b). 

For (c), Lemma~\ref{v2:lemma:transposition-natural}(b) for $\chi$ and Lemma~\ref{v2:lemma:transposition-can} for $I$ give a commutative diagram of chain maps
\[
\xymatrix@C=4pc{
 F \ar[r]^-{t(\overline\gamma)} \ar[d]_-{\can^I_F} & D_{R_{\Q/\Z}}(F) \ar[d]^-{\chi_*} \\
 D_I^2(F) \ar[r]^-{D_I(\gamma_I)} & D_I(F).
}
\]
As in (b), the left map represents the derived canonical map. So the bottom composite represents
\[
 \mathbb D_{R_{\Q/\Z}}(\gamma_I)\circ\can^{R_{\Q/\Z}}_F=t(\overline{\gamma_I}),
\]
by Lemma~\ref{v2:lemma:derived-transposition-can}. This proves the first claim, and it gives $t(\overline\gamma)_{\mathbb D}=\delta_*\circ t(\overline{\gamma_I})$. Since $\delta$ commutes with the involutions, Lemma~\ref{v2:lemma:derived-transposition-natural}(b) gives
\[
 \delta_*\circ t(\overline{\gamma_I})=t(\overline{\delta_*\circ\gamma_I})=t(\overline{\gamma_{\mathbb D}}).\qedhere
\]
\end{proof}

For $\Z_n$-tame complexes, the chain-level double dual is itself a model, and the canonical map is an isomorphism already at the chain level.

\begin{lemma}
\label{v2:lemma:Zn-can}
Let $F$ be a $\Z_n$-tame fusion complex.
\begin{enumerate}[(a)]
\item The chain map
\[
 \chi_*\colon D^2_{R_{\Q/\Z}}(F)\longrightarrow D_I\bigl(D_{R_{\Q/\Z}}(F)\bigr) \simeq \mathbb D^2_{R_{\Q/\Z}}(F)
\]
is a quasi-isomorphism. So $D^2_{R_{\Q/\Z}}(F)$ is a model for $\mathbb D^2_{R_{\Q/\Z}}(F)$, and the chain map $\can^{R_{\Q/\Z}}_F$ represents the derived canonical map.
\item The chain map $\can^{R_{\Q/\Z}}_F\colon F\to D^2_{R_{\Q/\Z}}(F)$ is an isomorphism of complexes.
\end{enumerate}
\end{lemma}

\begin{proof}
For (a), $D_{R_{\Q/\Z}}(F)$ is $\Z_n$-tame by Lemma~\ref{v2:lemma:RQZ}(c). So Theorem~\ref{v2:thm:tame-dual} applies, and the second claim follows from Proposition~\ref{v2:prop:tame-dictionary}(b).

For (b), fix $i$ and a basis $e_1,\dots,e_m$ of the $R_n$-module $F_i$. By Lemma~\ref{v2:lemma:RQZ}(c),
\[
 D_{R_{\Q/\Z}}(F)_{-i}=\Hom_{R_n}(\overline{F_i},R_n).
\]
It is free with basis $e^1,\dots,e^m$, where $e^a(\overline{e_b})=\delta_{ab}$. In the same way, $D^2_{R_{\Q/\Z}}(F)_i$ is free with basis $f_1,\dots,f_m$, where $f_b(\overline{e^a})=\delta_{ab}$. By Lemma~\ref{v2:lemma:ev-can-formulas},
\[
 \can^{R_{\Q/\Z}}_F(e_b)(\overline{e^a})=(-1)^ic\bigl(\overline{\delta_{ab}}\bigr)=(-1)^i\delta_{ab}.
\]
So $\can^{R_{\Q/\Z}}_F(e_b)=(-1)^if_b$, and $\can^{R_{\Q/\Z}}_F$ is an isomorphism in degree $i$.
\end{proof}

If $F$ is $\Z$-tame and nonzero, then $D_{R_{\Q/\Z}}(F)$ is not tame, and $\can^{R_{\Q/\Z}}_F$ is not an isomorphism of complexes. Nevertheless, the chain-level double dual is still a model, and the canonical map is a quasi-isomorphism.

\begin{lemma}
\label{v2:lemma:Z-can}
Let $F$ be a $\Z$-tame fusion complex.
\begin{enumerate}[(a)]
\item The chain map
\[
 \chi_*\colon D^2_{R_{\Q/\Z}}(F)\longrightarrow D_I\bigl(D_{R_{\Q/\Z}}(F)\bigr) \simeq \mathbb D^2_{R_{\Q/\Z}}(F)
\]
is a quasi-isomorphism. So $D^2_{R_{\Q/\Z}}(F)$ is a model for $\mathbb D^2_{R_{\Q/\Z}}(F)$, and the chain map $\can^{R_{\Q/\Z}}_F$ represents the derived canonical map.
\item The chain map $\can^{R_{\Q/\Z}}_F\colon F\to D^2_{R_{\Q/\Z}}(F)$ is a quasi-isomorphism.
\end{enumerate}
\end{lemma}

\begin{proof}
For (a), we first show that $\Ext^i_R(R_{\Q/\Z},R_{\Q/\Z})=0$ for $i>0$. Since $R$ is free over $\Z$, $R_{\Q/\Z}=R\otimes_\Z\Q/\Z$ is also the derived tensor product, so by Remark~\ref{rem:new-derived-tensor-hom-rings},
\[
 \RHom_R(R_{\Q/\Z},R_{\Q/\Z})\simeq\RHom_\Z(\Q/\Z,R_{\Q/\Z}).
\]
As an abelian group, $R_{\Q/\Z}$ is a direct sum of copies of $\Q/\Z$, so it is divisible, hence injective \cite[Corollary 2.3.2]{Weibel_1994}. So the right-hand side has no homology in negative degrees.

Now fix $k$, and let $m$ be the rank of $F_{-k}$. Since $\overline R\cong R$ and $c\colon\overline{R_{\Q/\Z}}\to R_{\Q/\Z}$ is an isomorphism,
\[
 \overline{D_{R_{\Q/\Z}}(F)_k}=\overline{\Hom_R(\overline{F_{-k}},R_{\Q/\Z})}\cong R_{\Q/\Z}^m.
\]
So $\Ext^i_R\bigl(\overline{D_{R_{\Q/\Z}}(F)_k},R_{\Q/\Z}\bigr)=0$ for $i>0$. Because $R_{\Q/\Z}$
has finite injective dimension by Lemma~\ref{v2:lemma:RQZ}(a), Theorem~\ref{thm:dualityisderived} applies to $\overline{D_{R_{\Q/\Z}}(F)}$ with $T=R_{\Q/\Z}$, and gives the first claim. The second claim follows from Proposition~\ref{v2:prop:tame-dictionary}(b).

For (b), $F$ is bounded with finitely generated torsion homology, so it lies in $\D^b_f(R)$ and is $\Q$-acyclic. So the derived canonical map is an isomorphism by Corollary~\ref{cor:QZ-biduality}. By (a), $\can^{R_{\Q/\Z}}_F$ represents it, so it is a quasi-isomorphism.
\end{proof}

\subsection{Changes of coefficients}

The first part of the paper uses $R_{\R/\Z}$-duality where $R_{\R/\Z}=R\otimes_\Z\R/\Z$, with the involution induced by that of $R$. We show that for tame complexes this gives the same duality. Let 
\[j\colon R_{\Q/\Z}\to R_{\R/\Z}\] be the inclusion.

\begin{lemma}
\label{v2:lemma:real-coefficients}
Let $F$ be a tame fusion complex.
\begin{enumerate}[(a)]
\item Let $V$ be a $\Q$-vector space complement to $\Q$ in $\R$, and let $R_V=R\otimes_\Z V$. Then
\[
 R_{\R/\Z}=j(R_{\Q/\Z})\oplus R_V,
\]
compatibly with the involutions. The complex $D_{R_V}(F)$ is zero if $F$ is $\Z_n$-tame, and contractible if $F$ is $\Z$-tame.
\item The chain map $j_*\colon D_{R_{\Q/\Z}}(F)\to D_{R_{\R/\Z}}(F)$ is a quasi-isomorphism, and it is an isomorphism if $F$ is $\Z_n$-tame. In particular, $D_{R_{\R/\Z}}(F)\simeq\mathbb D_{R[1]}(F)$.
\item If $\alpha\colon G\to F$ is a quasi-isomorphism between tame fusion complexes, then so is $D_{R_{\R/\Z}}(\alpha)$.

\item If $F$ is $\Z$-tame, then $\can^{R_{\R/\Z}}_F\colon F\to D^2_{R_{\R/\Z}}(F)$ is a quasi-isomorphism.
\end{enumerate}
\end{lemma}

\begin{proof}
For (a), we have $\R=\Q\oplus V$ as $\Q$-vector spaces, and $\Z\subseteq\Q$. So
\[
 \R/\Z=\Q/\Z\oplus V.
\]
Tensoring with $R$ gives $R_{\R/\Z}=j(R_{\Q/\Z})\oplus R_V$. The splitting commutes with the involutions, since they act only on the factor $R$. Hence
\[
 D_{R_{\R/\Z}}(F)=j_*D_{R_{\Q/\Z}}(F)\oplus D_{R_V}(F).
\]
\begin{itemize}
\item If $F$ is $\Z_n$-tame, each $F_i$ is $n$-torsion and $R_V$ is $\Z$-torsion-free. So $D_{R_V}(F)=0$.
\item If $F$ is $\Z$-tame, then since $R_V$ is an $R_\Q$-module, extension of scalars gives
\[
 D_{R_V}(F)\cong\Hom_{R_\Q}\bigl(\overline F\otimes_\Z\Q,R_V\bigr).
\]
The complex $\overline F\otimes_\Z\Q$ is a bounded complex of projective $R_\Q$-modules, and it is acyclic because $F$ is $\Q$-acyclic. So it is contractible, and hence so is $D_{R_V}(F)$.
\end{itemize}
Part (b) follows from (a) and Theorem~\ref{v2:thm:tame-dual}. Part (c) follows from (b), the naturality of $j_*$, and Corollary~\ref{v2:cor:tame-dual}(a).

For (d), Corollary~\ref{v2:cor:can-identities}(c) for $\psi=j$ gives
\[
 D_{R_{\R/\Z}}(j_*)\circ\can^{R_{\R/\Z}}_F=j_*\circ\can^{R_{\Q/\Z}}_F
\]
as chain maps $F\to\Hom_R\bigl(\overline{D_{R_{\Q/\Z}}(F)},R_{\R/\Z}\bigr)$. The terms of $D_{R_{\Q/\Z}}(F)$ are $\Z$-torsion and $R_V$ is $\Z$-torsion-free, so by (a), the map $j_*$ on the right is an isomorphism. By (a), $j_*\colon D_{R_{\Q/\Z}}(F)\to D_{R_{\R/\Z}}(F)$ is the inclusion of a summand with contractible complement, hence a homotopy equivalence, and so $D_{R_{\R/\Z}}(j_*)$ is a homotopy equivalence by Lemma~\ref{v2:lemma:dual-invariance}(a). Since $\can^{R_{\Q/\Z}}_F$ is a quasi-isomorphism by Lemma~\ref{v2:lemma:Z-can}(b), so is $\can^{R_{\R/\Z}}_F$.
\end{proof}

\subsection{Detection complexes}

The detection complex of a tame fusion complex $F$ is $F\otimes_R\widehat R$, where $\widehat R=\Hom_\Z(R,\Z)$ is the module of formal Laurent series of Appendix~\ref{subsec:notation}. It is defined at the chain level. To see that it is a derived object, we need the analogue of Theorem~\ref{thm:dualityisderived} for tensor products.

\begin{proposition}
\label{v2:prop:tor-acyclic}
Let $G$ be an $R$-module, and let $F$ be a possibly unbounded complex such that $\Tor^R_k(F_i,G)=0$ for every $i\in\Z$ and every $k>0$. Then the natural map
\[
 F\otimes^{\mathrm L}_R G\longrightarrow F\otimes_R G
\]
is an isomorphism in $\D(R)$.
\end{proposition}

\begin{proof}
Since $R$ has global dimension $d+1$, there is a projective resolution $p\colon P\to G$ with $P_j=0$ for $j\notin[0,d+1]$. A bounded complex of projective modules is $K$-flat. So $F\otimes_R P$ is a model for $F\otimes^{\mathrm L}_R G$, and the natural map is represented by $1\otimes p$. By Lemma~\ref{lem:new-cone-hom-tensor}, its cone is $F\otimes_R C$, where $C=\Cone(p)$ is bounded and acyclic.
\begin{itemize}
\item For each $i$, the complex $F_i\otimes_R C$ is acyclic, since the homology of $F_i\otimes_R P$ is $\Tor^R_*(F_i,G)$.
\item The complex $F\otimes_R C$ is the total complex of the double complex $F_i\otimes_R C_j$, which has only finitely many nonzero entries in each total degree.
\end{itemize}
As in the proof of Theorem~\ref{thm:dualityisderived}, filtering by $i$ shows that $F\otimes_R C$ is acyclic.
\end{proof}

\begin{proposition}
\label{v2:prop:tame-derived-tensor}
Let $F$ be a tame fusion complex. Then the natural map
\[
 F\otimes^{\mathrm L}_R\widehat R\longrightarrow F\otimes_R\widehat R
\]
is an isomorphism in $\D(R)$. In particular, a quasi-isomorphism between tame fusion complexes induces a quasi-isomorphism between their detection complexes.
\end{proposition}

\begin{proof}
By Proposition~\ref{v2:prop:tor-acyclic}, it suffices to show that $\Tor^R_k(R_n,\widehat R)=0$ for $k>0$. The resolution 
\[0\to R\xrightarrow{n}R\to R_n\to0\] shows that $\Tor^R_k(R_n,-)=0$ for $k\geq2$, and that $\Tor^R_1(R_n,\widehat R)$ is the $n$-torsion of $\widehat R$. This is zero because $\widehat R$ is $\Z$-torsion-free. The last claim holds because $-\otimes^{\mathrm L}_R\widehat R$ sends quasi-isomorphisms to isomorphisms, and the natural map is natural in $F$.
\end{proof}

\section{Tame BFCs}
\label{v2:app:tame-BFC}
We now turn to braidings on tame fusion complexes. The goal of this section is to describe tame BFCs, up to equivalence, by antisymmetric forms in the derived category. The key observation is that braidings are the antisymmetric cycles for the transposition on $\Hom_R\bigl(F,D_{R_{\Q/\Z}}(F)\bigr)$.

\subsection{Braidings as antisymmetric cycles}

Let $F$ be a tame fusion complex. By Definition~\ref{v2:def:transposition}, the transposition is an isomorphism of complexes
\[
 t\colon\overline{\Hom_R\bigl(F,D_{R_{\Q/\Z}}(F)\bigr)}\longrightarrow\Hom_R\bigl(F,D_{R_{\Q/\Z}}(F)\bigr).
\]
Since $c$ is an involution of $R_{\Q/\Z}$, we have $t\bigl(\overline{t(\overline\varphi)}\bigr)=\varphi$, by Lemma~\ref{v2:lemma:transposition-twice}. Since $t$ is a chain map,
\[
 \Hom_R\bigl(F,D_{R_{\Q/\Z}}(F)\bigr)^-=\bigl\{\varphi\in\Hom_R\bigl(F,D_{R_{\Q/\Z}}(F)\bigr)\bigm|t(\overline\varphi)=-\varphi\bigr\}
\]
is a subcomplex of abelian groups. Its elements are called \emph{antisymmetric}.

\begin{defn}
\label{v2:def:braiding}
Let $F$ be a tame fusion complex.
\begin{itemize}
\item A \emph{braiding} on $F$ is a degree $0$ cycle $\lambda$ of $\Hom_R(F,D_{R_{\Q/\Z}}(F))^-$ which is constant-alternating in degree zero. That is, $\lambda\colon F\to D_{R_{\Q/\Z}}(F)$ is a chain map with $t(\overline\lambda)=-\lambda$, and the constant coefficient of $\lambda(a)(\overline a)$ vanishes for every $a\in F_0$.
\item A \emph{tame BFC} (Definition~\ref{defn:BFC}) is a pair $(F,\lambda)$ of a tame fusion complex and a braiding on it.
\item An \emph{antisymmetric homotopy} is an element of $\Hom_R(F,D_{R_{\Q/\Z}}(F))^-$ of degree $1$.
\end{itemize}
\end{defn}

These are the braidings and antisymmetric homotopies of the paper (Definition~\ref{def:antisymmetric-homotopy}).

Since $R_{\Q/\Z}$ is concentrated in degree $0$, the transposition pairs opposite components. So an antisymmetric element is determined by half of its components. This is what makes the inductive constructions below work.

\begin{lemma}
\label{v2:lemma:reflection}
Let $F$ be a tame fusion complex, and let $\varphi\in \Hom_R(F,D_{R_{\Q/\Z}}(F))$ have degree $k$.
\begin{enumerate}[(a)]
\item The component $t(\overline\varphi)_j$ depends only on $\varphi_{-j-k}$.
\item If $\varphi$ is antisymmetric and $\varphi_j=0$ for all $j$ with $2j\geq-k$, then $\varphi=0$.
\item Let $\varphi_j\colon F_j\to D_{R_{\Q/\Z}}(F)_{j+k}$ be given for all $j$ with $2j\geq-k$. If $k$ is even, assume that $\varphi_{-k/2}$ is antisymmetric. Then there is a unique antisymmetric $\varphi$ of degree $k$ with these components.
\end{enumerate}
\end{lemma}

\begin{proof}
For (a), let $x\in F_i$ and $y\in F_j$. By Lemma~\ref{v2:lemma:transposition-formula}, $t(\overline\varphi)(y)(\overline x)$ is determined by $\varphi(x)(\overline y)$. This element lies in degree $i+j+k$ of $R_{\Q/\Z}$, so it is zero unless $i=-j-k$.

By (a), antisymmetry says that $\varphi_j=-t(\overline\varphi)_j$ is determined by $\varphi_{-j-k}$. The involution $j\mapsto-j-k$ exchanges the integers with $2j>-k$ and those with $2j<-k$, and fixes $-k/2$ when $k$ is even. This gives (b). For (c), define the components with $2j<-k$ by $\varphi_j=-t(\overline\varphi)_j$. Since $\varphi\mapsto t(\overline\varphi)$ is an involution, the resulting $\varphi$ is antisymmetric.
\end{proof}

\begin{lemma}
\label{v2:lemma:constant-alternating}
If $h$ is an antisymmetric homotopy on a tame fusion complex $F$, then $d h$ is a braiding on $F$.
\end{lemma}

\begin{proof}
Since $\Hom_R(F,D_{R_{\Q/\Z}}(F))^-$ is a subcomplex, $d h$ is an antisymmetric cycle of degree $0$. Let $a\in F_0$, and let $z=h(d a)(\overline a)$. By \eqref{eq:new-hom-complex} and antisymmetry of $h$,
\[
 (d h)(a)(\overline a)=h(a)(\overline{d a})+z=z-c(\overline z).
\]
Its constant coefficient vanishes because $c$ preserves constant coefficients.
\end{proof}

Braidings pull back along chain maps.

\begin{lemma}
\label{v2:lemma:pullback-braiding}
Let $(F,\lambda)$ be a tame BFC, and let $\alpha\colon G\to F$ be a chain map from a tame fusion complex $G$. Then $D_{R_{\Q/\Z}}(\alpha)\circ\lambda\circ\alpha$ is a braiding on $G$.
\end{lemma}

\begin{proof}
It is a chain map. It is antisymmetric, since by Lemma~\ref{v2:lemma:transposition-natural}(a),
\[
 t\bigl(\overline{D_{R_{\Q/\Z}}(\alpha)\circ\lambda\circ\alpha}\bigr)=D_{R_{\Q/\Z}}(\alpha)\circ t(\overline\lambda)\circ\alpha=-D_{R_{\Q/\Z}}(\alpha)\circ\lambda\circ\alpha.
\]
It is constant-alternating, since its value at $(a,\overline a)$ for $a\in G_0$ is $\lambda(\alpha a)(\overline{\alpha a})$.
\end{proof}

The paper sometimes describes braidings as forms. This is the same data.

\begin{remark}
\label{v2:rmk:BFC-as-forms}
By Remark~\ref{v2:rmk:forms}, a map $\varphi\in \Hom_R(F,D_{R_{\Q/\Z}}(F))$ of degree $k$ is the same as a sesquilinear form $\Lambda\colon\overline F\otimes_RF\to R_{\Q/\Z}$ of degree $k$, and $\varphi$ is antisymmetric if and only if $\Lambda$ is.
For a chain map $\alpha\colon G\to F$, the map $D_{R_{\Q/\Z}}(\alpha)\circ\varphi\circ\alpha$ corresponds to $\Lambda\circ(\overline\alpha\otimes\alpha)$. So a tame BFC is a pair $(F,\Lambda)$, where $\Lambda$ is an antisymmetric form of degree $0$ which is a chain map and is constant-alternating on $F_0$. Antisymmetric homotopies are antisymmetric forms of degree $1$.
\end{remark}

\subsection{The derived braiding}

A tame BFC $(F,\lambda)$ has a \emph{derived braiding}
\[
 \lambda_{\mathbb D}=\nu_F\circ\lambda\colon F\longrightarrow\mathbb D_{R[1]}(F)
\]
in $\D(R)$, as a special case of the definition in Section~\ref{subsection:canonical-maps}. By Proposition~\ref{v2:prop:tame-dictionary}(a), it depends only on the chain homotopy class of $\lambda$, and it is compatible with pullbacks along chain maps.

\begin{proposition}
\label{v2:prop:derived-braiding}
Let $(F,\lambda)$ be a tame BFC.
\begin{enumerate}[(a)]
\item The derived braiding is antisymmetric, that is,
\[
 t(\overline{\lambda_{\mathbb D}})=-\lambda_{\mathbb D},
\]
where $t$ is the derived transposition for $T=R[1]$.
\item The BFC $(F,\lambda)$ is nondegenerate if and only if $H_0(F)=0$ and $\lambda_{\mathbb D}$ is an isomorphism.
\end{enumerate}
\end{proposition}

\begin{proof}
Part (a) is Proposition~\ref{v2:prop:tame-dictionary}(c) for $\gamma=\lambda$, since $t(\overline\lambda)=-\lambda$. Part (b) holds because $\nu_F$ is an isomorphism.
\end{proof}

\subsection{Commutators and braidings}

The paper constructs braidings from their degree zero component.

\begin{defn}
\label{v2:def:commutator}
A \emph{commutator} on a tame fusion complex $F$ is a map
\[
 u_0\colon F_0\longrightarrow D_{R_{\Q/\Z}}(F)_0
\]
which is antisymmetric and constant-alternating, and such that $u_0\circ d$ takes values in the cycles of $D_{R_{\Q/\Z}}(F)$, that is,
\[
 u_0(d x)(\overline{d y})=0\qquad\text{for all }x,y\in F_1.
\]
\end{defn}

The degree zero component of a braiding is a commutator.

Both results below build the components of a braiding or a homotopy one degree at a time. At each step, we must lift a map into cycles to a map into chains. Properness makes this possible.
\begin{itemize}
\item The modules $F_j$ are projective over $R$, or over $R_n$ if $F$ is $\Z_n$-tame. In the second case, $D_{R_{\Q/\Z}}(F)$ is a complex of $R_n$-modules, by Lemma~\ref{v2:lemma:RQZ}(c). So maps out of $F_j$ lift along surjections between the modules of $D_{R_{\Q/\Z}}(F)$.
\item If $F$ is proper, then $D_{R_{\Q/\Z}}(F)$ has no homology in positive degrees, by Corollary~\ref{v2:cor:tame-dual}. So, for $i\geq1$, every map from $F_j$ to the cycles of $D_{R_{\Q/\Z}}(F)_i$ is of the form $d\circ\psi$ for some $\psi\colon F_j\to D_{R_{\Q/\Z}}(F)_{i+1}$.
\end{itemize}

\begin{lemma}
\label{v2:lemma:commutator-extension}
Let $F$ be a proper tame fusion complex, and let $u_0$ be a commutator on $F$.
\begin{enumerate}[(a)]
\item Let $u_1\colon F_1\to D_{R_{\Q/\Z}}(F)_1$ be an $R$-linear map with $d\circ u_1=u_0\circ d$. Then there is a braiding $\lambda$ on $F$ with $\lambda_0=u_0$ and $\lambda_1=u_1$. Its component $\lambda_{-1}$ is given by
\[
 \lambda_{-1}(y)(\overline x)=c\bigl(\overline{u_1(x)(\overline y)}\bigr).
\]
\item Such a map $u_1$ exists if and only if $u_0\circ d$ takes values in the boundaries of $D_{R_{\Q/\Z}}(F)$. This holds if $F$ is complete.
\end{enumerate}
\end{lemma}

\begin{proof}
For (a), set $\lambda_0=u_0$ and $\lambda_1=u_1$. Suppose that $\lambda_0,\dots,\lambda_i$ are constructed, with $i\geq1$ and $d\lambda_j=\lambda_{j-1}d$ for $1\leq j\leq i$. Then
\[
 d\lambda_id=\lambda_{i-1}dd=0,
\]
so, by exactness, there is $\lambda_{i+1}$ with $d\lambda_{i+1}=\lambda_id$. By Lemma~\ref{v2:lemma:reflection}(c), these components extend to an antisymmetric map $\lambda$ of degree $0$. Its differential $d\lambda$ is antisymmetric, and it vanishes in degrees $\geq1$ by construction. So $d\lambda=0$ by Lemma~\ref{v2:lemma:reflection}(b), that is, $\lambda$ is a chain map. It is constant-alternating in degree zero because $u_0$ is. The formula for $\lambda_{-1}$ is the antisymmetry condition for $x\in F_1$ and $y\in F_{-1}$.

For (b), if $u_0\circ d$ takes values in the boundaries, then $u_1$ exists because $F_1$ is projective. Conversely, $u_0\circ d=d\circ u_1$ takes values in the boundaries. If $F$ is complete, then $D_{R_{\Q/\Z}}(F)$ also has no homology in degree $0$, by Corollary~\ref{v2:cor:tame-dual}. So the cycles are boundaries.
\end{proof}

In the main-text conventions, the physical pairings $\Omega$ and $\Xi$ give $u_0=-\Omega_R$ and $u_1=\Xi_L$, so $\lambda_0=u_0$ and $\lambda_1=u_1$. 

In the paper, the condition $d\circ u_1=u_0\circ d$, that is,
\[
 u_1(x)(\overline{d a})=u_0(d x)(\overline a)\qquad\text{for }x\in F_1,\ a\in F_0,
\]
is the compatibility of the stabilizer--excitation pairing with the commutator (Section~\ref{subsec:bfc1}). Equivalently, $\overline{\Xi(x,d a)}=-\Omega(a,d x)$, or $\Xi(x,d a)=\Omega(d x,a)$.

\begin{lemma}
\label{v2:lemma:braiding-uniqueness}
Let $F$ be a proper tame fusion complex. Any two braidings $\lambda,\lambda'$ on $F$ with $\lambda_0=\lambda'_0$ are antisymmetrically homotopic, by a homotopy $h$ with $h_0=h_{-1}=0$.
\end{lemma}

\begin{proof}
Let $\gamma=\lambda'-\lambda$. It is an antisymmetric chain map with $\gamma_0=0$. Set $h_0=0$. Since $d\gamma_1=\gamma_0d=0$, there is $h_1$ with $\gamma_1=d h_1$. For $i\geq2$, suppose that $h_1,\dots,h_{i-1}$ are constructed with
\[
 \gamma_j=d h_j+h_{j-1}d\qquad\text{for }1\leq j<i.
\]
Then $d(\gamma_i-h_{i-1}d)=\gamma_{i-1}d-d h_{i-1}d=h_{i-2}dd=0$, so there is $h_i$ with $\gamma_i=d h_i+h_{i-1}d$. By Lemma~\ref{v2:lemma:reflection}(c), the $h_i$ extend to an antisymmetric map $h$ of degree $1$, and $h_{-1}=0$ because $h_0=0$. Then $\gamma-d h$ is antisymmetric, and it vanishes in degrees $\geq0$. So $\gamma=d h$ by Lemma~\ref{v2:lemma:reflection}(b).
\end{proof}

\begin{corollary}
\label{v2:cor:complete-commutator}
Let $F$ be a complete tame fusion complex, and let $u_0$ be a commutator on $F$. Then there is a braiding $\lambda$ on $F$ with $\lambda_0=u_0$, and it is unique up to antisymmetric homotopy.
\end{corollary}

\begin{proof}
This follows from Lemma~\ref{v2:lemma:commutator-extension} and Lemma~\ref{v2:lemma:braiding-uniqueness}.
\end{proof}

We note that the assumption that $F$ is complete is essential in Corollary~\ref{v2:cor:complete-commutator}, since the proof invokes exactness at $D_{\RQZ}(F)_0$. It is therefore natural to seek an explicit example of a commutator on an incomplete fusion complex that cannot be lifted to a braiding. \begin{example}\label{v2:ex:nonextendible-commutator}
Take $d=1$ and $R=\Z[x^{\pm1}]$.
The following fusion complex $F$, together with the commutator $u_0:F_0\to D_{\RQZ}(F)_0$, provides such an example:
\begin{equation}
    \begin{tikzcd}[ampersand replacement=\&]
        0\&R^2\&[6em]R^4\&[6em]R^2\&0\\[6ex]
        0\&R_{\Q/\Z}^2\&R_{\Q/\Z}^4\&R_{\Q/\Z}^2\&0
        \arrow[from=1-1, to=1-2]
        \arrow[from=1-2, to=1-3,"{\begin{psmallmatrix}2&0\\0&2\\1&-x\\-1&1\end{psmallmatrix}}"]
        \arrow[from=1-3, to=1-4,"{\begin{psmallmatrix}-1&x&2&0\\1&-1&0&2\end{psmallmatrix}}"]
        \arrow[from=1-4, to=1-5]
        \arrow[from=2-1, to=2-2]
        \arrow[from=2-2, to=2-3,"{\begin{psmallmatrix}-1&1\\\ol x&-1\\2&0\\0&2\end{psmallmatrix}}"']
        \arrow[from=2-3, to=2-4,"{\begin{psmallmatrix}-2&0&-1&1\\0&-2&\ol x&-1\end{psmallmatrix}}"']
        \arrow[from=2-4, to=2-5]
        \arrow[from=1-2, to=2-2,"{\begin{psmallmatrix}A&B\\C&D\end{psmallmatrix}}"]
        \arrow[from=1-3, to=2-3,"{u_0=\frac{1}{4}\begin{psmallmatrix}0&1&0&0\\-1&0&0&0\\0&0&0&0\\0&0&0&0\end{psmallmatrix}}"]
        \arrow[from=1-4, to=2-4,"{\begin{psmallmatrix}\ol A& \ol C \\ \ol B &\ol D\end{psmallmatrix}}"]
    \end{tikzcd}
\end{equation}
The differential $d_1$ is injective, $\ker d_0=\im d_1$, and $\coker d_0\cong R/(2,x-1)$, so this is a fusion complex with $H_{-1}(F)\cong\Z_2$. The displayed $u_0$ is constant-alternating, and $d_1^*u_0 d_1=0$ in $\RQZ$, so it is a commutator. Here $A,B,C,D\in R_{\Q/\Z}$ parametrize a hypothetical lift $\lambda:F\to D_{\RQZ}(F)$ of $u_0$. We now show that no such chain map exists. Observe that
\begin{equation}
    u_0\circ d_1=\frac{1}{2}\begin{pmatrix}0&1\\-1&0\\0&0\\0&0\end{pmatrix},\qquad
    d_1^{D_{\RQZ}(F)}\circ\lambda_1=\begin{pmatrix}C-A&D-B\\A\ol x-C&B\ol x-D\\2A&2B\\2C&2D\end{pmatrix}.
\end{equation}
Commutativity of the diagram requires $A=C$, and hence $A(\ol x-1)=-\frac{1}{2}.$ However, this equation has no solution in $R_{\Q/\Z}$.
Therefore, no such braiding $\lambda$ exists.

Let us explain the construction of this example. The fusion complex $F$ is obtained by applying a two-fold coarse-graining to the free resolution $B$ of $S=\Z_2$ constructed in Section~\ref{subsec:example1D}. Consequently, $F$ likewise describes a fully mobile particle with a $\Z_2$ fusion rule, now with a two-site unit cell. As discussed in Section~\ref{subsec:example1D}, the one-dimensional fusion statistics of a 1D particle are encoded by the commutation relations between neighboring hoppers. The physical commutator $\Omega_R=-u_0$ is chosen so that the commutation phase alternates from site to site between $\pm i$ and $1$. 
In other words, the particle alternates spatially between trivial and nontrivial one-dimensional
fusion statistics. This behavior is manifestly unphysical, since these statistics 
should not 
depend on position. The obstruction to lifting $u_0$ to a braiding reflects precisely this inconsistency.
\end{example}

\subsection{Pauli fusion complexes}

Pauli codes give the main examples of commutators. Write $D=D_{R_{\Q/\Z}}$. A Pauli fusion complex $F$ has $F_0=R_n^{2k}$, with the standard symplectic form $\Omega\colon F_0\to D(F_0)$, and $F_{-i}=D(F_i)$ for $i>0$. Its differentials satisfy $d_0=D(d_1)\circ\Omega$ and $d_{-i}=(-1)^{i+1}D(d_{i+1})$ for $i>0$. Its associated Pauli code is given by $P=F_0$ and $L=\im d_1$.

\begin{proposition}
\label{v2:prop:Pauli-fc-associated-code}
If two Pauli fusion complexes have the same associated Pauli code, then they are chain homotopy equivalent.
\end{proposition}

\begin{proof}
Let $F$ and $G$ be Pauli fusion complexes with the same associated code, so that $F_0=G_0=P$ and $\im d_1^F=\im d_1^G=L$. Let $F_+$ and $G_+$ be the brutal truncations of $F$ and $G$ to degrees $\geq1$, with differential out of degree $1$ set to zero. Via $d_1$, they are resolutions of $L$ by finitely generated free $R_n$-modules. For chain maps $u\colon F_+\to G_+$ and $v\colon G_+\to F_+$ with $d_1^Gu_1=d_1^F$ and $d_1^Fv_1=d_1^G$, let $\Phi(u,v)\colon F\to G$ have components $u_i$ for $i>0$, $\id_P$ for $i=0$, and $D(v_{-i})$ for $i<0$. The defining relations of a Pauli fusion complex show that $\Phi(u,v)$ is a chain map, and $\Phi(u',v')\circ\Phi(u,v)=\Phi(u'u,vv')$.

By the comparison theorem, there are such $f\colon F_+\to G_+$ and $g\colon G_+\to F_+$, and maps $h_i\colon F_i\to F_{i+1}$ for $i\geq1$ with $gf-\id=d h+hd$, where $h_0=0$. Then $\Phi(g,f)\circ\Phi(f,g)=\Phi(gf,gf)$, and $\Phi(gf,gf)-\id=d H+Hd$ for
\[
 H_i=h_i\ (i>0),\qquad H_0=0,\qquad H_i=(-1)^iD(h_{-i-1})\ (i<0).
\]
By symmetry, $\Phi(f,g)\circ\Phi(g,f)$ is also homotopic to the identity.
\end{proof}

\subsection{Equivalence of tame BFCs}

Tame BFCs are compared by quasi-isomorphisms that preserve the braidings, possibly up to antisymmetric homotopy.

\begin{defn}
\label{v2:def:quasi-isometry}
Let $(F,\lambda)$ and $(G,\mu)$ be tame BFCs, and let $\alpha\colon G\to F$ be a quasi-isomorphism.
\begin{itemize}
\item The map $\alpha$ is a \emph{quasi-isometry} from $(G,\mu)$ to $(F,\lambda)$ if
\[
 D_{R_{\Q/\Z}}(\alpha)\circ\lambda\circ\alpha=\mu.
\]
Then $(G,\mu)$ and $(F,\lambda)$ are \emph{quasi-isometric}.
\item The map $\alpha$ is a \emph{homotopy quasi-isometry} if $D_{R_{\Q/\Z}}(\alpha)\circ\lambda\circ\alpha-\mu=d h$ for an antisymmetric homotopy $h$ on $G$.
\item The tame BFCs $(F,\lambda)$ and $(G,\mu)$ are \emph{equivalent} if they are related by a zigzag
\begin{equation}
\label{v2:eq:bfc-zigzag}
 (F,\lambda)\longrightarrow(C^1,\tau_1)\longleftarrow(C^2,\tau_2)\longrightarrow\cdots\longleftarrow(G,\mu)
\end{equation}
of homotopy quasi-isometries between tame BFCs.
\end{itemize}
\end{defn}

These are the quasi-isometries and the equivalence of the paper (Definition~\ref{def:quasi-isometry} and Definition~\ref{def:tame-bfc-equivalence}). In the paper, the zigzag \eqref{v2:eq:bfc-zigzag} is drawn as the diagram \eqref{eqn:bfc-equiv-diagram}, whose squares commute up to antisymmetric homotopy.

We show that the homotopies are not needed. The idea is to absorb a homotopy into a larger complex, by adding a contractible summand.

\begin{lemma}
\label{v2:lemma:hyperbolic-zigzag}
Let $(F,\lambda)$ be a tame BFC, let $k\in \Hom_R(F,D_{R_{\Q/\Z}}(F))$ have degree $1$, and let $\gamma=d k$. Then
\[
 \lambda'=\lambda+\gamma-t(\overline\gamma)
\]
is a braiding on $F$, and there are quasi-isometries $(F,\lambda)\to(K,\tau)\leftarrow(F,\lambda')$ for some tame BFC $(K,\tau)$.
\end{lemma}

\begin{proof}
Let $C=\Cone(\id_F)$, let $\iota\colon F\to C$ be the inclusion $y\mapsto(y,0)$, and let $K=F\oplus C$. Since $C$ is contractible and tame of the same type as $F$, the complex $K$ is a tame fusion complex, and the projection $K\to F$ is a quasi-isomorphism. Since $\gamma=d k$ 
and the differential on $C$ is $d(y, x) = (dy + x, -dx)$,
the map
\[
 \beta\colon C\longrightarrow D_{R_{\Q/\Z}}(F),\qquad\beta(y,x)=\gamma(y)+k(x),
\]
is a chain map with $\beta\iota=\gamma$. Using $D_{R_{\Q/\Z}}(K)=D_{R_{\Q/\Z}}(F)\oplus D_{R_{\Q/\Z}}(C)$, let
\[
 \tau=\begin{pmatrix}\lambda&\beta\\-t(\overline\beta)&0\end{pmatrix}\colon K\longrightarrow D_{R_{\Q/\Z}}(K).
\]
\begin{itemize}
\item The transposition exchanges the two off-diagonal blocks, so $t(\overline\tau)=-\tau$ by Lemma~\ref{v2:lemma:transposition-twice}.
\item It is constant-alternating in degree zero. Indeed, for $(a,b)\in K_0$, the difference $\tau(a,b)(\overline{(a,b)})-\lambda(a)(\overline a)$ equals $z-c(\overline z)$ with $z=\beta(b)(\overline a)$, and its constant coefficient vanishes.
\end{itemize}
So $(K,\tau)$ is a tame BFC. The maps $j_0=(\id,0)$ and $j_1=(\id,\iota)$ from $F$ to $K$ are sections of the projection $K\to F$, hence quasi-isomorphisms. We have $D_{R_{\Q/\Z}}(j_0)\circ\tau\circ j_0=\lambda$ and
\[
 D_{R_{\Q/\Z}}(j_1)\circ\tau\circ j_1=\lambda+\beta\iota-t(\overline{\beta\iota})=\lambda'.
\]
In particular, $\lambda'$ is a braiding, by Lemma~\ref{v2:lemma:pullback-braiding}.
\end{proof}

To apply the lemma to an equivalence of BFCs, we use that an antisymmetric homotopy $h$ is even,
meaning it is of the form $\gamma - t(\bar \gamma)$ for some $\gamma$. 

\begin{proposition}
\label{v2:prop:homotopy-conjugation}
Equivalent tame BFCs are related by a zigzag \eqref{v2:eq:bfc-zigzag} of quasi-isometries.
\end{proposition}

\begin{proof}
A homotopy quasi-isometry $s\colon(C',\tau')\to(C,\tau)$ factors as the quasi-isometry
\[
 \bigl(C',D_{R_{\Q/\Z}}(s)\circ\tau\circ s\bigr)\longrightarrow(C,\tau),
\]
followed by an antisymmetric homotopy on $C'$. So it suffices to treat two braidings $\lambda,\lambda'$ on $F$ with $\lambda'-\lambda=d h$ and $t(\overline h)=-h$.

Let $e\colon F\to F$ be the projection onto the terms of odd degree, and let $k=h\circ e$. 
Clearly, $k - t(\ol k) = h$. 
Since $t$ is a chain map, $\gamma=d k$ satisfies $\gamma-t(\overline\gamma)=d h=\lambda'-\lambda$, and Lemma~\ref{v2:lemma:hyperbolic-zigzag} applies.
\end{proof}

\subsection{Odd fusion order}

At odd fusion order, tame BFCs up to equivalence are the same as antisymmetric morphisms in the derived category. The reason is that $2$ is invertible on such complexes in the derived category, so the derived antisymmetry can be made strict by averaging.

\begin{defn}
\label{v2:def:isometric}
Let $F,G\in\FC(R)$, and let $\varphi\colon F\to\mathbb D_{R[1]}(F)$ and $\psi\colon G\to\mathbb D_{R[1]}(G)$ be morphisms in $\D(R)$. An \emph{isometry} from $(G,\psi)$ to $(F,\varphi)$ is an isomorphism $\alpha\colon G\to F$ in $\D(R)$ with
\[
 \alpha^*\varphi=\mathbb D_{R[1]}(\alpha)\circ\varphi\circ\alpha=\psi.
\]
If one exists, $(F,\varphi)$ and $(G,\psi)$ are \emph{isometric}.
\end{defn}

\begin{lemma}
\label{v2:lemma:equivalent-isometric}
If the tame BFCs $(F,\lambda)$ and $(G,\mu)$ are equivalent, then $(F,\lambda_{\mathbb D})$ and $(G,\mu_{\mathbb D})$ are isometric.
\end{lemma}

\begin{proof}
Consider a homotopy quasi-isometry $f\colon(C,\tau)\to(C',\tau')$ in the zigzag \eqref{v2:eq:bfc-zigzag}. Then $\tau$ and $D_{R_{\Q/\Z}}(f)\circ\tau'\circ f$ are antisymmetrically homotopic. By Proposition~\ref{v2:prop:tame-dictionary}(a),
\[
 \tau_{\mathbb D}=f^*\tau'_{\mathbb D},
\]
and $f$ is an isomorphism in $\D(R)$. Composing along the zigzag gives the claim.
\end{proof}

\begin{lemma}
\label{v2:lemma:isometric-equivalent}
Let $(F,\lambda)$ and $(G,\mu)$ be tame BFCs, and assume that $G$ has odd fusion order. If $(F,\lambda_{\mathbb D})$ and $(G,\mu_{\mathbb D})$ are isometric, then $(F,\lambda)$ and $(G,\mu)$ are equivalent.
\end{lemma}

\begin{proof}
Let $\alpha\colon G\to F$ be an isomorphism in $\D(R)$ with $\alpha^*\lambda_{\mathbb D}=\mu_{\mathbb D}$.

We first reduce to the case where $G$ is $\Z$-tame. By Lemma~\ref{v2:lemma:tame-in-FC}(c), there is a $\Z$-tame complex $P$ isomorphic to $G$ in $\D(R)$. Since $P$ is $K$-projective, this isomorphism is represented by a quasi-isomorphism $p\colon P\to G$. By Lemma~\ref{v2:lemma:pullback-braiding}, $(P,D_{R_{\Q/\Z}}(p)\circ\mu\circ p)$ is a tame BFC, and it is quasi-isometric to $(G,\mu)$. By Proposition~\ref{v2:prop:tame-dictionary}(a), we may replace $(G,\mu)$ by it, and $\alpha$ by $\alpha p$. Then $\alpha$ is represented by a quasi-isomorphism of complexes, which we also call $\alpha$.

Now let $\beta=D_{R_{\Q/\Z}}(\alpha)\circ\lambda\circ\alpha$. By
Proposition~\ref{v2:prop:tame-dictionary}(a), $\beta_{\mathbb D}=\mu_{\mathbb D}$. Since $\nu_G$ is
an isomorphism and $G$ is $K$-projective, $\beta-\mu=d k$ for some $k\in \Hom_R(G,D_{R_{\Q/\Z}}(G))$
of degree $1$. Since $\beta-\mu$ is antisymmetric and $t$ is a chain map, we have
\[
 d\bigl(2(k-t(\overline k))\bigr)=4(\beta-\mu).
\]
So $h=2\bigl(k-t(\overline k)\bigr)$ is an antisymmetric homotopy from $4\mu$ to $4\beta$. Since $G$ has odd fusion order, $2\id_G$ is a quasi-isomorphism, and
\[
 D_{R_{\Q/\Z}}(2\id_G)\circ\mu\circ2\id_G=4\mu.
\]
So $2\id_G$ is a quasi-isometry from $(G,4\mu)$ to $(G,\mu)$, and likewise from $(G,4\beta)$ to $(G,\beta)$. Together with the homotopy $h$ and the quasi-isometry $\alpha$, this gives the zigzag
\[
 (F,\lambda)\xleftarrow{\ \alpha\ }(G,\beta)\xleftarrow{\ 2\id\ }(G,4\beta)\xrightarrow{\ \id\ }(G,4\mu)\xrightarrow{\ 2\id\ }(G,\mu),
\]
where the middle map is a homotopy quasi-isometry. So $(F,\lambda)$ and $(G,\mu)$ are equivalent.
\end{proof}

\begin{lemma}
\label{v2:lemma:odd-realization}
Let $F\in\FC(R)$ have odd fusion order, and let $\varphi\colon F\to\mathbb D_{R[1]}(F)$ be an antisymmetric morphism in $\D(R)$. Then $(F,\varphi)$ is isometric to $(P,\lambda_{\mathbb D})$ for some $\Z$-tame BFC $(P,\lambda)$.
\end{lemma}

\begin{proof}
By Lemma~\ref{v2:lemma:tame-in-FC}(c), there are a $\Z$-tame complex $P$ and an isomorphism $\alpha\colon P\to F$ in $\D(R)$. The morphism $\varphi'=\mathbb D_{R[1]}(\alpha)\circ\varphi\circ\alpha$ is again antisymmetric, by Lemma~\ref{v2:lemma:derived-transposition-natural}(a).

Since $P$ has odd fusion order, $2\id_P$ is an isomorphism in $\D(R)$, so we can halve $\varphi'$. Let $\psi\colon P\to\mathbb D_{R[1]}(P)$ be the morphism with $2\psi=\varphi'$. It is antisymmetric, since $2t(\overline\psi)=-2\psi$ and $2$ is invertible on morphisms out of $P$. Since $P$ is $K$-projective and $\nu_P$ is an isomorphism, $\psi=\kappa_{\mathbb D}$ for some chain map $\kappa\colon P\to D_{R_{\Q/\Z}}(P)$. Let
\[
 \lambda=\kappa-t(\overline\kappa).
\]
Then $t(\overline\lambda)=-\lambda$, and by Proposition~\ref{v2:prop:tame-dictionary}(c),
\[
 \lambda_{\mathbb D}=\psi-t(\overline\psi)=2\psi=\varphi'.
\]
For $a\in P_0$, we have $\lambda(a)(\overline a)=z-c(\overline z)$ with $z=\kappa(a)(\overline a)$, so $\lambda$ is constant-alternating. So $(P,\lambda)$ is a $\Z$-tame BFC, and $\alpha$ is an isometry from $(P,\lambda_{\mathbb D})$  to $(F,\varphi)$.
\end{proof}

\begin{theorem}
\label{v2:thm:odd-tame-derived-BFC}
Consider the following two kinds of objects:
\begin{itemize}
\item tame BFCs $(F,\lambda)$ of odd fusion order, up to equivalence;
\item pairs $(F,\varphi)$, where $F\in\FC(R)$ has odd fusion order and $\varphi\colon F\to\mathbb D_{R[1]}(F)$ is an antisymmetric morphism in $\D(R)$, up to isometry.
\end{itemize}
The assignment $(F,\lambda)\mapsto(F,\lambda_{\mathbb D})$ is a bijection from the first set of classes to the second. Every class of tame BFCs contains a $\Z$-tame BFC. Nondegenerate tame BFCs correspond to the pairs with $H_0(F)=0$ and $\varphi$ an isomorphism.
\end{theorem}

\begin{proof}
The map is well defined, since $\lambda_{\mathbb D}$ is antisymmetric by Proposition~\ref{v2:prop:derived-braiding}(a), and equivalent tame BFCs have isometric derived braidings by Lemma~\ref{v2:lemma:equivalent-isometric}. It is injective by Lemma~\ref{v2:lemma:isometric-equivalent}, and surjective, with $\Z$-tame representatives, by Lemma~\ref{v2:lemma:odd-realization}. The last claim follows from Proposition~\ref{v2:prop:derived-braiding}(b), since its conditions are invariant under isometry.
\end{proof}

\begin{remark}\label{rem:2rem}
The theory of quadratic and (anti)symmetric bilinear forms is much simpler when $2$ is invertible. In this case, quadratic forms and (anti)symmetric bilinear forms are equivalent concepts, the identification being given by a map from quadratic forms to (anti)symmetric bilinear forms called \emph{polarization}.
When $2$ is not invertible, however, these concepts no longer agree: polarization factors through a number of other possible so-called Poincar\'e structures, and the derived theory is much richer. In fact, a modern treatment of Hermitian forms when $2$ is not invertible was only recently given in \cite{CDHHLMNNS_2023_I}, and its framework uses $\infty$-categories rather than the classical approach to derived algebra to which we have limited ourselves in this paper.
\end{remark}

\subsection{Changes of coefficients}

Finally, we compare rational and real coefficients. A \emph{real tame BFC} is defined as a tame BFC, with $R_{\R/\Z}$ in place of $R_{\Q/\Z}$, and equivalence of real tame BFCs is defined in the same way. The results of this section hold verbatim for real tame BFCs, using Lemma~\ref{v2:lemma:real-coefficients}.

\begin{proposition}
\label{v2:thm:real-rational-BFC-equivalence}
The inclusion $j\colon R_{\Q/\Z}\to R_{\R/\Z}$ induces a bijection from equivalence classes of tame BFCs to equivalence classes of real tame BFCs.
\end{proposition}

\begin{proof}
Choose a splitting $R_{\R/\Z}=j(R_{\Q/\Z})\oplus R_V$ as in Lemma~\ref{v2:lemma:real-coefficients}(a), and let $p\colon R_{\R/\Z}\to R_{\Q/\Z}$ be the projection, so that $pj=\id$. The maps $j$ and $p$ are obtained by tensoring $R$ with homomorphisms between $\Q/\Z$ and $\R/\Z$. So they are $R$-linear, commute with the involutions, and preserve constant coefficients. Hence the induced maps $j_*$ and $p_*$ on duals have the following properties.
\begin{itemize}
\item They are natural, and they commute with the differentials.
\item They commute with transpositions, by Lemma~\ref{v2:lemma:transposition-natural}.
\item They preserve constant-alternation.
\end{itemize}
So they send tame BFCs to tame BFCs, and homotopy quasi-isometries to homotopy quasi-isometries. Thus both induce maps on equivalence classes, and the map induced by $j$ is injective because $p_*j_*=\id$.

For surjectivity, let $(F,\lambda)$ be a real tame BFC. Then $\Delta=\lambda-j_*p_*\lambda$ is an antisymmetric chain map from $F$ to $D_{R_V}(F)$. By Lemma~\ref{v2:lemma:real-coefficients}(a), $D_{R_V}(F)$ is zero or contractible. So $\Hom_R\bigl(F,D_{R_V}(F)\bigr)$ is acyclic, and $\Delta=d k$ for some $k$ of degree $1$. Since $R_V$ is a $\Q$-vector space, we can average:
\[
 h=\tfrac12\bigl(k-t(\overline k)\bigr),
 \qquad
 d h=\tfrac12\bigl(\Delta-t(\overline\Delta)\bigr)=\Delta.
\]
So $h$, viewed as a map to $D_{R_{\R/\Z}}(F)$, is an antisymmetric homotopy from $j_*p_*\lambda$ to $\lambda$. Hence $(F,\lambda)$ is equivalent to the image of the tame BFC $(F,p_*\lambda)$.
\end{proof}

\section{Pointlike fusion complexes and mutual statistics}
\label{v2:app:pointlike}

This section contains the results on pointlike fusion complexes used in Section~\ref{subsec:tpe} and in the proof of Theorem~\ref{thm:Sb-invariant}.  At the level of abstraction we present here, we do not need to assume that $2$ is invertible to state various results. However, for the applications of the paper, when $2$ is not invertible, the theory presented here will not suffice for classifications of pointlike theories. In technical terms, Poincar\'e structures are extremely subtle when $2$ is not a unit and the treatment presented here is not sufficient.

\begin{convention}
Throughout, we assume that $\cR$ satisfies Convention~\ref{conv:fusion-ring}. In this section we also assume $d\geq1$.
For $\Q$-acyclic $F\in\D^p(\cR)$, the map $\delta_*\colon\mathbb D_{\cR_{\Q/\Z}}(F)\to\mathbb D_{\cR[1]}(F)$ is a natural isomorphism compatible with the canonical maps. We write
\[
 \mathbb D=\mathbb D_{\cR_{\Q/\Z}}
 \qquad\text{and}\qquad
 \can=\can^{\cR_{\Q/\Z}},
\]
and we identify $\mathbb D_{\cR[1]}$ with $\mathbb D$ via $\delta_*$. 
\end{convention}

\subsection{Pointlike fusion complexes}

\begin{defn}
Let
\begin{itemize}
\item
 $\FC_{p}(\cR)\subset\D^p(\cR)$ be the full subcategory of $\Q$-acyclic complexes with homology concentrated in degree $-1$. We call this the category of \emph{pointlike fusion complexes}. 
 \item $\mathcal{M}_{p}(\cR)$ be the category of finitely generated $\Z$-torsion $\cR$-modules.
 \end{itemize}
\end{defn}

\begin{lemma}
\label{v2:lemma:pointlike-H}
The categories  $\FC_{p}(\cR)\subset\D^p(\cR)$ and  $\mathcal{M}_{p}(\cR)$  are related as follows:
\begin{enumerate}[(a)]
\item For $S\in\mathcal{M}_{p}(\cR)$, the complex $S[-1]$ lies in $\FC_{p}(\cR)$.
\item The functors
\begin{align*}
 H_{-1}&\colon \FC_{p}(\cR)\longrightarrow \mathcal{M}_{p}(\cR),
 \end{align*}
 and
 \begin{align*}
 \iota=(-)[-1]&\colon \mathcal{M}_{p}(\cR)\longrightarrow \FC_{p}(\cR)
\end{align*}
are inverse equivalences. 
More precisely, $H_{-1}\circ\iota=\id$, and there is a natural isomorphism
\[
 \epsilon_F\colon \iota(H_{-1}(F))=H_{-1}(F)[-1] \xlongrightarrow{\cong} F.
\]
This is the unique isomorphism in $\D(\cR)$ which induces the identity on $H_{-1}$. In particular, for a module $S$, the isomorphism $\epsilon_{S[-1]}$ is the identity of $S[-1]$.
\end{enumerate}
\end{lemma}

\begin{proof}
For (a), $S[-1]$ is perfect by Lemma~\ref{lemma:bounded-replacement}(c), and it is $\Q$-acyclic. For (b), by Lemma~\ref{lemma:single-degree}, a morphism between complexes with homology concentrated in degree $-1$ is determined by its effect on $H_{-1}$, and every homomorphism on $H_{-1}$ arises. So $H_{-1}$ is fully faithful. For $F\in\FC_{p}(\cR)$, let 
\[\epsilon_F\colon H_{-1}(F)[-1]\to F\] 
be the unique morphism in $\D(\cR)$ which induces the identity on $H_{-1}$. It is a quasi-isomorphism, hence an isomorphism. It is natural in $F$: for a morphism $f\colon F\to G$, the two composites $f\circ\epsilon_F$ and $\epsilon_G\circ H_{-1}(f)[-1]$ both induce $H_{-1}(f)$ on $H_{-1}$, so they are equal. Finally, for a module $S$, the identity of $S[-1]$ induces the identity on $H_{-1}$, so it is $\epsilon_{S[-1]}$ by uniqueness.
\end{proof}

\begin{remark}
\label{v2:lemma:pointlike-zigzag}
Let $F\in\FC_{p}(\cR)$. The isomorphism $\epsilon_F$ is represented by the zigzag of quasi-isomorphisms
\[
 \xymatrix{H_{-1}(F)[-1] & \ar[l]_-\simeq \tau_{\geq -1}F \ar[r]^-\simeq & F,}
\]
whose left map is the quotient map $\ker(d\colon F_{-1}\to F_{-2})\to H_{-1}(F)$ in degree $-1$. Indeed, both maps induce the identity on $H_{-1}$, so the zigzag is $\epsilon_F$ by uniqueness.
\end{remark}

\begin{corollary}
\label{v2:cor:S-complete-invt}
Let $F,G\in\FC_{p}(\cR)$, and let $\alpha\colon H_{-1}(F)\to H_{-1}(G)$ be an isomorphism. Then there is a unique isomorphism $F\to G$ in $\D(\cR)$ inducing $\alpha$. In particular, the isomorphism class of $H_{-1}(F)$ is a complete invariant of objects of $\FC_{p}(\cR)$.
\end{corollary}

\begin{proof}
This holds because $H_{-1}$ is fully faithful, by Lemma~\ref{v2:lemma:pointlike-H}.
\end{proof}

We now turn to totally pointlike complexes.
\begin{defn}
\label{v2:def:totally-pointlike}
We define totally pointlike complexes and modules as follows.
\begin{itemize}
\item An object $F\in\FC_{p}(\cR)$ is \emph{totally pointlike} if $\mathbb D(F)$ also has homology concentrated in degree $-1$. We write $\FC_{tp}(\cR)$ for the full subcategory of $\D(\cR)$ on the totally pointlike fusion complexes.
\item A module $S\in\mathcal{M}_{p}(\cR)$ is \emph{totally pointlike} if $S[-1]$ is. We write $\mathcal{M}_{tp}(\cR)$ for the full subcategory of $\Mod(\cR)$ on the totally pointlike modules.
\end{itemize}
\end{defn}
\begin{remark}
\label{v2:rmk:self-dual-tp}
Note that for $F\in\FC_{p}(\cR)$, if $F\cong\mathbb D(F)$ in $\D(\cR)$, then $F$ is totally pointlike.
\end{remark}

\begin{lemma}
\label{v2:lemma:tp-modules}
\begin{enumerate}[(a)]
\item A module $S\in\mathcal{M}_{p}(\cR)$ is totally pointlike if and only if
\[
 \Ext^k_\cR(\overline S,\cR_{\Q/\Z})=0\quad\text{for all }k\neq2.
\]
\item The equivalence of Lemma~\ref{v2:lemma:pointlike-H} restricts to inverse equivalences
\[
 H_{-1}\colon\FC_{tp}(\cR)\longrightarrow\mathcal{M}_{tp}(\cR)
\]
and
\[
 \iota\colon\mathcal{M}_{tp}(\cR)\longrightarrow\FC_{tp}(\cR).
\]
\item A nonzero module $S\in\mathcal{M}_{p}(\cR)$ is totally pointlike if and only if it is Cohen--Macaulay of pure dimension $d-2$.
\end{enumerate}
\end{lemma}

\begin{proof}
For (a), by Lemma~\ref{lemma:QZ-duality} and Lemma~\ref{lemma:torsion-dual-ext} with $j=-1$,
\[
 H_k\mathbb D(S[-1])\cong\Ext^{1-k}_\cR(\overline S,\cR_{\Q/\Z})
\]
for all $k$. For (b), let $F\in\FC_{p}(\cR)$. Then $F\cong H_{-1}(F)[-1]$ by Lemma~\ref{v2:lemma:pointlike-H}, so $F$ is totally pointlike if and only if $H_{-1}(F)$ is. 
Part (c) follows exactly as in the proof of Theorem~\ref{theorem:totally-pointlike}.
\end{proof}

\subsection{Categories with duality}
We now state a few formal results about categories equipped with duality.
\begin{defn}
\label{v2:def:cat-duality}
An \emph{additive category with duality} is a triple $(\cC,\mathbb D_\cC,\can^\cC)$, where $\cC$ is an additive category, $\mathbb D_\cC\colon\cC^\op\to\cC$ is an additive functor, and $\can^\cC\colon\id_\cC\Rightarrow\mathbb D_\cC^2$ is a natural isomorphism such that, for every $X\in\cC$, the diagram
\[
\xymatrix@C=4pc{
 \mathbb D_\cC(X) \ar[r]^-{\can^\cC_{\mathbb D_\cC(X)}} \ar[dr]_-{\id} & \mathbb D_\cC^3(X) \ar[d]^-{\mathbb D_\cC(\can^\cC_X)} \\
 & \mathbb D_\cC(X)
}
\]
commutes, that is, $\mathbb D_\cC(\can^\cC_X)\circ\can^\cC_{\mathbb D_\cC(X)}=\id_{\mathbb D_\cC(X)}$.
\end{defn}

\begin{example}
\label{v2:ex:tp-duality}
By Theorem~\ref{thm:fusion-duality}, the triple
\[
 (\D^p(\cR),\mathbb D_{\cR[1]},\can^{\cR[1]})
\]
is an additive category with duality. The category $\FC_{tp}(\cR)$ is closed under $\mathbb D$. Indeed, if $F$ is totally pointlike, then $\mathbb D(F)$ is a $\Q$-acyclic perfect complex with homology concentrated in degree $-1$, and $\mathbb D^2(F)\cong F$. So
\[
 (\FC_{tp}(\cR),\mathbb D,\can)
\]
is also an additive category with duality.
\end{example}

\begin{defn}
\label{v2:def:transpose}
Let $(\cC,\mathbb D_\cC,\can^\cC)$ be an additive category with duality, and let $\lambda\colon X\to\mathbb D_\cC(X)$ be a morphism.
\begin{itemize}
\item The \emph{transpose} of $\lambda$ is
\[
 \lambda^\dagger=\mathbb D_\cC(\lambda)\circ\can^\cC_X\colon X\longrightarrow\mathbb D_\cC(X).
\]
\item The morphism $\lambda$ is \emph{antisymmetric} if $\lambda^\dagger=-\lambda$.
\item For a morphism $q\colon X'\to X$, the \emph{pullback} of $\lambda$ is
\[
 q^*\lambda=\mathbb D_\cC(q)\circ\lambda\circ q\colon X'\longrightarrow\mathbb D_\cC(X').
\]
\end{itemize}
\end{defn}

For the duality $\mathbb D$, the transpose is the derived transposition: $\lambda^\dagger=t(\overline\lambda)$, by Lemma~\ref{v2:lemma:derived-transposition-can}.

\begin{lemma}
\label{v2:lemma:transpose}
Let $\lambda\colon X\to\mathbb D_\cC(X)$ and $q\colon X'\to X$ be morphisms. Then
\[
 (\lambda^\dagger)^\dagger=\lambda
 \qquad\text{and}\qquad
 (q^*\lambda)^\dagger=q^*(\lambda^\dagger).
\]
In particular, pullback preserves antisymmetry.
\end{lemma}

\begin{proof}
For the first identity, naturality of $\can^\cC$ gives $\mathbb D_\cC^2(\lambda)\circ\can^\cC_X=\can^\cC_{\mathbb D_\cC(X)}\circ\lambda$. So
\[
 (\lambda^\dagger)^\dagger=\mathbb D_\cC(\can^\cC_X)\circ\mathbb D_\cC^2(\lambda)\circ\can^\cC_X=\mathbb D_\cC(\can^\cC_X)\circ\can^\cC_{\mathbb D_\cC(X)}\circ\lambda=\lambda.
\]
For the second, naturality of $\can^\cC$ gives $\mathbb D_\cC^2(q)\circ\can^\cC_{X'}=\can^\cC_X\circ q$. So
\[
 (q^*\lambda)^\dagger=\mathbb D_\cC(q)\circ\mathbb D_\cC(\lambda)\circ\mathbb D_\cC^2(q)\circ\can^\cC_{X'}=\mathbb D_\cC(q)\circ\mathbb D_\cC(\lambda)\circ\can^\cC_X\circ q=q^*(\lambda^\dagger).\qedhere
\]
\end{proof}

Recall that a groupoid $\mathcal{G}$ is a category all of whose morphisms are isomorphisms. The components $\pi_0\mathcal{G}$ of the groupoid are the isomorphism classes of objects in $\mathcal{G}$. An equivalence of groupoids $\mathcal{G} \to \mathcal{G}'$ is an equivalence of categories. It induces a bijection  $\pi_0\mathcal{G} \to  \pi_0\mathcal{G}'$. 
\begin{defn}
\label{v2:def:Pn}
The \emph{groupoid of nondegenerate antisymmetric forms} $\mathrm{Pn}(\cC)$ of an additive category with duality has the following objects and morphisms.
\begin{itemize}
\item An object is a pair $(X,\lambda)$, where $\lambda\colon X\to\mathbb D_\cC(X)$ is an antisymmetric isomorphism.
\item A morphism $q\colon(X',\lambda')\to(X,\lambda)$ is an isomorphism $q\colon X'\to X$ with $q^*\lambda=\lambda'$.
\end{itemize}
That is, an object $(X,\lambda)$ of $\mathrm{Pn}(\cC)$ is a nondegenerate antisymmetric bilinear form, and a morphism of $\mathrm{Pn}(\cC)$ is an isometry. The set of components $\pi_0\mathrm{Pn}(\cC)$ consists of the isometry classes of nondegenerate antisymmetric bilinear forms in $\cC$.
\end{defn}

\begin{example}
\label{v2:ex:Pn-FC}
Consider the category with duality $(\FC_{tp}(\cR),\mathbb D,\can)$ of Example~\ref{v2:ex:tp-duality}. The groupoid $\mathrm{Pn}(\FC_{tp}(\cR))$ is as follows.
\begin{itemize}
\item An object is a pair $(F,\gamma)$, where $F$ is a totally pointlike complex and $\gamma\colon F\to\mathbb D(F)$ is an isomorphism in $\D(\cR)$ such that
\[
 \gamma^\dagger=\mathbb D(\gamma)\circ\can_F=-\gamma.
\]
\item A morphism $q\colon(F',\gamma')\to(F,\gamma)$ is an isomorphism $q\colon F'\to F$ in $\D(\cR)$ such that
\[
 q^*\gamma=\mathbb D(q)\circ\gamma\circ q=\gamma'.
\]
\end{itemize}
\end{example}

\begin{defn}
\label{v2:def:duality-equivalence}
Let $(\cC,\mathbb D_\cC,\can^\cC)$ and $(\cD,\mathbb D_\cD,\can^\cD)$ be additive categories with duality. An \emph{equivalence of categories with duality} is a pair
\[
 (f,\eta)\colon(\cC,\mathbb D_\cC,\can^\cC)\longrightarrow(\cD,\mathbb D_\cD,\can^\cD)
\]
with the following properties.
\begin{itemize}
\item The functor $f\colon\cC\to\cD$ is an additive equivalence.
\item The transformation $\eta\colon f\circ\mathbb D_\cC\Rightarrow\mathbb D_\cD\circ f^\op$ is a natural isomorphism.
\item For every $X\in\cC$, the diagram
\begin{equation}
\xymatrix@C=4pc{
 f(X) \ar[r]^-{f(\can^\cC_X)} \ar[d]_-{\can^\cD_{f(X)}} & f\bigl(\mathbb D_\cC^2(X)\bigr) \ar[d]^-{\eta_{\mathbb D_\cC(X)}} \\
 \mathbb D_\cD^2\bigl(f(X)\bigr) \ar[r]^-{\mathbb D_\cD(\eta_X)} & \mathbb D_\cD\bigl(f(\mathbb D_\cC(X))\bigr)
}
\label{v2:eq:duality-square}
\end{equation}
commutes, that is, $\eta_{\mathbb D_\cC(X)}\circ f(\can^\cC_X)=\mathbb D_\cD(\eta_X)\circ\can^\cD_{f(X)}$.
\end{itemize}
\end{defn}

The duality on the target can be transported from the source.

\begin{lemma}
\label{v2:lemma:transport}
Let $(\cC,\mathbb D_\cC,\can^\cC)$ be an additive category with duality, and let $\cD$ be an additive category. Assume we are given the following.
\begin{itemize}
\item An additive equivalence $f\colon\cC\to\cD$.
\item An additive functor $\mathbb D_\cD\colon\cD^\op\to\cD$.
\item A natural isomorphism $\eta\colon f\circ\mathbb D_\cC\Rightarrow\mathbb D_\cD\circ f^\op$.
\item A natural transformation $\can^\cD\colon\id_\cD\Rightarrow\mathbb D_\cD^2$ such that \eqref{v2:eq:duality-square} commutes for every $X\in\cC$.
\end{itemize}
Then $(\cD,\mathbb D_\cD,\can^\cD)$ is an additive category with duality, and $(f,\eta)$ is an equivalence of categories with duality.
\end{lemma}

\begin{proof}
We drop the superscripts on $\can$ and the subscripts on $\mathbb D$. By \eqref{v2:eq:duality-square},
\[
 \can_{fX}=\mathbb D(\eta_X)^{-1}\circ\eta_{\mathbb D X}\circ f(\can_X),
\]
so $\can_{fX}$ is an isomorphism. Every object of $\cD$ is isomorphic to some $fX$, so $\can$ is a natural isomorphism. It remains to verify the identity $\mathbb D(\can_Y)\circ\can_{\mathbb DY}=\id_{\mathbb DY}$ for $Y\in\cD$. Both sides are natural in $Y$, and every object of $\cD$ is isomorphic to some $fX$, so it suffices to take $Y=fX$.
Since $\eta_X\colon f(\mathbb DX)\to\mathbb D(fX)$ is an isomorphism, it suffices to show 
\[\mathbb D(\can_{fX})\circ\can_{\mathbb D(fX)}\circ\eta_X=\eta_X.\] 
By naturality of $\can$, and by \eqref{v2:eq:duality-square} at $X$ and at $\mathbb DX$,
\begin{align*}
 \mathbb D(\can_{fX})\circ\can_{\mathbb D(fX)}\circ\eta_X
 &=\mathbb D(\can_{fX})\circ\mathbb D^2(\eta_X)\circ\can_{f(\mathbb DX)}\\
 &=\mathbb D\bigl(f(\can_X)\bigr)\circ\mathbb D(\eta_{\mathbb DX})\circ\mathbb D^2(\eta_X)^{-1}\circ\mathbb D^2(\eta_X)\circ\mathbb D(\eta_{\mathbb DX})^{-1}\circ\eta_{\mathbb D^2X}\circ f(\can_{\mathbb DX})\\
 &=\mathbb D\bigl(f(\can_X)\bigr)\circ\eta_{\mathbb D^2X}\circ f(\can_{\mathbb DX}).
\end{align*}
By naturality of $\eta$ with respect to $\can_X\colon X\to\mathbb D^2X$, we have $\mathbb D\bigl(f(\can_X)\bigr)\circ\eta_{\mathbb D^2X}=\eta_X\circ f\bigl(\mathbb D(\can_X)\bigr)$. So, using the identity in $\cC$, this is
\[
 \eta_X\circ f\bigl(\mathbb D(\can_X)\circ\can_{\mathbb DX}\bigr)=\eta_X\circ f(\id_{\mathbb DX})=\eta_X,
\]
as required.
\end{proof}

\begin{proposition}
\label{v2:prop:Pn-equivalence}
Let $(f,\eta)\colon(\cC,\mathbb D_\cC,\can^\cC)\to(\cD,\mathbb D_\cD,\can^\cD)$ be an equivalence of categories with duality. For a morphism $\lambda\colon X\to\mathbb D_\cC(X)$, let
\[
 \beta_X(\lambda)=\eta_X\circ f(\lambda)\colon f(X)\longrightarrow\mathbb D_\cD\bigl(f(X)\bigr).
\]
\begin{enumerate}[(a)]
\item The map 
\[
 \beta_X\colon\Hom_\cC\bigl(X,\mathbb D_\cC(X)\bigr)\longrightarrow\Hom_\cD\bigl(f(X),\mathbb D_\cD(f(X))\bigr)
\]
 is a bijection, and $\lambda$ is an isomorphism if and only if $\beta_X(\lambda)$ is.
\item For every morphism $q\colon X'\to X$, we have $\beta_{X'}(q^*\lambda)=f(q)^*\beta_X(\lambda)$.
\item We have $\beta_X(\lambda^\dagger)=\beta_X(\lambda)^\dagger$. In particular, $\lambda$ is antisymmetric if and only if $\beta_X(\lambda)$ is.
\item The functor $(X,\lambda)\mapsto\bigl(f(X),\beta_X(\lambda)\bigr)$, $q\mapsto f(q)$, is an equivalence of groupoids $\mathrm{Pn}(\cC)\to\mathrm{Pn}(\cD)$.
\end{enumerate}
\end{proposition}

\begin{proof}
Part (a) holds because $f$ is fully faithful and $\eta_X$ is an isomorphism. 
For (b), by functoriality of $f$,
\[
 \beta_{X'}(q^*\lambda)=\eta_{X'}\circ f\bigl(\mathbb D_\cC(q)\circ\lambda\circ q\bigr)=\eta_{X'}\circ f\bigl(\mathbb D_\cC(q)\bigr)\circ f(\lambda)\circ f(q).
\]
Naturality of $\eta$ with respect to $q\colon X'\to X$ gives $\eta_{X'}\circ f\bigl(\mathbb D_\cC(q)\bigr)=\mathbb D_\cD\bigl(f(q)\bigr)\circ\eta_X$. So
\[
 \beta_{X'}(q^*\lambda)=\mathbb D_\cD\bigl(f(q)\bigr)\circ\eta_X\circ f(\lambda)\circ f(q)=\mathbb D_\cD\bigl(f(q)\bigr)\circ\beta_X(\lambda)\circ f(q)=f(q)^*\beta_X(\lambda).
\]
For (c), by the definition of the transpose, by \eqref{v2:eq:duality-square}, and by naturality of $\eta$ with respect to $\lambda$,
\begin{align*}
 \beta_X(\lambda)^\dagger
 &=\mathbb D_\cD\bigl(f(\lambda)\bigr)\circ\mathbb D_\cD(\eta_X)\circ\can^\cD_{fX}\\
 &=\mathbb D_\cD\bigl(f(\lambda)\bigr)\circ\eta_{\mathbb D_\cC X}\circ f(\can^\cC_X)\\
 &=\eta_X\circ f\bigl(\mathbb D_\cC(\lambda)\circ\can^\cC_X\bigr)=\beta_X(\lambda^\dagger).
\end{align*}
For the third line, we used that naturality of $\eta$ with respect to $\lambda\colon X\to\mathbb D_\cC X$ gives
\[\mathbb D_\cD\bigl(f(\lambda)\bigr)\circ\eta_{\mathbb D_\cC X}=\eta_X\circ f\bigl(\mathbb D_\cC(\lambda)\bigr)\colon f(\mathbb D_\cC^2X)\to\mathbb D_\cD(fX).\]
For the second claim, $\lambda$ is antisymmetric exactly when $\lambda^\dagger=-\lambda$. Since $\beta_X$ is injective, this holds if and only if $\beta_X(\lambda^\dagger)=\beta_X(-\lambda)$. By the first claim and additivity of $\beta_X$, this equation is $\beta_X(\lambda)^\dagger=-\beta_X(\lambda)$, which says that $\beta_X(\lambda)$ is antisymmetric.

For (d), parts (a) and (c) show that $\beta_X(\lambda)$ is an antisymmetric isomorphism whenever $\lambda$ is, and part (b) shows that an isometry $q\colon(X',\lambda')\to(X,\lambda)$ is sent to an isometry $f(q)\colon\bigl(f(X'),\beta_{X'}(\lambda')\bigr)\to\bigl(f(X),\beta_X(\lambda)\bigr)$. So the functor is well defined.

The functor is faithful because $f$ is. For fullness, let $p\colon f(X')\to f(X)$ be an isometry from $\beta_{X'}(\lambda')$ to $\beta_X(\lambda)$. Since $f$ is an equivalence, $p=f(q)$ for a unique isomorphism $q\colon X'\to X$. By (b), $\beta_{X'}(q^*\lambda)=f(q)^*\beta_X(\lambda)=\beta_{X'}(\lambda')$, and the injectivity in (a) gives $q^*\lambda=\lambda'$. So $q$ is an isometry.

For essential surjectivity, let $(Y,\mu)\in\mathrm{Pn}(\cD)$. Since $f$ is essentially surjective, there is an isomorphism $u\colon f(X)\to Y$ for some $X\in\cC$. The pullback $u^*\mu\colon f(X)\to\mathbb D_\cD(f(X))$ is an isomorphism, and it is antisymmetric by Lemma~\ref{v2:lemma:transpose}. By (a), $u^*\mu=\beta_X(\lambda)$ for an isomorphism $\lambda$, and $\lambda$ is antisymmetric by (c). Then $u$ is an isometry from $\bigl(f(X),\beta_X(\lambda)\bigr)$ to $(Y,\mu)$.
\end{proof}

\begin{example}
\label{v2:ex:change-of-coefficients}
Let $\cC=\FC_{tp}(\cR)$. The connecting map $\delta\colon \cR_{\Q/\Z}\to \cR[1]$ of \eqref{eq:ZQZ-triangle} commutes with the involutions. So 
\[(\id,\delta_*)\colon(\cC,\mathbb D_{\cR_{\Q/\Z}},\can^{\cR_{\Q/\Z}})\to(\cC,\mathbb D_{\cR[1]},\can^{\cR[1]})\] 
is an equivalence of categories with duality. 
In particular, $\gamma$ is antisymmetric if and only if $\delta_*\gamma$ is, and $\gamma\mapsto\delta_*\gamma$ gives an equivalence of groupoids
\[\mathrm{Pn}(\cC,\mathbb D_{\cR_{\Q/\Z}})\simeq\mathrm{Pn}(\cC,\mathbb D_{\cR[1]}).\]
\end{example}

\subsection{Statistical duality and mutual statistics}

\begin{defn}
\label{v2:def:sD}
The \emph{statistical dual} of an $\cR$-module $S$ is $S^\vee=H_{-1}\bigl(\mathbb D(S[-1])\bigr)$, and of a morphism $f$ of $\cR$-modules is $f^\vee=H_{-1}\bigl(\mathbb D(f[-1])\bigr)$. This defines a functor $(-)^\vee\colon\Mod(\cR)^\op\to\Mod(\cR)$.
For $F\in\FC_{p}(\cR)$, let
\[
 \alpha_F=H_{-1}\bigl(\mathbb D(\epsilon_F)\bigr)\colon H_{-1}\mathbb D(F)\longrightarrow H_{-1}(F)^\vee.
\]
The \emph{statistical canonical map} of $S\in\mathcal{M}_{tp}(\cR)$ is
\[
 \scan_S=\alpha_{\mathbb D(\iota S)}\circ H_{-1}\bigl(\can_{\iota S}\bigr)\colon S\longrightarrow S^{\vee\vee}.
\]
\end{defn}

\begin{lemma}
\label{v2:lemma:sD-basic}
\begin{enumerate}[(a)]
\item The statistical dual restricts to a functor 
\[(-)^\vee \colon \mathcal{M}_{tp}(\cR)^\op\to\mathcal{M}_{tp}(\cR).\]
\item The maps $\alpha_F$ form a natural isomorphism 
\[\alpha\colon H_{-1}\circ\mathbb D\Rightarrow(-)^\vee\circ H_{-1}\] 
of functors $\FC_{tp}(\cR)^\op\to\mathcal{M}_{tp}(\cR)$.
\item For $S\in \mathcal{M}_p$, there is an isomorphism
$ S^\vee\cong\Ext^2_\cR(\overline S,\cR_{\Q/\Z})$,
natural in $S$.
\end{enumerate}
\end{lemma}

\begin{proof}
Part (a) follows from Example~\ref{v2:ex:tp-duality} and Lemma~\ref{v2:lemma:tp-modules}(b). For (b), $\epsilon_F$ is an isomorphism, natural in $F$, so $\alpha_F=H_{-1}(\mathbb D(\epsilon_F))$ is an isomorphism, natural in $F$. Part (c) follows from Lemma~\ref{lemma:QZ-duality} and Lemma~\ref{lemma:torsion-dual-ext} with $j=k=-1$.
\end{proof}

\begin{theorem}
\label{v2:thm:pointlike-duality}
The pair $(H_{-1},\alpha)$ is an equivalence of categories with duality
\[
 (\FC_{tp}(\cR),\mathbb D,\can)\xrightarrow{\ \simeq\ }(\mathcal{M}_{tp}(\cR),(-)^\vee,\scan).
\]
In particular, $(\mathcal{M}_{tp}(\cR),(-)^\vee,\scan)$ is an additive category with duality, so that $\scan$ is a natural isomorphism, and for every $S\in\mathcal{M}_{tp}(\cR)$ the diagram
\[
\xymatrix@C=4pc{
 S^\vee \ar[r]^-{\scan_{S^\vee}} \ar[dr]_-{\id} & S^{\vee\vee\vee} \ar[d]^-{(\scan_S)^\vee} \\
 & S^\vee
}
\]
commutes, that is, $(\scan_S)^\vee\circ\scan_{S^\vee}=\id_{S^\vee}$.
\end{theorem}

\begin{proof}
By Example~\ref{v2:ex:tp-duality}, $(\FC_{tp}(\cR),\mathbb D,\can)$ is an additive category with duality, and by Lemma~\ref{v2:lemma:tp-modules}(b), $H_{-1}$ is an equivalence. By Lemma~\ref{v2:lemma:transport}, it suffices to show that $\scan$ is natural and that, for every $F\in\FC_{tp}(\cR)$, the diagram
\[
\xymatrix@C=4pc{
 H_{-1}(F) \ar[r]^-{H_{-1}(\can_F)} \ar[d]_-{\scan_{H_{-1}(F)}} & H_{-1}\bigl(\mathbb D^2F\bigr) \ar[d]^-{\alpha_{\mathbb D(F)}} \\
 H_{-1}(F)^{\vee\vee} \ar[r]^-{\alpha_F^\vee} & \bigl(H_{-1}\mathbb D(F)\bigr)^\vee
}
\]
commutes. The map $\scan$ is natural, since it is a composite of natural transformations. For $F=\iota S$, we have $\alpha_{\iota S}=\id$ since $\epsilon_{\iota S}=\id$, and the diagram commutes by the definition of $\scan_S$. For general $F$, both composites in the diagram are natural transformations $H_{-1}\Rightarrow(-)^\vee\circ H_{-1}\circ\mathbb D$. Naturality with respect to the isomorphism $\epsilon_F\colon\iota H_{-1}(F)\to F$, which satisfies $H_{-1}(\epsilon_F)=\id$, shows that the diagram for $F$ is obtained from the diagram for $\iota H_{-1}(F)$ by postcomposing both composites with the same isomorphism.
\end{proof}

\begin{defn}
A \emph{theory of mutual statistics} is an object $(S,b)$ of $\mathrm{Pn}(\mathcal{M}_{tp}(\cR))$.  An \emph{isometry} is a morphism of $\mathrm{Pn}(\mathcal{M}_{tp}(\cR))$. Explicitly:
\begin{itemize}
\item A theory of mutual statistics is a pair $(S,b)$, where $S\in\mathcal{M}_{tp}(\cR)$ and $b\colon S\to S^\vee$ is an isomorphism of $\cR$-modules such that
\[
b^\dagger=b^\vee\circ\scan_S=-b.
\]
\item An isometry $h\colon(S',b')\to(S,b)$ is an isomorphism of $\cR$-modules $h\colon S'\to S$ such that
\[
 h^*b=h^\vee\circ b\circ h=b'.
\]
\end{itemize}
\end{defn}

For $F\in\FC_{tp}(\cR)$ and a morphism $\gamma\colon F\to\mathbb D(F)$ in $\D(\cR)$, let
\[
 b_\gamma=\alpha_F\circ H_{-1}(\gamma)\colon H_{-1}(F)\longrightarrow H_{-1}(F)^\vee.
\]
In the notation of Proposition~\ref{v2:prop:Pn-equivalence}, applied to the equivalence $(H_{-1},\alpha)$ of Theorem~\ref{v2:thm:pointlike-duality}, we have 
\[b_\gamma=\beta_F(\gamma).\] 
The main consequence is the following.

\begin{corollary}
\label{v2:cor:tms}
The functor $(F,\gamma)\mapsto\bigl(H_{-1}(F),b_\gamma\bigr)$ is an equivalence of groupoids
\[
 \mathrm{Pn}\bigl(\FC_{tp}(\cR),\mathbb D\bigr)\xrightarrow{\ \simeq\ }\mathrm{Pn}(\mathcal{M}_{tp}(\cR),(-)^\vee).
\]
 In particular, isometry classes of pairs $(F,\varphi)$, with $F\in\FC_{tp}(\cR)$ and $\varphi\colon F\to\mathbb D(F)$ an antisymmetric isomorphism, correspond bijectively to isometry classes of theories of mutual statistics.
\end{corollary}

\begin{proof}
The equivalence is Proposition~\ref{v2:prop:Pn-equivalence}(d) for the equivalence $(H_{-1},\alpha)$ of Theorem~\ref{v2:thm:pointlike-duality}. An equivalence of groupoids induces a bijection on isomorphism classes, which gives the last statement.
\end{proof}

\subsection{Tame models and the statistical dual}
\label{v2:subsec:pointlike-tame}

From now on, $\cR=R$.

\begin{lemma}
\label{v2:lemma:tp-tame}
Let $F$ be a pointlike tame fusion complex. Then $F$ is totally pointlike if and only if the chain-level dual $D_{R_{\Q/\Z}}(F)$ has homology concentrated in degree $-1$.
\end{lemma}

\begin{proof}
By Corollary~\ref{v2:cor:tame-dual}(b),
$ H_kD_{R_{\Q/\Z}}(F)\cong H_k\mathbb D(F)
$
for all $k$.
\end{proof}

The main text computes the statistical duality with the model $D_I$ of Lemma~\ref{v2:lemma:involutive-resolution}. The following proposition identifies these computations with the definitions of Definition~\ref{v2:def:sD}. In particular, Definition~\ref{v2:def:sD} agrees with the definition of $S^\vee$ in Section~\ref{subsec:tpe}.

\begin{proposition}
\label{v2:prop:sD-model}
Let $S\in\mathcal{M}_{tp}(R)$, and let $F\in\FC_{tp}(R)$ with $H_{-1}(F)=S$.
\begin{enumerate}[(a)]
\item The statistical dual is
\[
 S^\vee=H_{-1}\bigl(D_I(S[-1])\bigr),
\]
and $f^\vee=H_{-1}\bigl(D_I(f[-1])\bigr)$ for a morphism $f$ of $\mathcal{M}_{tp}(R)$.
\item The isomorphism $\alpha_F$ is induced on $H_{-1}$ by the zigzag
\[
 D_I(F)\longrightarrow D_I(\tau_{\geq-1}F)\longleftarrow D_I(S[-1]).
\]
\item The statistical canonical map $\scan_S$ is induced on $H_{-1}$ by the zigzag
\[
 S[-1]\xrightarrow{\can^I_{S[-1]}}D_I^2(S[-1])\longrightarrow D_I\bigl(\tau_{\geq-1}D_I(S[-1])\bigr)\longleftarrow D_I\bigl(S^\vee[-1]\bigr).
\]
\end{enumerate}
\end{proposition}

\begin{proof}
Since $I$ is $K$-injective, $D_I(X)$ is a model for $\mathbb D(X)$ for every complex $X$. Part (a) follows, since $S^\vee=H_{-1}\bigl(\mathbb D(S[-1])\bigr)$. For (b), recall that $\alpha_F=H_{-1}(\mathbb D(\epsilon_F))$. By Remark~\ref{v2:lemma:pointlike-zigzag}, $\epsilon_F$ is represented by the zigzag $S[-1]\leftarrow\tau_{\geq-1}F\to F$. Applying $D_I$ gives the zigzag in (b). For (c), by definition,
\[
 \scan_S=\alpha_{\mathbb D(\iota S)}\circ H_{-1}\bigl(\can_{\iota S}\bigr).
\]
The morphism $\can_{\iota S}$ is represented by $\can^I_{S[-1]}$, by Lemma~\ref{v2:lemma:derived-models}(b) and naturality of $\can^I$. Applying (b) to the model $D_I(S[-1])$ of $\mathbb D(\iota S)$ gives the rest of the zigzag.
\end{proof}

We now compare the homological braiding with the correspondence of Corollary~\ref{v2:cor:tms}. Let $(F,\lambda)$ be a pointlike tame BFC, and let $S=H_{-1}(F)$. The braiding $\lambda\colon F\to D_{R_{\Q/\Z}}(F)$ gives the chain map
\[
 \lambda_I=\chi_*\circ\lambda\colon F\longrightarrow D_I(F).
\]
Since $D_I(F)$ is a model for $\mathbb D(F)$, it represents a morphism $F\to\mathbb D(F)$ in $\D(R)$, which we also write $\lambda_I$.

\begin{remark}
Up to the equivalence $\delta_*$, the morphism $\lambda_I$ is the derived braiding $\lambda_{\mathbb D}=\nu_F\circ\lambda$ of Section~\ref{v2:app:tame-BFC}, so the results there apply to it. In particular, $\lambda_I$ is antisymmetric by Lemma~\ref{v2:lemma:chi-antisymmetric}.
\end{remark}

By Proposition~\ref{v2:prop:sD-model}(b), the isomorphism $a_F$ of Definition~\ref{def:homological-braiding} is
$ a_F=\alpha_F\circ H_{-1}(\chi_*)$.
Hence the homological braiding of $(F,\lambda)$ is
$ b=a_F\circ H_{-1}(\lambda)=b_{\lambda_I}$.
If $(F,\lambda)$ is nondegenerate, $(S,b)$ is therefore the image of $(F,\lambda_I)$ under the correspondence of Corollary~\ref{v2:cor:tms}.

\begin{corollary}
\label{v2:cor:tame-tms}
Let $F$ be a pointlike tame fusion complex, with $S=H_{-1}(F)$.
\begin{enumerate}[(a)]
\item Assume that $F$ is totally pointlike. Let $\gamma\colon F\to D_{R_{\Q/\Z}}(F)$ be a chain map, which need not be antisymmetric. Then the diagram
\[
\xymatrix@C=10pc{
 S \ar[r]^-{H_{-1}(D_{R_{\Q/\Z}}(\gamma)\circ\can^{R_{\Q/\Z}}_F)} \ar[d]_-{\scan_S} & H_{-1}D_{R_{\Q/\Z}}(F) \ar[d]^-{a_F} \\
 S^{\vee\vee} \ar[r]^-{(a_F\circ H_{-1}(\gamma))^\vee} & S^\vee
}
\]
commutes, that is,
\[
 a_F\circ H_{-1}\bigl(D_{R_{\Q/\Z}}(\gamma)\circ\can^{R_{\Q/\Z}}_F\bigr)=\bigl(a_F\circ H_{-1}(\gamma)\bigr)^\vee \circ\scan_S.
\]
\item If $(F,\lambda)$ is a tame BFC with $F$ totally pointlike, then its homological braiding $b$ is antisymmetric.
\item If $(F,\lambda)$ is a nondegenerate tame BFC, then $F$ is totally pointlike, and $(F,\lambda_I)$ is an object of $\mathrm{Pn}(\FC_{tp}(R),\mathbb D)$. In particular, $(S,b)$ is a theory of mutual statistics.
\end{enumerate}
\end{corollary}

\begin{proof}
For (a), let $\gamma_I=\chi_*\circ\gamma$, so that $a_F\circ H_{-1}(\gamma)=b_{\gamma_I}$. 
By Lemma~\ref{v2:lemma:transposition-can}, $t(\overline\gamma)=D_{R_{\Q/\Z}}(\gamma)\circ\can^{R_{\Q/\Z}}_F$, where $t$ is the transposition of Definition~\ref{v2:def:transposition} for $T=R_{\Q/\Z}$.
By Proposition~\ref{v2:prop:tame-dictionary}(c), the chain map $\chi_*\circ t(\overline\gamma)$ 
represents $t(\overline{\gamma_I})=\gamma_I^\dagger$ (Lemma~\ref{v2:lemma:derived-transposition-can}).
So, by Proposition~\ref{v2:prop:Pn-equivalence}(c),
\[
 a_F\circ H_{-1}\bigl(t(\overline\gamma)\bigr)=b_{\gamma_I^\dagger}=(b_{\gamma_I})^\dagger=\bigl(a_F\circ H_{-1}(\gamma)\bigr)^\vee\circ\scan_S.
\]
Part (b) is (a) for $\gamma=\lambda$, since $t(\overline\lambda)=-\lambda$. For (c), $\lambda_I$ is an isomorphism by Proposition~\ref{v2:prop:derived-braiding}(b), and it is antisymmetric by Lemma~\ref{v2:lemma:chi-antisymmetric}. So $F$ is totally pointlike by Remark~\ref{v2:rmk:self-dual-tp}, and $(F,\lambda_I)$ is an object of $\mathrm{Pn}(\FC_{tp}(R),\mathbb D)$. Its image $(S,b)$ under Corollary~\ref{v2:cor:tms} is a theory of mutual statistics.
\end{proof}

At odd fusion order, the theory of mutual statistics is a complete invariant.

\begin{corollary}
\label{v2:cor:odd-tms}
Let $(F,\lambda)$ and $(G,\mu)$ be nondegenerate pointlike tame BFCs, with theories of mutual statistics $(S,b)$ and $(S',b')$.
\begin{enumerate}[(a)]
\item The theories $(S,b)$ and $(S',b')$ are isometric if and only if $(F,\lambda_I)$ and $(G,\mu_I)$ are isometric. 
\item If $(F,\lambda)$ and $(G,\mu)$ are equivalent  in the sense of Definition~\ref{def:tame-bfc-equivalence}, then $(S,b)$ and $(S',b')$ are isometric. The converse holds if $G$ has odd fusion order.
\item Every theory of mutual statistics of odd fusion order $n$ is isometric to the theory of mutual statistics of a nondegenerate pointlike $\Z$-tame BFC of fusion order $n$.
\end{enumerate}
In particular, this proves Theorem~\ref{thm:Sb-invariant}.
\end{corollary}

\begin{proof}
Part (a) holds because the functor of Corollary~\ref{v2:cor:tms} is an equivalence of groupoids, which sends $(F,\lambda_I)$ to $(S,b)$. Part (b) follows from (a), Lemma~\ref{v2:lemma:equivalent-isometric} and Lemma~\ref{v2:lemma:isometric-equivalent}.

For (c), let $(S_0,b_0)$ be a theory of mutual statistics of odd fusion order $n$. By Corollary~\ref{v2:cor:tms}, there is an object $(S_0[-1],\varphi)$ of $\mathrm{Pn}(\FC_{tp}(R),\mathbb D)$ whose image is $(S_0,b_0)$. By Lemma~\ref{v2:lemma:odd-realization}, there is a $\Z$-tame BFC $(P,\lambda)$ such that $(P,\lambda_I)$ is isometric to $(S_0[-1],\varphi)$. The complex $P$ is isomorphic to $S_0[-1]$, so it is pointlike of fusion order $n$. In particular, $H_0(P)=0$, and $\lambda_I$ is an isomorphism, so $(P,\lambda)$ is nondegenerate by Proposition~\ref{v2:prop:derived-braiding}(b). By Corollary~\ref{v2:cor:tms}, its theory of mutual statistics is isometric to $(S_0,b_0)$.
\end{proof}

\begin{remark}[Even fusion order]
\label{v2:rmk:even-order}
Recall from Remark~\ref{rem:2rem} that the theory of Hermitian forms is much more subtle when $2$ is not invertible. Indeed,
at even fusion order, even in the simpler pointlike case, pointlike theories of mutual statistics are not rich enough to describe phases.

More specifically, the arguments for Corollary~\ref{v2:cor:odd-tms}(b) and (c) break down in two places.

\emph{Uniqueness.} On a $\Z$-tame model, two braidings $\lambda$ and $\lambda'$ with $\lambda_I=\lambda'_I$ in $\D(R)$ differ by a null-homotopic antisymmetric chain map $\lambda'-\lambda=dk$, but an equivalence requires an antisymmetric homotopy. The proof of Lemma~\ref{v2:lemma:isometric-equivalent} uses $2\bigl(k-t(\overline k)\bigr)$, an antisymmetric homotopy between $4\lambda$ and $4\lambda'$, together with the quasi-isometry $2\id$, which needs multiplication by $2$ to be a quasi-isomorphism. For example, the semion and anti-semion BFCs have the same theory of mutual statistics, as discussed after Definition~\ref{def:tms}.

\emph{Existence.} On a $\Z$-tame model $F$, an object of $\mathrm{Pn}(\FC_{tp}(R))$ is represented by a chain map $\lambda\colon F\to D_{R_{\Q/\Z}}(F)$ (Theorem~\ref{v2:thm:tame-dual}) that is antisymmetric only up to homotopy. It is not known to us whether every such $\lambda$ is homotopic to a braiding, which is strictly antisymmetric and constant-alternating, as was the case at odd fusion order by Lemma~\ref{v2:lemma:odd-realization}.
\end{remark}

We end with a technical lemma on chain-level models of pointlike complexes.

\begin{lemma}
\label{v2:rmk:S-complete-invt-Zn-restated}
\label{v2:lem:truncated-gamma}
\begin{enumerate}[(a)]
\item Let $F$ be a complex of $R$-modules with $F_i=0$ for $i<-1$, and let $G$ be a complex of $R$-modules with homology concentrated in degree $-1$. Then every chain map $\gamma\colon F\to G$ factors uniquely as
\[
\xymatrix@C=3pc{
 F \ar[r]^-{\gamma} \ar@{.>}[d]_-{\tilde\gamma} & G \\
 \tau_{\geq-1}G \ar[ur]_-{\mathrm{incl}} &
}
\]
and $\gamma$ and $\tilde\gamma$ induce the same map on $H_{-1}$.
\item In (a), assume further that either each $F_i$ is a projective $R$-module, or each $F_i$ is a projective $R_n$-module and $G$ is a complex of $R_n$-modules. Then every homomorphism $H_{-1}(F)\to H_{-1}(G)$ is induced by a chain map $F\to G$, and two chain maps $\gamma,\gamma'\colon F\to G$ are chain homotopic if and only if they induce the same map on $H_{-1}$.
\item Let $F$ and $G$ be pointlike tame fusion complexes which are both $\Z_n$-tame, or both $\Z$-tame, and let $\alpha\colon H_{-1}(F)\to H_{-1}(G)$ be an isomorphism. Then the isomorphism of Corollary~\ref{v2:cor:S-complete-invt} is represented by a zigzag of quasi-isomorphisms $F\leftarrow P\rightarrow G$, where $P$ is a pointlike tame fusion complex of the same type.
\end{enumerate}
\end{lemma}

\begin{proof}
For (a), since $F_{-2}=0$, we have $d\gamma(x)=\gamma(d x)=0$ for $x\in F_{-1}$, so 
\[\gamma(F_{-1})\subseteq\ker(d\colon G_{-1}\to G_{-2})=(\tau_{\geq-1}G)_{-1}.\] 
In all other degrees, $(\tau_{\geq-1}G)_i$ is $G_i$ or zero. Uniqueness holds because the inclusion is injective, and the inclusion induces the identity on $H_{-1}$.

For (b), homotopic maps induce the same map on homology. Conversely, by (a) it suffices to show that the factored maps $\tilde\gamma$ and $\tilde\gamma'$ are homotopic, since composing a homotopy with the inclusion gives a homotopy of $\gamma$ and $\gamma'$. The complex $\tau_{\geq-1}G$ is exact in degrees $\geq0$, and the $F_i$ are projective over $R$, or over $R_n$ with $G$ a complex of $R_n$-modules. So the comparison theorem \cite[Theorem 2.2.6]{Weibel_1994} gives both the existence of a lift and the homotopy.

For (c), let $S=H_{-1}(F)$. Choose a resolution $P$ of $S$ by finitely generated free modules, indexed so that $P_i=0$ for $i<-1$. If $F$ is $\Z_n$-tame, we take $P$ over $R_n$. If $F$ is $\Z$-tame, we take $P$ over $R$ and of finite length, which is possible since $R$ has finite global dimension. By (b), there are chain maps $f\colon P\to F$ and $g\colon P\to G$ inducing $\id_S$ and $\alpha$. They are quasi-isomorphisms, and $gf^{-1}$ induces $\alpha$. So it is the isomorphism of Corollary~\ref{v2:cor:S-complete-invt}.
\end{proof}

\section{Coarsening}
\label{v2:app:coarsening}

Coarsening replaces the translation group by a subgroup of finite index by restriction of scalars. We first treat restriction of scalars and duality for general complexes, then specialize to fusion complexes, BFCs and statistical duality.

Let $k$ be a commutative ring with involution, let $\T\cong\Z^d$, and let $\T'\hookrightarrow\T$ have finite index. Set
\[
 A=k[\T],\qquad B=k[\T'],
\]
and let
\[
	\rho = \operatorname{Res}^A_B
\]
be the restriction of scalars functor from $A$-modules to $B$-modules, or the corresponding functor on complexes over those
rings.
The involutions on the group rings extend that of $k$ and send each translation to its inverse. For an $A$-module or complex $F$, we also write $\rho F=\operatorname{Res}^{A}_{B}(F)$. Restriction leaves the underlying abelian groups and differentials unchanged, but restricts the module action to $B$.

For a complex $T'$ of $B$-modules, its \emph{coinduced complex} is the complex of $A$-modules
\[
 T=\Hom_B(A,T'),\qquad (a f)(u)=f(ua),
\]
for $a,u \in A$ and $f \in \Hom_B(A, T')$.  If $T'$ has an involution, the induced involution on $T$ is given by
\[
 c_T(\overline f)(u)=c_{T'}\bigl(\overline{f(\overline u)}\bigr).
\]
Evaluation at $1$ is the counit of the adjunction; it
gives a $B$-linear chain map $\pi:\rho T\to T'$, $\pi(f)=f(1)$, compatible with these involutions.
In the following lemma, $D_T$ and $\mathbb D_T$ denote the ordinary and derived dualities over $A$, and $D'_{T'}$ and $\mathbb D'_{T'}$ those over $B$.

\begin{lemma}[Restriction of scalars and duality]
\label{v2:lemma:coarsening-general}
With the notation above, the following hold.
\begin{enumerate}[(a)]
\item Restriction of scalars is exact and faithful, commutes naturally with conjugation, shifts, homology and canonical truncations, and preserves quasi-isomorphisms. It preserves perfect complexes.
\item For every complex $F$ of $A$-modules, postcomposition with $\pi$ is a natural chain isomorphism
\[
 \beta_F:\rho D_T(F)\xrightarrow{\cong}D'_{T'}(\rho F).
\]
There is also a natural isomorphism in $\D(B)$
\[
 \boldsymbol\beta_F:\rho\mathbb D_T(F)
 \xrightarrow{\cong}\mathbb D'_{T'}(\rho F).
\]
Both comparisons are natural in the coefficient complex $T'$ as well as in $F$.
\item If $T'$ has an involution, these comparisons preserve canonical evaluation and transposition. Explicitly, the ordinary canonical maps satisfy
\[
 \beta_{D_T(F)}\circ\rho(\can^T_F)
 =D'_{T'}(\beta_F)\circ\can^{T'}_{\rho F},
\]
and, for any $\varphi\in\Hom_A(G,D_T(F))$,
\[
 t'\bigl(\overline{\beta_F\circ\rho\varphi}\bigr)
 =\beta_G\circ\rho\bigl(t(\overline\varphi)\bigr).
\]
The analogous identities hold for derived canonical maps and derived transposition.
\end{enumerate}
\end{lemma}

\begin{proof}
For (a), restriction does not change kernels, images or cokernels. Its compatibility with conjugation follows because the involution of $A$ restricts to that of $B$; the other compatibilities follow from the definitions. A set of representatives of $\T/\T'$ is a finite $B$-basis of $A$. Thus restriction takes finitely generated projective $A$-modules to finitely generated projective $B$-modules and preserves perfect complexes.

For (b), the restriction-coinduction adjunction gives
\[
	\rho
 \Hom_A\bigl(\overline F,\Hom_B(A,T')\bigr)
 \cong\Hom_B(\rho\overline F,T')
 =\Hom_B(\overline{\rho F},T').
\]
The adjunction map is $f\mapsto\pi\circ f$, is natural in both arguments, and commutes with the Hom differentials. 
This proves the comparison at the chain complex level. 

Choose a $K$-injective resolution $T'\to I'$. Coinduction is exact because $A$ is free over $B$, and $\Hom_B(A,I')$ is $K$-injective over $A$: for any acyclic $A$-complex $E$, adjunction identifies
\[
 \Hom_A\bigl(E,\Hom_B(A,I')\bigr)
 \cong\Hom_B(\rho E,I'),
\]
which is acyclic. Hence $T\to\Hom_B(A,I')$ is a $K$-injective resolution, and applying the ordinary comparison with $I'$ gives the derived comparison.

For (c), if $x\in F_i$ and $f\in D_T(F)_j$, the evaluation formulas of Lemma~\ref{v2:lemma:ev-can-formulas} give
\[
 \pi\bigl(\ev^T_F(\overline x\otimes f)\bigr)
 =\ev^{T'}_{\rho F}\bigl(\overline x\otimes\beta_F(f)\bigr).
\]
Since $\pi$ commutes with the coefficient involutions, the same formulas give the stated canonical-map identity. The transposition identity follows from Lemma~\ref{v2:lemma:transposition-formula}, with the same Koszul signs on both sides.

These identities also compute the derived ones. Indeed, restriction preserves $K$-projective complexes: if $P$ is $K$-projective and $E'$ is acyclic over $B$, then
\[
 \Hom_B(\rho P,E')
 \cong\Hom_A\bigl(P,\Hom_B(A,E')\bigr)
\]
is acyclic by exactness of coinduction. Thus one may use $K$-projective source models and the coinduced $K$-injective coefficient model above on both sides. Lift the coefficient involution as in Lemma~\ref{v2:lemma:involutive-model}; coinduction and evaluation at $1$ intertwine these lifts. The ordinary evaluation and transposition identities on these models give the derived identities.
\end{proof}

\begin{lemma}[Group-ring coefficients and successive coarsening]
\label{v2:lemma:coarsening}
Let $K$ be any complex of $k$-modules, with an involution when canonical maps or transpositions are considered. Set $T=A\otimes_k K$ and $T'=B\otimes_k K$. Projection onto translations in $\T'$ defines a $B$-linear chain map
\[
 \pi:T\longrightarrow T',\qquad
 \pi\Bigl(\sum_{g\in\T}g\otimes q_g\Bigr)
 =\sum_{g\in\T'}g\otimes q_g.
\]
There is a natural isomorphism
\[
 T\xrightarrow{\cong}\Hom_B(A,T'),\qquad
 q\longmapsto\bigl(a\mapsto\pi(aq)\bigr).
\]
Under this isomorphism, $\pi$ is evaluation at $1$. Thus all the duality comparisons of Lemma~\ref{v2:lemma:coarsening-general} apply, and $\pi$ preserves constant coefficients.

For successive finite-index subgroups $\T''\subseteq\T'\subseteq\T$, these comparisons compose: restriction directly to $k[\T'']$ agrees with successive restriction, and the direct duality comparison is the composite of the two comparisons, at both the ordinary and derived levels.
\end{lemma}

\begin{proof}
Choose coset representatives $g_1,\ldots,g_m$ for $\T/\T'$. Every $q\in T$ has a unique expression $q=\sum_i g_i^{-1}q_i$ with $q_i\in T'$, and $\pi(g_iq)=q_i$. These values determine an arbitrary $B$-linear map $A\to T'$, proving the isomorphism degreewise. The construction commutes with differentials, coefficient maps and involutions. The identity translation belongs to $\T'$, so its coefficient is unchanged. Finally, projections onto nested subgroups compose, as do their postcomposition maps on Hom complexes. The derived assertion follows either by resolving coefficients or by the natural restriction--coinduction adjunction.
\end{proof}

In particular, taking $k=\Z$ and $K=\Q/\Z$, $\Z[1]$ or $\Z_n$ gives the coefficient systems used in this paper. The comparisons also commute with the change from $R_{\Q/\Z}$ to $R[1]$ on $\Q$-acyclic complexes: coefficient projection gives a map between the short exact sequences $0\to R\to R_{\Q}\to R_{\Q/\Z}\to0$ and their coarsened counterparts, and hence commutes with the connecting morphisms.

We now specialize to $R=\Z[\T]$ and $R'=\Z[\T']$, and use $\pi$ for the projection $R_{\Q/\Z}\to R'_{\Q/\Z}$. Its postcomposition map on ordinary duals is denoted by $\pi_*$.

\begin{corollary}
\label{v2:cor:coarsening-fusion}
Restriction preserves derived fusion complexes, pointlike and totally pointlike complexes, and tame fusion complexes. For a tame fusion complex $F$,
\[
 H_iD_{R'_{\Q/\Z}}(\rho F)
 \cong\rho H_iD_{R_{\Q/\Z}}(F).
\]
Moreover, $\rho F$ is proper (resp.\ complete) if and only if $F$ is.
\end{corollary}

\begin{proof}
Restriction preserves perfect complexes and homology, including $\Z$-torsion and finite generation, so it preserves derived fusion complexes. Pointlikeness is a homology condition, and total pointlikeness also follows from the derived duality comparison. Both kinds of tameness are preserved because $R$ and $R_n$ are finitely generated free over $R'$ and $R'_n$, respectively. The displayed isomorphism follows from Lemma~\ref{v2:lemma:coarsening} and exactness of restriction. Since restriction reflects vanishing of modules, the properness and completeness assertions follow from Corollary~\ref{v2:cor:tame-dual}(b),(c).
\end{proof}

\begin{corollary}
\label{v2:cor:coarsening-BFC}
Let $\T'$, $R'$, $\rho$, $\pi$ and $\pi_*$ be as in Lemma~\ref{v2:lemma:coarsening}. If $(F,\lambda)$ is a tame BFC, then
\[
 \bigl(\rho F,\pi_*\circ\rho\lambda\bigr)
\]
is a tame BFC over $R'$. In particular, the coarsened braiding is constant-alternating. It is nondegenerate if $(F,\lambda)$ is. Equivalent tame BFCs give equivalent tame BFCs over $R'$.
\end{corollary}

\begin{proof}
For $\varphi\in\Hom_R\bigl(G,D_{R_{\Q/\Z}}(F)\bigr)$, let $\varphi'=\pi_*\circ\rho\varphi$, so that
\[
 \varphi'(x)(\overline y)=\pi\bigl(\varphi(x)(\overline y)\bigr).
\]
By Lemma~\ref{v2:lemma:coarsening}, the assignment $\varphi\mapsto\varphi'$ commutes with the differentials, with composition by chain maps, and with transpositions.

By Corollary~\ref{v2:cor:coarsening-fusion}, $\rho F$ is a tame fusion complex over $R'$, and $\lambda'=\pi_*\circ\rho\lambda$ is an antisymmetric chain map. For $a\in F_0$, the element $\lambda'(a)(\overline a)=\pi\bigl(\lambda(a)(\overline a)\bigr)$ has the same constant coefficient as $\lambda(a)(\overline a)$, because $1\in\T'$. So $(\rho F,\lambda')$ is a tame BFC.

Restriction preserves quasi-isomorphisms and $H_0$, and $\pi_*$ is an isomorphism, so nondegeneracy is preserved. Homotopy quasi-isometries are sent to homotopy quasi-isometries, since antisymmetric homotopies are sent to antisymmetric homotopies.
\end{proof}

\begin{lemma}[Coarsening and statistical duality]
\label{v2:lemma:coarsening-statistical}
For every $R$-module $S$, there is a natural isomorphism
\[
 \eta_S:\rho(S^\vee)\xrightarrow{\cong}(\rho S)^\vee,
\]
where the statistical duals (Definition~\ref{v2:def:sD}) on the two sides are taken over $R$ and $R'$, respectively.
Restriction preserves totally pointlike fusion modules, and on these modules the comparison respects statistical canonical maps:
\[
 \eta_{S^\vee}\circ\rho(\scan_S)
 =(\eta_S)^\vee\circ\scan_{\rho S}.
\]
Consequently, if $(S,b)$ is a theory of mutual statistics, then
\[
 (\rho S,\eta_S\circ\rho b)
\]
is a theory of mutual statistics over $R'$. The construction preserves isometries, agrees with the homological braiding of a coarsened tame pointlike BFC, and is compatible with successive coarsenings.
\end{lemma}

\begin{proof}
Combining exactness, shifts and derived duality gives
\begin{align*}
 \rho(S^\vee)
 &=\rho H_{-1}\mathbb D_{R_{\Q/\Z}}(S[-1])\\
 &\cong H_{-1}\mathbb D_{R'_{\Q/\Z}}((\rho S)[-1])
 =(\rho S)^\vee.
\end{align*}
This defines $\eta$ naturally for all modules. For totally pointlike fusion modules, preservation follows from Corollary~\ref{v2:cor:coarsening-fusion} applied to $S[-1]$.

Restriction commutes with the canonical truncation zigzag and its induced isomorphism $\epsilon_F:\iota H_{-1}(F)\to F$ for pointlike complexes. Thus the duality comparison intertwines the maps $\alpha_F$ of Definition~\ref{v2:def:sD}. Applying $H_{-1}$ to the canonical-map compatibility in Lemma~\ref{v2:lemma:coarsening-general}(c), and then using these $\alpha$ comparisons, gives the displayed identity for $\scan$. The change-of-coefficients compatibility in Lemma~\ref{v2:lemma:coarsening} identifies this computation with the appendix's $R[1]$ model of statistical duality.

Naturality of $\eta$ and the canonical-map identity imply that $b\mapsto\eta_S\circ\rho b$ preserves transposition and hence antisymmetry. It preserves isomorphisms and isometries as well. For a tame pointlike BFC, the same compatibility with the truncation zigzag and coefficient comparison identifies the resulting map with its coarsened homological braiding. Finally, successive coarsening is compatible by Lemma~\ref{v2:lemma:coarsening}.
\end{proof}

\section{Extended evaluation}
\label{app:extended-evaluation-proof}

\begin{proof}[Proof of Theorem~\ref{thm:extended-evaluation-comparison}]
Write $J=\hR^{\Q/\Z}$, and for any module $M$ write ${}_n M=\{x\in M:nx=0\}$. The Hom--tensor adjunction and the chain-map property of $\widehat{\ev}_F$ show that $\widehat{\ev}_{F,L}$ is a chain map.

We will use the coefficientwise isomorphism
\[
\rho_n:\hR/n\hR\xrightarrow{\cong}{}_n J,\qquad
\rho_n([\hat r])=\mu\bigl((1/n+\Z)\otimes\hat r\bigr).
\]
Indeed, on each coefficient this sends $[a]\in\Z/n\Z$ to $a/n+\Z$, identifying $\Z/n\Z$ with the subgroup of $\Q/\Z$ annihilated by $n$. Applied to unrestricted coefficient families, this is surjective with kernel $n\hR$ before passing to the quotient.

Suppose first that $F$ is $\Z_n$-tame. Choose a basis $a_1,\dots,a_m$ for $F_p$ and dual
basis $\alpha_1,\dots,\alpha_m\in D_{\RQZ}(F)_{-p} = \Hom_R(\overline{F_p},R_{\Q/\Z})$ satisfying $\alpha_j(\overline{a_k})=\delta_{jk}/n+\Z$.  Evaluation on these bases identifies the underlying abelian groups as follows:
\[
F_p\otimes_R\hR\cong(\hR/n\hR)^m,\qquad
\Hom_R(\overline{D_{\RQZ}(F)_{-p}},J)\cong ({}_n J)^m.
\]
Under these coordinates, the $j$th component of $\widehat{\ev}_{F,L}(\sum_k a_k\otimes\hat r_k)$ is
$(-1)^p\mu((1/n+\Z)\otimes\hat r_j)$.
By bijectivity of $\rho_n$, this is an isomorphism in every degree.  

Now suppose that $F$ is $\Z$-tame.  Introduce the coefficient module and map
\[
E=\Hom_R(\overline{\RQZ},J),\qquad
\eta:\hR\longrightarrow E,\qquad
\eta(\hat r)(\bar q)=\mu(\bar q\otimes\hat r).
\]
The ordinary scalar action on $E$ makes $\eta$ $R$-linear.  We claim that it is injective with uniquely divisible cokernel.  For every positive integer $n$, applying the exact functor $\Hom_R(\overline{(-)},J)$ to
\[
0\longrightarrow R_n\xrightarrow{[r]\mapsto(1/n+\Z)r}\RQZ
\xrightarrow{n}\RQZ\longrightarrow0
\]
gives
\[
0\longrightarrow E\xrightarrow{n}E
\longrightarrow\Hom_R(\overline{R_n},J)\longrightarrow0.
\]
Let $i_n : R_n \to \RQZ$ be the inclusion $[r]\mapsto(1/n+\Z)r$, and consider $D_J(i_n) \eta : \widehat R  \to
\Hom_{R}(\overline{R_n}, \widehat R^{\Q/\Z})$. 
	After evaluation at $1$, this sends $\hat r$ to $\rho_n([\hat r])$, with $\ker D_J(i_n) \eta = n \widehat R$. Intersecting these kernels over $n\geq1$, we see
\[
	\ker \eta = \bigcap_{n \geq 1} n \widehat R = 0.
\]
Hence $\eta$ is injective. Moreover it is an isomorphism modulo $n$, 
since restriction followed by evaluation at $1$ identifies the induced map with the isomorphism $[\hat r]\mapsto\rho_n([\hat r])$.
The preceding exact sequence shows that $E$ is $\Z$-torsion-free, since multiplication by every positive integer $n$ is injective. 
{Then $\hat R$ is also $\Z$-torsion-free, since it injects into $E$.}  
Surjectivity of $\eta$ modulo $n$ gives $E=\eta(\hR)+nE$, so multiplication by $n$ on $V:=E/\eta(\hR)$ is surjective. If $ne=\eta(\hat r)$, injectivity modulo $n$ gives $\hat r=n\hat s$; torsion-freeness of $E$ then gives $e=\eta(\hat s)$, proving injectivity on $V$.  Thus $V$ is naturally an $R_\Q$-module, where $R_\Q=R\otimes_\Z\Q$.

Finite freeness identifies the comparison map with tensoring $\eta$ with $F$.  Explicitly, the chain isomorphism
\[
\kappa_F:F\otimes_R E\longrightarrow\Hom_R(\overline{D_{\RQZ}(F)},J)
\]
is given for $a\in F_p$ and $\varphi\in D_{\RQZ}(F)_{-p}$ by
\[
\kappa_F(a\otimes h)(\overline\varphi)=(-1)^p h(\overline{\varphi(\bar a)}).
\]
Note that $\kappa_F$ is $R$-linear. 
For $\varphi\in D_{\RQZ}(F)_{1-p}$, the Hom differential gives
\[
(d\kappa_F(a\otimes h))(\overline\varphi)
=(-1)^{p-1}h\bigl(\overline{\varphi(\overline{d^F a})}\bigr)
=\kappa_F(d^F a\otimes h)(\overline\varphi),
\]
so $\kappa_F$ is a chain map; a basis in each degree verifies bijectivity.  Its composite with $\id_F\otimes\eta$ is exactly $\widehat{\ev}_{F,L}$.  We therefore have a short exact sequence of complexes
\[
0\longrightarrow F\otimes_R\hR
\xrightarrow{\widehat{\ev}_{F,L}}\Hom_R(\overline{D_{\RQZ}(F)},J)
\longrightarrow F\otimes_R V\longrightarrow0.
\]
Since $F$ has $\Z$-torsion homology, $F\otimes_R R_\Q$ is acyclic.  It is a bounded complex of free $R_\Q$-modules and hence contractible, so
\[
F\otimes_R V
\cong(F\otimes_R R_\Q)\otimes_{R_\Q}V
\]
is contractible as well.  This proves that $\widehat{\ev}_{F,L}$ is a quasi-isomorphism in the $\Z$-tame case.

Finally, injectivity of $J$ gives natural isomorphisms
\[
H_k\Hom_R(\overline{D_{\RQZ}(F)},J)
\cong\Hom_R(\overline{H_{-k}(D_{\RQZ}(F))},J), \qquad
\qquad k\in\Z.
\]
Together with the quasi-isomorphism just proved, these give Equation~\ref{eqn:detection-complex-homology}.
\end{proof}

\bibliography{bibliography}

\end{document}